\documentclass[aps,rmp,reprint,amsmath,amssymb,longbibliography]{revtex4-1}
\usepackage{xspace}
\usepackage{bm}
\usepackage{graphicx}
\usepackage{multirow}

\newcommand{\ov}[1]{\overline{#1}}

\newcommand{\dm}{\ensuremath{\Delta M}}
\newcommand{\dg}{\ensuremath{\Delta \Gamma}}

\newcommand{\f}{\ensuremath{f}}
\newcommand{\fb}{\ensuremath{\bar f}}

\newcommand{\Cpar}	{\ensuremath{C_{\f}}\xspace}
\newcommand{\Sbpar}	{\ensuremath{S_{\fb}}\xspace}
\newcommand{\Spar}	{\ensuremath{S_{\f}}\xspace}
\newcommand{\Dbpar}	{\ensuremath{{A_{\fb}^{\Delta\Gamma}}}\xspace}
\newcommand{\Dpar}	{\ensuremath{{A_{\f}^{\Delta\Gamma}}}\xspace}

\newcommand {\Bd} {\ensuremath{B^0}}
\newcommand {\Bs} {\ensuremath{B^0_s}}

\newcommand {\barBd} {\ensuremath{\bar{B}^0}}
\newcommand {\barBs} {\ensuremath{\bar{B}^0_s}}

\newcommand {\Dsminus} {\ensuremath{D^-_s}}
\newcommand {\Dsplus} {\ensuremath{D^+_s}}
\newcommand{\asls}{\ensuremath{a_{\rm sl}^s}}
\newcommand{\asld}{\ensuremath{a_{\rm sl}^d}}

\newcommand{\Km}{K^-}
\newcommand{\Bm}{B^-}
\newcommand{\Kp}{K^+}
\newcommand{\Ks}{K^0_S}
\newcommand{\Kstz}{K^{\star0}}
\newcommand{\Kstzb}{\bar{K}^{\star0}}
\newcommand{\pip}{\pi ^+}
\newcommand{\pim}{\pi ^-}
\newcommand{\Ds}{D_s}
\newcommand{\Dsp}{D_s^+}

\newcommand{\Dsm}{D_s^-}
\newcommand{\Dm}{D^-}
\newcommand{\neumb}{\bar{\nu}_\mu}
\newcommand{\mun}{\mu^-}
\newcommand{\epm}{e^+e^-}

\newcommand{\CP}{$CP$}

\def\betas{\beta _s}

\newcommand\jpsi     {\ensuremath{\mathrm{J}/\psi}}
\def\lhcb {LHCb\xspace}
\def\ux85 {UX85\xspace}

\def\babar  {BaBar\xspace}

\def\dzero  {D0\xspace}

\def\invfb{fb^{-1}}
\newcommand{\dms}{\ensuremath{\Delta M_{s}}\xspace}

\def\delpar{\delta_\parallel}
\def\delperp{\delta_\perp}

\def\aperpsq{|A_{\perp}|^2}

\def\azerosq{|A_{0}|^2}

\def\maglambda{|\lambda|}

\begin{document}

\title{CP violation in the $\Bs$ system}

\author{Marina Artuso}
       \affiliation{Department of Physics, Syracuse University,
                    Syracuse NY 13244, USA}
\author{Guennadi Borissov}
       \affiliation{Physics Department, Lancaster University,
                    Lancaster LA1 4YB, United Kingdom}
\author{Alexander Lenz}
       \affiliation{Institute for Particle Physics Phenomenology,
                    Durham University, South Road, Durham DH1 3LE,
                    United Kingdom}

\date{\today{}}

\begin{abstract}
We review experimental and theoretical studies of CP violation
in the $\Bs$ system.
Updated predictions for the mixing parameters of the
$\Bs$ mesons expected in the Standard Model (SM) are given, namely
the mass  difference  $\Delta M_s^{\rm SM} = (18.3 \pm 2.7)$ ps$^{-1}$,
the decay rate difference $\Delta \Gamma_s^{\rm SM} = (0.085 \pm 0.015)$ ps$^{-1}$, and
the flavour specific CP asymmetry $a_{\rm fs}^{s, \rm SM} = (2.22 \pm 0.27) \times
10^{-5}$ and the equivalent quantities in the $B^0$-sector.
Current experimental values of $\Delta M_s$
and $\Delta \Gamma_s$  agree with remarkable precision with theoretical expectations.
This agreement supports the applicability
of theoretical tools such as the Heavy Quark Expansion (HQE) to these decays.
%
%
CP violating studies in the $\Bs$ system provide essential information to test the SM expectations,
and to unveil possible contribution of the new physics (NP).
%
  NP effects on $\Delta M_s$  of the order of $15\%$ are still possible.
The CP phase $\phi_s$ due to CP violation in interference of decays and mixing
 can accommodate effects of the order of ${\cal O} (100 \%)$.
The semileptonic CP asymmetry $a_{\rm sl}^s$  due to CP violation in mixing
could still be a factor of 130 larger than
its robust SM expectation and thus provides a very clean observable for NP searches.
Theoretical improvements that are necessary to make full use of
the experimental precision are discussed.
\end{abstract}


\maketitle

\tableofcontents{}

\section{Introduction}
\label{intro}
The phenomenon of CP violation, 
discovered more than 50 years ago
\cite{Christenson:1964fg},
is an essential ingredient to explain the apparent imbalance between matter
and anti-matter in the Universe \cite{Sakharov:1967dj}.
Consequently, this topic attracts a lot of attention.
In the Standard Model (SM) \cite{Glashow:1961tr,Weinberg:1967tq,Salam:1968rm}
CP violation arises in the Yukawa-sector
via quark mixing and it is described by a complex parameter in the
Cabibbo-Kobayashi-Maskawa matrix
(CKM matrix) \cite{Cabibbo:1963yz,Kobayashi:1973fv}.
Intensive studies of CP violation, especially at the $e^+ e^-$ $B$ factories
(see e.g. \cite{Bevan:2014iga} for a comprehensive review),
provide convincing evidence that the main source of
CP violation is the phase in the CKM matrix.
More precisely, a vast body of measurements  performed in different
 experimental conditions, such as accelerators, energies of operation, and detectors, 
confirm the unitarity
of the CKM matrix, see \cite{Amhis:2014hma}.

The CKM phase accounts for all the observed CP violating phenomena, but it is too small to account for the abundance of matter in the Universe. Thus additional sources of CP violation must be found. A recent
discussion of this problem can be found in \cite{Bambi:2015mba}. Thus the quest for a broader understanding of CP violation is 
strongly motivated and may provide 
 hints on the path towards a more complete physics picture of the elementary particles and their interactions.

In particular, the study of CP violation in the $\Bs$ system
offers an excellent opportunity to uncover new physics (NP).
 SM predictions  for several $\Bs$ meson observables have achieved 
reasonable precision. In addition, SM CP violating effects 
are expected to be more highly suppressed  than in $\Bd$ meson decays.
Therefore, even a relatively small
contribution of new physics effects could be clearly visible
in the $\Bs$ system, see e.g. \cite{Dunietz:2000cr}. More precisely, the angle $\beta$\footnote{Instead of the notation $\alpha$, $\beta$ and $\gamma$
for the angles of the unitarity triangle, also $\phi_2$, $\phi_1$ and $\phi_3$ are commonly used.}
describing  CP violation in
interference of decay and mixing in the $\Bd$ system
is predicted to be of the order of $22^\circ$. The corresponding angle $\beta_s$ in the $\Bs$ system
is expected to be about $1^\circ$. Thus the sensitivity to new physics is potentially enhanced.
Unfortunately, the contribution of the so-called penguin effects  to the measured value of  $\beta_s$
can also be $\sim$1$^\circ$. 
Thus a more precise determination
of penguin contributions is mandatory \cite{Aaij:2014vda}.
On the contrary, solid conclusions about the existence of new physics could be drawn
by the investigation
of CP violation in mixing extracted from semileptonic charge asymmetries. Here, a measured value
of about two or three times the value of the SM predictions
would be an unambiguous signal for new physics.

The study of $\Bs$ mesons at  $e^+ e^-$ $B$-factories is
possible only by running at  the $\Upsilon(10860)$ center-of-mass energy. The typical center-of-mass
 energy of both the BaBar and Belle experiments corresponds to  the $\Upsilon(4S)$ mass, and is not sufficient to produce  $\Bs\barBs$ pairs. 
The Belle experiment took some data  at the higher energy
and obtained several interesting results, notably some branching fractions of  $\Bs$ decays,
see e.g. \cite{Agashe:2014kda}. However, their statistical accuracy is not sufficient to study CP violation observables.  Thus, the main
source of information on $\Bs$ mesons comes from hadron collider experiments
at the Tevatron (CDF, D0) and the LHC (ATLAS, CMS, LHCb). In particular LHCb, the first experiment designed to
study beauty and charm decays at the  LHC, has produced an impressive body of data on CP violation
in $\Bs - \barBs$ mixing and decay.

This paper aims at summarising the current experimental knowledge of $CP$ violation in the $\Bs$
system as well as the theoretical implication of these data. It is organised as follows.
Section \ref{Bs_system} describes the main properties of  the $\Bs$ system,
such as its mass and width difference, and the time evolution of the $\Bs$ system.
Section \ref{CPVmix} reports studies $CP$ violation in $\Bs$-$\barBs$ mixing.
CP violation in interference of $\Bs$ mixing and decay is discussed in Section \ref{CPVinter}
with a detailed review of penguin contributions.
Section \ref{CPVdecay} reviews studies of CP violation in  $\Bs$ decays, as well as methods
to derive the CKM angle $\gamma$ from $\Bs$ decays.
Section \ref{model} examines the data reported in this review in the context of NP searches.
Model independent constraints on NP contributions  inferred from
$\Bs$ data reported here are presented. Finally,  Section \ref{conclusion} gives an outlook to future developments.
In the appendix details of the numerical updates of the Standard Model predictions
for the mixing quantities are listed.

\section{The \boldmath{$\Bs$} system}
\label{Bs_system}
\subsection{Theory: Basic mixing quantities, time evolution of the \boldmath{$\Bs$} system and the HQE}
\subsubsection{Mixing observables}
\begin{figure*}
\includegraphics[width=0.90 \textwidth]{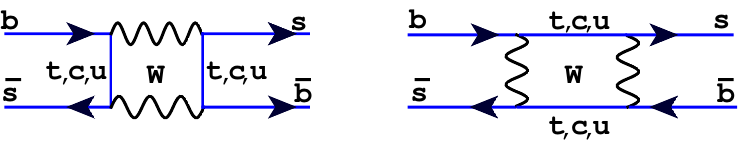}
\caption{\label{box} Standard Model  diagrams for the transition between $\Bs$ and $\barBs$ mesons.
The contribution of internal on-shell particles
(only the charm and the up quark can contribute) is denoted by
$\Gamma_{12}^s$; the contribution of internal off-shell particles
(all depicted particles can contribute) is denoted by $M_{12}^s$.
}
\end{figure*}

The quantum mechanical time evolution of a decaying particle $B$ with mass
$m_B$ and lifetime $\tau_B = 1 / \Gamma_B$ is given as
\begin{equation}
| B(t) \rangle = e^{-i m_B t - \frac{\Gamma_B}{2}t}  | B(0) \rangle \; ,
\end{equation}
where $\Gamma_B$ denotes the total decay width of the $B$ particle.
We now consider the system of neutral $\Bs$ mesons, defined by their
quark flavour content $|\Bs\rangle = |(\bar{b}s)\rangle$,  and their anti-particles,
$|\barBs \rangle = |(b \bar{s}) \rangle$. Its
time-evolution is described by this simple differential
equation for a two-state system:
\begin{equation}
i \frac{d}{dt} \left(
\begin{array}{c}
| \Bs (t) \rangle
\\
| \barBs(t) \rangle
\end{array}
\right)
= \left(
\hat{M}^s - \frac{i}{2} \hat{\Gamma}^s
\right)
\left(
\begin{array}{c}
| \Bs(t) \rangle
\\
| \barBs(t) \rangle
\end{array}
\right) \; .
\end{equation}
Naively one expects the diagonal entries of the $2\times 2$ matrix $\hat{M}^s$ to be equal to
the mass of the $\Bs$ meson, $M_{\Bs}$, the diagonal entries
of $\hat{\Gamma}^s$ to be equal to the decay rate of the $\Bs$ meson,
$\Gamma_s$ and all non-diagonal entries to vanish.
However, because of the weak interaction, the flavour eigenstate $\Bs$ can transform
into its anti-particle $\barBs$ and vice versa. This transition is
governed by the so-called box diagrams, depicted in Fig.~\ref{box}, and it gives rise
to the off-diagonal elements $M_{12}^s$ in $\hat{M}^s$ and $\Gamma_{12}^s$ in
$\hat{\Gamma}^s$.
These box diagrams include  contributions
from virtual internal particles, denoted by $M_{12}^s$ and contributions
from internal on-shell particles, denoted by $\Gamma_{12}^s$.
Only internal charm and up quarks are involved in $\Gamma_{12}^s$,
while $M_{12}^s$ is sensitive to all possible internal particles, and, in principle, also
to heavy new physics particles\footnote{There can also be new physics contributions to $\Gamma_{12}^s$,
for example,  by modified tree-level operators or by new $bs \tau \tau$-operators, as discussed in Section \ref{CPVmix}.}.
Due to the CKM structure both
$M_{12}^s$ and $\Gamma_{12}^s$ can be complex.
\begin{eqnarray}
M_{12}^s & = & |M_{12}^s| e^{i \phi_M}  \; ,
\\
\Gamma_{12}^s & = & | \Gamma_{12}^s | e^{i \phi_\Gamma} \; .
\end{eqnarray}
The CKM phases $\phi_M$ and $\phi_\Gamma$ are not physical, but depend on the
phase convention used in the CKM matrix.
Later on we will see that
\begin{equation}
e^{i \phi_M} = \frac{V_{ts}^* V_{tb}}{V_{ts} V_{tb}^*}.
\label{phiM}
\end{equation}
No such simple relation exists for $\phi_\Gamma$, because $\Gamma_{12}^s$ depends
 on three
different CKM structures in the Standard Model.
\\
In order to obtain the physical eigenstates of the mesons with a definite mass and decay rate,
the matrices  $\hat{M}^s$ and  $\hat{\Gamma}^s$ have to be diagonalised.
This gives the meson eigenstates $|B_{s,H} \rangle$ (H=heavy)
and $|B_{s,L} \rangle$ (L=light) as linear combinations of the flavour eigenstates:
\begin{eqnarray}
| B_{s,L} \rangle & = & p |\Bs  \rangle+ q |\barBs  \rangle\; ,
\\
| B_{s,H} \rangle & = & p |\Bs  \rangle- q |\barBs  \rangle\; ,
\end{eqnarray}
which are in general not orthogonal.
The complex coefficients $p$ and $q$ fulfill $|p|^2+|q|^2 = 1$
and the corresponding masses and decay rates of these states are denoted by
$M_L^s$, $M_H^s$ and  $\Gamma_L^s$, $\Gamma_H^s$.
The mass eigenstates of the $\Bs$ mesons are almost CP eigenstates. Using the same conventions
as e.g. \cite{Dunietz:2000cr}  for the CP properties and defining
\begin{equation}
CP | \Bs \rangle = - | \barBs \rangle \; ,
\label{CP}
\end{equation}
we get for the CP eigenstates of the $\Bs$ meson
\begin{eqnarray}
| B_s^{\rm even} \rangle & = & \frac{1}{\sqrt{2}} \left(
|\Bs  \rangle -  |\barBs  \rangle \right)
\; ,
\\
| B_s^{\rm odd} \rangle & = & \frac{1}{\sqrt{2}} \left(
|\Bs  \rangle +  |\barBs  \rangle \right)
\; .
\end{eqnarray}
In absence of CP violation in mixing, which is a very
small effect\footnote{CP violation in mixing is expected to be of the order of $2 \cdot 10^{-5}$ in the SM.}, 
the heavy eigenstate
is CP odd ($| B_{s,H} \rangle \approx | B_s^{\rm odd} \rangle $)
and the light one is CP even ($| B_{s,L} \rangle \approx | B_s^{\rm even} \rangle $) ,
in this case one has $p = 1/\sqrt{2}$ and  $q = -1/\sqrt{2}$.
\\
If we expand\footnote{Such an expansion does not hold in the charm system, because there
$\Delta \Gamma$ and $\Delta M$ are of a similar size.}  the eigenvalues
of $\hat{M}^s$ and $\hat{\Gamma}^s$ in powers of
$|\Gamma_{12}^s/M_{12}^s| \approx 5 \cdot 10^{-3}$ in the SM,
we can express the mass and decay rate differences as
\begin{eqnarray}
\Delta M_s      & := & M_H^s - M_L^s
\nonumber
\\
&= &
2 \left| M_{12}^s\right| \left( 1 -
\frac{\left|\Gamma_{12}^s\right|^2 \sin^2 \phi_{12}^s}{8 \left|M_{12}^s\right|^2}
+ ... \right)  ,
\label{DMdef}
\\
\Delta \Gamma_s & := & \Gamma_L^s - \Gamma_H^s
\nonumber
\\
&= &
2 \left| \Gamma_{12}^s\right| \cos \phi_{12}^s \!
\left( \! 1 \! +
\frac{\left|\Gamma_{12}^s\right|^2 \sin^2 \phi_{12}^s}{8 \left|M_{12}^s\right|^2}
+ \!  ... \! \right) \!  ,
\label{DGdef}
\end{eqnarray}
with the mixing phase
\begin{equation}
\phi_{12}^s := \arg \left(- \frac{M_{12}^s}{\Gamma_{12}^s} \right)
= \pi + \phi_M - \phi_\Gamma.
\label{phi12def}
\end{equation}
In contrast to $\phi_M$ and $\phi_\Gamma$,  this phase difference is 
physical.
We follow here the definition given in \cite{Bediaga:2012py}. In some references, for example \cite{Anikeev:2001rk,Lenz:2006hd},
$\phi_{12}^s$ is denoted as $\phi_s$. However, in the literature the notation
$\phi_s$ is  often used for different
quantities, also related to CP violation in interference.
We will define the phase that appears in interference in
Section \ref{CPVinter}.
The correction factor $1/8 \, |\Gamma_{12}^s/M_{12}^s|^2 \sin^2 \phi_{12}^s$
in Eq.~(\ref{DMdef}) and  Eq.~(\ref{DGdef}) is of the order of
$ 6 \cdot 10^{-11}$ in the Standard Model and the current
experimental bound for this factor is smaller than $5 \cdot 10^{-5}$, thus it
can be safely neglected. Diagonalisation of $\hat{M}^s$ and $\hat{\Gamma}^s$
gives also
\begin{eqnarray}
\frac{q}{p} & = & - e^{-i \phi_M}
\left[
1- \frac{1}{2} \frac{|\Gamma_{12}^s|  }{|M_{12}^s|  } \sin   \phi_{12}^s
+ {\cal O} \left(\frac{|\Gamma_{12}^s|^2}{|M_{12}^s|^2} \right)
 \right]
\nonumber
\\
&=&
- \frac{V_{ts} V_{tb}^*}{V_{ts}^* V_{tb}} \left[ 1- \frac{a_{\rm fs}^s}{2} \right]
+ {\cal O} \left(\frac{|\Gamma_{12}^s|^2}{|M_{12}^s|^2} \right)
\; ,
\label{pq}
\end{eqnarray}
with the notation
\begin{equation}
a_{\rm fs}^s = \frac{|\Gamma_{12}^s|  }{|M_{12}^s|  } \sin   \phi_{12}^s \; .
\label{abbrev}
\end{equation}
Later on, in Section \ref{CPVmix}, we will see that $a_{\rm fs}^s$ equals
the so-called flavour-specific CP asymmetry. From Eq.~(\ref{pq}) it follows also
that, in the absence of CP violation in mixing,  $q/p = -1$.
In Eq.~(\ref{pq}) again all terms of order $|\Gamma_{12}^s|^2/|M_{12}^s|^2 $ can be discarded,
many times also the term of order $a_{\rm fs}^s$ is not necessary.
\subsubsection{Time evolution of neutral mesons}
We now consider the time evolution of the flavour eigenstates of the
$\Bs$ mesons\footnote{A more detailed discussion of the $\Bs$ mixing system and its time
evolution can be found in e.g. \cite{Anikeev:2001rk}.}.
$ |\Bs (t)\rangle$ denotes a meson at time $t$ that was produced as a $\Bs$ meson at time
$t=0$. At a later time $t$, $ |\Bs (t)\rangle$ will have components both of
$|\Bs \rangle$ and $|\barBs \rangle$:
   \begin{eqnarray}
   |\Bs (t)\rangle       & = & g_+(t) |\Bs \rangle  + \frac{q}{p} g_-(t) |\barBs \rangle \; ,
   \label{evol1}
   \\
   |\barBs (t)\rangle & = & \frac{p}{q} g_-(t) |\Bs \rangle  + g_+(t) |\barBs \rangle  \; ,
   \label{evol2}
   \end{eqnarray}
   with the coefficients
   \begin{eqnarray}
   g_+ (t) & & =    e^{- i M_{s} t} e^{- \frac12 \Gamma_{s} t} \times
     \\ &&
    \left[ \cosh \frac{\Delta \Gamma_s t }{4} \cos \frac{\Delta M_s t}{2} -
         i \sinh \frac{\Delta \Gamma_s t }{4} \sin \frac{\Delta M_s t}{2}
   \right] \; ,
   \nonumber \\
   g_- (t) & & =    e^{- i M_{s} t} e^{- \frac12 \Gamma_s t}\times
    \\ &&
    \left[ - \sinh \frac{\Delta \Gamma_s t }{4} \cos \frac{\Delta M_s t}{2} +
           i \cosh \frac{\Delta \Gamma_s t }{4} \sin \frac{\Delta M_s t}{2}
   \right] \; .
   \nonumber
    \end{eqnarray}
   Here we used the averaged mass $M_{\Bs}$  and decay rate $\Gamma_s$:
   \begin{equation}
    M_{s} = \frac{M_H^s + M_L^s}{2} \,   ,
   \hspace{1cm}
   \Gamma_{s} = \frac{\Gamma_H^s + \Gamma_L^s}{2} \, .
   \end{equation}
Next we consider the time evolution of the decay rate for a $\Bs$ meson, that
was initially (at time $t=0$) tagged as a $\Bs$ flavour eigenstate into an arbitrary
final state $f$.
\begin{widetext}
\begin{eqnarray}
\Gamma \left[\Bs (t) \to f \right]  =  N_f \left|{\cal A}_f\right|^2
\left(1+|\lambda_f|^2 \right)
e^{-\Gamma t}
&&
\left\{
                                           \frac{\cosh \left( \frac{\Delta \Gamma_s}{2} t\right)}{2}
+ \frac{1-|\lambda_f|^2 }{1+|\lambda_f|^2 }\frac{\cos  \left( \Delta M_s t\right)}{2}
\right.
\nonumber
\\
&&
\left.
- \frac{2 \Re (\lambda_f )}{1+|\lambda_f|^2 }
  \frac{\sinh \left( \frac{\Delta \Gamma_s}{2} t\right)}{2}
- \frac{2 \Im (\lambda_f )}{1+|\lambda_f|^2 }
  \frac{\sin  \left( \Delta M_s t\right)}{2}
\right\}
\label{GammaBf} \; .
\end{eqnarray}
\end{widetext}
Here $N_f$ denotes a time-independent normalisation factor, which includes
phase space effects.
The decay amplitude describing the transition of the flavour eigenstate $\Bs$
in the final state $f$ is
denoted by ${\cal A}_f$; for the decay of a $\barBs$ state into $f$ we use the notation $\bar{\cal A}_f$:
\begin{equation}
{\cal A}_f = \langle f | {\cal H}_{eff} |\Bs \rangle \; ,
\hspace{1cm}
\bar{\cal A}_f = \langle f | {\cal H}_{eff} | \barBs \rangle .
\label{amplitude}
\end{equation}
The flavour changing weak quark transitions are described by an effective Hamiltonian
including also perturbative and non-perturbative QCD-effects. ${\cal H}_{eff}$ will
be described in more detail in Section \ref{CPVinter-theory}.
The amplitudes  ${\cal A}_f$  and $\bar{\cal A}_f$
are typically governed by hadronic effects and they are
very difficult to be
calculated reliably in theory.
In Section \ref{CPVinter-theory} it will also be shown that
CP symmetries are governed by a single quantity $\lambda_f$,
which is given by
\begin{equation}
\lambda_f = \frac{q}{p} \frac{\bar{\cal A}_f}{{\cal A}_f}
\approx - \frac{V_{ts} V_{tb}^*}{V_{ts}^* V_{tb}} \frac{\bar{ A}_f}{{ A}_f}
\left[ 1 - \frac{a_{\rm fs}^s}{2} \right]  \; .
\label{lambdaf}
\end{equation}
For the terms appearing on the right-hand-side of Eq.~(\ref{GammaBf}) the following
definitions are typically used
\begin{eqnarray}
{\cal A}_{\rm CP}^{\rm dir} & = & \frac{1-|\lambda_f|^2}{1+|\lambda_f|^2} \; ,
\label{Adir}
\\
{\cal A}_{\rm CP}^{\rm mix} & = & -\frac{2 \Im \left(\lambda_f \right)}{1+|\lambda_f|^2} \; ,
\label{Amix}
\\
{\cal A}_{\Delta \Gamma}    & = & -\frac{2 \Re \left(\lambda_f \right)}{1+|\lambda_f|^2} \; .
\label{AGamma}
\end{eqnarray}
${\cal A}_{\rm CP}^{\rm dir}$ describes effects related to direct CP violation, which is
described in Section \ref{CPVdecay}. This can be seen by neglecting CP violation in
mixing, i.e. assuming $|q/p|=1$ and considering the decay into a final state $f$, that is a
CP eigenstate, i.e. $\bar{f} = \eta_{CP} f$. With these assumptions we get
$|\lambda_f| = |\bar{\cal A}_{\bar{f}}|/|{\cal A}_f|$.
\break
A non-vanishing value for ${\cal A}_{\rm CP}^{\rm dir}$ is obtained for $|\lambda_f|\neq 1$
and this corresponds now to $|\bar{\cal A}_{\bar{f}}| \ne  |{\cal A}_f|$, which is
equivalent to direct CP violation.
${\cal A}_{\rm CP}^{\rm mix}$ encodes effects due to interference between mixing and decay,
which is discussed in Section \ref{CPVmix}
and $ {\cal A}_{\Delta \Gamma} $ is a  correction factor, due to a finite
value of the decay rate difference $\Delta \Gamma_s$; $ {\cal A}_{\Delta \Gamma} $
also appears in the definition of the effective lifetimes $\tau^{\rm eff}$~\footnote{The total lifetime
of the $\Bs$ mesons is defined as $\tau (\Bs) = 1/\Gamma_s = 2/(\Gamma_H^s + \Gamma_L^s)$. But, the
decay of a $\Bs$ meson is actually a superposition of a decay of a $B_H$ meson and a $B_L$ meson. Fitting
such a decay with only one exponential PDF leads to the effective lifetime, which differs from the total lifetime. \nopagebreak} :
\begin{eqnarray}
\tau^{\rm eff} & = & \tau_{\Bs}  \frac{1}{1- y_s^2}
\left(
\frac{1 + 2 {\cal A}_{\Delta \Gamma} y_s + y_s^2}{1 + {\cal A}_{\Delta \Gamma} y_s}
\right)
\end{eqnarray}
with
\begin{equation}
\tau_{\Bs}  = \frac{1}{\Gamma_{\Bs}} \; \, \hspace{1cm}
y_s = \frac{\Delta \Gamma_s}{2 \Gamma_{\Bs}} \; .
\end{equation}
Such lifetimes can also be used to determine $\Delta \Gamma_s$, 
examples of theoretical derivation can be found in \cite{Dunietz:1995cp,Hartkorn:1999ga,Dunietz:2000cr} and will be discussed in
Section \ref{Bs_system_exp}.
In general ${\cal A}_{\rm CP}^{\rm dir}$, ${\cal A}_{\rm CP}^{\rm mix}$ and $ {\cal A}_{\Delta \Gamma} $
are governed by  non-perturbative effects and there are no simple expressions
for these quantities in terms of basic Standard Model parameters.
These three quantities are, however,  not independent and the following relation holds
\begin{equation}
\left({\cal A}_{\rm CP}^{\rm dir}\right)^2
+
\left({\cal A}_{\rm CP}^{\rm mix}\right)^2
+
\left({\cal A}_{\Delta \Gamma}\right)^2
= 1 \; .
\end{equation}
Under certain circumstances, we get, however, simplified expressions for ${\cal A}_{\rm CP}^{\rm dir}$, ${\cal A}_{\rm CP}^{\rm mix}$ and $ {\cal A}_{\Delta \Gamma} $:
\begin{enumerate}
\item In the  case of flavour-specific decays that are discussed in Section \ref{CPVmix},
      we have $\bar{\cal A}_f = 0$ and thus $\lambda_f = 0$, hence we get
      \begin{eqnarray}
      {\cal A}_{\rm CP}^{\rm fs,dir} & = & 1 \; , \hspace{0.2cm}
      {\cal A}_{\rm CP}^{\rm fs,mix} = 0 \; , \hspace{0.2cm}
      {\cal A}_{\Delta \Gamma}^{\rm fs} = 0 \; ,
      \\
      \tau^{\rm fs, eff} & = & \tau_{\Bs}  \frac{1+ y_s^2}{1- y_s^2} \; .
      \label{taueff}
      \end{eqnarray}
\item In Section \ref{CPVinter} we will introduce so-called golden modes, which  have only one contributing CKM structure and one
      considers the decay into a CP eigenstate $f$.
      In that case we have $|\lambda_f| =1 $ and thus the simple relations
      \begin{equation}
      {\cal A}_{\rm CP}^{\rm dir}    = 0 \; , \hspace{0.2cm}
      {\cal A}_{\rm CP}^{\rm mix}    = - \Im (\lambda_f) \; , \hspace{0.2cm}
      {\cal A}_{\Delta \Gamma}       = - \Re (\lambda_f) \; .
      \label{relationsimple}
      \end{equation}
      Moreover the real and imaginary parts of $\lambda_f$ are now given by simple
      combinations of CKM elements, which will be discussed in Section \ref{CPVinter}.
\end{enumerate}
After discussing the decay of a $\Bs$ meson into the final state $f$, we
consider next the time evolution of the decay rate for a $\barBs$ meson
into the same final state $f$. It is given by
\begin{widetext}
\begin{eqnarray}
\Gamma \left[\barBs (t) \to f \right]  =  N_f \left|{\cal A}_f\right|^2
\left(1+|\lambda_f|^2 \right) (1+a_{\rm fs}^s)
e^{-\Gamma t}
&&
\left\{
                                           \frac{\cosh \left( \frac{\Delta \Gamma_s}{2} t\right)}{2}
- \frac{1-|\lambda_f|^2 }{1+|\lambda_f|^2 }\frac{\cos  \left( \Delta M_s t\right)}{2}
\right.
\nonumber
\\
&&
\left.
- \frac{2 \Re (\lambda_f )}{1+|\lambda_f|^2 }
  \frac{\sinh \left( \frac{\Delta \Gamma_s}{2} t\right)}{2}
+ \frac{2 \Im (\lambda_f )}{1+|\lambda_f|^2 }
  \frac{\sin  \left( \Delta M_s t\right)}{2}
\right\}
\label{GammabarBf} \; .
\end{eqnarray}
\end{widetext}
The common pre-factors, i.e. $N_f$ and $\left|{\cal A}_f\right|^2
\left(1+|\lambda_f|^2 \right)$, typically cancel in CP asymmetries and we do not need to know their
value. This is very advantageous because the hadronic quantity  ${\cal A}_f$ is notoriously
difficult to calculate.
Nevertheless, a dependence on the parameter $\lambda_f$ will still be left in CP asymmetries. As already stated,
in general this parameter cannot be calculated from first principles. 
Making, however,  some additional assumptions, like neglecting penguin effects, a theory prediction
for  $\lambda_f$ can be made, which enables  then an extraction
of fundamental standard model parameters (i.e. a combination of CKM elements)
from the measurement of a CP asymmetry.
\\
For completeness we also consider the decay of $\Bs$ and $\barBs$ mesons
into the CP conjugate of $f$, which will be
denoted by $\bar{f}$.
\begin{equation}
| \bar{f} \rangle = CP | f \rangle \; .
\label{barf}
\end{equation}
With the definitions
\begin{equation}
\bar{\cal A}_{\bar{f}} = \langle \bar{f} | {\cal H}_{eff} |\barBs \rangle \; ,
\hspace{1cm}
\lambda_{\bar{f}} = \frac{q}{p} \frac{\bar{\cal A}_{\bar{f}}}{{\cal A}_{\bar{f}}}
\end{equation}
and assuming $N_f = N_{\bar{f}}$ we get for the time evolution of~the decay rates
\begin{widetext}
\begin{eqnarray}
\Gamma \left[\Bs (t) \to \bar{f} \right]  =  N_f \left|\bar{\cal A}_{\bar f}\right|^2
\left(1+|\lambda_{\bar{f}}|^{-2} \right) (1-a_{\rm fs}^s)
e^{-\Gamma t}
&&
\left\{
                                           \frac{\cosh \left( \frac{\Delta \Gamma_s}{2} t\right)}{2}
- \frac{1-|\lambda_{\bar{f}}|^{-2} }{1+|\lambda_{\bar{f}}|^{-2} }\frac{\cos  \left( \Delta M_s t\right)}{2}
\right.
\nonumber
\\
&&
\left.
- \frac{2 \Re (\frac{1}{\lambda_{\bar{f}}} )}{1+|\lambda_{\bar{f}}|^{-2} }
  \frac{\sinh \left( \frac{\Delta \Gamma_s}{2} t\right)}{2}
+ \frac{2 \Im (\frac{1}{\lambda_{\bar{f}}} )}{1+|\lambda_{\bar{f}}|^{-2} }
  \frac{\sin  \left( \Delta M_s t\right)}{2}
\right\}
\label{GammaBbarf} \; ,
\\
\Gamma \left[\barBs (t) \to \bar{f} \right]  =  N_f \left|\bar{\cal A}_{\bar f}\right|^2
\left(1+|\lambda_{\bar{f}}|^{-2} \right)
e^{-\Gamma t}
&&
\left\{
                                           \frac{\cosh \left( \frac{\Delta \Gamma_s}{2} t\right)}{2}
+ \frac{1-|\lambda_{\bar{f}}|^{-2} }{1+|\lambda_{\bar{f}}|^{-2} }\frac{\cos  \left( \Delta M_s t\right)}{2}
\right.
\nonumber
\\
&&
\left.
- \frac{2 \Re (\frac{1}{\lambda_{\bar{f}}} )}{1+|\lambda_{\bar{f}}|^{-2} }
  \frac{\sinh \left( \frac{\Delta \Gamma_s}{2} t\right)}{2}
- \frac{2 \Im (\frac{1}{\lambda_{\bar{f}}} )}{1+|\lambda_{\bar{f}}|^{-2} }
  \frac{\sin  \left( \Delta M_s t\right)}{2}
\right\}
\label{GammabarBbarf} \; .
\end{eqnarray}
\end{widetext}
The above formulae can be used to extract the observables $\Delta M_s$, $\Delta \Gamma_s$
and $a_{\rm fs}^s$
from experiment, which can then be compared with the theory predictions.
According  to Eq.~(\ref{DMdef}) and Eq.~(\ref{DGdef}) these three observables
are related to the matrix elements $\Gamma_{12}^s$ and $M_{12}^s$, thus a Standard Model
calculation of the three mixing observables requires a calculation of the box diagrams in
Fig.~\ref{box}.
\subsubsection{Theoretical determination of $M_{12}^s$}
The calculation of the Standard Model value for $M_{12}^s$ is straight-forward.
In principle there are nine different combinations of internal quarks in the box diagrams, 
thus  we get
\begin{eqnarray}
M_{12}^s & \propto &
\lambda_u^2         F(u,u) + \lambda_u \lambda_c F(u,c) + \lambda_u \lambda_t F(u,t) +
\nonumber
\\ &&
\lambda_c \lambda_u F(c,u) + \lambda_c^2         F(c,c) + \lambda_c \lambda_t F(c,t) +
\nonumber\\ &&
\lambda_t \lambda_u F(t,u) + \lambda_t \lambda_c F(t,c) + \lambda_t^2         F(t,t) \, ,
\end{eqnarray}
with the CKM structures $\lambda_q = V_{qs}^* V_{qb}$. The functions $F(x,y)$ depend on
the masses of the internal quarks $x$ and $y$ normalised to the $W$ boson mass.
Using CKM unitarity, i.e. $\lambda_u+\lambda_c+\lambda_t = 0$, we get
\begin{eqnarray}
M_{12}^s & \propto &
\; \; \; \; \lambda_c^2 \left[F(c,c) - 2 F(u,c) + F(u,u) \right]
\nonumber \\ &&
+ 2 \lambda_c \lambda_t \left[ F(c,t) - F(u,t) - F(u,c) + F(u,u)\right]
\nonumber \\ &&
+\;  \lambda_t^2 \left[ F(t,t) - 2 F(u,t) + F(u,u) \right] \, .
\end{eqnarray}
From this equation one sees clearly the arising GIM cancellation \cite{Glashow:1970gm}
in all three terms: if all masses would be equal, each of the three terms would vanish.
Because of that also any constant term in the functions $F(x,y)$ cancels in $M_{12}^s$
and only the mass dependent terms will survive. An explicit calculation shows that
$F(x,y)$ grows strongly with the masses (see Eq.~(\ref{InamiLim})), thus there is a very severe GIM cancellation
in the first two terms ($m_u/M_W$ and $m_c/M_W$ can be very well be approximated by zero),
while the third term will give a sizable contribution ($m_t/M_W >1$). Since the
CKM structures have all a similar size ($\lambda_c \propto \lambda^4 \propto \lambda_t$,
with the Wolfenstein parameter $\lambda$ \cite{Wolfenstein:1983yz}) we get to a very
good approximation
\begin{eqnarray}
M_{12}^s & \propto &   \lambda_t^2 \left[ F(t,t) - 2 F(u,t) + F(u,u) \right]
\\
& \propto &   \lambda_t^2 S_0 \left( \frac{m_t^2}{M_W^2} \right) \; ,
\end{eqnarray}
where $S_0$ denotes the  Inami-Lim function  \cite{Inami:1980fz}:
\begin{equation}
S_0(x) = \frac{4x -11x^2 +x^3}{4 (1-x)^2} - \frac{3 x \ln x}{2 (1-x)^2}\, .
\label{InamiLim}
\end{equation}
In that respect it is sometimes stated that only the top quark contributes
to $M_{12}^s$. Formally the process of calculating $M_{12}^s$ can be viewed as
performing an operator product expansion (OPE) by integrating out the
heavy $W$ boson and the heavy top quark. Since both of these masses are far above
the hadronic scale and the $b$ quark mass, there is no doubt in the applicability
of the OPE. This will change in the discussion of $\Gamma_{12}^s$.
The complete calculation of $M_{12}^s$ yields
\begin{eqnarray}
M_{12}^s & = & \frac{G_F^2}{12 \pi^2}
\lambda_t^2 M_W^2 S_0(x_t)
B f_{B_s}^2  M_{\Bs} \hat{\eta }_B\, ,
\label{M12}
\end{eqnarray}
with simple pre-factors: the Fermi-constant $G_F$, the masses of the $W$ boson,
$M_W$, and of the $B_s$ meson, $M_{\Bs}$ and the normalisation factor $1/12\pi^2$.
As we have seen above there is only one CKM structure contributing
$ \lambda_t = V_{ts}^* V_{tb}$.
The CKM elements are the only place in Eq.~(\ref{M12}) where an imaginary part can arise.
By writing
\begin{equation}
\lambda_t^2 = |\lambda_t^2| \frac{\lambda_t}{\lambda_t^*} = |\lambda_t^2| e^{i \phi_M}
\end{equation}
we get the  explicit dependence of  the phase $\phi_M$ on CKM parameters, which was already stated
in Eq.~(\ref{phiM}).
As discussed above, the result of the 1-loop diagrams given in Fig.~\ref{box}
is denoted by the Inami-Lim function
$ S_0(x_t= (\bar{m}_t(\bar{m}_t))^2/M_W^2)$, where $\bar{m}_t(\bar{m}_t)$ is
the $\overline{MS}$-mass \cite{Bardeen:1978yd} of the top quark.
Perturbative 2-loop QCD corrections are compressed in the factor
$\hat{\eta }_B \approx 0.84$, they have been calculated by
\cite{Buras:1990fn}.
Performing the calculation of $M_{12}^s$ one gets a spinor operator
for each external quark in the box diagram. Together with the arising
Dirac matrices they form the four quark $\Delta B=2$ operator
\begin{eqnarray}
Q & = & \bar s^\alpha \gamma_\mu (1- \gamma_5) b^\alpha
        \times
        \bar s^\beta  \gamma^\mu (1- \gamma_5) b^\beta \; .
      \label{Q}
\end{eqnarray}
$\alpha$ and $\beta$ are the colour indices of the $b$ and $s$ quark spinors.
All hadronic effects that describe the binding of the quarks into meson states
as well as the non-perturbative QCD effects contributing to the
transition of the $\Bs$ meson into the $\barBs$ meson and vice versa
are encoded in the hadronic matrix element of the operator $Q$.
The hadronic matrix element\footnote{Throughout this review we will
use the conventional relativistic normalisation for
the $\Bs$ meson states, i.e. $\langle \bar{B}_s^0 | \Bs \rangle = 2 E V$
($E$: energy, $V$: volume).} of this operator
is parametrised in terms of a decay constant $f_{B_s}$ and a bag parameter
$B$:
\begin{eqnarray}
\langle Q \rangle & \equiv &      \langle \bar{B}_s^0|Q |\Bs \rangle
       =
      \frac{8}{3}  M_{\Bs}^2  f_{B_s}^2 B (\mu) \; ,
\label{ME1}
\end{eqnarray}
The factor $8/3 = 2 (1 + 1/N_c)$ stems from the colour structure.
It ensures that the bag parameter $B$ obtains the value one in vacuum
insertion approximation\footnote{The matrix element in Eq.~(\ref{ME1}) can be rewritten, 
by inserting a complete set of states between the two currents of the operator $Q$, 
given in Eq.~(\ref{Q}). Next this expression is equated to the contribution of the vacuum state only times
 a correction factor $B$ (bag factor), that corrects for the neglect of all higher states in the sum.
Setting the bag parameter to one, corresponds to the vacuum insertion approximation. Many lattice evaluations
show, that this assumption seems to be very well justified \cite{Bazavov:2016nty}.
The remaining matrix elements of the form $\langle \Bs |  \bar s^\alpha \gamma_\mu (1- \gamma_5) b^\alpha | 0 \rangle $
are proportioanl to $f_{B_s} p_\mu$, where $p_\mu$ is the four-momentum of the $\Bs$ meson.}.
We also indicated the renormalisation scale dependence
of the bag parameter; in our analysis we take $\mu = m_b$. 
\\
Sometimes a different notation for the QCD corrections and the bag parameter
is used in the literature
(e.g. by the Flavour Lattice Averaging Group (FLAG): \cite{Aoki:2013ldr}),
$(\eta_B, \hat{B})$ instead of $(\hat{\eta}_B, B)$ with
\begin{eqnarray}
\hat{\eta}_B B & =: & \eta_B \hat{B} 
\\
& = & \eta_B   \alpha_s(\mu)^{-\frac{6}{23}} 
\left[ 1 + \frac{\alpha_s(\mu)}{4 \pi} \frac{5165}{3174} \right] B
\; ,
\\
\hat{B} & = & 1.51599 \, B \; .
\end{eqnarray}
The parameter $\hat{B}$ has the advantage
of being renormalisation scale and scheme independent.
\\
A commonly used Standard Model prediction of $\Delta M_s$ was given by \cite{Lenz:2011ti}
\begin{equation}
\Delta M_s^{\rm SM, 2011} = \left(17.3 \pm 2.6 \right)\; \mbox{ps}^{-1} \; .
\label{DeltaM2011}
\end{equation}
Using the most recent numerical inputs ($G_F$, $M_W$, $M_{B_s}$ and $m_b$ from the
Particle Data Group (PDG)
\cite{Agashe:2014kda},
the top quark mass from \cite{ATLAS:2014wva},
the non-perturbative parameters from FLAG (web-update of \cite{Aoki:2013ldr})
and CKM elements from the CKMfitter group [ web-update of \cite{Charles:2004jd} ],
[similar values can be taken from the UTfit group \cite{Bona:2006ah}],
we predict the mass difference of the neutral $\Bs$ mesons to be
\begin{equation}
\Delta M_s^{\rm SM, 2015} = \left(18.3 \pm 2.7 \right)
\; \mbox{ps}^{-1} \; .
\label{DeltaM2015}
\end{equation}
Here the dominant uncertainty comes  from the lattice predictions for the non-perturbative
parameters  $B$ and $ f_{B_s}$, giving a relative error of $14 \%$.
This input did not change
compared to the 2011 prediction from \cite{Lenz:2011ti}.
The uncertainty in the CKM elements
contributes about $5\%$ to the error budget. The CKM parameters were determined assuming
unitarity of the $3 \times 3$ CKM matrix. For some  new physics models
this assumption might have to be given up, leading to larger CKM uncertainties.
The uncertainties due to ${m}_t$, ${m}_b$ and $\alpha_s$
can be safely neglected at the current stage.
A detailed discussion of the input parameters and the
error budget is given in Appendix~\ref{app:error}.
\\
There is, however, a word of caution:
in the above theory prediction (\ref{DeltaM2015})
we use the non-perturbative value from FLAG
$f_{B_s} \sqrt{B} = 216 \pm 15 \;\mbox{MeV}$ \footnote{This value is derived from the FLAG
value of $f_{B_s} \sqrt{\hat{B}}$. It is by accident equal to the value of
$f_{B_d} \sqrt{\hat{B}}$ quoted from FLAG.}
(with $N_f = 2+1$ active flavours in the lattice simulations). However,
only one number -- from the HPQCD Collaboration \cite{Gamiz:2009ku} --
is included in the FLAG average.
It would of course be advantageous to have more numbers from different collaborations
and there are currently some more (mostly preliminary) numbers on the market:
\begin{eqnarray}
f_{B_s} \sqrt{B} \approx 200 \; {\rm MeV} & \Rightarrow &
          \Delta M_s^{\rm HPQCD} \approx  15.7 \; \mbox{ps}^{-1}  ,
\label{DeltaMHPQCD}
\\
f_{B_s} \sqrt{B}\approx 211  \; {\rm MeV} & \Rightarrow &
          \Delta M_s^{\rm ETMC} \approx  17.4 \; \mbox{ps}^{-1}  ,
\label{DeltaMETMC}
\\
f_{B_s} \sqrt{B} \approx 227 \; {\rm MeV} & \Rightarrow &
          \Delta M_s^{\rm Fermilab} \approx\  20.2 \; \mbox{ps}^{-1}  .
\label{DeltaMFermi}
\end{eqnarray}
HPQCD updated their results in \cite{Dowdall:2014qka} and for our numerical estimate in
Eq.~(\ref{DeltaMHPQCD}) we had to read off the numbers from Fig.~3
in their proceedings \cite{Dowdall:2014qka}. Their investigations suggest a
possible error
of about $5\%$ for $f_{B_s}^2 B$ in the near future, which would be a major improvement.
The ETMC number stems from
\cite{Carrasco:2013zta}, it is obtained with only two active flavours in the lattice simulation.
The Fermilab-MILC number is an update for the LATTICE 2015
conference of \cite{Bouchard:2011xj}\footnote{During the refereeing process for this review,
Fermilab-MILC presented final results in \cite{Bazavov:2016nty}. The numerical effect of these new 
inputs on mixing observables was studied in \cite{Jubb:2016mvq}.}.
The range of the  above numbers seems to be nicely covered by the current FLAG
average, but it would of course be very interesting to have final numbers and an average
for the values given in Eq.~(\ref{DeltaMHPQCD}), Eq.~(\ref{DeltaMETMC}) and Eq.~(\ref{DeltaMFermi}).
There is  also a large value from RBC-UKQCD presented at LATTICE 2015, $f_{B_s} \sqrt{B} = 262 \; {\rm MeV} $
(update of \cite{Aoki:2014nga}).
However, this number is obtained in the static limit and currently missing $1/m_b$
corrections are expected to be very sizable. Thus we do not give a value of $\Delta M_s$ for this lattice value.
For our numerical analysis, we only use the value from FLAG.
In summary, an uncertainty of about $\pm 5\%$
might be feasible for the theory prediction of $\Delta M_s$ taking future lattice improvements into account.
\subsubsection{Heavy Quark Expansion}
\begin{figure}[t]
      \includegraphics[width=0.350 \textwidth,angle=270]{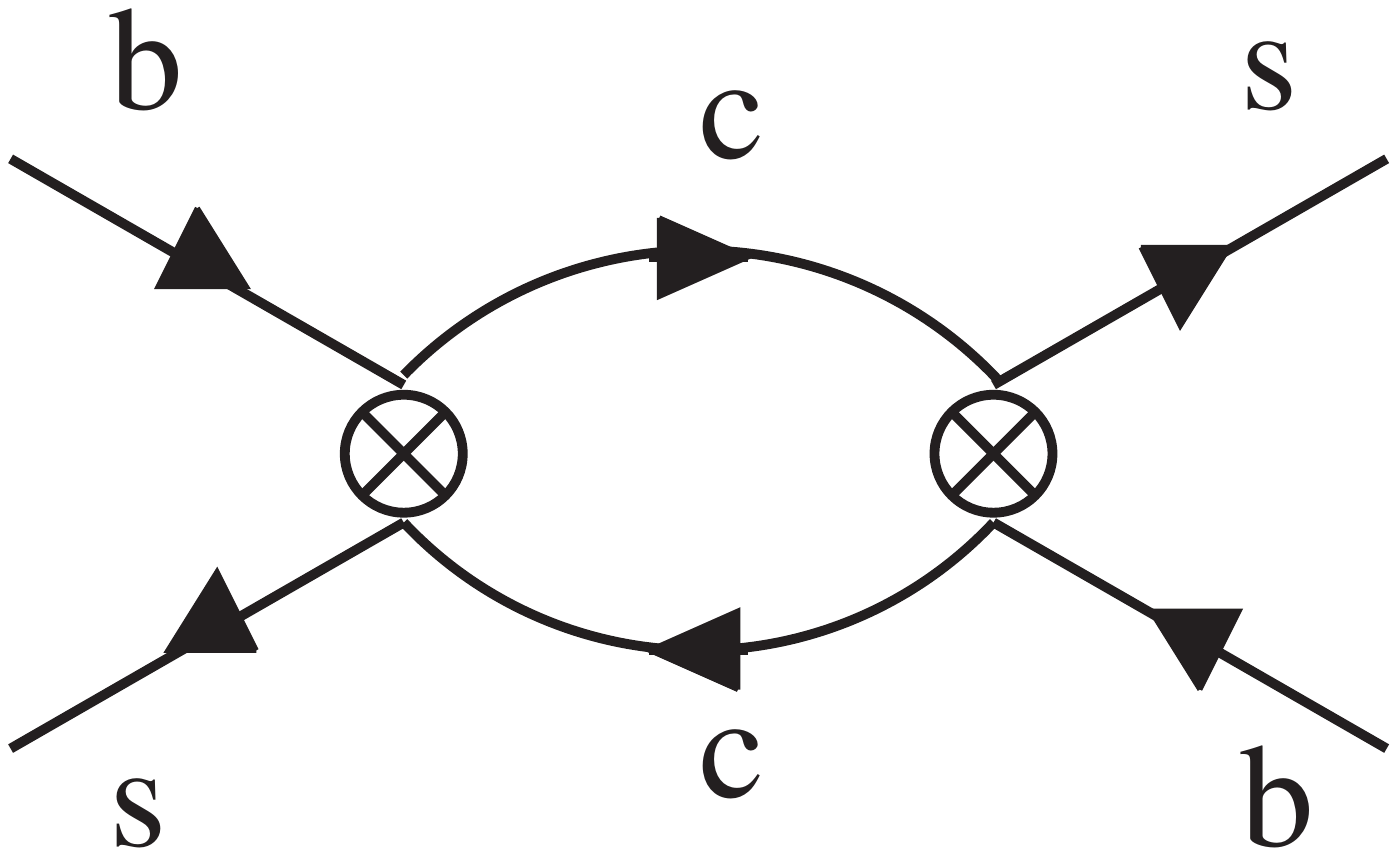}
      \caption{To $\Gamma_{12}^s$ only the box diagrams with internal up and charm
               quarks are contributing in the Standard Model, see Fig.~\ref{box}.
               Integrating out the heavy $W$ boson, we are left with a bi-local object, which
               is shown here for internal charm and anti charm quarks.}
      \label{bilocal}
      \end{figure}

\begin{figure*}[t]
      \includegraphics[width=0.95 \textwidth,angle=0]{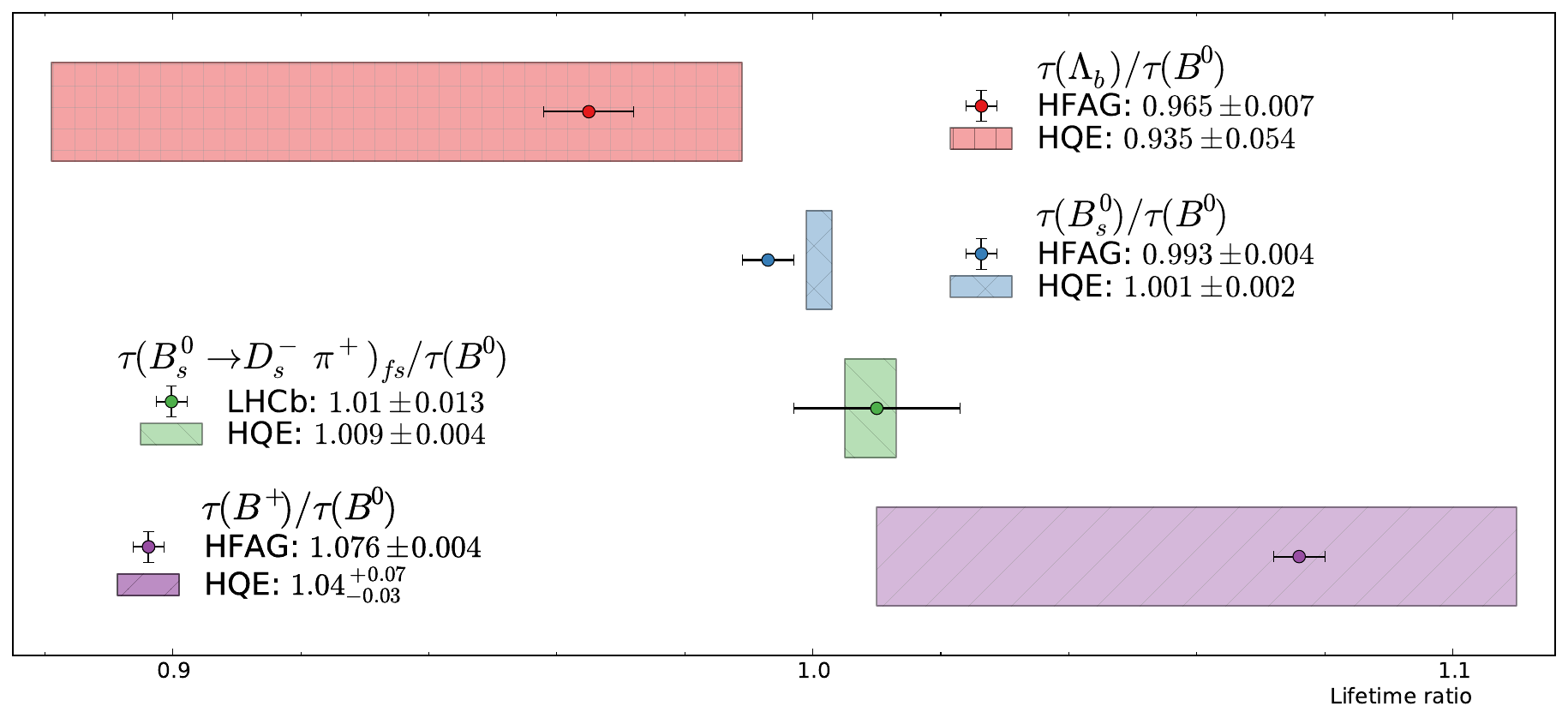}
      \caption{Comparison of HQE predictions for lifetime ratios of
      heavy hadrons with experimental values.
      The theory values are taken from \cite{Lenz:2014jha}. Experimental numbers are taken
      from (fall 2014) HFAG \cite{Amhis:2014hma}.}
      \label{lifetime}
      \end{figure*}

The calculation of the decay rate difference $\Delta \Gamma_s$ is more involved.
In the box diagrams depicted in Fig.~\ref{box}, we have to take into account now only the
internal up and charm quarks. Integrating out all heavy particles (in this case only the
$W$ boson) we are not left with a local $\Delta B =2$ operator as in the case of
$M_{12}^s$, but with a bi-local object depicted in Fig.~\ref{bilocal}.
To get to the level of local operators, which is needed for being able to make a theory
prediction, a second operator product expansion is required.
The second OPE relies on the smallness
 of the parameter $\Lambda / m_b$, where $\Lambda$
is expected to be of the order of the hadronic scale $\Lambda_{\rm QCD}$ and $m_b$ is the $b$ quark mass.
More precisely the HQE is an expansion in $\Lambda$ normalised to
the momentum release of the decay given by
$\sqrt{M_i^2 - M_f^2}$, with the initial mass $M_i$ and the final state masses $M_f$.
For massless final states an expansion in $\Lambda/m_b$ is generally expected to
converge, while for a transition like $b \to c \bar{c} s$ it is not a priori clear,
whether $\Lambda/ \sqrt{m_b^2 - 4 m_c^2}$ is small enough to get a converging series.  Thus the validity of this
so-called heavy quark expansion (HQE) has to be tested by comparisons of experiment and
theory.
The formulation of the HQE is based on work by
Voloshin and Shifman in \cite{Khoze:1983yp}, \cite{Shifman:1984wx}, \cite{Bigi:1991ir},
\cite{Blok:1992hw}, \cite{Bigi:1992su}, \cite{Blok:1992he}
and in detail described in \cite{Lenz:2014jha}\footnote{See e.g. \cite{Bigi:1987in} and 
\cite{Bigi:1997fj}, for  early reviews.}.
The HQE applies also for lifetimes and totally inclusive decay rates of  heavy hadrons.
Historically there had been several discrepancies between experiment and
theory that questioned the validity of the HQE:
\begin{itemize}
\item In the mid-nineties the {\it missing charm puzzle}
      (see e.g. \cite{Lenz:2000kv} for a brief review), a disagreement between
      experiment and theory about the
      average number of charm quarks produced per $b$-decay, was a hot topic.
      A possible interpretation could be  new physics, but a violation of quark hadron duality, i.e.
      a violation of the validity of the HQE, was also considered to solve
      this discrepancy, in particular in the decay $b \to c \bar{c} s$.
      This issue has now been resolved, by more precise data
      and improved theory predictions
      (see \cite{Krinner:2013cja}), leading to a nice agreement between
      experiment and theory within uncertainties.
\item For a long time the measured $\Lambda_b$ lifetime was considerably
      shorter than its predicted value (according to estimates of the HQE - see e.g. \cite{Bigi:1997fj,Voloshin:2000zc}). 
      This issue has been resolved by recent measurements, mostly by the
      LHCb Collaboration (\cite{Aaij:2013oha,Aaij:2014owa,Aaij:2014zyy})
      but also by the Tevatron experiments (e.g. \cite{Aaltonen:2014wfa}).
      The history of the {\it $\Lambda_b$ lifetime} - HFAG quoted
      2003 a value of $\tau_{\Lambda_b}^{\rm HFAG \; 2003} = (1.229 \pm 0.080)$ ps,
      which is about 3 standard deviations away from the 2015 average
      of  $\tau_{\Lambda_b}^{\rm HFAG \; 2015} = (1.466 \pm 0.010)$ ps -
      and also (sometimes embarrassing) theoretical attempts to
      obtain low theory values are discussed in detail in the review of \cite{Lenz:2014jha}.
      The low experimental values reported in the early measurements are
      mostly determined using semileptonic decays with an undetectable
      neutrino \cite{Stone:2014pra}, while new measurements use non-leptonic decays with xsfully reconstructed
     final states. The huge range in the
      theory predictions for the $\Lambda_b$ lifetime stems from our missing knowledge
      about the size of the hadronic
      matrix elements. Some theory groups tried to create some extraordinary large
      enhancements of these matrix elements in order to describe the experimental data,
      while other groups, including, for example, Bigi and Uraltsev,  stuck to theory estimates
      that were in conflict with the old measurements, but agree perfectly with
      the new ones.
      The current status of lifetimes is depicted in Fig.~\ref{lifetime}.
      No lifetime
      puzzle exists anymore.
      The theoretical precision is strongly limited by a lack of
      up-to-date values for the arising
      non-perturbative parameters. For the $\Lambda_b$-baryon the most recent
      lattice numbers stem from
      1999 \cite{DiPierro:1999tb} and for the $B$ mesons the
      most recent numbers are from 2001
      \cite{Becirevic:2001fy}.
      This lack of theoretical investigations limits also our current knowledge about
      the intrinsic precision of the HQE.
\item Since $\Delta \Gamma_s$ is dominated by a $ b \to c \bar{c} s$ transition,
      the applicability of the HQE was in particular questioned for
      $\Delta \Gamma_s$, see e.g. \cite{Ligeti:2010ia} and the discussion in
      \cite{Lenz:2011zz} and the references therein.
      In the last years this was also related
      to the unexpected measurement of a large value of the
      di-muon asymmetry by the D0 collaboration
      \cite{Abazov:2013uma,Abazov:2011yk,Abazov:2010hj,Abazov:2010hv}.
      In 2012 the issue of $\Delta \Gamma_s$ was solved experimentally by a
      direct measurement of this
      quantity by the LHCb Collaboration. The current HFAG \cite{Amhis:2014hma} average,
      combining values from LHCb, ATLAS, CMS, D0 and CDF, is in perfect agreement
      with the HQE prediction from \cite{Lenz:2011ti}, which is based on on the calculations
      of
      \cite{Lenz:2006hd,Beneke:2003az,Ciuchini:2003ww,Beneke:1998sy}. This will be discussed
      in detail below.
\end{itemize}
All in all the HQE has been experimentally proven to be very successful and
one could try next to test its applicability also for charm-physics,
see e.g. \cite{Lenz:2013aua,Bobrowski:2010xg}
for some recent investigations.
Charm studies would be very helpful for assessing  the intrinsic uncertainties
of the HQE. Having more confidence in the validity of HQE, it can now also
be applied to quantities that are sensitive
to new physics, in particular to the semileptonic CP asymmetries, which will be
discussed in Section \ref{CPVmix}. A very recent study of the possible size of duality violating effects (i.e. deviations from
the HQE expectations) can be found in \cite{Jubb:2016mvq}.
\subsubsection{Theoretical determination of $\Gamma_{12}^s$}
\begin{figure*}
\includegraphics[width=0.50 \textwidth,angle=270]{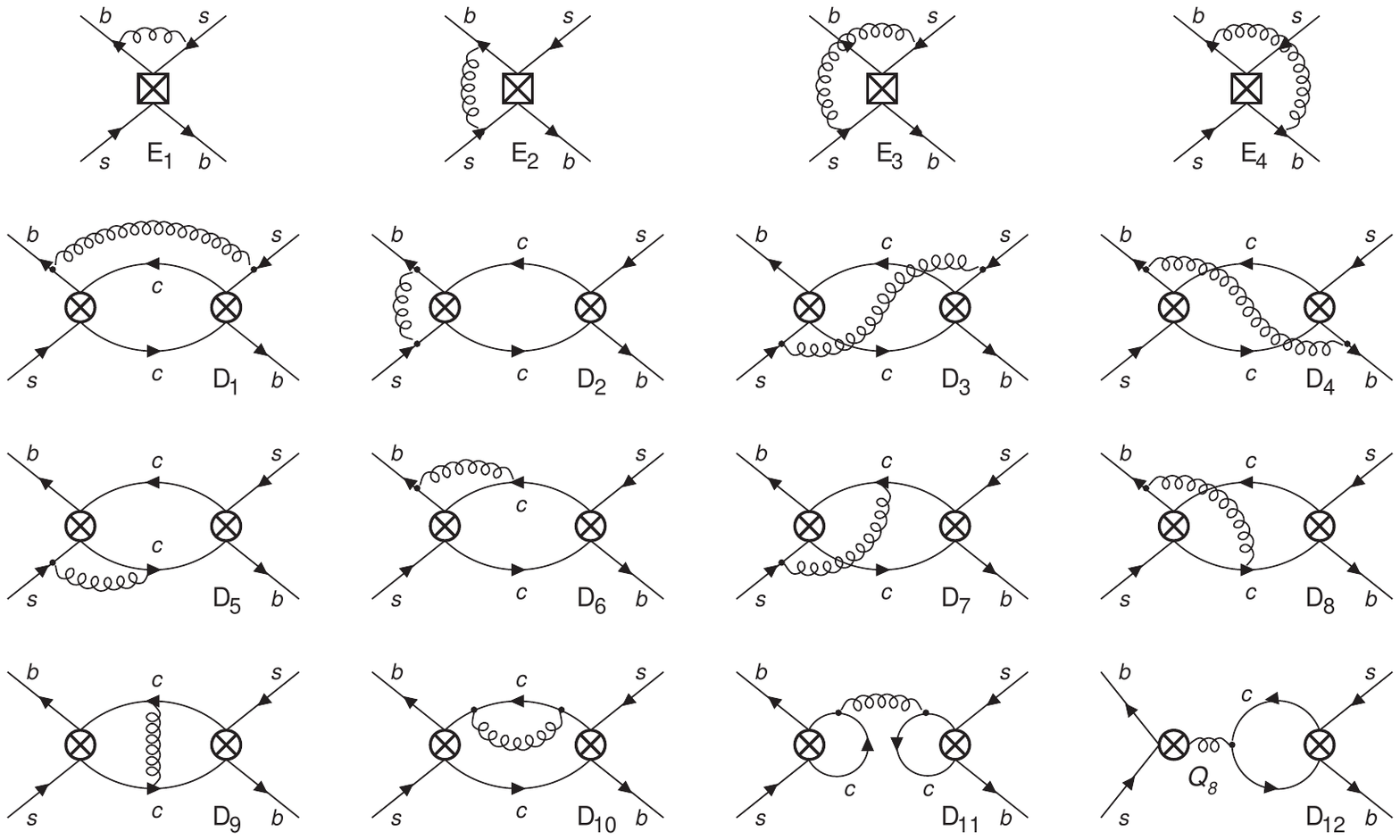}
\caption{\label{nloQCD} Standard Model diagrams contributing to $\Gamma_{12}^s$ at NLO-QCD,
i.e. $\Gamma_3^{s,(1)}$.
For obtaining the NLO-QCD Wilson coefficients one has to calculate one-loop corrections to the
$\Delta B=2$ operators (E1-E4) and also two-loop corrections to the double insertion of  $\Delta B=1$ operators
(D1-D12). An explicit
cancellation of all infra-red singularities in the matching was shown by
\cite{Beneke:1998sy} and later by \cite{Beneke:2003az} and \cite{Ciuchini:2003ww}.
Such an IR-safety is crucial for the consistency of the HQE. The next future steps
will be the determination of  $\Gamma_4^{s,(1)}$ and $\Gamma_3^{s,(2)}$. For that one has to
take into account in the above diagrams a non-vanishing strange quark momentum and one has
to add a further gluon in the above diagrams.}
\end{figure*}

According to the HQE, the off-diagonal element $\Gamma_{12}^s$ of the $\Bs$ mixing matrix
can be expanded as a power series in the inverse of
the heavy $b$-quark mass $m_b$ and the strong coupling $\alpha_s$:
\begin{equation}
\Gamma_{12}^s = \frac{\Lambda^3}{m_b^3}
\left(\Gamma_3^{s,(0)} + \frac{\alpha_s}{4 \pi} \Gamma_3^{s,(1)} + ... \right)
+
\frac{\Lambda^4}{m_b^4}
\left(\Gamma_4^{s,(0)} + ... \right) + ... \, .
\label{HQE}
\end{equation}
$\Lambda$ denotes a hadronic scale, which is assumed to be of the order of $\Lambda_{QCD}$,
but its
actual value has to be determined by a non-perturbative calculation.
Each of the $\Gamma_i^{s,(j)}$ is a product of perturbative Wilson coefficients
and non-perturbative matrix elements.
In $\Gamma_3^s$ these matrix elements arise from dimension 6 four quark operators,
in $\Gamma_4^s$ from dimension 7 operators and so on.
\\
The leading term in Eq.~(\ref{HQE}), $\Gamma_3^{s,(0)}$, was calculated already quite
long ago by
\cite{Ellis:1977uk}, \cite{Hagelin:1981zk}, \cite{Franco:1981ea}, \cite{Chau:1982da},
\cite{Buras:1984pq} and \cite{Khoze:1986fa}.
Here three different 4 quark operators arise; besides $Q$ from Eq.~(\ref{Q}) these are
\begin{eqnarray}
Q_S & = & \bar s^\alpha (1 + \gamma_5) b^\alpha
          \times
          \bar s^\beta  (1 + \gamma_5) b^\beta \; ,
\\
\tilde{Q}_S & = & \bar s^\alpha (1 + \gamma_5) b^\beta
                 \times
                 \bar s^\beta  (1 + \gamma_5) b^\alpha \; .
\end{eqnarray}
The general structure of the leading term $\Gamma_{3}^s$ has three ($uc = cu$) different CKM contributions
\begin{eqnarray}
\Gamma_{3}^s & = & -
\sum \limits_{x=u,c}
\sum \limits_{y=u,c}
\lambda_x \lambda_y
\Gamma_{12}^{s,xy}
\label{HQE2}
\end{eqnarray}
and each factor $\Gamma_{12}^{s,xy}$ has contributions of the three operators
$Q$, $Q_S$ and $\tilde{Q}_S$
\begin{eqnarray}
\Gamma_{12}^{s,xy}
& = &
\Gamma_{xy}^{s,Q} \langle Q \rangle
+
\Gamma_{xy}^{s,{Q}_S} \langle Q_S \rangle
+
\Gamma_{xy}^{s,\tilde{Q}_S} \langle \tilde{Q}_S \rangle
\label{HQE3} \; .
\end{eqnarray}
The matrix elements of the newly arising  operators are typically
parameterised as
\begin{eqnarray}
      \langle Q_S \rangle & \equiv & \langle \bar{B_s}|Q_S |B_s \rangle
       =
      - \frac{5}{3}  M_{\Bs}^2 f_{B_s}^2
      B_S' \; ,
      \\
      \langle \tilde{Q}_S  \rangle
      & \equiv &
      \langle \bar{B_s}|\tilde{Q}_S |B_s \rangle
      =
       \frac{1}{3}  M_{\Bs}^2 f_{B_s}^2
      \tilde{B}_S' \; ,
\end{eqnarray}
with the modified bag parameters
\begin{equation}
B_X' =
\frac{M_{\Bs}^2}{\left[\bar{m}_b(\bar{m}_b) + \bar{m}_s(\bar{m}_b)\right]^2} \, B_X
\approx 1.57706 \; B_X \; .
\end{equation}
In the vacuum insertion approximation, the unmodified bag parameters are equal to one.
More reliable values can be obtained by using non-perturbative methods like
QCD sum rules\footnote{A QCD sum rule determination of $\langle Q \rangle$ is given e.g.
in
\cite{Korner:2003zk}. However, we will not use the number obtained there in our analysis.}
or lattice QCD.
$Q$, $Q_S$ and $\tilde{Q}_S$ were determined by several lattice groups, who actually determined all
five operators of the so-called SUSY basis\footnote{In the Standard Model only $Q$ contributes to
$\Delta M_s$, while in extensions of the Standard Model additional contributions of new operators
can appear. The whole set of these operators is called SUSY-basis and typically denoted by
$O_1$...$O_5$. It turns out, however, that all these five operators are also needed for a
precise standard
model prediction of $\Delta \Gamma_s$.}.
\cite{Becirevic:2001xt},
\cite{Carrasco:2013zta} and
\cite{Dowdall:2014qka}
use the notation $O_1$, $O_2$ and $O_3$ for these three operators:
\begin{equation}
Q            \equiv  O_1 \; ,
Q_S          \equiv  O_2 \; ,
\tilde{Q}_S  \equiv  O_3 \; .
\end{equation}
In the case of  \cite{Bouchard:2011xj} there is also an additional factor $4$ present.
\begin{equation}
Q            \equiv  4 O_1 \; ,
Q_S          \equiv  4 O_2 \; ,
\tilde{Q}_S  \equiv  4 O_3 \; .
\end{equation}
\cite{Becirevic:2001xt} and \cite{Carrasco:2013zta}
use the same definitions of the bag parameters as we do
\begin{equation}
B           \equiv B_1 \; ,
B_s         \equiv B_2 \; ,
\tilde{B}_S \equiv B_3 \; ,
\end{equation}
while \cite{Dowdall:2014qka})  and \cite{Bouchard:2011xj} use the modified bag parameters
\begin{equation}
B           \equiv B_1 \; ,
B_s'         \equiv B_2 \; ,
\tilde{B}_S' \equiv B_3 \; .
\end{equation}
It was found, that  these three operators are not independent (see e.g. \cite{Beneke:1996gn})
and that the following relation holds
\begin{equation}
R_0 = Q_S + \alpha_1 \tilde{Q}_S + \frac{\alpha_2}{2} Q = 0 + {\cal O} \left(\frac{\Lambda}{m_b} \right) \; ,
\label{relation}
\end{equation}
with the coefficients
(obtained in \cite{Beneke:1998sy} using the renormalisation scheme described there)
\begin{eqnarray}
\alpha_1 & = & 1 +  \frac{\alpha_s(\mu)}{3 \pi} \left( 12 \ln \frac{\mu}{m_b} + 6\right) \; ,
\\
\alpha_2 & = & 1 + \frac{\alpha_s(\mu)}{3 \pi} \left( 6 \ln \frac{\mu}{m_b} + \frac{13}{2} \right) \; .
\end{eqnarray}
With the help of  Eq.~(\ref{relation}) one can substitute one of the three operators;
historically $\tilde{Q}_S$ was
eliminated, obtaining
\begin{eqnarray}
\Gamma_{12}^{s,xy}
& = &
\left[ \Gamma_{xy}^{s,Q}   - \frac12 \frac{\alpha_2}{\alpha_1}  \Gamma_{xy}^{s,\tilde{Q}_S} \right]
\langle Q \rangle
+
\nonumber
\\
&&
\left[ \Gamma_{xy}^{s,Q_S} - \frac{1}{\alpha_1}  \Gamma_{xy}^{s,\tilde{Q}_S} \right]
\langle Q_S \rangle
+ {\cal O} \left(\frac{\Lambda}{m_b} \right) \; ,
\end{eqnarray}
which was denoted in the literature as
\begin{eqnarray}
\Gamma_{12}^{s,xy} & = & \frac{G_F^2 m_b^2}{24 \pi M_{B_s}}
\left[ G^{s,xy}   \langle  Q \rangle
      -G_S^{s,xy} \langle Q_S \rangle
\right] + \Gamma_{12, \frac{1}{m_b}}^{s,xy}
\label{Gammaxy}
\\
 & = & \frac{G_F^2 m_b^2 f_{B_s}^2 \! M_{\Bs}}{24 \pi } \!
\left[ \frac83 G^{s,xy}   B + \frac53 G_S^{s,xy} B_S'
\right] \! + \Gamma_{12, \frac{1}{m_b}}^{xy} \; ,
\nonumber
\end{eqnarray}
where the Wilson coefficients $G^{s,xy}$ and $G_S^{s,xy}$ contain
the result of the calculation of the box diagrams with internal
on-shell up and/or charm quarks; $xy \in \{uu,uc,cc\}$.
Neglecting the mass of the charm quark and penguin contributions,
$G^{s,xy}$ and $G_S^{s,xy}$ read in LO-QCD
\begin{eqnarray}
 G^{s,xy} & = & 3 C_1^2 + 2 C_1 C_2 + \frac12 C_2^2
\; ,
\\
 G_S^{s,xy} & = & - \left( 3 C_1^2 + 2 C_1 C_2 - C_2^2 \right)
\; ,
\end{eqnarray}
where $C_{1,2}$ denote the $\Delta B =1$ Wilson coefficients of the effective Hamiltonian
describing $b$ quark decays (in our notation
$C_2$ corresponds to the colour allowed operator).
Early LO-QCD estimates of $G^{s,xy}$ and $G_S ^{s,xy}$ can be found in
\cite{Ellis:1977uk}, \cite{Hagelin:1981zk}, \cite{Franco:1981ea}, \cite{Chau:1982da},
\cite{Buras:1984pq} and \cite{Khoze:1986fa}.
NLO QCD corrections, i.e. $\Gamma_3^{s,(1)}$ in Eq.~(\ref{HQE}),
were done for the first time in \cite{Beneke:1998sy}, they turned out
to be quite large. This work was also a proof of the IR-safety of the HQE by direct
calculation. The corresponding NLO-QCD diagrams are shown in Fig.~\ref{nloQCD}.
General arguments for such a proof were given already in the seminal paper
of \cite{Bigi:1991ir}, which resolved the theoretical issues that were prohibiting a systematic
expansion in the inverse of the heavy $b$ quark mass.
Five years later the calculation of the QCD corrections was
confirmed and also sub-leading CKM structures were included
by \cite{Beneke:2003az} and \cite{Ciuchini:2003ww}. In these papers the full expressions for
$G^{s,xy}$ and $G_S^{s,xy}$  are given; they also include contributions from the QCD penguin
operators $Q_1$-$Q_6$ and the chromo-magnetic penguin operator $Q_8$.
\cite{Beneke:2002rj} found that the use of $\bar{m}_c(\bar{m}_b)$ (charm mass at the bottom mass scale)
instead of $\bar{m}_c(\bar{m}_c)$, sums up large logs of the form
$m_c^2/m_b^2 \ln m_c^2/m_b^2 $ to all orders; we will thus use
the parameter $\bar{z}$ in our numerical analysis, given by
\begin{equation}
\bar{z} = \left(\frac{\bar{m}_c(\bar{m}_b)}{\bar{m}_b(\bar{m}_b)} \right)^2 \; .
\label{zbardef}
\end{equation}
In Eq.~(\ref{Gammaxy}) the term $\Gamma_{12, 1/m_b}^{s,xy}$ denotes  sub-leading $1/m_b$
corrections to $\Gamma_{12}^s$ - in
Eq.~(\ref{HQE}) these terms were called $\Gamma_4^{s,(0)}$. Such sub-leading $1/m_b$ corrections
were first calculated by \cite{Beneke:1996gn}
and they also turned out to be quite sizable. 
The operators arising in  $\Gamma_4^{s,(0)}$ are of dimension 7 (e.g.
four quark operators with one derivative),
they are denoted by $R_0$, $R_1$, $R_2$ and $R_3$,
as well as the colour-rearranged counterparts $\tilde{R}_1$, $\tilde{R}_2$ and $\tilde{R}_3$, see
e.g. \cite{Lenz:2006hd} for more details.
The operators $R_0$, $R_1$ and $\tilde{R}_1$ can be reduced to four quark operators
(see e.g. the definition of $R_0$  in Eq.~(\ref{relation})) and thus they can be studied
with current lattice technologies; their results can be deduced from
\cite{Becirevic:2001xt},
\cite{Bouchard:2011xj},
\cite{Carrasco:2013zta}
and
\cite{Dowdall:2014qka}, who were calculating the full five-dimensional SUSY basis of
$\Delta B = 2$ operators. All of those five independent 
operators ($Q$, $Q_S$, $\tilde{Q}_S$, $R_1$ and $\tilde{R}_1$) 
contribute to $\Gamma_{12}^s$.
The genuine dimension 7 operators $R_2$, $R_3$, $\tilde{R}_2$ and $\tilde{R}_3$ are
considerably more complicated.
For the corresponding matrix elements currently no lattice determination is available,
so we have to rely on vacuum insertion approximation, i.e. the bag parameters
$B_{R_2}$, $B_{R_3}$, $B_{\tilde{R}_2}$ and
$B_{\tilde{R}_3}$ are set to one.
First steps towards a non-perturbative determination of these matrix
elements within the framework of QCD sum rules have been done by
\cite{Mannel:2007am,Mannel:2011zza}. Here a more complete study would be very desirable,
because - as will be seen below - these parameters give currently the dominant
uncertainty to $\Gamma_{12}^s$.
\\
The precision of the theory prediction can be further improved by using ratios of theoretical expressions
and by choosing an optimal operator basis:
\begin{itemize}
\item $\Gamma_{12}^s$ depends on $f_{Bs}^2 B$, which is currently
      not very well-known. Thus, it might be advantageous to consider
      the ratio $\Gamma_{12}^s/M_{12}^s$, where the decay constant cancels.
      One gets from this ratio
      \begin{equation}
      {\rm Re} \left(  \frac{\Gamma_{12}^s}{M_{12}^s} \right)
       =
       - \frac{\Delta \Gamma_s}{\Delta M_s} \; ,
      \; \; \; \; \;
      {\rm Im} \left(  \frac{\Gamma_{12}^s}{M_{12}^s} \right)
       = a_{fs}^s \; .
      \end{equation}
      The ratio $\Gamma_{12}^s/M_{12}^s$ can be further modified by using
      the CKM unitarity ($\lambda_u + \lambda_c + \lambda_t = 0$):
      \begin{eqnarray}
      - \frac{\Gamma_{12}^s}{M_{12}^s}  & = & \frac{     \lambda_c^2         \Gamma_{12}^{s,cc}
                         + 2 \lambda_c \lambda_u \Gamma_{12}^{s,uc}
                         +   \lambda_u^2         \Gamma_{12}^{s,uu}}{\lambda_t^2 \tilde{M}_{12}^s}
      \\
       & = &  \frac{\Gamma_{12}^{s,cc}}{\tilde{M}_{12}^s}
        + 2 \frac{\lambda_u}{\lambda_t}
          \frac{\Gamma_{12}^{s,cc} -\Gamma_{12}^{s,uc}}{\tilde{M}_{12}^s}
      \nonumber \\
      &&
        + \left(\frac{\lambda_u}{\lambda_t}\right)^2
          \frac{\Gamma_{12}^{s,cc} -2 \Gamma_{12}^{s,uc}+\Gamma_{12}^{s,uu} }{\tilde{M}_{12}^s}
      \label{ratio1}
      \\
      & = & - 10^{-4}
       \left[ c + a \frac{\lambda_u}{\lambda_t} + b\left(\frac{\lambda_u}{\lambda_t}\right)^2
      \right] \; ,
      \label{ratio2}
      \end{eqnarray}
      where $\tilde{M}_{12}^s$ is defined in such a way that only the CKM-dependence of $M_{12}^s$
      in Eq.~(\ref{M12}) is split off. Eq.~(\ref{ratio2}) introduces the $a$, $b$ and $c$ notation
      of \cite{Beneke:2003az}.
      In the ratios $\Gamma_{12}^{s,xy} / \tilde{M}_{12}^s$ - which are the building
      blocks of the parameters $a$, $b$ and $c$  - many quantities cancel,
      in particular the decay constant $f_{B_s}$, the mass of the $B_s$ meson and the Fermi
      constant.
      We get
     \begin{equation}
     \frac{\Gamma_{12}^{s,xy}}{\tilde{M}_{12}^s} = \frac{\pi m_b^2
        \left[ 8 G^{s,xy} + 5 G_S^{s,xy} \frac{B_S'}{B} +
        {\cal O} \left(\frac{1}{m_b}\right)\right]}{6 M_W S_0(x_t) \hat{\eta}_B}
        \; .
      \label{ratio_anatomy}
     \end{equation}
     Now the first term in Eq.~(\ref{ratio_anatomy}), proportional to $ G^{s,xy}$ is completely free of any non-perturbative
    contribution. It can be completely determined in perturbative QCD. Because of all these
    cancellations $a$, $b$ and $c$ are theoretically quite clean and they are also almost identical
    for $B_d$ and $B_s$ mesons, except
    for differences in the  primed bag factors and in the $1/m_b$ corrections.
    The way of writing $\Gamma_{12}^s/M_{12}^s$ in Eq.~(\ref{ratio1}) and Eq.~(\ref{ratio2}) can be viewed
    as a Taylor expansion in the small ratio of CKM parameters,  $\lambda_u/\lambda_t$, for which we get
    the following numerical values
    \begin{eqnarray}
    \frac{\lambda_u}{\lambda_t} & = & -8.0486  \cdot 10^{-3} + 1.81082 \cdot 10^{-2}I \; ,
    \label{lambdaut}
    \\
    \left( \frac{\lambda_u}{\lambda_t} \right)^2& = &
      -2.63126 \cdot 10^{-4} - 2.91491 \cdot 10^{-4} I \; .
    \end{eqnarray}
    Moreover a pronounced GIM (\cite{Glashow:1970gm})  cancellation is arising
    in the coefficients $a$ and $b$ in Eq.~(\ref{ratio2}). With the newest input parameters described
    in Appendix \ref{app:input}, we get for the numerical values of $a$, $b$ and $c$:
    \begin{eqnarray}
    c & =  & -48.0 \pm 8.3 \; \; (-49.5 \pm 8.5)
        \; ,
        \\
    a & = &  + 12.3 \pm 1.4\; \; (+11.7 \pm 1.3)
        \; ,
        \\
    b & = & + 0.79 \pm 0.12\; \; (+0.24 \pm 0.06)
        \; .
    \end{eqnarray}
    The numbers in brackets denote the corresponding values for the $B^0$ system.
    Putting all this together, we see that the real part of $\Gamma_{12}^s/M_{12}^s$ is absolutely
    dominated by the coefficient $c$, while for the imaginary party only $a$ and to a
    lesser extent $b$ are contributing.
    We get
    \begin{eqnarray}
    \Re \left( \frac{\Gamma_{12}^s}{M_{12}^s} \right) & = &10^{-4} \left(
     c +  a \Re \left[\frac{\lambda_u}{\lambda_t}                \right]
       + b \Re \left[\frac{\lambda_u^2}{\lambda_t^2} \right]
     \right)
     \nonumber
     \\
   \Rightarrow \frac{\Delta \Gamma_s}{\Delta M_s} & \approx & - 10^{-4} c \; ,
      \\
     \Im \left( \frac{\Gamma_{12}^s}{M_{12}^s} \right) & = & 10^{-4} \left(
         a \Im \left[\frac{\lambda_u}{\lambda_t}                \right]
       + b \Im \left[\frac{\lambda_u^2}{\lambda_t^2} \right]
       \right)
       \nonumber
      \\
        \Rightarrow  a_{\rm fs}^s &\approx&
      10^{-4}  a \Im \left[\frac{\lambda_u}{\lambda_t}                \right]
      \; .
      \end{eqnarray}
     So for a determination of only  $\Delta \Gamma_s$ (or also $\Delta \Gamma_d$)
     to a good approximation the first term of Eq.~(\ref{ratio1})
     - or equivalently the coefficient $c$ - is sufficient.
\item Unfortunately it turned out after the calculation of the NLO-QCD and the sub-leading
      $1/m_b$ corrections that $\Delta \Gamma_s$ is not very well-behaved (see
      \cite{Lenz:2004nx}): all corrections are quite large and they have the same sign.
      Surprisingly this problem could be solved to a large extent by using $Q$ and $\tilde Q_S$
      as the two independent operators instead of $Q$ and $Q_S$,
     which is just a change of the operator basis, see \cite{Lenz:2006hd}.
      As an illustration of the improvement
      we discuss the real part of the ratio $\Gamma_{12}^s/M_{12}^s$ and split up the
      terms according to Eq.~(\ref{ratio_anatomy}). We leave only the ratio of
      bag parameters as free parameters, while we else insert all Standard Model parameters
      according to the values given in the appendix. We get now for
      $\Delta \Gamma_s / \Delta M_s$ in the old (operators $Q$ and $Q_S$)
      and the new basis (operators $Q$ and $\tilde{Q}_S$):
      \begin{eqnarray}
      \frac{\Delta \Gamma_s}{\Delta M_s}^{\mbox{Old}} & = &
      10^{-4} \cdot
      \left[ 2.6  + 69.7 \frac{ B_S}{B}  - 24.3  \frac{B_R}{B}
      \right]  ,
      \\
      \frac{\Delta \Gamma_s}{\Delta M_s}^{\mbox{New}} & = &
      10^{-4} \cdot
      \left[ 44.8  + 16.4 \frac{ \tilde{B}_S}{B}  - 13.0  \frac{B_R}{B}
      \right]  ,
      \end{eqnarray}
      where $B_R$ is an abbreviation for all seven bag parameters of the dimension 7 operators.
      In the old basis the first term, which has no dependence on non-perturbative lattice parameters,
      is almost negligible. The second term, that depends on the ratio of the
      matrix elements of the operators  $Q_S$ and $Q$ is
      by far dominant and the third term, that describes $1/m_b$ corrections
      gives an important negative contribution.
      In the new basis the first term, being completely free of any non-perturbative uncertainties,
      is numerical dominant. The second term is sub-leading and the $1/m_b$ corrections became
      smaller and undesired cancellations therein are less pronounced. Thus the second
      formulation has a much weaker
      dependence on the badly known bag parameters, also on the dimension seven ones.
      If all bag parameters were known precisely, then such a change of basis has no
      effect, but since $B_R$ is unknown and the ratios $B_S'/B$ and $\tilde{B}_S'/B$ are
      much less known compared to the exact value one (stemming from $B/B)$,
      now a basis, where the coefficients of $B_R/B$ and $\tilde{B}_S'/B$ are small, gives
      results with a much better theoretical control.
      For more details we refer the reader to \cite{Lenz:2006hd}.
\end{itemize}
$1/m_b$ corrections for the sub-leading CKM structures
in $\Gamma_{12}^s$ \cite{Dighe:2001gc}
and $1/m_b^2$ corrections for $\Delta \Gamma_s$ \cite{Badin:2007bv}
were also determined; their numerical effect is small.
A commonly used Standard Model prediction for $\Delta \Gamma_s$ was given by \cite{Lenz:2011ti}
\begin{equation}
\label{DeltaG2011}
\Delta \Gamma_s^{\rm SM, 2011} = \left( 0.087 \pm 0.021 \right) \; \mbox{ps}^{-1} \; .
\end{equation}
We take the most recent numerical inputs from the following sources:
$G_F$, $M_W$, $M_{B_s}$ and $m_b$ from the PDG \cite{Agashe:2014kda},
the top quark mass from \cite{ATLAS:2014wva},
the non-perturbative parameters from FLAG (web-update of \cite{Aoki:2013ldr})
and $\tilde{B}_S/B$, $B_{R_0}$, $B_{R_1}$ and   $B_{\tilde{R}_1}$ from
\cite{Becirevic:2001xt},
\cite{Bouchard:2011xj},
\cite{Carrasco:2013zta}
and
\cite{Dowdall:2014qka}
and CKM elements from CKMfitter (web-update of \cite{Charles:2004jd})
 - similar values can be taken from UTfit \cite{Bona:2006ah}.
With these new values we  predict the decay rate difference of the neutral $B_s$ mesons to be
\begin{equation}
\Delta \Gamma_s^{\rm SM, 2015} = \left(  0.088 \pm 0.020 \right)\; \mbox{ps}^{-1} \; .
\label{DeltaG2015}
\end{equation}
The dominant uncertainty stems from the dimension 7 bag parameter
$B_{R_2}$ (about 15$\%$), closely followed by  $f_{B_s} \sqrt{B}$ (about 14 $\%$)
and the renormalisation scale dependence, which contributes about $8\%$ to the error budget.
A detailed listing of all the contributing uncertainties can be found in Appendix \ref{app:error}.
In order to reduce the theory uncertainty to a value between $5\%$ and $10 \%$, a non-perturbative
determination of $B_{R_2}$, a calculation of NNLO-QCD corrections (denoted by $\Gamma_3^{s,(2)}$
in Eq.~(\ref{HQE}) , a first step in this direction, has been done by \cite{Asatrian:2012tp}
and by $\Gamma_4^{s,(1)}$)
and more precise
values of  the matrix elements of the operators $Q$, $Q_S$ and $\tilde{Q}_S$ are mandatory.
All of this seems to be feasible in the next few years.
\\
In the discussion of the dimuon asymmetry in Section \ref{CPVmix}
we will also need several mixing quantities
from the $B^0$ sector. Their calculation within the Standard Model is analogous to the one
in the $\Bs$ sector. We present here numerical updates of the predictions given in
\cite{Lenz:2011ti}. The input parameters are identical to the ones in  the $\Bs$ system, except
$f_{B_d} \sqrt{B}$, $\tilde{B}_S/B$, $M_{\Bd}$ and $m_d$, which can found in the same literature
as the values for the $\Bs$ system.
Our  new predictions are
\begin{eqnarray}
\Delta M_d^{\rm SM, 2015} & = & ( 0.528 \pm  0.078 ) \;  \mbox{ps}^{-1} \; ,
\label{DMd_sm}
\\
\Delta \Gamma_d^{\rm SM, 2015} & = & ( 2.61 \pm  0.59 ) \cdot 10^{-3} \;  \mbox{ps}^{-1} \; ,
\label{DGd_sm}
\\
\left( \frac{\Delta \Gamma_d}{\Gamma_d} \right)^{\rm SM, 2015} & = & ( 3.97 \pm  0.90 )  \cdot 10^{-3} \; ,
\label{DGGd_sm}
\\
\Re \left( \frac{\Gamma_{12}^d}{M_{12}^d} \right)^{\rm SM, 2015} & = &
(-49.4 \pm 8.5 ) \cdot 10^{-4} \; .
\label{G12M12d_sm}
\end{eqnarray}
A detailed error analysis is given in Appendix \ref{app:error}.

\subsection{Experiment: Mass and decay rate difference $\Delta M_s$
                        and $\Delta \Gamma_s$}
\label{Bs_system_exp}


Experimental studies of $\Delta M_s$ and $\Delta \Gamma_s$
and their comparison with the theoretical predictions of
Eq.(\ref{DeltaM2015}) and Eq.(\ref{DeltaG2015})
constitute an important SM test. In addition, $\Delta M_s$ together with
the mass difference $\Delta M_d$ of the $\Bd$ meson can be used
to evaluate the ratio of the CKM parameters
$|V_{ts} / V_{td}|$.  These elements are not likely to be measurable
with high precision in tree-level decays involving a top-quark, because the top quark is too short-lived to
form a hadron (see
e.g. \cite{Agashe:2014kda}), but
the ratio between $\Delta M_d$ and $\Delta M_s$ provides a theoretically
clean and precise constraint. Using the results discussed below, and
un-quenched lattice calculations,
Ref.~\cite{Agashe:2014kda} quotes
\begin{equation}
\left|\frac{V_{td}}{V_{ts}}\right|=0.216\pm 0.001\pm 0.011,
\end{equation}
where the first error stems from experiment and the second from theory.
Therefore, the measurement of $\Delta M_s$, although
not directly related to CP violation, contributes significantly
to the test of the unitarity of the CKM matrix \cite{Amhis:2014hma}.

The measurement of $\Delta M_s$ and $\Delta \Gamma_s$ eluded experimentalists
for a very long time.
A relatively large value of $|V_{ts}|$ results in a high oscillation frequency of
$\Bs$ mesons and numerous
transitions from particle to anti-particle during its lifetime. Therefore,
a high precision of the proper decay length measurement is required to be sensitive
to $\Delta M_s$.
On the other side, the measurement of
$\Delta \Gamma_s$ is also challenging because
$\Delta \Gamma_s / \Gamma_s = \mathcal{O}(10\%)$.

The measurement of $\Delta M_s$ was attempted by many experiments during more than 20 years;
 the
CDF collaboration at Fermilab first succeeded to perform it with a statistical significance
exceeding five
standard deviations \cite{Abulencia:2006ze}.

From a technical point of view, the measurement of $\Delta M_s$ requires these essential
components:
\begin{itemize}
\item identification of the flavor of the $\Bs$ meson
 at the time of production;
\item identification of the flavor of the $\Bs$ meson when it decays;
\item measurement of its proper lifetime.
\end{itemize}

\begin{figure}[tbh]
  \includegraphics[width=0.45 \textwidth,angle=0]{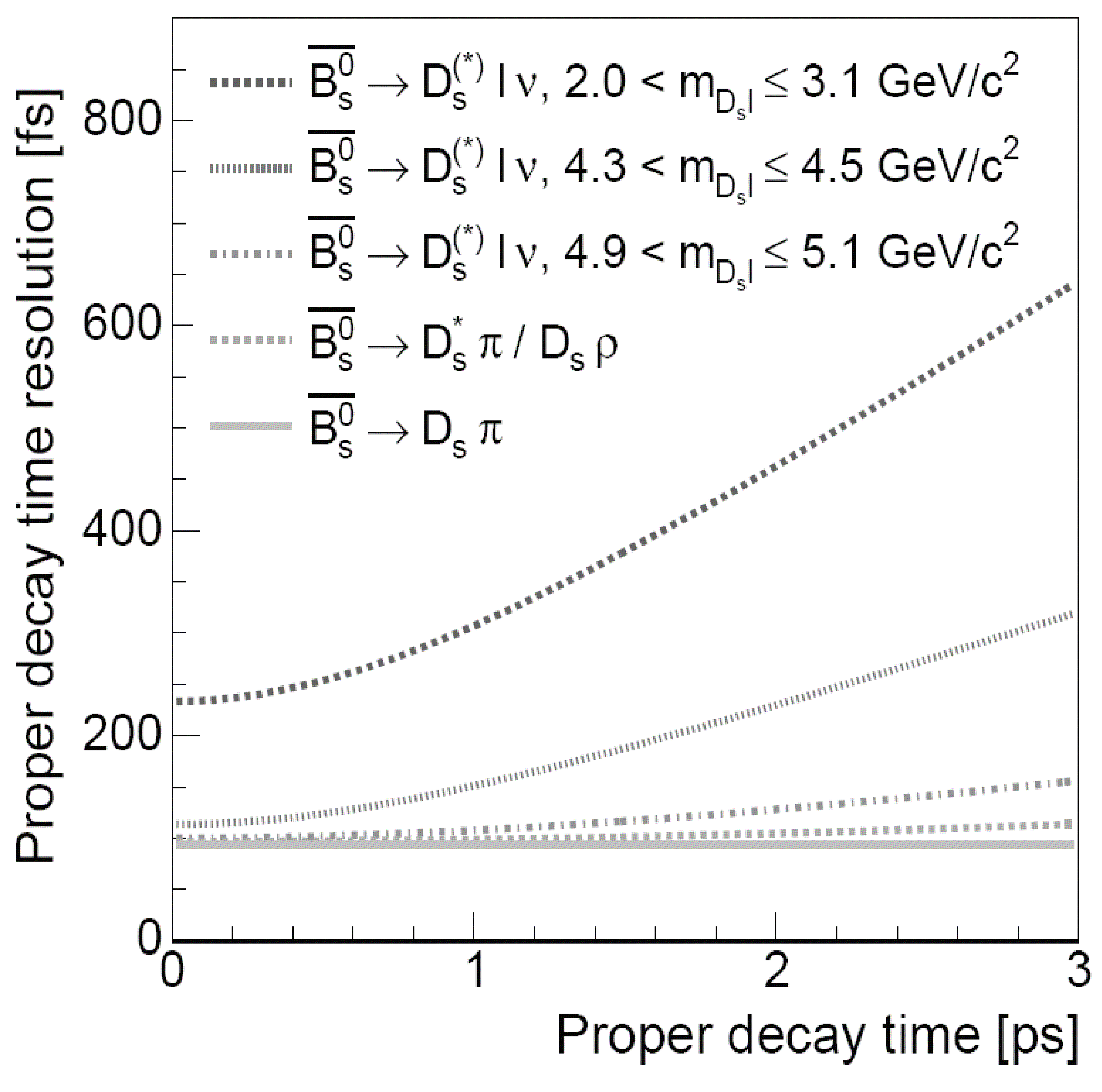}
  \caption{The proper decay time resolution measured by the CDF collaboration. The
  plot is taken
  from Ref. \cite{Abulencia:2006ze}. }
  \label{cdf-dms-resolution}
\end{figure}

To measure the final state of the $\Bs$ meson decay, a flavour-specific transition is used.
The simplest flavour-specific state
is the semileptonic decay $\Bs \to \Dsminus \mu^+ \nu_\mu$ since the muon usually
provides an excellent possibility
for an efficient selection of such decays during both the data taking and the
subsequent analysis. However,
the precision of the proper lifetime measurement in this decay mode is rather poor
because of the missing neutrino
taking some part of the $\Bs$ momentum. Figure \ref{cdf-dms-resolution} shows the
proper decay time
resolution for different decay modes as a function of the $\Bs$ proper decay time
in the CDF measurement.
The resolution in the
semileptonic decay channel deteriorates very quickly with the increase of the proper time.
Therefore, the ability of an experiment to reconstruct hadronic
$\Bs$ decays such as $\Bs \to \Dsminus \pi^+$ plays a crucial role in the $\Delta M_s$
measurement.

The identification of the $\Bs$ initial state, also known as the
{\it initial flavour tagging} (IFT),
was first developed and used at hadron colliders by the
CDF \cite{Abulencia:2006ze} and D0 \cite{Abazov:2006qp} experiments at the Tevatron.
In the LHCb implementation of the IFT
\cite{Aaij:2012mu,Aaij:2013oba,Aaij:2015pha}, the special capabilities
of the detector, such as the particle identification and efficient reconstruction
of secondary decays,
are extensively used.

Technically, the IFT is divided into opposite-side (OS) and same-side (SS) tagging.
At LHC, where the gluon splitting dominates the $b \bar b$ production and the $b$ quarks are
considerably boosted, the ``opposite-side'' is actually not ``opposite'' at all.
Therefore, the naming of the two IFT methods is nowadays
largely historical and does not reflect the actual topology of the $b \bar b$ events.
The OS tagging is based on the correlation of the flavours of two produced $B$ hadrons,
while the SS tagging exploits the correlation of the flavour of the $\Bs$ meson and the charge of additional
particles produced
in the hadronisation of the initial $b$ quark.
The performance of the IFT is quantified by the {\it tagging power} $P$, which is expressed
as $P = \epsilon (1- 2 w)^2$, where $\epsilon$ is the tagging efficiency and $w$ is
the wrong-tag
probability. The tagging power multiplied by the total number of events in the analysis
corresponds
to the effective statistics used to measure $\Delta M_s$.

The performance of the IFT in different experiments is presented in Table \ref{tab1}. It
includes the results of
the ATLAS \cite{Aad:2016tdj} and CMS \cite{Khachatryan:2015nza} collaborations, who
use the IFT
for the measurement of CP violation.  It can be seen that the tagging power never exceeds
few percents
meaning that a large statistics should be collected to obtain the significant
measurement of $\Delta M_s$. In general, the tagging power improves with a better understanding of the underlying event and with the
refinement of multi-variate tagging methods.

\begin{table}[htb]
  \begin{center}
    \begin{tabular}{l c l l}
      \hline
      Experiment & method & $P$ (\%) & Ref.  \\
      \hline
      CDF  & OS & $1.8 \pm 0.1$    & \cite{Abulencia:2006ze}\\
      CDF & SS & 3.7$\pm$0.9 & \cite{Abulencia:2006ze}    \\
      D0  & OS & $2.48 \pm 0.22 $  &  \cite{Abazov:2006qp} \\
      LHCb & OS & $2.55 \pm 0.14$   & \cite{Aaij:2013oba}\\
      LHCb & SS & $1.26 \pm 0.17$   & \cite{Aaij:2013oba}\\
      ATLAS & OS & $1.49 \pm 0.02$  &  \cite{Aad:2016tdj}\\
      CMS  & OS & $1.307 \pm 0.032$ &  \cite{Khachatryan:2015nza}\\
       \hline
    \end{tabular}

    \caption{Performance of the initial flavour tagging in different experiments.
             The numbers shown correspond to the same-side (SS) or the
             opposite-side (OS) tagging power ($P$).
             The uncertainty shown is the combination of the statistical and systematic
             uncertainties.
             In general, the same-side flavour tagging depends upon the mode being investigated.
             CDF finds a SS tagging power of $P=(4.8\pm 1.2) \%$ \cite{Abulencia:2006ze}
             in the semileptonic decay sample.
             }
    \label{tab1}
  \end{center}
\end{table}

\begin{figure}[tbh]
  \includegraphics[width=0.45 \textwidth,angle=0]{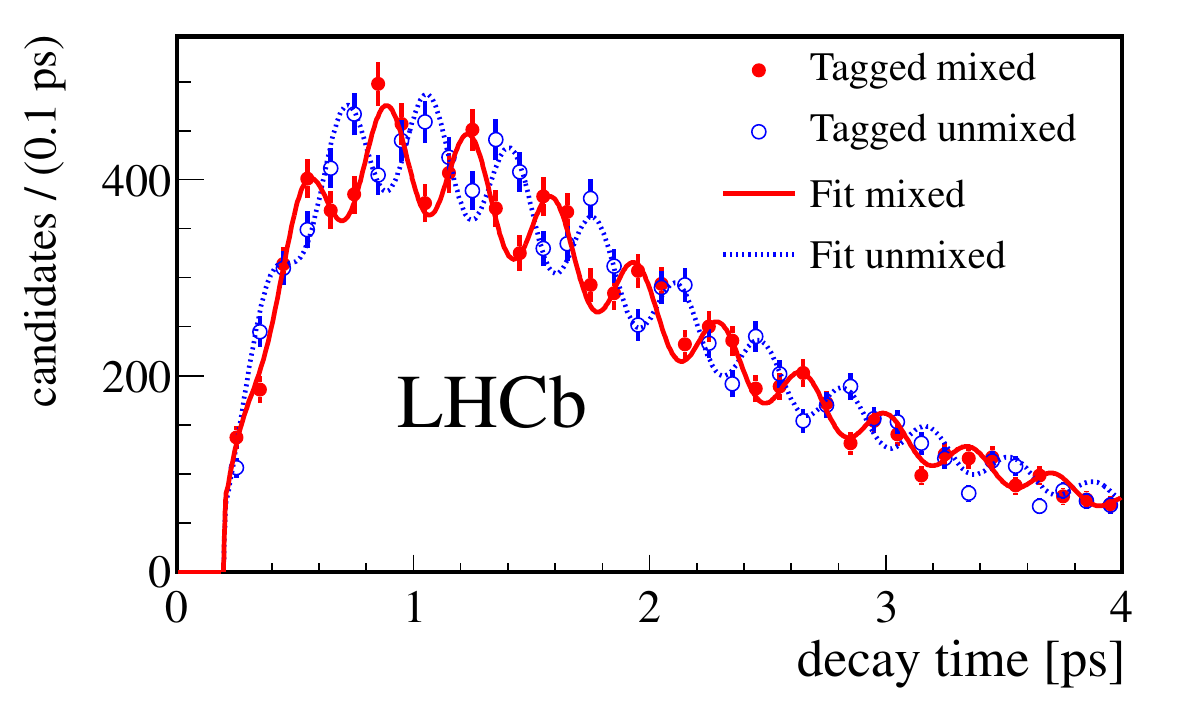}
  \caption{Proper decay time distribution for the selected $\Bs$ decays
candidates tagged as mixed (different flavour at decay and production; red,
continuous line) or unmixed (same flavour at decay and production; blue, dotted
line). The data and the fit projections are plotted in a signal window around the
reconstructed $\Bs$ mass of $5.32 - 5.55$ GeV/$c^2$. The plot is taken
  from Ref. \cite{Aaij:2013mpa}. }
  \label{lhcb-dms}
\end{figure}

The period of oscillation of the $\Bs$ meson corresponding to $\Delta M_s = 17.76$ ps$^{-1}$ is
$T = 2 \pi / \Delta M_s \simeq 350$ fs. To measure it reliably and thus extract $\Delta M_s$, the precision
of the proper lifetime measurement should be at least four times better. The precision of the
proper decay length measurement in the CDF experiment was about 100 fs, while for the LHCb experiment
it is about 44 fs. This excellent performance together with large statistics collected by the LHCb
experiment in  the LHC Run I results in a much better precision of the $\Delta M_s$ measurement.
They also succeeded to obtain a clear oscillation pattern in the proper decay length distribution, which
is shown in Fig. \ref{lhcb-dms}.

The first double sided bound at 90 \% C.L on the $\Delta M_s$ value was obtained by the
D0 collaboration \cite{Abazov:2006dm}.
Soon after that the CDF collaboration reported the actual measurement of this quantity \cite{Abulencia:2006ze}
\begin{equation}
\Delta M_s^{\rm{CDF}} = 17.77 \pm 0.10 \mbox{(stat)} \pm 0.07 \mbox{(syst) ps}^{-1}.
\end{equation}
Later, the LHCb collaboration performed the most precise single-experiment measurement of $\Delta M_s$ \cite{Aaij:2013mpa}
\begin{equation}
\Delta M_s^{\rm{LHCb}} = 17.768 \pm 0.023 \mbox{(stat)} \pm 0.006 \mbox{(syst) ps}^{-1}.
\end{equation}
The combination of all $\Delta M_s$ measurements by the HFAG \cite{Amhis:2014hma} gives
\begin{equation}
\Delta M_s^{\rm{HFAG \; 2015}} = 17.757 \pm 0.021~\mbox{ps}^{-1}.
\label{DeltaMExp}
\end{equation}

The currently most precise measurement of  $\Delta \Gamma_s$
consists in the simultaneous study of the proper decay length and angular distributions of the
decay $\Bs \to \jpsi K^+ K^-$ which mainly includes the $\Bs \to \jpsi \phi$ final state.
For simplicity, this study is denoted as $\Bs \to \jpsi \phi$ channel in the following
discussion, although it should be remembered that the addition of the non-resonant contribution
is required for an appropriate analysis of data.
Both the CP-even and CP-odd $\Bs$ states contribute in this decay mode
and therefore its properties are sensitive to both the $\Bs$ width difference and the phase
$\phi_s$ (defined in Section \ref{CPVinter-theory})
describing CP violation in the interference of decay and mixing.

All collider experiments at the Tevatron and LHC perform the measurement of $\Delta \Gamma_s$
in the $\Bs \to \jpsi \phi$ decay. The first results were obtained by the CDF \cite{Aaltonen:2012ie}
and D0 \cite{Abazov:2011ry} collaborations,
who largely developed the measurement technique.
The ATLAS \cite{Aad:2016tdj}, CMS \cite{Khachatryan:2015nza} and LHCb \cite{Aaij:2014zsa}
collaborations continue this study at LHC, where a significantly larger statistics is collected
and much more data are expected in the future.

As for $\Delta M_s$, the measurement of $\Delta \Gamma_s$ in $\Bs \to \jpsi \phi$ decay
requires  IFT and the proper decay length of the $\Bs$ meson. In addition, the study of the
angular distributions of the $\Bs$ decay products is needed. This is the reason  why this analysis
is sensitive to the quality of the data description by the simulation. All
experiments succeed in achieving an excellent understanding of their detectors.
%

The measurements
of $\Delta \Gamma_s$ using $\jpsi\phi (\Kp\Km)$  are summarised in Table \ref{tab2}. It also includes the world average
value obtained by the HFAG  \cite{Amhis:2014hma}, which is found to be
\begin{equation}
\label{dgs-jpsiphi}
\Delta \Gamma_s = 0.079 \pm 0.006 \; \mbox{ps}^{-1}~~(\Bs \to \jpsi \phi) \; .
\end{equation}

\begin{table*}[t]
  \begin{center}
    \begin{tabular}{lllll}
      \hline
      Exp. & \  $\Delta \Gamma_s$ (ps$^{-1}$) & $\Gamma _s$ (ps$^{-1}$)& Ref.\\
      \hline
CDF &  $0.068 \pm 0.026 \pm 0.009$   & $0.654 \pm 0.008 \pm 0.004$ & \cite{Aaltonen:2012ie}\\
D0   & $0.163^{+0.065}_{-0.064}$  & 0.693$^{+0.018}_{-0.017}$ & \cite{Abazov:2011ry} \\
ATLAS&   $0.083\pm 0.011 \pm 0.007$ &0.677$\pm0.003\pm 0.003$ & \cite{Aad:2016tdj} \\
CMS  &$0.095\pm 0.013\pm 0.007$ & 0.6704$\pm 0.0043\pm 0.0051$&\cite{Khachatryan:2015nza} \\
LHCb & $0.0805\pm 0.0091\pm 0.0033$  &0.6603$\pm 0.0027\pm 0.0015$ &\cite{Aaij:2014zsa}  \\\hline
HFAG 2015 & $0.079 \pm 0.006$ & 0.6649$\pm$0.0022 & \cite{Amhis:2014hma}\\
       \hline
    \end{tabular}
    \caption{Measurements of $\Delta \Gamma_s$ in $\Bs \to \jpsi \phi$ decay. The last
    line gives the world average value obtained by HFAG. }
    \label{tab2}
  \end{center}
\end{table*}


An alternative approach to determine $\Delta \Gamma_s$ relies upon the direct measurement of the effective lifetime
of $ \Bs$ decays
to pure $CP$ eigenstates. The extraction of $\Delta \Gamma_s$ with this method is discussed in detail in Ref.\cite{Fleischer:2011cw}.

To first order in $y_s\equiv \Delta \Gamma_s/(2\Gamma_s)$, we have \cite{Amhis:2014hma}

\begin{eqnarray}
\tau _{single}(\Bs\to CP-{\rm even})\approx \frac{1}{\Gamma _L}\left( 1+\frac{(\phi _s)^2 y_s}{2}\right),\\
\tau _{single}(\Bs\to CP-{\rm odd})\approx \frac{1}{\Gamma _H}\left( 1-\frac{(\phi _s)^2 y_s}{2}\right),
\end{eqnarray}
where $\tau _{single}$ is the effective lifetime of the $\Bs$ decaying to a specific $CP$-eigenstate state $f$.
This formula assumes that ${\cal A}^{\Delta \Gamma}_{CP-EVEN}=\cos{\phi_s}$ and ${\cal A}^{\Delta \Gamma}_{CP-ODD}=-\cos{\phi_s}$, where the  mixing angle $\phi_s$ will
be defined in Section \ref{CPVinter-theory}.
Thus, the decay width measured in the CP-even final state, such as $\Bs \to K^+ K^-$ and $\Bs \to D^+_s D^-_s$,
is approximately equal to $1/\Gamma_{L}(s)$. Similarly, the CP-odd decay modes $\Bs \to \jpsi K_s^0$
and $\Bs \to \jpsi f_0(980)$ provide measurements of $1/\Gamma_{H}(s)$, thus $\Delta \Gamma_s$
can be obtained as the difference of these two quantities. There are several subtleties that need to be taken into account when using
this method to measure $\Delta \Gamma_s$. For example, the decays $\Bs\to\Kp\Km$ and $\Bs\to\jpsi\Ks$ may suffer from $CP$ violation due
to interfering tree and loop amplitudes.  Thus Ref.~\cite{Amhis:2014hma} uses only the effective lifetimes obtained for $\Dsp\Dsm$ ($CP$-even),
and $\jpsi f_0$, $\jpsi \pi\pi$ ($CP$-odd) decays to obtain
\begin{eqnarray}
\tau_{single}(\Bs\to CP-{\rm even})& = &1.379\pm0.031\ {\rm ps}\\
\tau_{single}(\Bs\to CP-{\rm odd}) &=& 1.656\pm0.033\ {\rm ps}.
\end{eqnarray}
Table \ref{tab3}
summarises the current values as well as the average values of $1/\Gamma_{L}^s$ and $1/\Gamma_{H}^s$ reported
in Ref.~\cite{Amhis:2014hma}. Note that the effective lifetimes measured in $\Bs\to\Kp\Km$ abd $\Bs\to\jpsi\Ks$ have not been used in these averages because of the
difficulty in quantifying the penguin contribution in these modes. These effective lifetimes correspond to
\begin{equation}
\label{dg-direct}
\Delta \Gamma _s=0.121\pm 0.020.
\end{equation}
This value is higher by two standard deviations than the one shown in Eq.~(\ref{dgs-jpsiphi}).
However, this difference should be considered with caution.
The value in Eq.~(\ref{dg-direct}) is
obtained with theoretical assumptions and external input on weak phases and hadronic parameters.

\begin{figure}[h]
  \includegraphics[width=0.45 \textwidth,angle=0]{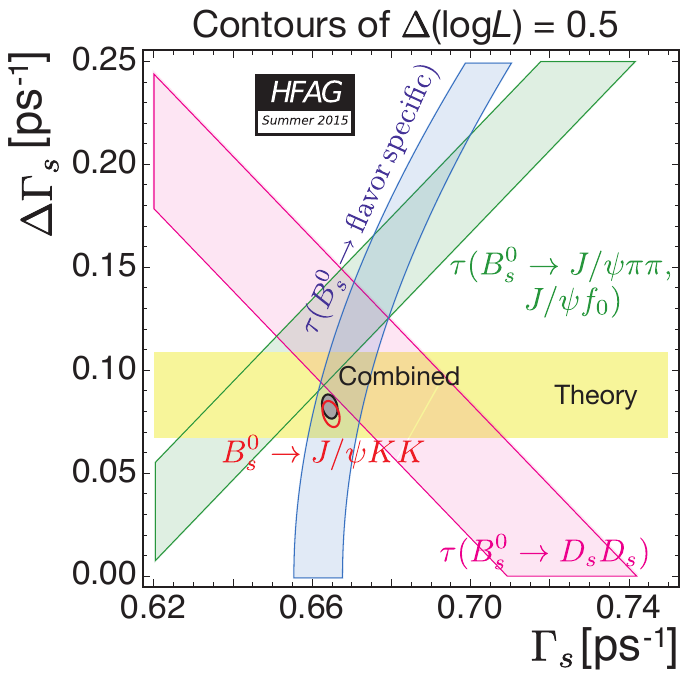}
  \caption{The average of all the $\Bs \to \jpsi \phi$ and $\Bs \to \jpsi K^+ K^-$
  results is shown as the red contour, and
  the constraints given by the effective lifetime measurements of $\Bs$
  to flavour-specific (see Eq.(\ref{taueff})), pure CP-odd
  and pure CP-even final states are shown as the blue, green and purple bands, respectively.
  The average taking all constraints into account is shown as the grey-filled contour.
  The yellow band is the theory prediction given in Eq.(\ref{DeltaG2011})
  that assumes no new physics in $\Bs$ mixing. The plot is taken from \cite{Amhis:2014hma}.
  }
  \label{hfag-dgs}
\end{figure}

Using these data in conjunction with the $\jpsi\phi (\Kp\Km)$
determinations of $\Delta \Gamma _s$, the current experimental average is \cite{Amhis:2014hma}
\begin{eqnarray}
\Delta \Gamma_s^{\rm HFAG \; 2015} & = & 0.083 \pm 0.006 \; \mbox{ps}^{-1} \; .
\label{DeltaGExp}
\end{eqnarray}
The comparison of different lifetime measurements of CP eigenstates, which can be used to extract  $\Delta \Gamma_s$ is presented in Fig. \ref{hfag-dgs}.

\begin{table*}[t]
  \begin{center}
    \begin{tabular}{l l l l}
      \hline
      Quantity & Source & Channel & result (ps) \\
      \hline
      \multirow{3}{*}{$1/\Gamma_{L}^s$}  & LHCb\cite{Aaij:2014fia} & $\Bs \to K^+ K^-$     & $1.407 \pm 0.016 \pm 0.007$     \\
                                        & LHCb\cite{Aaij:2013bvd} & $\Bs \to D^+_s D^-_s$ & $1.379 \pm 0.026 \pm 0.017$     \\
      \hline
      \multirow{4}{*}{$1/\Gamma_{H}^s$}  & CDF \cite{Aaltonen:2011nk}  & $\Bs \to \jpsi f_0(980)$ & $1.70^{+0.12}_{-0.11} \pm 0.03$     \\
                                        & LHCb \cite{Aaij:2012nta} & $\Bs \to \jpsi f_0(980)$ & $1.700 \pm 0.040 \pm 0.026$     \\
                                        & LHCb\cite{Aaij:2013eia} & $\Bs \to \jpsi  K_s^0$    & $1.75 \pm 0.12 \pm 0.07$     \\
                                        & LHCb\cite{Aaij:2013oba}   &$\Bs \to \jpsi \pip\pim$ &1.652$\pm$0.024$\pm$0.024 \\
      \hline
    \end{tabular}
    \caption{The $\Bs$ width difference can be extracted from lifetime measurements in different channels with
             a definite CP quantum number.}
    \label{tab3}
  \end{center}
\end{table*}

%

At the end of this section we would like to compare the experimental and theoretical numbers
for the mass difference and the decay rate difference. For the experimental value of the mass difference
we take the value from  Eq.~(\ref{DeltaMExp}) and for the value of the decay width difference we take
Eq.~(\ref{DeltaGExp})
For the theory value, we take the more precise prediction of the ratio $\Delta \Gamma_s/\Delta M_s$.
We find a very good agreement for experiment
and theory
      \begin{eqnarray}
       \frac{ \left( \Delta \Gamma_s / \Delta M_s \right)^{\rm Exp} }
            { \left( \Delta \Gamma_s / \Delta M_s \right)^{\rm SM} }
      & = &
      \frac{0.00467 (1 \pm 0.072 )}
           {0.00481 (1 \pm 0.173)}
      \\
      & = & 0.97  \pm 0.07 \pm 0.17 \; .
      \end{eqnarray}
In the last line the first error is the experimental and the second the theoretical.
This results proves that the heavy quark expansion is working in the $B$-sector with a precision of at
least $20 \%$, also for the decay channel
$b \to c \bar{c} s$, which seems to be most sensitive to violations of quark hadron duality.
Assuming that there are no new physics effects in $\Delta M_s$ and taking into account
that the ratio $\Delta \Gamma_s / \Delta M_s$ is theoretically cleaner than $\Delta \Gamma_s$ alone,
we get an improved prediction for $\Delta\Gamma_s$
\begin{equation}
\Delta \Gamma_s^{\rm SM, 2015b} = \left( \frac{\Delta \Gamma_s}{\Delta M_s} \right)^{\rm SM} \! \! \! \!
       \cdot \,
     \Delta M_s^{\rm Exp} = 0.085 \pm 0.015 \; \mbox{ps}^{-1} .
     \end{equation}
This is the most precise theory value for $\Delta \Gamma_s$ that can currently be obtained. In future this
theory uncertainty might be improved by a factor of up to three, as explained in 
Section~\ref{Bs_system} .

\section{CP violation in mixing}
\label{CPVmix}
\subsection{Theory:  HQE}
CP violation in mixing is described by the weak  mixing phase $\phi_{12}^s$
defined in Eq.(\ref{phi12def}). It can be measured directly via CP asymmetries
of so-called flavour specific decays.
A flavour specific decay $\Bs \to f$ is defined by the following properties:
\begin{itemize}
\item The decays { $\barBs \to f$} and {$\Bs \to \bar{f}$} are forbidden. This reads
      in our notation 
      \begin{equation}
      \bar{\cal A}_f = 0 = {\cal A}_{\bar{f}} \;
      \end{equation}
      and thus
      \begin{equation}
      \lambda_f = 0 = \frac{1}{\lambda}_{\bar{f}} \; .
      \end{equation}
      Hence the time evolution of these decays is quite simple, compared to the 
      general case.
\item No direct CP violation arises in the decay, i.e.
      { $|\langle f       | {\cal H}_{eff} | \Bs \rangle| =
         | \langle\bar{f} | {\cal H}_{eff} | \barBs\rangle|  $},
      which again reads in our notation
      \begin{equation}
      |{\cal A}_f |= |\bar{\cal A}_{\bar{f}}| \; .
      \end{equation}
\end{itemize}
Examples for such decays are e.g.
$\Bs \to D_s^- \pi^+$ or  $\Bs \to X l \nu$ - therefore the corresponding 
asymmetries in semileptonic decays are also called
{\it semileptonic CP asymmetries}.
The CP asymmetry for flavour specific decays is defined as
\begin{eqnarray}
 a_{\rm fs}^s & = & 
\frac{\Gamma\left(\barBs(t) \to f \right) - \Gamma\left(\Bs(t) \to \bar{f} \right)}
     {\Gamma\left(\barBs(t) \to f \right) + \Gamma\left(\Bs(t) \to \bar{f} \right)}  
\equiv a_{\rm sl}^s \; .
\end{eqnarray}
Inserting the time evolution of the $\Bs$ mesons - given in 
Eq.(\ref{GammabarBf}) and  
Eq.(\ref{GammaBbarf}) - the flavour specific CP asymmetry
$a_{\rm fs}^s$ can be further simplified\footnote{This result was already used in 
Eq.(\ref{abbrev}).} as
\begin{eqnarray}
a_{\rm fs}^s
&= &
 - 2 \left( \left| \frac{q}{p}\right| - 1 \right)
\nonumber \\
& = &  \Im \left(\frac{\Gamma_{12}^s}{M_{12}^s} \right) =
\left| \frac{\Gamma_{12}^s}{M_{12}^s} \right| \sin \phi^s_{12}  \; .
\end{eqnarray}
For the SM prediction of the flavour specific asymmetries 
we can now simply use our determination of the ratio of the matrix elements 
$M_{12}^s$ and $\Gamma_{12}^s$ 
from the previous section, in particular
we need only the coefficient $a$ ($b$ gives only a small correction) 
defined in Eq.(\ref{ratio2}) to get:
\begin{eqnarray} 
a_{fs}^s & \approx & 
\Im \left(\frac{\lambda_u}{\lambda_t} \right) \cdot a \cdot 10^{-4} \; .
\end{eqnarray} 
The coefficient $a$ was given given by the difference of the internal charm-charm loop
and the internal up-charm loop.
Using the exact expression for $\Im \left( \Gamma_{12}^s / M_{12}^s \right)$ 
the Standard Model prediction of $a_{\rm fs}^s$ was given by \cite{Lenz:2011ti}
\begin{equation}
a_{\rm fs}^{s, \rm SM, 2011} = (1.9 \pm 0.3) \cdot 10^{-5} \; .
\end{equation}
With the most recent numerical inputs ($G_F$, $M_W$, $M_{B_s}$ and $m_b$ from the PDG
\cite{Agashe:2014kda},
the top quark mass from \cite{ATLAS:2014wva},
the non-perturbative parameters from FLAG (web-update of \cite{Aoki:2013ldr} in Summer 2015)
and $\tilde{B}_S/B$, $B_{R_0}$, $B_{R_1}$ and   $B_{\tilde{R}_1}$ from
\cite{Becirevic:2001xt},
\cite{Bouchard:2011xj},
\cite{Carrasco:2013zta}
and
\cite{Dowdall:2014qka}
and CKM elements from CKMfitter (web-update of \cite{Charles:2004jd} in Summer 2015)
( similar values can be taken from UTfit \cite{Bona:2006ah} )
we predict the flavour specific CP asymmetries of the neutral $\Bs$ mesons to be
\begin{equation}
\label{asls_sm}
a_{\rm fs}^{s,\rm SM, 2015} = (2.22  \pm 0.27) \cdot 10^{-5} \; .
\end{equation}
The dominant uncertainty stems from the renormalisation scale dependence, with 
$9\%$, followed by the CKM dependence with $5\%$ and the charm quark mass dependence
with  $ 4\%$. A detailed discussion of the uncertainties is given in Appendix \ref{app:error}.
Because of this small value and the proven validity of the HQE, 
the flavour specific asymmetries represent a nice null test, as
any sizable experimental deviation from the prediction in Eq.(\ref{asls_sm}) is a clear 
indication for new physics, see \cite{Jubb:2016mvq} for a more detailed discussion of this point.
\\
In addition we obtain the SM  prediction for the mixing phase
$\phi_{12}^s$:
\begin{eqnarray}
\phi_{12}^{s, {\rm SM}, 2015} & = & (4.6 \pm 1.2) \cdot 10^{-3} \; \mbox{rad}
\\
& = & 0.26^\circ \pm 0.07^\circ \; .
\end{eqnarray}
In the discussion of the dimuon asymmetry 
below we  also need the semileptonic CP asymmetry from the $B^0$ sector. 
Its calculation within the SM is analogous to the one of $a_{\rm sl}^s$. 
We update the predictions given in
\cite{Lenz:2011ti}, by using the same input parameters as for the $\Bs$-system,
except using $M_{B^0}$, $m_d$ and $\tilde{B}_S/B$.
We get as  new Standard Model values
\begin{eqnarray}
a_{\rm fs}^{d,\rm SM, 2015} & = & (-4.7  \pm 0.6) \cdot 10^{-4} \; ,
\label{asld_sm}
\\
\phi_{12}^{d, {\rm SM}, 2015} & = & (-0.096 \pm 0.025) \; \mbox{rad}
\label{phi12d_sm}
\nonumber
\\
                              & = & -5.5^\circ \pm 1.4^\circ \; .
\end{eqnarray}
A more detailed analysis of the uncertainties can be found in Appendix \ref{app:error}.
\begin{figure*}[t]
\includegraphics[width =0.85  \textwidth]{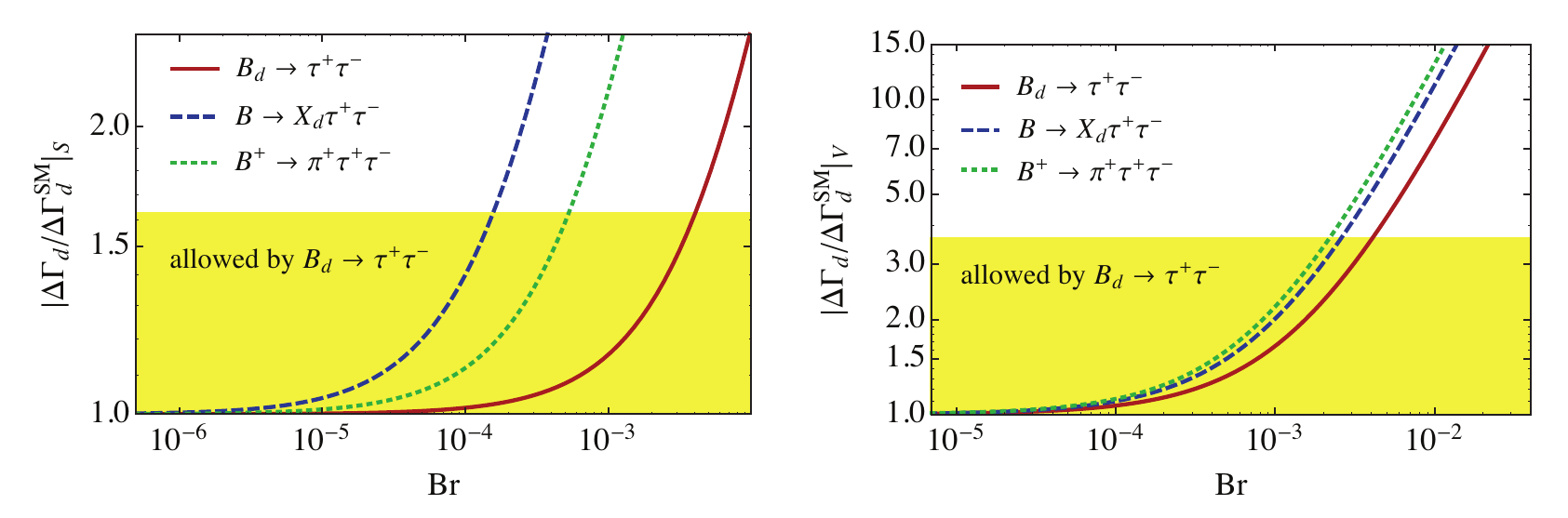}
\caption{\label{bdtautau}
The possible enhancement factor of $\Delta \Gamma_d$ by new scalar (l.h.s.) or vector
(r.h.s) $bd \tau \tau$ operators are indicated by the yellow region. In the case of a 
scalar operator $\Delta \Gamma_d$ can still be enhanced to about 1.6 times of the SM 
values. In the case of a 
vector operator $\Delta \Gamma_d$ can even be enhanced to about 3.5 times of the SM 
values. More precise bounds on $\Bd \to \tau^+ \tau^-$,  $\Bd \to X_d\tau^+ \tau^-$ and $B^+ \to \pi^+\tau^+ \tau^-$ could further shrink the allowed enhancement factor.
The relation between the bounds  $\Bd \to \tau^+ \tau^-$,  $\Bd \to X_d\tau^+ \tau^-$ and $B^+ \to \pi^+\tau^+ \tau^-$ and the possible enhancement factor of $\Delta
\Gamma_d$ is given by the red, blue and green line.}
\end{figure*}
Measurements of the  dimuon asymmetry triggered a lot of interest in $\Bd$ 
and $\Bs$ mixing, because early measurements seemed to indicate large new physics effects
\cite{Abazov:2013uma,Abazov:2011yk,Abazov:2010hj,Abazov:2010hv}.
Originally, the dimuon asymmetry $A_{CP}$
was considered to be given by a linear combination of the semileptonic
CP asymmetries in the $\Bd$ and the $\Bs$ system (see e.g.
\cite{Abazov:2011yk,Abazov:2010hj,Abazov:2010hv})
\begin{equation}
A_{CP} = C_d a_{\rm sl}^d + C_s a_{\rm sl}^s \; ,
\end{equation}
with $C_d$ and $C_s$ being roughly equal.
The large deviation of the measured value of $A_{CP}$ from the calculated
values of the linear combination of  $a_{\rm sl}^d$ and $a_{\rm sl}^s$
seemed to be a hint for large new physics effects in the semileptonic
CP asymmetries.
In 2013 Borissov and Hoeneisen \cite{Borissov:2013wwa} 
found that there is actually also an additional contribution from indirect CP violation.
This led to the following new interpretation (also used in 
\cite{Abazov:2013uma})
\begin{equation}
A_{CP} = C_d a_{\rm sl}^d + C_s a_{\rm sl}^s + C_{\Delta \Gamma_d} 
\frac{\Delta \Gamma_d}{\Gamma_d}\; .
\end{equation}
Because of the small value of $\Delta \Gamma_d$ in the SM, see 
Eq.(\ref{DGd_sm}) and Eq.(\ref{DGGd_sm})
the additional term did not solve the discrepancy.
It was  pointed out \cite{Nierste:2014ckm}, that the relation should be 
further modified to
\begin{equation}
A_{CP} = C_d a_{\rm sl}^d + C_s a_{\rm sl}^s + \alpha C_{\Delta \Gamma_d} 
\frac{\Delta \Gamma_d}{\Gamma_d}\; ,
\end{equation}
where $\alpha \leq 1/2$.
An interesting feature of this new interpretation is that a large enhancement
of $\Delta \Gamma_d$ by new physics effects could explain the experimental value of 
the dimuon asymmetry, while huge enhancements of the semileptonic CP asymmetries 
are disfavoured by direct measurements, see next section.
The investigation of \cite{Bobeth:2014rda} has further shown that 
enhancements of $\Delta \Gamma_d$ by several hundred per cent are not excluded
by any other experimental constraint - 
such an enhancement could bring the dimuon asymmetry in agreement with experiment. 
One possible enhancement mechanism would be e.g. new $bd\tau\tau$ transitions. 
Since two tau leptons are lighter than a $\Bd$ meson such a new operator could 
contribute to $\Gamma_{12}^d$.
This possibility can be tested by investigating $bd\tau\tau$ transitions directly.
In Fig. \ref{bdtautau} we show the possible enhancement of $\Delta \Gamma_d$
due to new scalar (l.h.s.)
and due to new vector (r.h.s.) $bd\tau\tau$ operators. Currently enhancements
within the yellow regions are allowed. 
In the case of vector operators $\Delta \Gamma_d$ can be enhanced to about 3.5 the SM
value of $\Delta \Gamma_d$.
The connection between a direct
measurement of or a bound on $\Bd \to \tau^+ \tau^-$ is given by the red line. From  
Fig. \ref{bdtautau} one can read off that a bound on $\Bd \to \tau^+ \tau^-$
of the order of $10^{-3}$ would limit the enhancement of $\Delta \Gamma_d$ to 
about $15 \%$ of the SM value in the case of scalar new physics operators
and to
about $50 \%$ of the SM value in the case of scalar new physics operators.
Similar relations between a possible enhancement of $\Delta \Gamma_d$
and a direct search for $\Bd \to X_d\tau^+ \tau^-$ and $B^+ \to \pi^+\tau^+ \tau^-$
are indicated by the blue line and the green line.
\\
Another enhancement mechanism would be new physics effects in tree-level decays,
which are typically neglected. Such studies were performed systematically in
\cite{Bobeth:2014rda,Bobeth:2014rra,Brod:2014bfa} and could also lead to sizable 
enhancements of $\Delta \Gamma_d$. Here a more precise measurement of 
$\Delta \Gamma_d$ would of course be very helpful.

\subsection{Experiment: Semi-leptonic asymmetries $a_{sl}^s$ and $a_{sl}^d$,
                    the di-muon asymmetry}
The measurement of the flavour-specific charge asymmetry is conceptually simple. Essentially, it is given by the
 asymmetry between  flavour-specific decays $\Bs \to f$ and $\barBs \to \bar f$.
As the expected value of the asymmetry is tiny, great care needs to be taken to assess any potential source
of asymmetry, for example, production dynamics, background sources, or detection asymmetry.
The final state typically used for this measurement is
 the semi-leptonic decay $\Bs \to D_s^{(*)-} \mu^+ \nu$ where the notation $(*)$
denotes the production of either $\Dsminus$, $D_s^{*-}$, or $D_{sJ}$ states.
The published results consider only the decay $D_s \to \phi \pi$ with $\phi \to K^+ K^-$. 
The initial flavor of the $\Bs$ meson is not determined
and the measured quantity is
\begin{equation}
A_{\rm meas} = \frac{N(\Dsminus \mu^+) - N(\Dsplus \mu^-)} {N(\Dsminus \mu^+) +  N(\Dsplus \mu^-)},
\end{equation}
where  $N(f)$ ($f = \Dsminus \mu^+$ or $\Dsplus \mu^-$) is the number of reconstructed
events in the final state $f$.
It can be expressed as
\begin{eqnarray}
N(f) \propto \int_{0}^{+\infty} [ & & \sigma(\Bs)\Gamma(\Bs(t) \to f) + \nonumber \\
& &    \sigma(\barBs) \Gamma(\barBs(t) \to f) ] \epsilon(f, t) dt.
\end{eqnarray}
This expression takes into account the absence of the
initial flavour tagging, the possible difference in the production cross-sections
$\sigma(\Bs)$ and $\sigma(\barBs)$,
and time dependent reconstruction efficiency $\epsilon(f,t)$ of the final state $f$.
The most important  instrumental charge-asymmetries  are 
related to differences between $\mu^+$ -$\mu^-$, and $\pip$-$\pim$ detection efficiencies.
The two opposite-charge kaons from $\phi$ decay 
have almost the same momentum spectrum, and thus charge-dependent detection effects do not influence the
measured asymmetry.

Using the expressions of the time evolution of $\Bs$ mesons, assuming that the ratio
of the reconstruction efficiencies $r_\epsilon \equiv \epsilon(\Dsminus \mu^+, t) / \epsilon(\Dsplus \mu^-, t)$
does not depend on time, and neglecting the second order terms, the semi-leptonic charge asymmetry $\asls$ is
related to $A_{\rm meas}$ as
\begin{eqnarray}
\label{ameas}
A_{\rm meas} & = & \frac{\asls}{2} - \frac{1-r_\epsilon}{2}+ \left ( a_{\rm P} - \frac{\asls}{2} \right) I \\
I  & \equiv & \frac{\int_{0}^{+\infty} e^{-\Gamma_s t} \cos (\Delta M_s t) \epsilon(t) dt }
                             {\int_{0}^{+\infty} e^{-\Gamma_s t} \cosh (\Delta \Gamma_s t /2) \epsilon(t) dt}. \nonumber
\end{eqnarray}
Here $a_{\rm P}$ is the production asymmetry of the $\Bs$ meson defined as
\begin{equation}
a_{\rm P} = \frac{\sigma(\Bs) - \sigma(\barBs)}{\sigma(\Bs) + \sigma(\barBs)}.
\end{equation}

The asymmetry $a_{\rm P}$ is zero at a $p \bar p$ collider, while it does not exceed a few percent for the
$\Bs$ production at LHC  (see \cite{Norrbin:2000jy, Aaij:2013iua, Aaij:2012jw}).
Because of the large value of $\Delta M_s$, the value of $I$ is about 0.2\%. As a result, the
value of the third term in Eq.~(\ref{ameas}) is of the order of $10^{-4}$ and can be safely neglected.
Thus, the main experimental task in the measurement of the $\asls$ is the determination of $r_\epsilon$.

\begin{figure}[tbh]
  \includegraphics[width=0.45 \textwidth,angle=0]{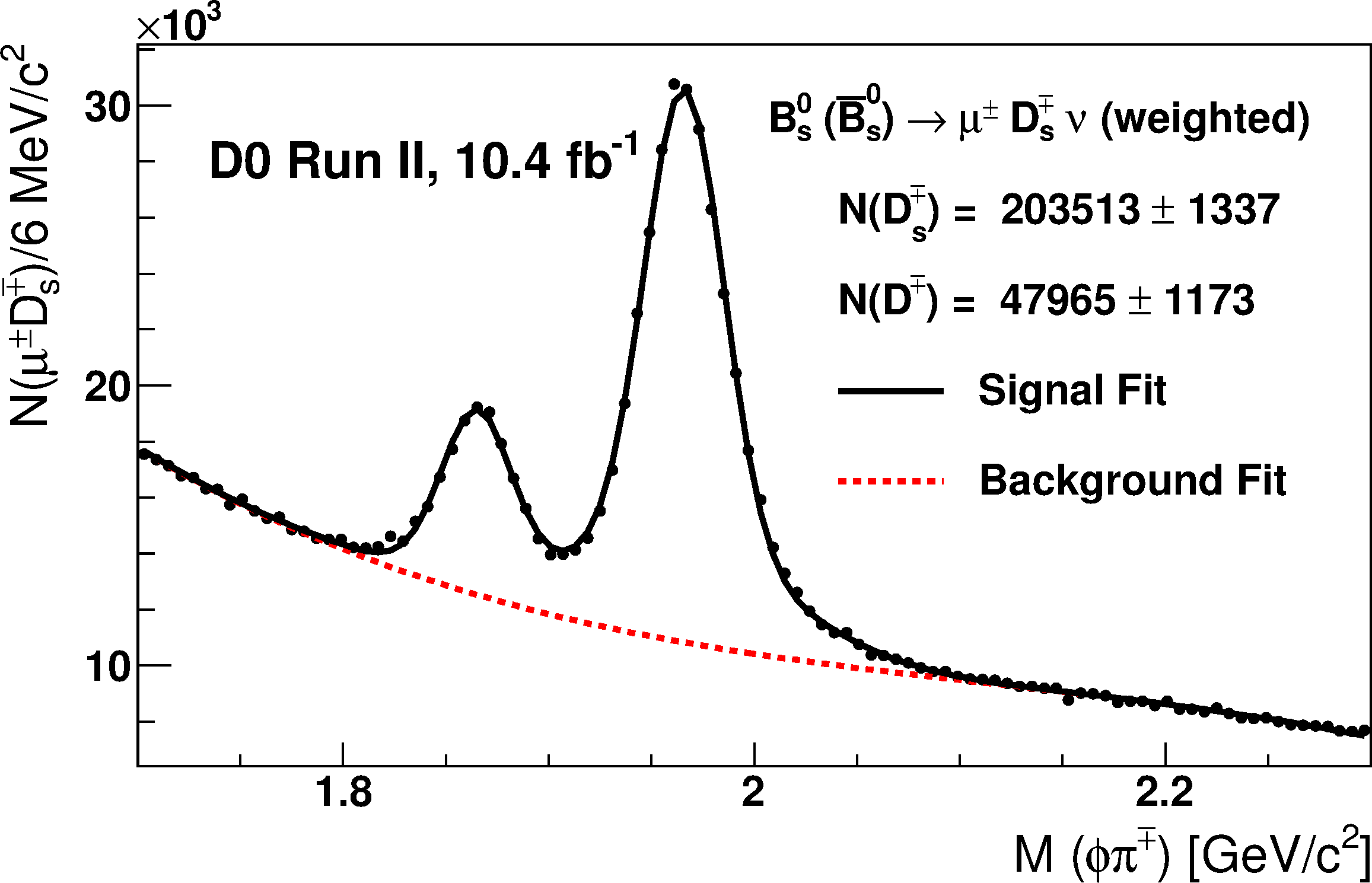}
  \caption{
  The weighted $K^+ K^- \pi^\mp$ invariant mass distribution for the $\mu \phi \pi^\mp$
  sample. The solid line represents the result of the fit and the dashed line shows
  the background parametrisation. The lower mass peak is due to the decay $D^\mp \to \phi \pi^\mp$
  and the second peak is due to the $D_s^\mp$ meson decay. Note the suppressed zero on the vertical axis. The plot is taken from \cite{Abazov:2012zz}.
  }
  \label{d0-dsmu}
\end{figure}

\begin{figure}[tbh]
  \includegraphics[width=0.45 \textwidth,angle=0]{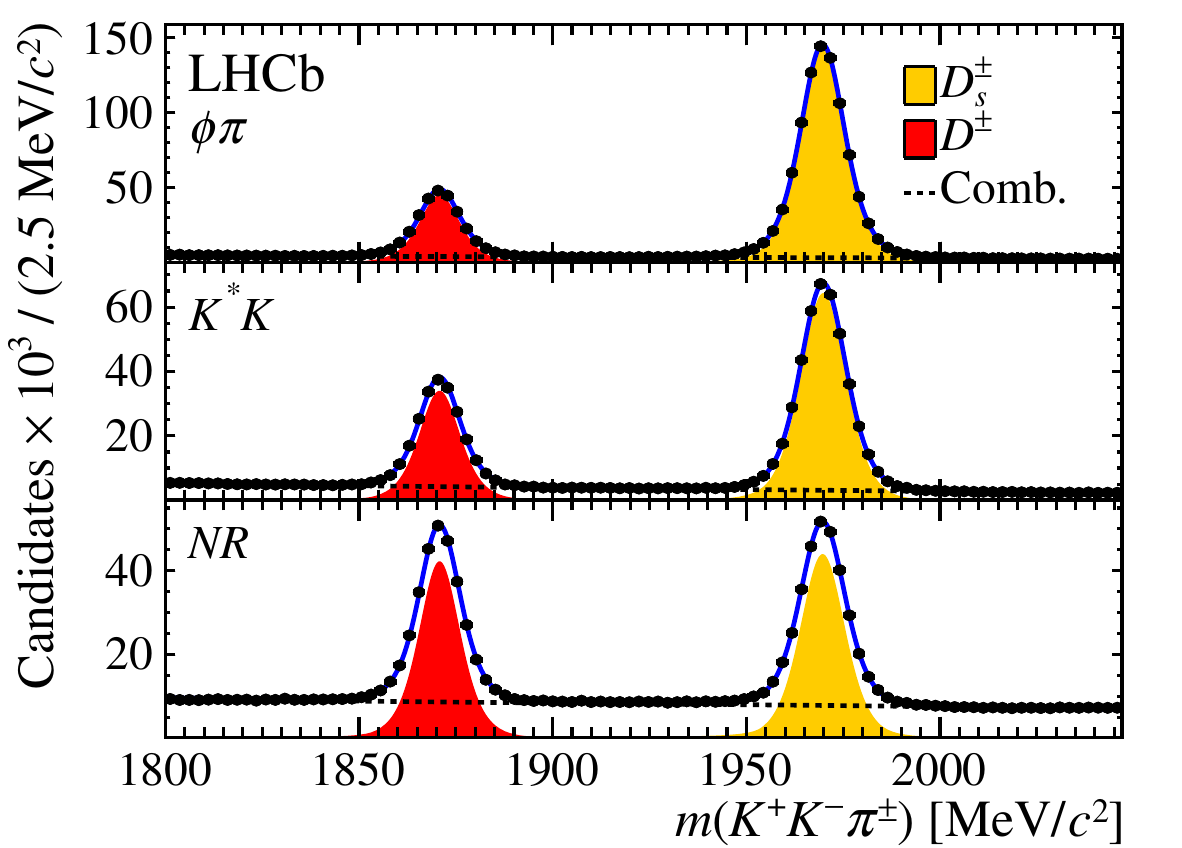}
  \caption{
  Invariant mass distributions of $K^+ K^- \pi^+$ and (b) $K^+ K^- \pi^-$ in the three Dalitz regions studied in the LHCb analysis, summed over both magnet polarities and data taking periods. Overlaid are the results of the fits, with signal and combinatorial background components as indicated in the legend. The plot is taken from \cite{Aaij:2016yze} .
 }
  \label{lhcb-dsmu}
\end{figure}

Measurements  of the asymmetry $\asls$ have been reported by the
D0 \cite{Abazov:2012zz} and
LHCb \cite{Aaij:2013gta} collaborations.  Both D0 and LHCb  collected large
statistics using semi-leptonic $\Bs$ decays. The number
of reconstructed signal events in the D0 measurement is $215763 \pm 1467$. The
corresponding $\mu^\pm  \phi \pi^\mp$ mass distribution is shown in Fig. \ref{d0-dsmu}. Recently, LHCb updated the measurement of $\asls$, using three fb$^{-1}$  and including all
the possible $\Ds$ decays to the $\Kp\Km\pip$ final state.
The corresponding
mass distribution is shown in Fig. \ref{lhcb-dsmu}.

\begin{figure}[tbh]
  \includegraphics[width=0.45 \textwidth,angle=0]{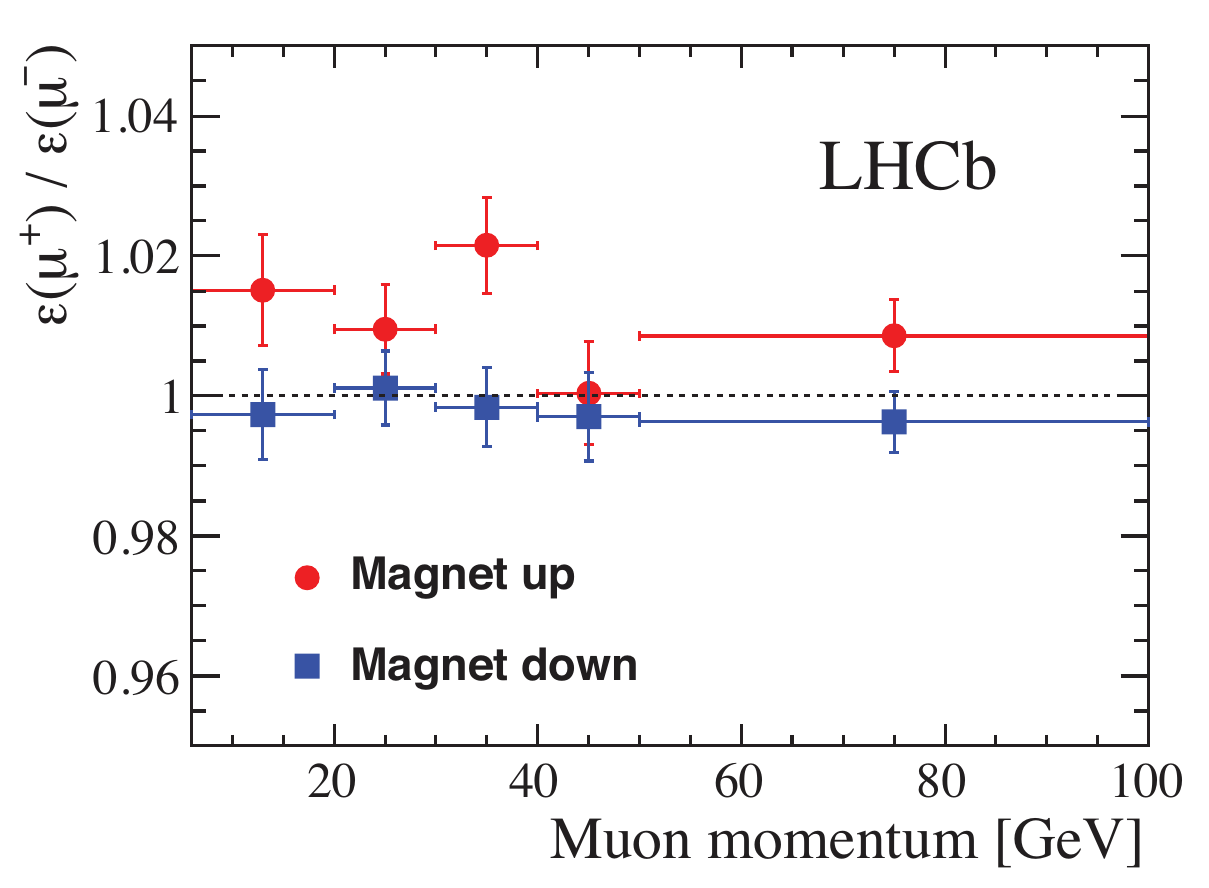}
  \caption{Relative muon efficiency as a function of the muon momentum.
  The plot is taken from \cite{Aaij:2013gta} .
  }
  \label{lhcb-reps}
\end{figure}

The important feature of both experiments is a regular
reversal of the magnet polarities. In the D0 experiment, the polarities of the toroidal and solenoidal magnetic
fields \cite{Abazov:2005pn} were reversed on average every two weeks so that the four solenoid-toroid polarity combinations are exposed to
approximately the same integrated luminosity. D0 reported only results averaged over all the magnet polarities.
The 1 fb$^{-1}$ LHCb sample comprises approximately 40\% of data taken with the magnetic field up, oriented along
the positive $y$-axis in the LHCb coordinate system,  and the rest with the opposite down polarity.  The 2 fb$^{-1}$ sample comprises equal
amounts of data with the two magnet polarities. LHCb analyses
data with magnetic field up and down separately, to allow  a quantitative assessment of charge-dependent
asymmetries. Figure \ref{lhcb-reps}
shows their measurement of the ratio $r_\epsilon$  for two magnet polarities and the two data sets.
 It can be seen
that the majority of the detection asymmetry changes sign with the reversal of the magnet polarity, and thus the final average of the two samples is much less sensitive to detection asymmetry.

The resulting values of $\asls$ obtained by the two experiments as well as their  average are
\begin{eqnarray}
{\asls}^{\rm{D0}}   & = & (-1.12 \pm 0.74 \pm 0.17) \times 10^{-2} \\
{\asls}^{\rm{LHCb}} & = &( +0.39 \pm 0.26 \pm 0.20 )\times 10^{-2}\\
{\asls}^{\rm{average}} & = &( +0.17 \pm 0.30)\times 10^{-2} .
\end{eqnarray}
Both results are consistent with the Standard Model expectation (\ref{asls_sm}), albeit the uncertainty is still
a factor of about 130 larger than the central value in the Standard Model.

The Babar, Belle, D0, and LHCb collaborations perform the independent measurement of the asymmetry $\asld$.
Their results are summarised in  Table~\ref{tab:asld}. The world average value of $\asld$ is
\begin{equation}
\asld(\mbox{HFAG}) = 0.0001 \pm 0.0020.
\end{equation}

\begin{table}
\begin{tabular}{lll}
\hline
Experiment & measured $\asld$ (\%) & Ref. \\\hline
\lhcb  $D^{(\star)} \mu\nu X$ & $-0.02 \pm 0.19 \pm 0.30$         & \cite{Aaij:2014nxa} \\
\dzero \hspace{0.35cm} $D^{(\star)} \mu\nu X$ & $+0.68 \pm 0.45 \pm 0.14$         &\cite{Abazov:2012hha} \\
\babar $D^{\star}  \ell\nu X$ & $+0.29 \pm 0.84^{+1.88}_{-1.61}$ & \cite{Lees:2013sua}\\
\babar $\ell\ell$             & $-0.39 \pm 0.35 \pm 0.19$      &\cite{Lees:2014kep}
 \\\hline
\end{tabular}
\caption{Most recent measurements of the $CP$ violation parameter $\asld$.}
\label{tab:asld}
\end{table}

The D0 experiment also reports a complementary measurement related to the semi-leptonic asymmetries of $\Bs$
and $\Bd$ mesons \cite{Abazov:2013uma}. It performs the simultaneous study of
the inclusive semi-leptonic charge asymmetry
and of the like-sign di-muon charge asymmetry. These quantities are defined as
\begin{eqnarray}
a & = & \frac{n^+ - n^-}{n^+ + n^-}, \\
A & = & \frac{N^{++} - N^{--}}{N^{++} + N^{--}}.
\end{eqnarray}
Here $n^+$ and $n^-$ are the number of events with the reconstructed positive or negative muon, respectively.
 $N^{++}$ and $N^{--}$ are the number of events with two positive or two negative muons, respectively.
  The asymmetries $a$ and $A$ are cast as
   \begin{eqnarray}
 a & = & a_{\rm CP} + a _{\rm bkg}, \\
 A & = & A_{\rm CP} + A_{\rm bkg}.
 \end{eqnarray}
 Here $a_{\rm CP}$ and $A_{\rm CP}$ are the asymmetries due to the genuine CP-violating processes, such as
 CP violation in mixing of $\Bd$ and $\Bs$ mesons. The asymmetries $a_{\rm bkg}$ and $A_{\rm bkg}$
 are produced by the background processes not related to  CP violation.
 The main source of these asymmetries is the difference in the interaction cross-section of the positive
 and negative charged particles with the detector material.
 The main challenge in the D0 analysis is the accurate estimate of the background asymmetries
 $a_{\rm bkg}$ and $A_{\rm bkg}$  and the extraction of the values of $a_{\rm CP}$ and $A_{\rm CP}$.

 The asymmetries $a_{\rm CP}$ and $A_{\rm CP}$ depend on both $\asld$ and $\asls$. Since the oscillation
 period of $\Bd$ and $\Bs$ mesons is significantly different, the contribution of $\asld$ and $\asls$
 strongly depends on the decay time of collected $B$ mesons. This decay time is not measured
 in the inclusive analysis. Instead, the D0 experiment measures the asymmetries $a_{\rm CP}$ and $A_{\rm CP}$
 in  sub-samples containing the muons with different muon impact parameter.
 The division into the sub-samples according to the muon impact parameter is used to estimate the contribution of $\asld$ and $\asls$.
 In addition, the asymmetry $A_{\rm CP}$ is sensitive to the width difference $\Delta \Gamma_d$
of $\Bd$ meson (see \cite{Borissov:2013wwa})
 and this quantity is also obtained in the D0 analysis. Their result is
 \begin{eqnarray}
 \asld & = & (-0.62 \pm 0.43) \times 10^{-2}, \\
 \asls & = & (-0.82 \pm 0.99) \times 10^{-2}, \\
 \frac{\Delta \Gamma_d}{\Gamma_d} & = & (+0.50 \pm 1.38) \times 10^{-2}.
 \end{eqnarray}
 The correlation between the fitted parameters are
 \begin{equation}
 \rho_{d,s} = -0.61,~~~~\rho_{d,\Delta \Gamma} = -0.03,~~~~\rho_{s, \Delta \Gamma} = +0.66.
 \end{equation}
Although the central values of all three quantities are consistent with the SM prediction within the uncertainties,
a deviation from the SM prediction by 3 standard deviations is reported
 because of the correlation between these observables.

\begin{figure}[h]
  \includegraphics[width=0.45 \textwidth,angle=0]{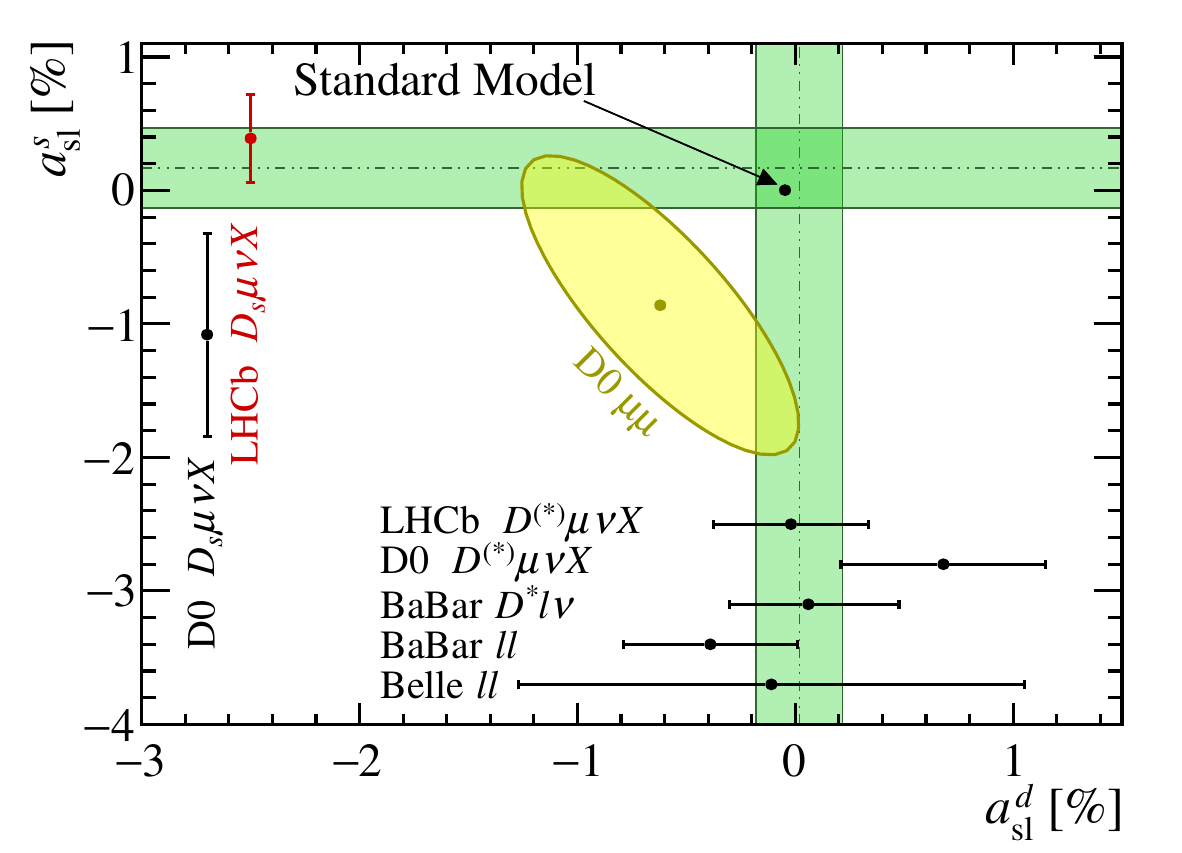}
  \caption{Overview of measurements in the $\asld$ - $\asls$ plane.
Direct measurements of $\asls$ and $\asld$ listed in Tab.~\ref{tab:asld}
 ($\Bd$ average as the vertical band, $\Bs$ average as the horizontal band, D0 di-muon result as the yellow ellipse). The black point close to (0; 0) is the
Standard Model prediction reported in this paper with error bars multiplied by 10.
The plot is taken from Ref.\cite{Aaij:2016yze}.}
  \label{hfag-asl}
\end{figure}

The world knowledge of $CP$ violating parameters in $\Bs$ and $\Bd$ mixing is
summarised in Fig. \ref{hfag-asl}, that shows that the individual measurements of $\asld$ and
$\asls$ are consistent with the Standard Model. Only the \dzero\ di-muon
result suggests a deviation from Standard Model expectations in $CP$ violation in neutral $B^0$ oscillations.
Since this measurement is inclusive, other unknown effects not directly related to  CP violation
in $\Bs$ mixing could contribute to it.

The LHCb experiment is currently taking data and is expected to collect
an additional sample of $\sim$6 fb$^{-1}$ in the current LHC run, and at least 50 fb$^{-1}$
with an upgraded detector to be installed in 2020. 
Moreover Belle II will start taking data in a time scale comparable to the expected start of the LHCb upgraded detector.
Therefore the prospects for increased precision in $CP$ violating asymmetries in neutral $B$ meson decays
are excellent.

\section{CP violation in interference}
\label{CPVinter}
\subsection{Theory}
\label{CPVinter-theory}
In this section we discuss CP violating effects that arise from interference
between mixing and decay, which is also called {\it mixing-induced CP violation}.
Therefore we consider a final state $f$ in which in principle both 
the $\Bs$-meson
and the $\barBs$-meson can decay. The corresponding decay amplitudes
will be denoted by ${\cal A}_f$ and $\bar{\cal A}_f$, defined in Eq.(\ref{amplitude}).
These amplitudes can have contributions from different CKM structures;
their general structure looks like
\begin{eqnarray}
    {\cal A}_f & = & \sum \limits_j {\cal A}_j e^{i (\phi_j^{\rm strong} + \phi_j^{\rm CKM})} \; ,
\end{eqnarray}
where $j$ sums over the different CKM contributions, $\phi_j^{\rm CKM}$ denotes the corresponding
CKM phase and ${\cal A}_j e^{i \phi_j^{\rm strong}}$ encodes the whole non-perturbative physics
as well as the moduli of the CKM-elements.
The calculation of the strong amplitudes and phases from first principles is a non-trivial problem,
for which a general solution has not yet been developed.
Currently several working tools are available in order to investigate this
non-perturbative problem:
QCD factorisation (QCDF; e.g. \cite{Beneke:1999br,Beneke:2000ry,Beneke:2001ev,Beneke:2003zv}),
Soft Collinear Effective Theory (SCET; e.g. \cite{Bauer:2000yr,Bauer:2001yt,Bauer:2004tj}),
light cone sum rules (LCSR; e.g. \cite{Balitsky:1989ry,Khodjamirian:2000mi,Khodjamirian:2003eq})
and  perturbative QCD (pQCD; e.g. \cite{Li:1994iu,Yeh:1997rq}).
\\
Considering  the CP conjugate decay $\barBs \to \bar{f}$,  one finds
\begin{eqnarray}
\bar{\cal A}_{\bar{f}} & = & - \sum \limits_j {\cal A}_j e^{i (\phi_j^{\rm strong} - \phi_j^{\rm CKM})} \; ,
\end{eqnarray}
so only the CKM phase has changed its sign, while the strong amplitude and the strong phase remain
unmodified. The overall sign is due to the CP properties of the $\Bs$-mesons, defined in 
Eq.({\ref{CP}) and $\bar{f}$ defined in Eq.(\ref{barf}).
\\
In some CP asymmetries  the hadronic amplitudes cancel to a good approximation in the 
ratios of decay rates. The corresponding decay modes are the so-called {\it gold-plated modes},
which  were introduced e.g. by \cite{Carter:1980tk} and \cite{Bigi:1981qs}.
Later on we will see that gold-plated modes will appear, when the decay amplitude is  governed by
a single CKM structure. This could be the case in a decay like $\barBs \to \jpsi \phi$, 
which is governed
on quark-level by a $ b \to c \bar{c} s$-transition. Such a transition has a large 
tree-level contribution
and a suppressed penguin contribution, see Fig. \ref{penguincont}.
To a good first approximation the penguins can 
be neglected and we have 
a gold-plated mode, with a precise relation of the corresponding CP asymmetry to fundamental 
Standard Model parameters,
including the CKM-couplings. In view of the dramatically increased experimental precision in recent years
it turns out, however, that it is necessary to investigate the possible size of penguin effects, the 
so-called {\it penguin pollution}. This will be discussed below.
\\
Let us go back to the general case, and consider the following time-dependent CP asymmetry
for a $\Bs \to f$ transition without any approximations concerning the structure of the decay amplitude:
\begin{equation}
A_{CP,f}(t) = \frac{\Gamma \left( \barBs(t) \to f \right) - \Gamma \left( \Bs(t) \to f \right)}
              {\Gamma \left( \barBs(t) \to f \right) + \Gamma \left( \Bs(t) \to f \right)}
\; .
\label{CPasyminter}
\end{equation}
Inserting the time evolution given in Eq.(\ref{GammaBf}) and Eq.(\ref{GammabarBf}) one finds \footnote{A more
detailed derivation can be found in \cite{Anikeev:2001rk}.}
\begin{equation}
A_{CP,f}(t)  = - \frac{{\cal A}_{\rm CP}^{\rm dir} \cos (\Delta M_s t) + {\cal A}_{\rm CP}^{\rm mix}\sin (\Delta M_s t)}
              {                   \cosh (\frac{\Delta \Gamma_s t}{2}) 
               + {\cal A}_{\Delta \Gamma}\sinh (\frac{\Delta \Gamma_s t}{2})
              }
+ {\cal O} \left(a_{fs}^s \right)
\; ,
\label{CPasyminter2}
\end{equation}
with ${\cal A}_{\rm CP}^{\rm dir}$, ${\cal A}_{\rm CP}^{\rm mix}$ and $ {\cal A}_{\Delta \Gamma} $
being defined in Eq.(\ref{Adir}), Eq.(\ref{Amix}) and Eq.(\ref{AGamma}).
We can rewrite two of those definitions as
\begin{eqnarray}
{\cal A}_{\rm CP}^{\rm mix} & = & -\frac{2 | \lambda_f |}{1+|\lambda_f|^2} \sin \left[ \arg(\lambda_f) \right]
=  +\frac{2 | \lambda_f |}{1+|\lambda_f|^2} \sin \left[ \phi_s \right]
\; ,
\nonumber
\\
\label{Amix2}
\\
{\cal A}_{\Delta \Gamma}    & = & -\frac{2 | \lambda_f |}{1+|\lambda_f|^2} \cos \left[ \arg(\lambda_f) \right]
=  -\frac{2 | \lambda_f |}{1+|\lambda_f|^2} \cos \left[ \phi_s \right]
\; ,
\nonumber
\\
\label{AGamma2}
\end{eqnarray}
with the phase $\phi_s$ to be defined as
\begin{eqnarray}
\phi_s & = & - \arg(\lambda_f)= 
         - \arg \left( \frac{q}{p} \frac{\bar{\cal A}_f}{{\cal A}_f} \right)
\\
& = & - \pi + \phi_M - \arg \left( \frac{\bar{\cal A}_f}{{\cal A}_f} \right) \; .
\end{eqnarray}
This is the most general definition of the phase that appears in interference.
However, in this form a measurement of $\phi_s$ does not enable us to connect
the phase with fundamental parameters of the underlying theory. To do so, we either find some
simplifications for the decay amplitudes or we have to evaluate the ratio of amplitudes 
non-perturbatively. Before discussing a particular simplification, we note
that sometimes  a different notation ($S_f$ for the coefficient that is arising in 
Eq.(\ref{CPasyminter2}) with
the term $\sin (\Delta M_st)$ and  $C_f$ or $A_f$ for the coefficient that is arising with
the term $\cos (\Delta M_st)$ - up to signs)
is used
\begin{eqnarray}
  -A_f = C_f  & \equiv &{\cal A}_{\rm CP}^{\rm dir} \; ,
\\
- S_f  &\equiv & {\cal A}_{\rm CP}^{\rm mix}  \; .
\end{eqnarray}
BaBar uses $C_f$ and Belle $A_f$. 
Expanding the hyperbolic functions in Eq.(\ref{CPasyminter2}) for small arguments, i.e.
small decay rate differences and/or short times,
we can express the time-dependent CP asymmetry $A_{CP,f}(t)$ as
\begin{eqnarray}
A_{CP,f}(t)  &  \approx &  \frac{S_f\sin (\Delta M_s t)-C_f \cos (\Delta M_s t) }
              {1 + {\cal A}_{\Delta \Gamma}   \frac{\Delta \Gamma_s}{2 \Gamma_s} \frac{t}{\tau_s} +
   \frac12 \left(\frac{\Delta \Gamma_s}{2 \Gamma_s} \frac{t}{\tau_s}\right)^2}
\; .
\end{eqnarray}
This formula holds in general, and no approximation on the corresponding decay amplitudes 
has been made yet.
In this general case,  the quantities ${\cal A}_{\rm CP}^{\rm dir}$, ${\cal A}_{\rm CP}^{\rm mix}$ 
and $ {\cal A}_{\Delta \Gamma} $ are unknown hadronic contributions that are very difficult to be determined
in theory.
\\
In the following we discuss the simplified case of the gold-plated modes.
Here we consider the final state $f$ to be a CP eigenstate, i.e. $ f = f_{\rm CP} = \eta_{\rm CP} \bar{f}$
and we assume that only one CKM structure is contributing to the decay amplitude - by e.g.
neglecting penguins. In this special case we get
\begin{eqnarray}
{\cal A}_{f_{\rm CP}}               & = & {\cal A}_j e^{i (\phi_j^{\rm strong} + \phi_j^{\rm CKM})} \; ,
\\
\bar{\cal A}_{f_{\rm CP}} & = & \eta_{\rm CP} \bar{\cal A}_{{\bar{f}}_{\rm CP}} 
                       =   - \eta_{\rm CP} {\cal A}_j e^{i (\phi_j^{\rm strong} - \phi_j^{\rm CKM})} \; ,
\\
\Rightarrow
\frac{\bar{\cal A}_{f_{\rm CP}}}{{\cal A}_{f_{\rm CP}}} & = &  - \eta_{\rm CP} e^{-2 i \phi_j^{\rm CKM}} \; .
\label{ratioA}
\end{eqnarray}
So in the case of gold-plated modes all hadronic
uncertainties cancel exactly in
the ratio of the two decay amplitudes in Eq.(\ref{ratioA}) 
and one is left with a pure weak CKM phase.
Thus the parameter $\lambda_f$, which triggers the CP asymmetries is given by 
\begin{eqnarray}
\lambda_{f_{\rm CP}} & = & \frac{q}{p} \frac{\bar{\cal A}_{f_{\rm CP}}}{{\cal A}_{f_ {\rm CP}}}
 = 
 \eta_{CP} \frac{V_{ts} V_{tb}^*}{V_{ts}^* V_{tb}} e^{-2 i \phi_j^{\rm CKM}} \; .
\end{eqnarray}
Therefore we have in the case of only one contributing CKM structure $|\lambda_{f_{\rm CP}}| = 1$ and thus
\begin{eqnarray}
{\cal A}_{\rm CP}^{\rm dir} & = & 0\; ,
\\
{\cal A}_{\rm CP}^{\rm mix} & = & + \sin \left(\phi_s \right) \; ,
\\
{\cal A}_{\Delta \Gamma}    & = & - \cos \left(\phi_s \right)  \; ,
\end{eqnarray}
leading to the simplified formula for the asymmetry
\begin{equation}
A_{CP,f}(t)  \approx  \frac{\sin \phi_s \sin (\Delta M_s t)}
              {\cos \phi_s   \sinh (\frac{\Delta \Gamma_s t}{2}) -
                             \cosh (\frac{\Delta \Gamma_s t}{2}) 
              }
\; .
\label{CPasyminter3}
\end{equation}
This formula holds in the case of only one contributing CKM-structure to the 
whole decay amplitude and the final state being a CP eigenstate.
\\
If the corresponding decay is triggered e.g. by a $ b \to c \bar{c} s$ transition on quark-level,
as in the case of $\Bs \to \jpsi \phi$,  we get
\begin{eqnarray}
\phi_s
& = & - \arg \left(  \eta_{CP} \frac{V_{ts}   V_{tb}^*}{V_{ts}^* V_{tb}  } 
                               \frac{V_{cs}^* V_{cb}  }{V_{cs}   V_{cb}^*}
\right)
\; .
\end{eqnarray}
Thus a measurement of the mixing phase $\phi_s$ gives us direct information
about the phases, i.e. the amount of CP violation, of the CKM elements.
If in addition the final state has a CP eigenvalue $\eta_{CP} =+1$, then we get
\begin{eqnarray}
\phi_s
& = &  - 2 \beta_s
\; ,
\label{phisbetas}
\end{eqnarray}
with the commonly used notation
\begin{eqnarray}
\beta_s & = &- \arg \left[ - \frac{V_{ts}^* V_{tb}}{V_{cs}^* V_{cb}}\right]
\label{betas}
\\
   & = & 0.0183 \pm 0.0010  =   (1.05 \pm 0.05 )^\circ \; .
\end{eqnarray}
Here we used a definition for $\beta_s$ that ensures that its numerical value is positive; sometimes
a different sign is used.
\begin{figure*}
\includegraphics[width=0.45 \textwidth,angle=0]{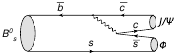}
\hfill
\includegraphics[width=0.45 \textwidth,angle=0]{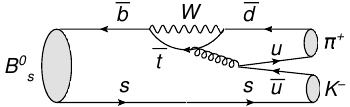}

\vspace{0.6cm}

\includegraphics[width=0.45 \textwidth,angle=0]{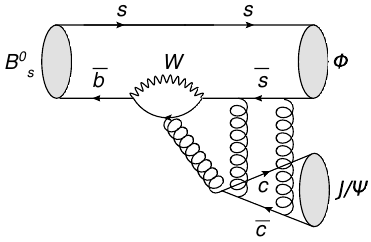}
\hfill
\includegraphics[width=0.45 \textwidth,angle=0]{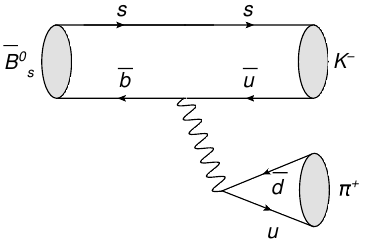}
\caption{\label{penguincont} Different decay topologies contributing to the decays
                         $\Bs \to \jpsi \phi$ (l.h.s.) and $\Bs \to K^- \pi^+$ (r.h.s.). 
                        The top row shows the colour suppressed topologies (in the case
                        of $\Bs \to \jpsi \phi$ this is the tree-level contribution and in
                        the case of  $\Bs \to K^- \pi^+$ this is the penguin contribution) and 
                        the lower row shows the colour allowed topologies (in the case
                        of $\Bs \to \jpsi \phi$ this is now the penguin  contribution and in
                        the case of  $\Bs \to K^- \pi^+$ this is the tree-level contribution). Since 
                        the $\jpsi$-meson is colour neutral, we have to add additional gluons
                        for the penguin contribution to $\Bs \to \jpsi \phi$.}
\end{figure*}
If there is only a modest experimental precision available,  penguins can be neglected, to a first approximation,
for the tree-level dominated $b \to c \bar{c} s$ decays like $B_s \to \jpsi \phi$
penguins, and we can use simplified formulae like Eq.(\ref{phisbetas}).
\\
However, we will see below that the current experimental precision in the determination of $\phi_s$ is
of the order of $\pm 2^\circ$, which equals the SM expectation of  $\phi_s^{\rm SM} = (2.1 \pm 0.1)^\circ$. 
In view of  this high experimental precision, it seems mandatory to
determine the possible size of penguin contributions,
in order to make  profound statements about  new physics effects in these CP asymmetries.
\\
Let us examine the general expression for the decay amplitude without neglecting penguins. 
Examples for decays with both tree-level and penguin contributions 
are e.g. $B_s \to \jpsi \phi$ or $B_s \to K^- \pi^+$. 
The former is governed on quark-level by a $b \to c \bar{c} s$-transition, and the latter by a 
$b \to u \bar{u} d$-transition. The tree-level components and penguin contributions to these
decays are shown in Fig. \ref{penguincont}.
\\
A naive dimensional estimate (size of CKM couplings, number of strong couplings and colour counting)
gives a small penguin contribution in the case of $B_s \to \jpsi \phi$ and a large 
penguin contribution in the case $B_s \to K^- \pi^+$. Thus $B_s \to \jpsi \phi$ is a prime candidate
for a gold-plated  mode, while in the case of  $B_s \to K^- \pi^+$ direct CP violation, i.e. CP
violation directly in the decay might be visible; this will be further discussed in Section 
\ref{CPVdecay}.
\\
To become more quantitative, we take a closer look at the general structure of the 
decay amplitude of a
$b \to c \bar{c} s$-transition. Using the effective Hamiltonian for $\Delta B = 1$
transitions (see e.g. \cite{Buras:1998raa} for a nice introduction) we get for the amplitude.
\begin{equation}
{\cal A}_f \left( \Bs \longrightarrow f \right)
=
\langle f | {\cal H}_{eff} | \Bs \rangle \; ,
\end{equation}
with the effective SM Hamiltonian for $b \to c \bar c s$ transitions
\begin{eqnarray}
 {\cal H}_{eff.} = &\frac{G_F}{\sqrt{2}}&
\left[  \lambda_u \left( C_1 Q_1^u + C_2 Q_2^u \right)
      + \lambda_c \left( C_1 Q_1^c + C_2 Q_2^c \right)
\right.
\nonumber
\\
  && \left.    + \lambda_t \sum \limits_{i=3}^6 C_i Q_i
\right] + h.c. \;.
\end{eqnarray}
The CKM structure is given as before by $\lambda_q := V_{qb} V_{qs}^*$;  
the decay $b \to c \bar c s$
proceeds via the current-current operators $Q_1^c, Q_2^c$ and the QCD penguin operators $Q_3,...,Q_6$.
$C_1, ... , C_6$ are the corresponding Wilson coefficients.
When the current-current operators $Q_1^u, Q_2^u$ are inserted in a penguin diagram in the effective
theory, they also contribute to $b \to c \bar c s$.
Electro-weak penguins are neglected.
\\
Therefore we have the following structure of the amplitude
 ${\cal A}_f \left( \Bs \longrightarrow f \right)$ 
\begin{eqnarray}
{\cal A}_f = &\frac{G_F}{\sqrt{2}} &  \left[ 
  \lambda_u    \sum \limits_{i = 1,2} C_i  \langle Q_i^u \rangle^P 
+ \lambda_c    \sum \limits_{i = 1,2} C_i  \langle Q_i^c \rangle^{T+P} 
\right. \nonumber \\
&& \left.
+ \lambda_t    \sum \limits_{i = 3}^6 C_i  \langle Q_i \rangle^T 
\right] \; .
\end{eqnarray}
$\langle Q \rangle^T$ denotes the tree-level insertion of the local operator $Q$, 
$\langle Q \rangle^P$ denotes the insertion of the operator $Q$ in a penguin diagram.
Using further the unitarity of the CKM matrix, $\lambda_t = - \lambda_u - \lambda_c$, 
we can rewrite the amplitude in a form where only two different CKM structures are appearing.
\begin{eqnarray}
{\cal A}_f & = & \frac{G_F}{\sqrt{2}} \lambda_c  
\left[
  \sum \limits_{i = 1,2} C_i \langle Q_i^c \rangle^{T+P} 
- \sum \limits_{i = 3}^6 C_i \langle Q_i \rangle^T 
\right.
\nonumber
\\
&+& \left. 
\frac{\lambda_u}{\lambda_c} \left( 
  \sum \limits_{i = 1,2} C_i  \langle Q_i^u \rangle^P 
- \sum \limits_{i = 3}^6 C_i \langle Q_i \rangle^T \right)
\right]
\\
& = & {\cal A}^{\rm Tree}_f + {\cal A}^{\rm Peng}_f \; .
\label{amplitude_2}
\end{eqnarray} 
In the last line we have defined separately a tree-level amplitude and a penguin amplitude.
They are  given by
\begin{eqnarray}
{\cal A}^{\rm Tree}_f & = & \frac{G_F}{\sqrt{2}} \lambda_c  
\left[
  \sum \limits_{i = 1,2} C_i \langle Q_i^c \rangle^{T+P} 
- \sum \limits_{i = 3}^6 C_i \langle Q_i \rangle^T \right]
\nonumber
\\
& = & \left|{{\cal A}}^{\rm Tree}_f\right| e^{i \left[\phi^{\rm QCD}_{\rm Tree} + \arg(\lambda_c)
           \right]}  \; ,
\label{amplitude_3}
\\
{\cal A}^{\rm Peng}_f  & = &  \frac{G_F}{\sqrt{2}} \lambda_u
\left[
  \sum \limits_{i = 1,2} C_i  \langle Q_i^u \rangle^P 
- \sum \limits_{i = 3}^6 C_i \langle Q_i \rangle^T 
\right] 
\nonumber
\\
& = & \left|{{\cal A}}^{\rm Peng}_f\right| e^{i \left[\phi^{\rm QCD}_{\rm Peng} + \arg(\lambda_u)
           \right]} \; .
\label{amplitude_4}
\end{eqnarray}
Here we  split up the amplitudes into their modulus and their phase. Sometimes it advantageous
to split off the explicit dependence on the modulus of the  CKM structure:
\begin{eqnarray}
\left|{{\cal A}}^{\rm Tree}_f\right| 
& = &
\frac{G_F}{\sqrt{2}} \left|\lambda_c \right| \left|\tilde{{\cal A}}^{\rm Tree}_f\right| 
\; ,
\\
\left|{{\cal A}}^{\rm Peng}_f\right|  
& = & 
\frac{G_F}{\sqrt{2}} \left|\lambda_u \right| \left|\tilde{{\cal A}}^{\rm Peng}_f\right| 
\; .
\end{eqnarray}
The strong amplitudes and the strong phases are in principle unknown. A first naive estimate of the
size of the modulus can be done by investigating, what $\Delta B=1$ Wilson coefficients are contributing.
In the case of $\Bs \to \jpsi \phi$ the tree-level amplitude is enhanced by the CKM-elements in 
$\lambda_c$ and the tree-level contribution of  the large Wilson coefficients $C_1$ and $C_2$; 
the penguin amplitude is suppressed by $\lambda_u$ and further either by small 
penguin Wilson-coefficients $C_{3...6}$ or by a loop.
\\
In general, i.e. without any approximations concerning the size of the hadronic effects,
we get the ratio of decay amplitudes
\begin{eqnarray}
\frac{\bar{\cal A}_{\bar{f}} }
     {    {\cal A}_f         }    
& = &  - e^{-2 i \arg(\lambda_c)} \left[ \frac{1 + r e^{- i \arg \left(\frac{\lambda_u}{\lambda_c}\right)}}
                                              {1 + r e^{+ i \arg \left(\frac{\lambda_u}{\lambda_c}\right)}}\right]
\label{ratioamplitude}
\end{eqnarray}
with $r$ being defined as
\begin{equation}
r = \left|\frac{\lambda_u}{\lambda_c} \right|
    \left|\frac{\tilde{{\cal A}}^{\rm Peng}_f}{\tilde{{\cal A}}^{\rm Tree}_f} \right|
    e^{i (\phi^{\rm QCD}_{\rm Peng} - \phi^{\rm QCD}_{\rm Tree})} \;.
\label{r}
\end{equation}
In the case of $\Bs \to \jpsi$ the CKM part of $r$ is very small, it is given by 
$|\lambda_u/\lambda_c| \approx 0.02 $.
The hadronic part of $r$ is a non-perturbative
quantity that can currently not be calculated from first principles. Before we turn to
some quantitative investigations in the literature, we have a look at a naive estimate:
$\tilde{{\cal A}}^{\rm Peng}_f$ and $\tilde{{\cal A}}^{\rm Tree}_f$ contain Wilson coefficients
from the effective Hamiltonian. The penguin Wilson coefficients $|C_{3,...,6}|$ are typically 
smaller than 0.04, therefore one can neglect them in comparison to the Wilson coefficient $C_2 \approx 1$, 
see e.g. \cite{Buchalla:1995vs} for numerical values.
Thus we are left with the tree-level insertion of the operator $Q_2$ in the case of 
$\tilde{{\cal A}}^{\rm Tree}_f$ and with the penguin insertion of the operator $Q_2$ in the case of 
$\tilde{{\cal A}}^{\rm Peng}_f$. 
Since we do not know the relative size, of these two, we take the analogy of inclusive $b$-quark
decays as a first indication of its size.
For the inclusive decay $b \to c \bar{c} s$
it was found
(see e.g. \cite{Bagan:1995yf,Lenz:1997aa,Krinner:2013cja})
that  $\langle Q \rangle^P \leq 0.05 \langle Q \rangle^T$. 
Taking this value as an indication for the size of
$\tilde{{\cal A}}^{\rm Peng}_f / \tilde{{\cal A}}^{\rm Tree}_f$
we get an estimate of $r$ of about $|r| \approx 0.001$. 
One should be aware, however, that this very naive 
estimate can easily be off by a factor of 10 and we also cannot quantify the size 
of the strong phase in this approach.
Using the same methods for the decay $\Bs \to K^- \pi^+$ we would get a value of 
$r^{\Bs \to K^- \pi^+}$ of about 0.1, so roughly 100 times larger than in the case
of $\Bs \to \jpsi \phi$. $\Bs \to K^- \pi^+$ is thus  a prime candidate for decays where we are looking 
for large penguin effects, e.g. if we want to measure a direct CP asymmetry in the
$\Bs$ system. Our naive estimate does not take into account that these two channels proceed 
via different topologies; hence the factor 100 might have to be modified considerably.
\\
Nevertheless it seems that $r$ is a small number in the case of $\Bs \to \jpsi \phi$ and we can make 
a Taylor expansion in Eq.(\ref{ratioamplitude}) to obtain
\begin{eqnarray}
\frac{\bar{\cal A}_{\bar{f}} }
     {    {\cal A}_f         }    
& \approx & - e^{-2 i \arg(\lambda_c)} \left[
1 - 2 i r \sin \left( \arg \left( \frac{\lambda_u}{\lambda_c} \right)\right) \right] \; .
\label{ratioamplitude2}
\end{eqnarray}
Investigating further Eq.(\ref{ratioamplitude2}) or Eq.(\ref{ratioamplitude}), 
we see that the first term on the r.h.s. gives 
rise to $-2 \beta_s$ in the CP asymmetry in Eq.(\ref{CPasyminter3}). 
The second term (proportional to $r$), 
corresponds to the SM penguin pollution, which we denote by $\delta^{\rm Peng, SM}$.
Therefore the experimentally measured phase $\phi_s$ has two contributions in the 
Standard Model:
\begin{eqnarray}
\phi_s & = & -2 \beta_s + \delta^{\rm Peng, SM} \, ,
\label{pollution}
\end{eqnarray}
where the Standard Model penguin is given by
\begin{eqnarray}
e^{i\delta^{\rm Peng, SM}} & \approx & 1 - 2 i r \sin \left[ \arg \left( \frac{\lambda_u}{\lambda_c}
                                                                  \right)
                                                      \right] 
e^{i (\phi^{\rm QCD}_{\rm Peng} - \phi^{\rm QCD}_{\rm Tree})} \; .
\nonumber
\\
\end{eqnarray}
Inserting the above approximations for $\Bs \to \jpsi \phi$ we get as a very rough estimate
of the penguin pollution:
\begin{eqnarray}
e^{i\delta^{\rm Peng, SM}}
& \approx & 1-  0.002 i e^{i (\phi^{\rm QCD}_{\rm Peng} - \phi^{\rm QCD}_{\rm Tree})}
\nonumber
\\
\Rightarrow \delta^{\rm Peng, SM} & \leq & \pm 0.002 = \pm 0.1^\circ \; .
\end{eqnarray}
Thus very naively we expect a penguin pollution of at most $\pm 0.1^\circ$ in the case of 
$\Bs \to \jpsi \phi$.
This very rough estimate could, however, be 
easily modified by a factor of e.g. 10, due to non-perturbative effects and then we 
would be close to the current experimental uncertainties. 
Thus more theoretical work has to be done to quantify the size of penguin contributions.
There are now several strategies to achieve this point:
\begin{enumerate}
\item Measure $\phi_s$ in different decay channels: assuming that penguins are negligible, different
      determinations should give the same value for the mixing phase.
      Until now we have focused on the  extraction of the phase $\phi_s$ 
      from the decay $\Bs \to \jpsi \phi$. This final state is an admixture of CP-even and CP-odd 
      components. To extract information on $\Delta \Gamma_s$ and $\phi_s$ an angular analysis
      is required, see the discussion in Section \ref{Bs_system_exp} and Section \ref{CPV_inter_exp}
      or the early references: \cite{Dighe:1995pd,Dighe:1998vk,Dunietz:2000cr}.
      Moreover the $\jpsi \phi (\to K^+ K^-)$ final state can be investigated for non-resonant 
      $K^+ K^-$ contributions
      in order to increase the statistics.
      The phase $\phi_s$ has also been determined in different $b \to c \bar{c} s$-channels, like
      $\Bs \to \jpsi \pi^+ \pi^-$  
       (including   $\Bs \to \jpsi f_0$, 
      see e.g. \cite{Stone:2008ak,Colangelo:2010wg,Fleischer:2011au,Zhang:2012zk,Stone:2013eaa}),
      $\Bs \to \jpsi \eta^{(')}$  
      (see e.g. \cite{Dunietz:2000cr,DiDonato:2011kr,Fleischer:2011ib})
      and $\Bs \to D_s^{(*)+} D_s^{(*)-}$  
      (see e.g. \cite{Dunietz:2000cr,Fleischer:2007zn})
      as a cross-check.
\begin{figure*}
\begin{center}
\includegraphics[width=0.9 \textwidth,angle=0]{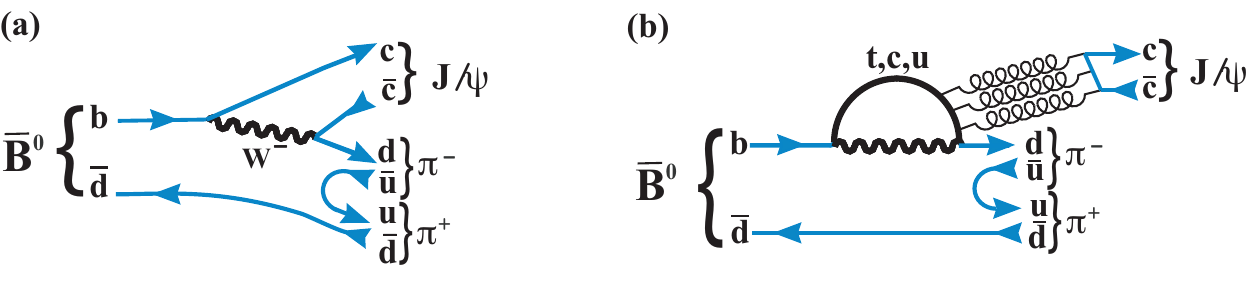}
\end{center}
\caption{\small (a) Tree level and (b) penguin diagram for $\barBd$ decays into $\jpsi \pi^+\pi^-$, which contains also $\barBd \to \jpsi \rho^0$.}
\label{FeyPenguin}
\end{figure*}
      Here $\Bs \to \jpsi f_0$ is a CP-odd final state,
      $ \jpsi \eta^{(')}$ and $ D_s^{+} D_s^{-}$ are CP-even final states,
      and $D_s^{*+} D_s^{*-}$ is again an admixture of different CP components.
      Getting different values for $\phi_s$ from different decay modes 
      would point towards different and large penguin contributions in the individual channels.
      The different experimental results will be discussed in the next section: they show
      no significant deviations within the current experimental uncertainties, but 
      there is also plenty of space left for some sizable differences. 
\item Measure the phase $\phi_s$ for different polarisations of the final states 
      in $\Bs \to \jpsi \phi$: potential
      differences might originate from penguins, which in general contribute differently to
      different polarisations, see \cite{Fleischer:1999zi,Faller:2008gt}. 
      Such an analysis was done in \cite{Aaij:2014zsa}
      and within the current experimental uncertainties 
      no hint for a polarisation dependence of $\phi_s$ was found:
      \begin{eqnarray}
      \phi_{s,||}    - \phi_{s,0} & = & - \left( 1.03 \pm 2.46 \pm 0.52 \right)^\circ \; ,
      \\
      \phi_{s,\perp} - \phi_{s,0} & = & - \left( 0.80 \pm 2.01 \pm 0.34 \right)^\circ \; .
      \end{eqnarray}
      On the other hand one sees that effects of the order of $2^\circ$, which would be as large as the 
      whole SM prediction for $\phi_s$, are not ruled out yet. 
      A further discussion of this result was done by \cite{DeBruyn:2014oga}.
\item Compare the decay $\Bs \to \jpsi \phi$ to a decay with a similar hadronic structure, but a
      CKM enhanced penguin contribution:
      differences in the phase $\phi_s$ extracted from $\Bs \to \jpsi \phi$ and from the new decay 
      might then give experimental hints for the size of the penguin contribution.
      \\
      Exchanging the $s$-quark line in Fig.\ref{penguincont} with a $d$-quark line
      one arrives at decays like
      $\Bs \to \jpsi K_S$ (see e.g. \cite{Fleischer:1999nz})
      or 
      $\Bs \to \jpsi \bar{K}^*(892)$ (see e.g. \cite{Dunietz:2000cr,Faller:2008gt,DeBruyn:2014oga}). 
      In the first decay there is only one vector particle in the final state, while the latter case
      we have two ($\bar{K}^*(892)$ is a vector meson) as in the case of $\Bs \to \jpsi \phi$.
      Thus we consider here only the decay $\Bs \to \jpsi \bar{K}^*(892)$.
      \\ 
      The analogous modes in  $\Bd$ decays are
      $\Bd \to \jpsi K_S \Leftrightarrow \Bd \to \jpsi \pi^0$. The extraction of penguin pollution
      via this relation was discussed in e.g. 
      \cite{Ciuchini:2005mg,Faller:2008zc,Ciuchini:2011kd}.
      To get an idea of the  size of the penguin uncertainties, we note that 
      \cite{Ciuchini:2011kd} found a 
      possible Standard Model penguin pollution of about $\pm 1.1^\circ$ in the gold-plated mode
      $\Bd \to \jpsi K_S$. 
      \\
      Coming back to the $\Bs$ system, the relative size of the penguin contributions in the decays $\Bs \to \jpsi K_S$ 
      and $\Bs \to \jpsi \bar{K}^*(892)$ ,
      compared to the tree-level components,
      are larger by a factor of about $1/\lambda^2 \approx 25$  than in
      $\Bs \to \jpsi \phi$. This enhancement of the penguin contribution might manifest itself
      in different values for the extracted values of the phase $\phi_s$.
      A disadvantage of these decays is that they are more difficult to measure, because they
      proceed on quark-level via $b\to c \bar{c}d$, whose branching ratio is suppressed by a 
      factor of about 
      $\lambda^2 \approx 1/25$ compared to $\Bs \to \jpsi \phi$. 
      This is the reason, why the CP asymmetries in $\Bs \to \jpsi K_S$ 
      and the one in $\Bs \to \jpsi \bar{K}^*(892)$ have only been determined recently with large
      uncertainties by \cite{Aaij:2015tza} and by \cite{Aaij:2015mea}.
      The corresponding branching  ratios have  been
      measured earlier by the LHCb Collaboration \cite{Aaij:2012nh,Aaij:2012di}.
      \cite{DeBruyn:2014oga} discuss some strategies to extract the size of penguin pollution
      without having the full knowledge about these CP asymmetries.
      A further drawback of this method is that the size of the hadronic effects due 
      to the exchange of a $\phi$-meson with 
      a $\bar{K}^*(892)$-meson cannot be quantified from first principles.
      Finally  there are also penguin annihilation and weak exchange topologies contributing
      to $\Bs \to \jpsi \phi$, 
      that are not present in the  $\Bs \to \jpsi \bar{K}^*(892)$ case, see e.g. \cite{Faller:2008gt}. 
      Whether it is justified to neglect such contributions can e.g. be tested by decays
      like $\Bd \to \jpsi \phi$ that proceed only via weak-exchange and annihilation topologies.
      Experimental constraints on $\Bd \to \jpsi \phi$ from
      Belle \cite{Liu:2008bta}, BaBar \cite{Lees:2014lra} and LHCb \cite{Aaij:2013mtm}
      indicate, however,  no unusual enhancement of annihilation or weak exchange contributions.
\item Compare the decay $\Bs \to \jpsi \phi$ with a decay which is related to it via a symmetry of QCD:
      having now a symmetry might add confidence in obtaining some control over the effect 
      of exchanging the initial and final states mesons with other mesons.
      Such a symmetry is the flavour symmetry $SU(3)_F$, i.e. a symmetry of QCD under the exchange of
      $u$-, $d$- and $s$-quarks. The application of these symmetry is quite widespread, see e.g.
      \cite{Fleischer:1999nz,Ciuchini:2005mg,Faller:2008gt,Ciuchini:2011kd,Jung:2012mp,Bhattacharya:2012ph,DeBruyn:2014oga,Ligeti:2015yma}
      for some examples related to $B$-meson decays.
      Again a word of caution: it is currently not clear how well the $SU(3)_F$-symmetry 
      is working and how large the corrections are, see e.g.
      \cite{Faller:2008gt,Frings:2015eva} for some critical comments. On the other hand
      a comparison of experimental data finds that $SU(3)_F$ might work quite well, for some of these
      decay channels, see e.g. \cite{DeBruyn:2014oga}.
      \\
      A sub-group of  $SU(3)_F$, which is supposed to work particularly well, is
      the so-called {\it U-spin symmetry}, 
      i.e. the invariance of QCD under the exchange of the $s$-quark with a $d$-quark, 
      see e.g. \cite{Fleischer:1999nz,Fleischer:1999zi,DeBruyn:2014oga}.
      Substituting the $s$- and $\bar{s}$-quark on the l.h.s. of Fig.\ref{penguincont} with down-type
      quarks one gets, e.g. \cite{Fleischer:1999zi}.
      \begin{eqnarray}
      \Bs \to \jpsi \phi & \Leftrightarrow & \Bd \to \jpsi \rho^0 \; , \;  \jpsi \pi^0 \; . 
      \end{eqnarray}
      The decay  $ \Bd \to \jpsi \rho^0$ has also enhanced penguin contributions and a similar
      structure as $\Bd \to \jpsi \pi^0$ and $\Bs \to \jpsi K_S$; 
      tree and penguin contributions to $\Bd \to \jpsi \pi^+ \pi^-$, which contains   
      $\Bd \to \jpsi \rho^0$
      are depicted in Fig.\ref{FeyPenguin}.
      $ \Bd \to \jpsi \rho^0$ 
      is discussed further e.g. by \cite{DeBruyn:2014oga} and there are also 
      first measurements of the mixing induced CP asymmetries by the LHCb Collaboration
      \cite{Aaij:2014vda}. In this decay we have again two vector mesons in the final state, as in the case 
      $\Bs \to \jpsi \phi$. Thus here we do  not consider the decay $\Bd \to \jpsi \pi^0$ any further. However 
      this decay
       gave important constraints on the penguin pollution in 
      $\Bd \to \jpsi K_S$ - as it was explained above.
      \\
      Applying U-spin symmetry to the $\Bd$ system one gets e.g.
      \begin{eqnarray}
      \Bd \to \jpsi K_S & \Leftrightarrow & \Bs \to \jpsi K_S \; .
      \end{eqnarray}
      The decay $\Bs \to \jpsi K_S$ was already mentioned above for estimating penguin uncertainties
      in $\Bs \to \jpsi \phi$. It is, however, much better suited for the decay $\Bd \to \jpsi K_S$, see
      \cite{Fleischer:1999nz,Faller:2008zc,DeBruyn:2014oga}. Further experimental studies of this decay 
      were performed by \cite{Aaij:2013eia}.
      \\
      Currently symmetry considerations put a quite strong bound on the penguin pollution;
      \cite{DeBruyn:2014oga} (see also \cite{Fleischer:2015mla}) get for the 
      decay $\Bs \to \jpsi \phi$ the following possible size of penguin pollution:
      \begin{equation}
      \delta^{\rm Peng, SM}_{\jpsi \phi}  = 
         \left[ 0.08 ^{+0.56}_{-0.72} (\rm stat)
                     ^{+0.15}_{-0.13} (\rm SU(3))
         \right]^\circ \; .
      \end{equation}
      This bound is currently dominated by  statistical uncertainties stemming from experiment
      and it will thus be getting stronger in the future by improved measurements, even without 
      theoretical improvements.
\item Investigate purely penguin induced decays: an example for a decay that has no tree-level
      contribution is $\Bs \to \phi \phi$, which is governed
      by a $ b \to s \bar{s} s$-quark level transition. 
      Traditionally such decays are considered to be
      most sensitive to new physics effects.
      The decay $\Bs \to \phi \phi$ has contributions from an $u$, $c$ and $t$ penguin. Its
      amplitude reads
      \begin{eqnarray}
        {\cal A}_f (\Bs \to \phi \phi) & &= \frac{G_F}{\sqrt{2}}   \left[ 
          \lambda_u    \sum \limits_{i = 1,2} C_i  \langle Q_i^u \rangle^P 
        \right. \ \\
        && \left.
          + \lambda_c    \sum \limits_{i = 1,2} C_i  \langle Q_i^c \rangle^{P} 
          + \lambda_t    \sum \limits_{i = 3}^6 C_i  \langle Q_i \rangle^T 
        \right] \; .
        \nonumber
      \end{eqnarray}
      Using again the unitarity of the CKM matrix, 
      we can rewrite the amplitude in a form where only two different CKM structures are appearing.
      \begin{eqnarray}
      {\cal A}_f & = & \frac{G_F}{\sqrt{2}} \lambda_c  
      \left[
       \sum \limits_{i = 1,2} C_i \langle Q_i^c \rangle^{P} 
       - \sum \limits_{i = 3}^6 C_i \langle Q_i \rangle^T 
      \right.
      \nonumber
      \\
      &+& \left. 
      \frac{\lambda_u}{\lambda_c} \left( 
       \sum \limits_{i = 1,2} C_i  \langle Q_i^u \rangle^P 
      - \sum \limits_{i = 3}^6 C_i \langle Q_i \rangle^T \right)
      \right]
     \end{eqnarray} 
     Neglecting the second term, proportional to $\lambda_u/\lambda_c$ we get the same result as
     in the case of the gold-plated mode $\Bs \to \jpsi \phi$: the measured mixing phase is 
     $\phi_s = - 2 \beta_s$.
     In the case of $\Bs \to \phi \phi$ this might, however, not be a very good approximation. Our
     leading term is now given by the difference of the charm penguins and the top penguins, 
     which will give a
     small contribution compared to the large tree-level term in the case of  $\Bs \to \jpsi \phi$.
     The sub-leading term is suppressed by $\lambda_u/\lambda_c$ , which is a small number, but
     the hadronic part is now the difference of up penguins and top penguins, which is of a similar size
     as leading term.
     Thus deviations of the measured phase in  $\Bs \to \phi \phi$ from $-2 \beta_s$
     might tell us something about unexpected non-perturbative enhancements of the up quark penguins 
     compared to the charm quark penguins.
     More advanced theory investigations can be found in
     \cite{Beneke:2006hg,Bartsch:2008ps,Cheng:2009mu,Datta:2012ky}. First measurements
     (see \cite{Aaij:2014kxa}) have still a sizable uncertainty, but they show no significant deviation
     of the mixing phase in   $\Bs \to \phi \phi$ from $-2 \beta_s$.
\item Try to do a calculation from first principles. Very recently penguin effects were estimated 
      in that manner by \cite{Frings:2015eva} by proofing the infrared safety of the penguin
      contributions in a factorisation approach. This study  suggests that penguin contributions 
      to $\phi_s$ in the case of $B_s \to J / \psi \phi$ should be smaller than about $1^\circ$.
      First steps in such a direction have been performed by \cite{Boos:2004xp} and they were pioneered
      by \cite{Bander:1979px}. In the framework of pQCD this was attempted recently by \cite{Liu:2013nea}. 
\end{enumerate}
Most of the current investigations point toward a maximal size of SM  penguin effects
of about $\pm 1^\circ$,
which is unfortunately very close to the current experimental precision of about
$\pm 2^\circ$. Thus more theoretical work has to be done in that direction. Note that the LHCb constraint from
the study of the decay $\Bd\to\jpsi\rho$ \cite{Aaij:2014vda} gives a limit on penguin effects at about  $1^\circ$.

\subsection{Experiment}
\label{CPV_inter_exp}
The experimental study of the CP-violating phase $\phi_s$ has been pursued
vigorously and considerable experimental progress has been achieved. The main channels
used are  $\Bs\to\jpsi  h^+h^-$, where the  $h^+h^-$ system in general may
comprise many states with different angular momenta. Many studies focus on the
``golden mode,''  $\Bs \to \jpsi \phi$,
discussed in Section \ref{Bs_system_exp},
which also contains the references to the latest experimental results. The analysis
of this final state provides the constraint on both $\Delta \Gamma_s$ and $\phi_s$
and is therefore presented as a two-dimensional confidence level contour.

The determination of $\phi_s$ requires the CP-even and CP-odd components
to be disentangled by analysing the differential distribution $d\Gamma/dtd\Omega$,
where $\Omega\equiv(\cos{\theta _h},\cos{\theta _\mu, \chi}$,  defined as (a)
$\theta_h$ is the angle between the $h^+$ direction in the $h^+h^-$ rest frame
with respect to the direction of the $h^+h^-$ pair in the $\Bs$ rest frame,
(b) $\theta_\mu$ is the angle between the $\mu ^+$ direction in the \jpsi\
frame with respect to the $\jpsi$ direction in the $\Bs$ rest frame, and (c)
$\chi$ is the angle between the $\jpsi$ and the $h^+h^-$, as shown  in
Fig.~\ref{fig:helangles}.
 \begin{figure}[tbh]
   \includegraphics[width=0.45 \textwidth,angle=0]{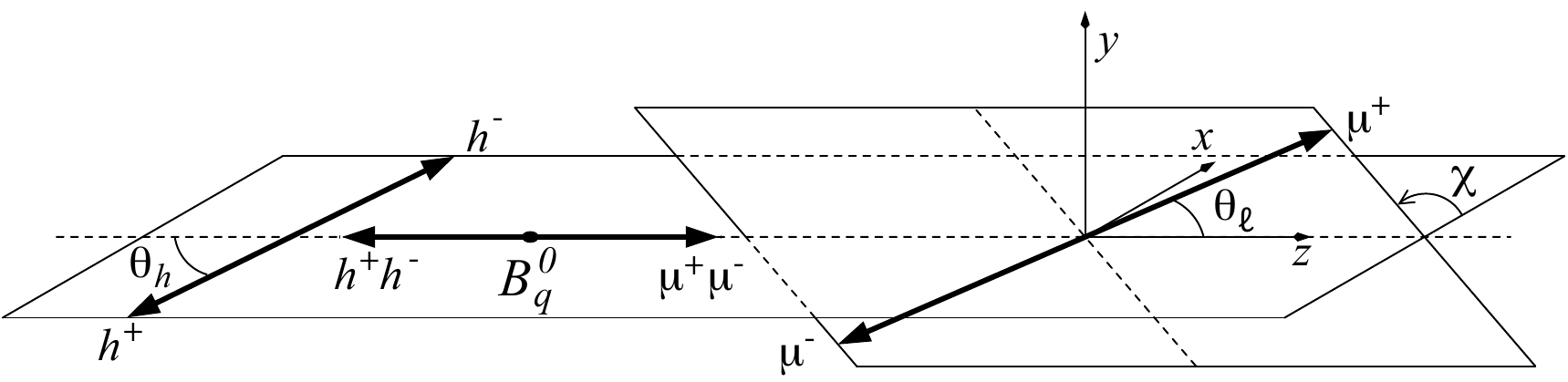}
   \caption{Definition of the helicity angles. For details see text.
   The plot is taken from \cite{Zhang:2012zk}. In this figure the angle 
   $\theta_\mu$ is denoted as $\theta_l$.
   }
   \label{fig:helangles}
\end{figure}

 The decay $\Bs\to \jpsi \Kp\Km$ proceeds predominantly via $\Bs\to \jpsi \phi$ with
the $\phi$ meson decaying subsequently to $\Kp\Km$. In this case, the $\Bs$ decays
into two vector particles, and the $\Kp\Km$ pair is in a P-wave configuration. The
final state is then the superposition of CP-even and CP-odd states, depending
upon the relative orbital angular momentum of the $\jpsi$ and the $\phi$. The
same final state can be produced also with $\Kp\Km$ pairs in an S-wave configuration, as pointed
out by Stone and Zhang \cite{Stone:2008ak}.
This S-wave component is CP-odd.

 The decay width can be expressed in terms of four time-dependent complex amplitudes 
$A_i(t)$. Three of them arise from the P-wave configuration, and, correspond to the 
relative orientation of the linear polarisation vectors of the \jpsi\  and $\phi$ 
mesons, ($0,\perp,\|$) ( see \cite{Aaij:2014nxa}), and one of them corresponds to the S-wave 
configuration. The distributions of decay angles and time for a $\Bs$ meson produced 
at time $t=0$ can be expressed in terms of ten terms,  corresponding to the four 
polarisation amplitudes and their interference terms.  The expressions for the 
decay rate $d\Gamma(\Bs)/dtd\Omega$ is invariant under the transformation
 \begin{equation}
(\phi _s,\Delta\Gamma _s,\delta _0,\delta _{\|},\delta _{\perp},\delta _S)
\to 
(\pi-\phi _s,-\Delta\Gamma _s,-\delta _{\|},\pi-\delta _{\perp},-\delta _S).
\end{equation}
Here, the convention $\delta_0=0$ is chosen.
Thus in principle there is a two-fold ambiguity in the results. This is removed 
by performing fits in bins of $m_{hh}$, see \cite{Xie:2009fs}.
 Thus the LHCb collaboration performs the fit to the distribution $dn/dtd\Omega$ 
in bins of $m_{hh}$ to resolve this ambiguity. The projections of the decay time 
and angular distributions obtained from the analysis of the 3 $\invfb$ LHCb data 
set is shown in Fig.~\ref{fig:lhcbphis}, and the corresponding fit parameters are 
summarised in Table~\ref{tab:phislhcb}. Note that the mixing parameter $\dms$ is 
not constrained from other measurements in this fit and is consistent with world averages.

\begin{figure}
\centering{
  \includegraphics[width=0.40\textwidth]{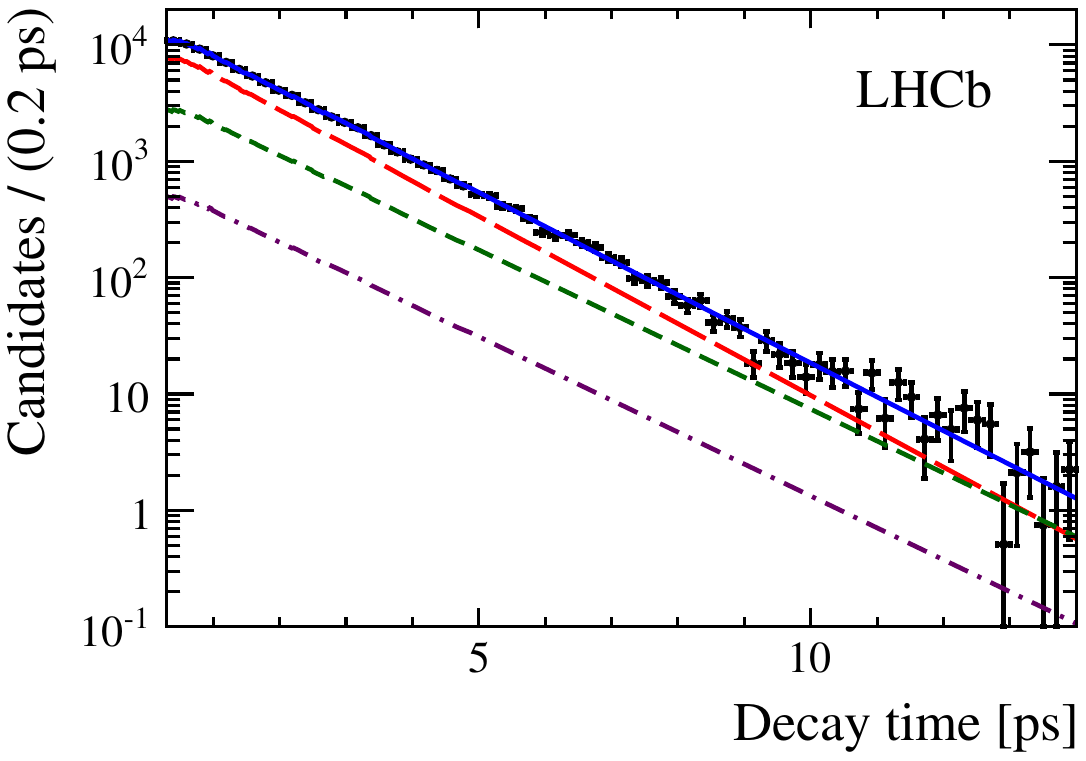}
  \includegraphics[width=0.40\textwidth]{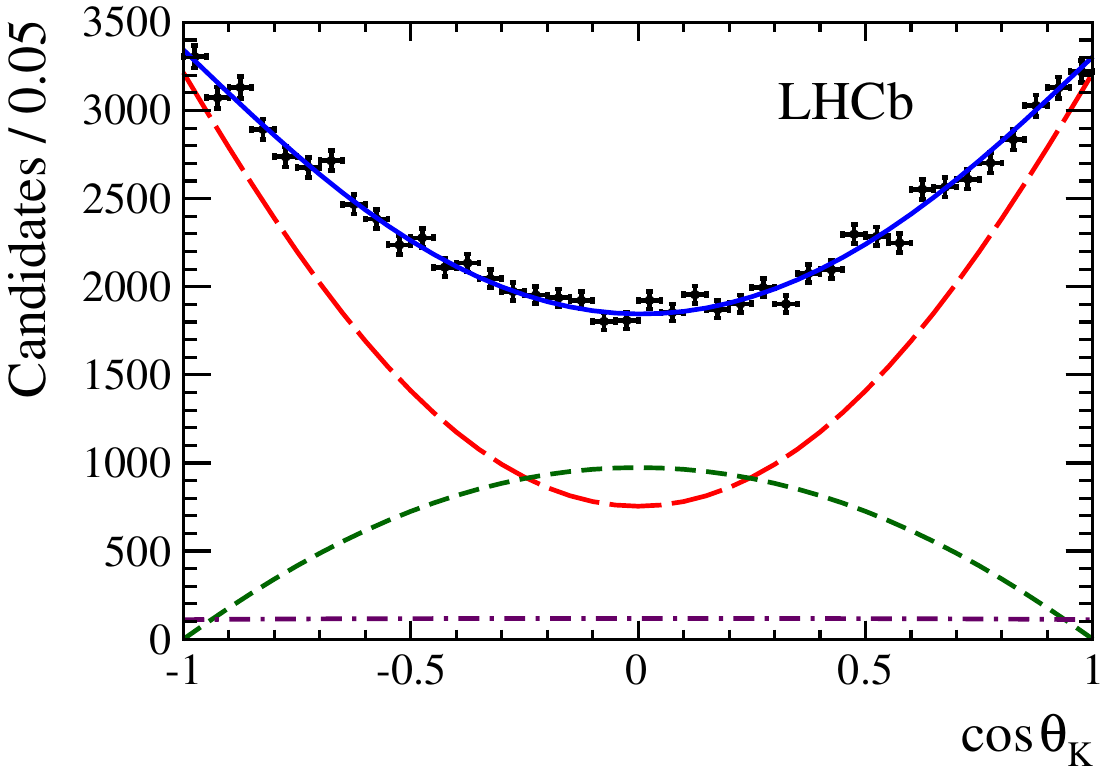}\\
  \includegraphics[width=0.40\textwidth]{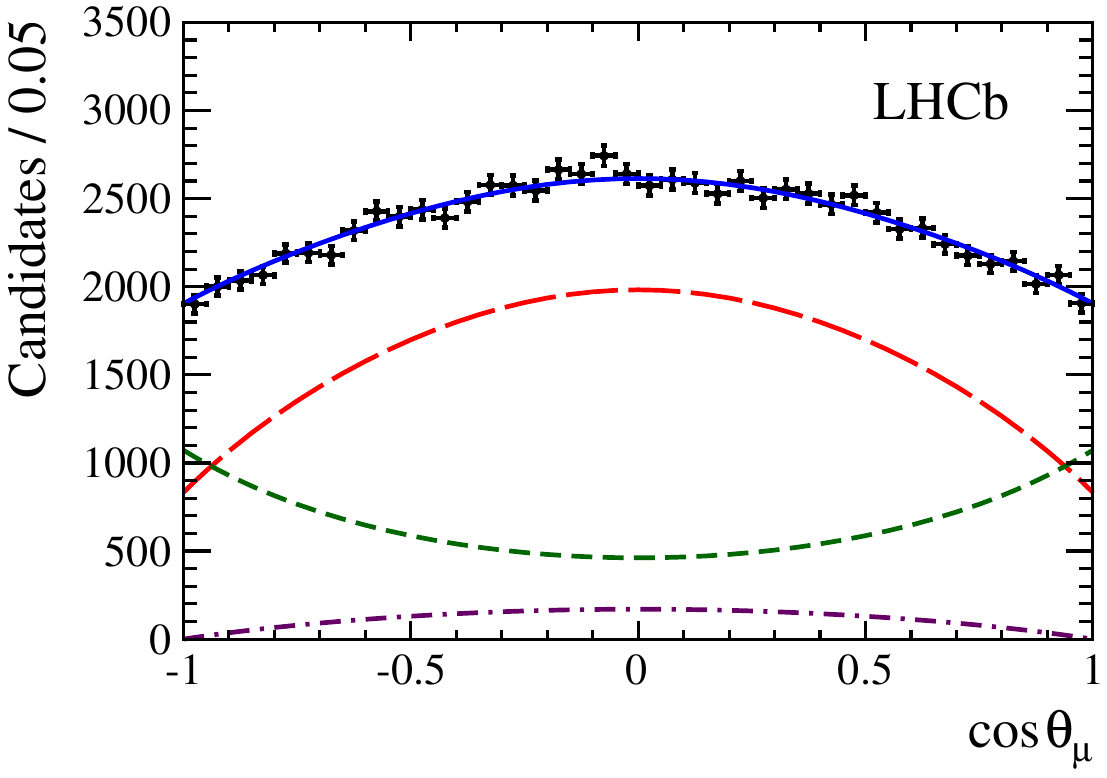}
  \includegraphics[width=0.40\textwidth]{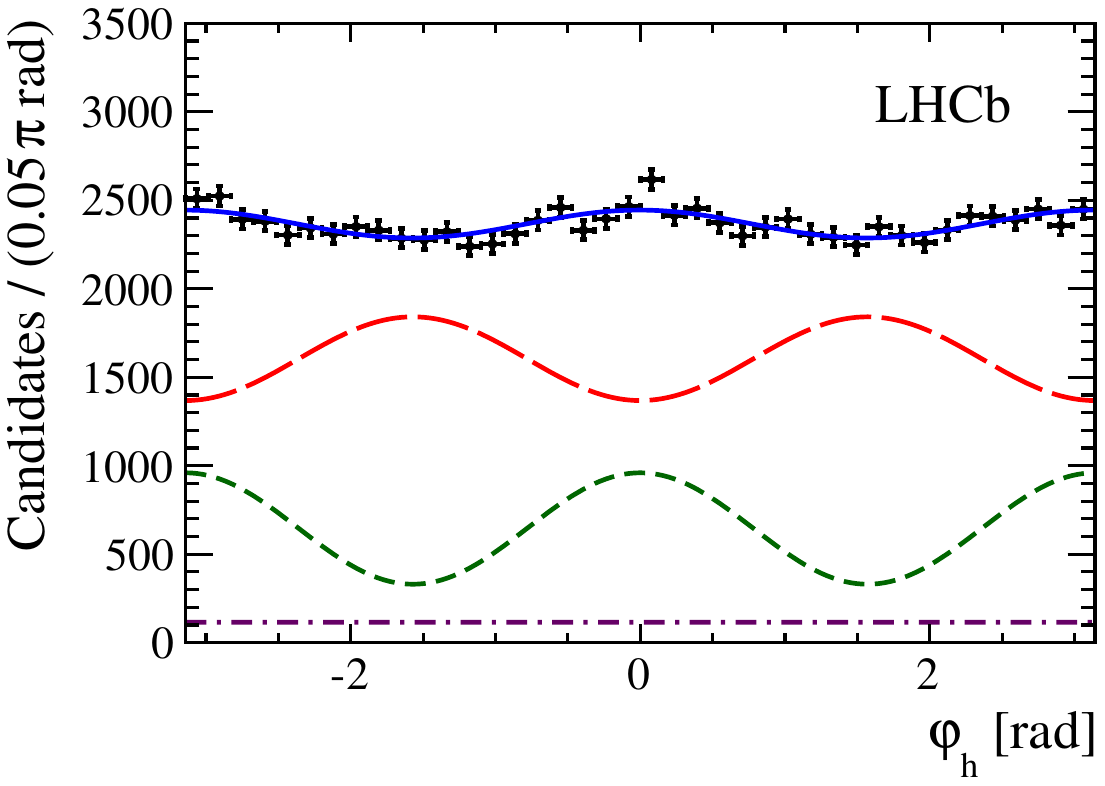}
}
 \caption{
Decay-time and helicity-angle distributions for $\Bs\to\jpsi K^+K^-$
decays (data points) with the one-dimensional fit projections overlaid.
The solid blue line shows the total signal contribution, which is composed
of CP-even (long-dashed red), CP-odd (short-dashed green) and S-wave (dotted-dashed purple) contributions. The figure is taken from Ref.~\cite{Aaij:2014zsa}.
}
\label{fig:lhcbphis}
\end{figure}

\begin{table}[t]
\caption{\small Values of the principal physics parameters determined from the LHCb 
                polarisation-independent analysis of $\Bs\to\jpsi\phi$ in \cite{Aaij:2014zsa}.}
\begin{center}
    \begin{tabular}{l c}
                                Parameter    &     Value\\
     \hline
      \rule{0mm}{4.5mm}  $\Gamma _s$ $[\rm ps^{-1}]$       & $0.6603 \pm 0.0027    \pm  0.0015$  \\
      \rule{0mm}{4.5mm}  $\Delta\Gamma _s$ $[\rm ps^{-1}]$ & $0.0805 \pm 0.0091    \pm  0.0032$   \\
      \rule{0mm}{4.5mm}  $\aperpsq$       		   & $0.2504 \pm 0.0049    \pm  0.0036$  \\
      \rule{0mm}{4.5mm}  $\azerosq$        		   & $0.5241 \pm 0.0034    \pm  0.0067$  \\
      \rule{0mm}{4.5mm}  $\delpar$~[rad]   	           & $3.26 \ ^{+0.10\ +0.06}_{-0.17\ -0.07}\ \ \, $ \\
      \rule{0mm}{4.5mm}  $\delperp$~[rad]  	           & $3.08 \ ^{+0.14}_{-0.15} \pm 0.06$    \\
      \rule{0mm}{4.5mm}  $\phi_s$~[rad]      	           & $-0.058 \pm 0.049     \pm  0.006$   \\
      \rule{0mm}{4.5mm}  $\maglambda$    	           & $\phantom{+}0.964     \pm  0.019  \pm  0.007$   \\
      \rule{0mm}{4.5mm}  $\dms$   $[\rm ps^{-1}]$  	   & $17.711\  ^{+0.055}_{-0.057} \pm0.011$   \\
     \hline
    \end{tabular}
\end{center}
\label{tab:phislhcb}
\end{table}

This decay mode has been studied also in the general purpose detectors at the
Tevatron  \cite{Aaltonen:2012ie,Abazov:2011ry}
and LHC  \cite{Aad:2014cqa,Khachatryan:2015nza}. 
The analysis method is similar to the one described before. 
Fig.~\ref{fig:cmsfits} shows the fit projections obtained with the recent CMS measurements 
reported in Ref.~\cite{Khachatryan:2015nza}. 

\begin{figure}[hbt]
\centering{
\includegraphics[width=0.3\textwidth]{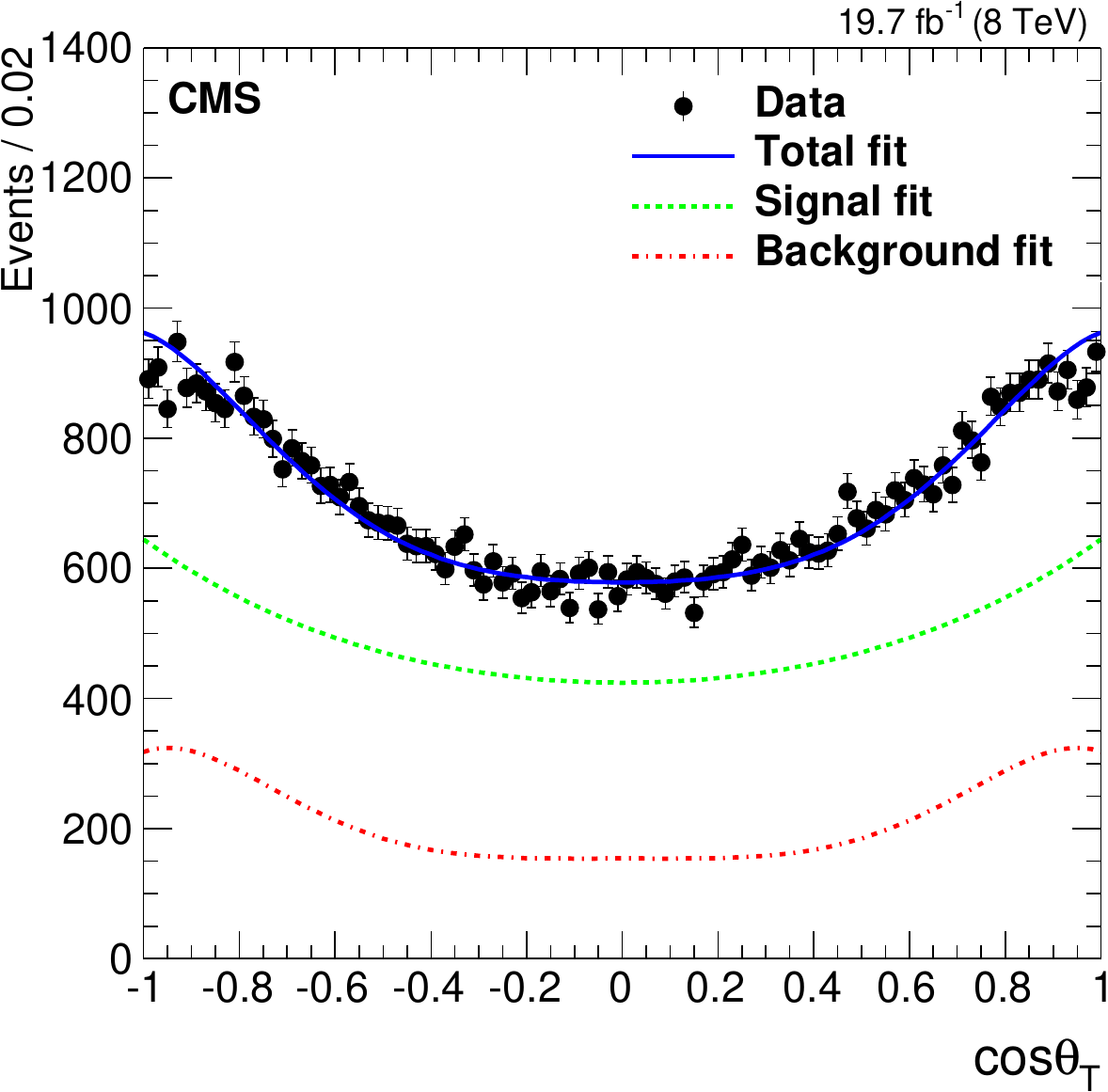}
\includegraphics[width=0.3\textwidth]{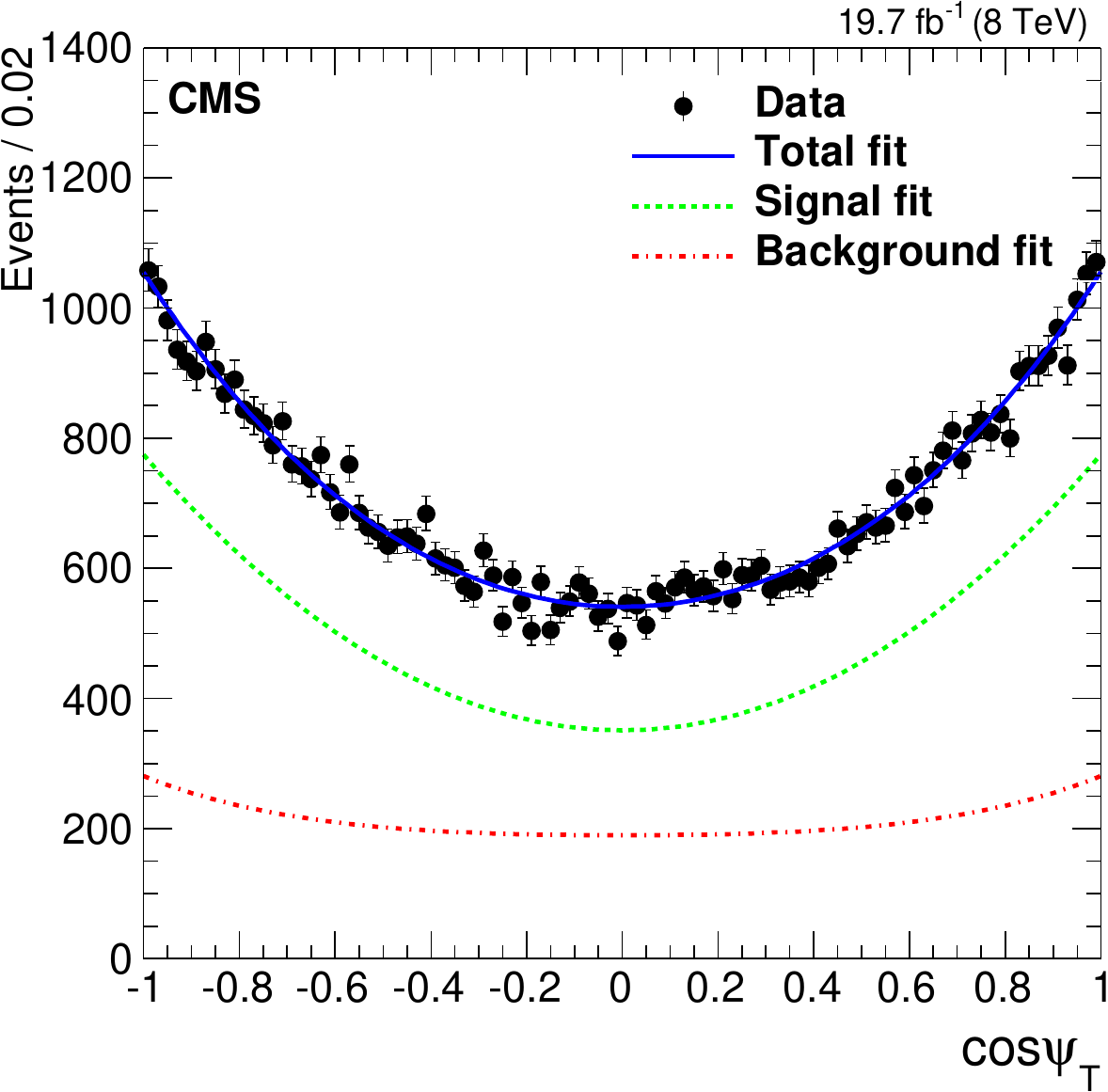}
\includegraphics[width=0.3\textwidth]{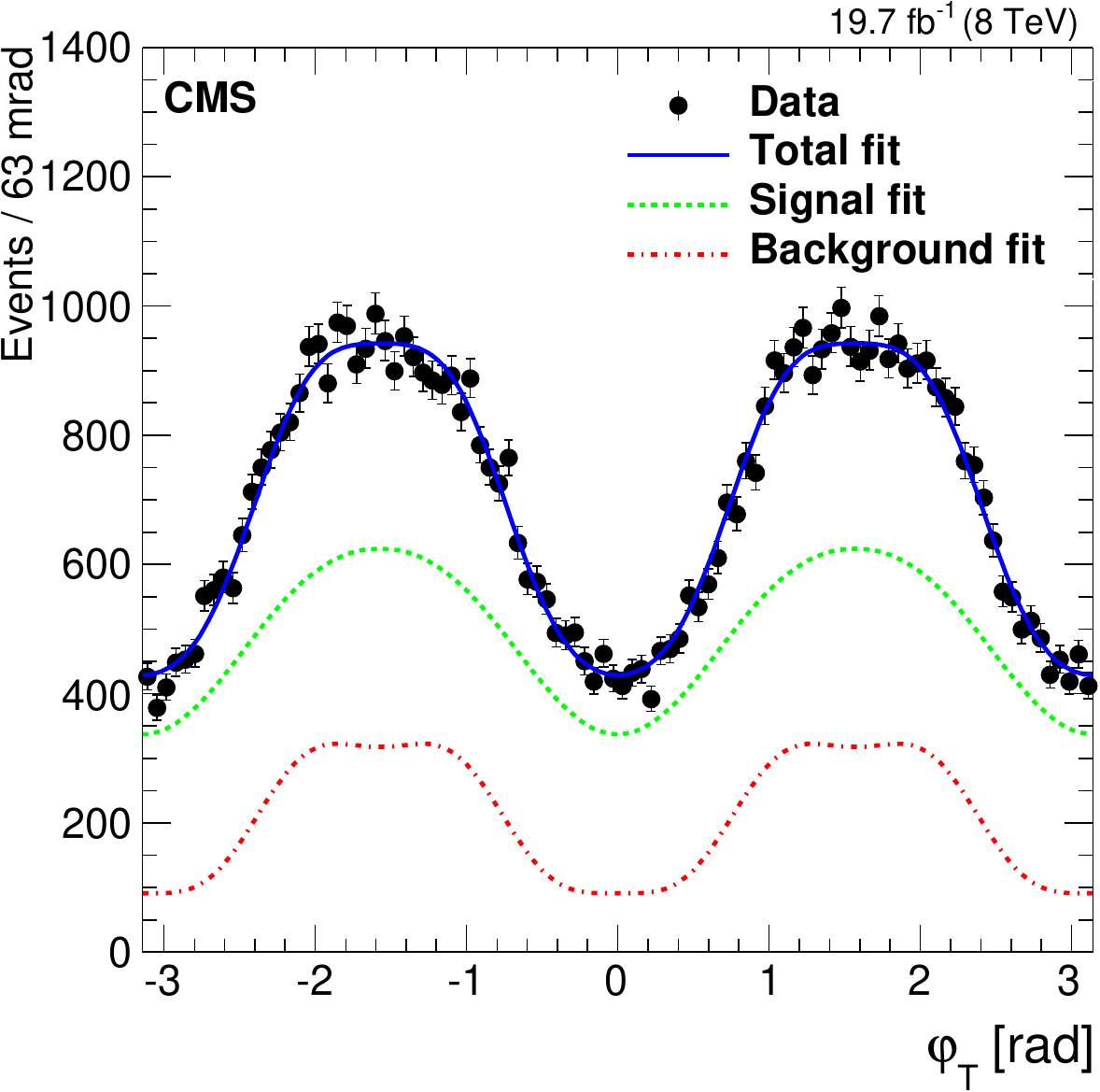}}
\caption{The angular distributions ($\cos\theta_\mathrm{T}$, $\cos\psi_\mathrm{T}$, $\varphi_\mathrm{T}$) of the \Bs{} candidates from data (solid markers). The angles $\theta_T $ and $\psi_T$ are the polar and azimuthal
angles, respectively, of the $\mu ^+$ in the rest frame of the $\jpsi$, where the $x$ axis is defined by the
direction of the $\phi$ meson in the $\jpsi$ rest frame, and the $x-y$ plane is defined by the decay
plane of the $\phi\to\Kp\Km$ decay. The helicity angle $\psi_\mathrm{T}$ is the angle of the $\Kp$ in the $\phi$ rest
frame with respect to the negative $\jpsi$ momentum direction. The solid line is the result of the fit, the dashed line is the signal fit, and the dot-dashed line is the background fit. The figure is taken from Ref.~\cite{Khachatryan:2015nza}.}
\label{fig:cmsfits}
\end{figure}

Another channel (see \cite{Stone:2008ak}) was recognised to provide complementary information 
on $\phi_s$, namely $\Bs\to\jpsi f_0$, with $f_0\to\pip\pim$.  The original appeal of this 
mode is that it was assumed to be predominantly an S-wave decay, and thus not in need of 
the complex multidimensional fit just described. The study of the Dalitz plot of 
$\Bs\to\jpsi\pip\pim$ \cite{LHCb:2012ae, Aaij:2014siy} revealed a more complex resonant structure.   
A combination of five resonant states are required to described the data (see \cite{Aaij:2014siy}):  
$f_0(980)$,  $f_0(1500)$, $f_0(1790)$, $f_0(1270)$, and  $f^\prime_2(1525)$. The data are 
compatible with no additional non-resonant (NR) component, as well as a combination of
 these 5 resonances plus significant NR component, with a fit fraction of (5.9$\pm$1.4)\%.  
The latter solution is shown in Fig.~\ref{5R-1NR}. Thus the most recent study of CP 
violation in $\Bs\to\jpsi\pip\pim$ uses the formalism developed in Ref.~\cite{Zhang:2012zk}. 
Their approach is to couple the three body Dalitz formalism applied to the final 
state $\jpsi\pip\pim$ with the time-dependent CP-violation analysis, by splitting 
the final state into CP-even and CP-odd components.  They perform an unbinned 
maximum likelihood fit to the $\jpsi\pip\pim$ invariant mass $m$, the decay time $t$, 
the di-pion invariant mass, the three helicity angles $\Omega$, along with flavor 
information of the decay hadron, namely whether it was produced as a $\Bs$ or $\barBs$. 
Assuming the absence of direct CP violation, the result is
$$\phi _s=75\pm 67\pm 8\ {\rm mrad},$$
while allowing for direct CP violation they obtain
$$\phi _s=70\pm 68\pm 8\ {\rm mrad}, |\lambda|=0.89\pm0.05\pm 0.01.$$

Another channel that provides an independent constraint on $\phi _s$, investigated by the LHCb experiment, 
is $\Bs\to \Dsp\Dsm$. This decay mode is particularly appealing because it is a CP even final state and, including two pseudo-scalar mesons in the final state, does not require an angular analysis. 
They obtain $\phi _s=(0.02\pm 0.17\pm 0.02)$ rad, see \cite{Aaij:2014ywt}.

\begin{figure}[!t]
      \includegraphics[width=3.7in]{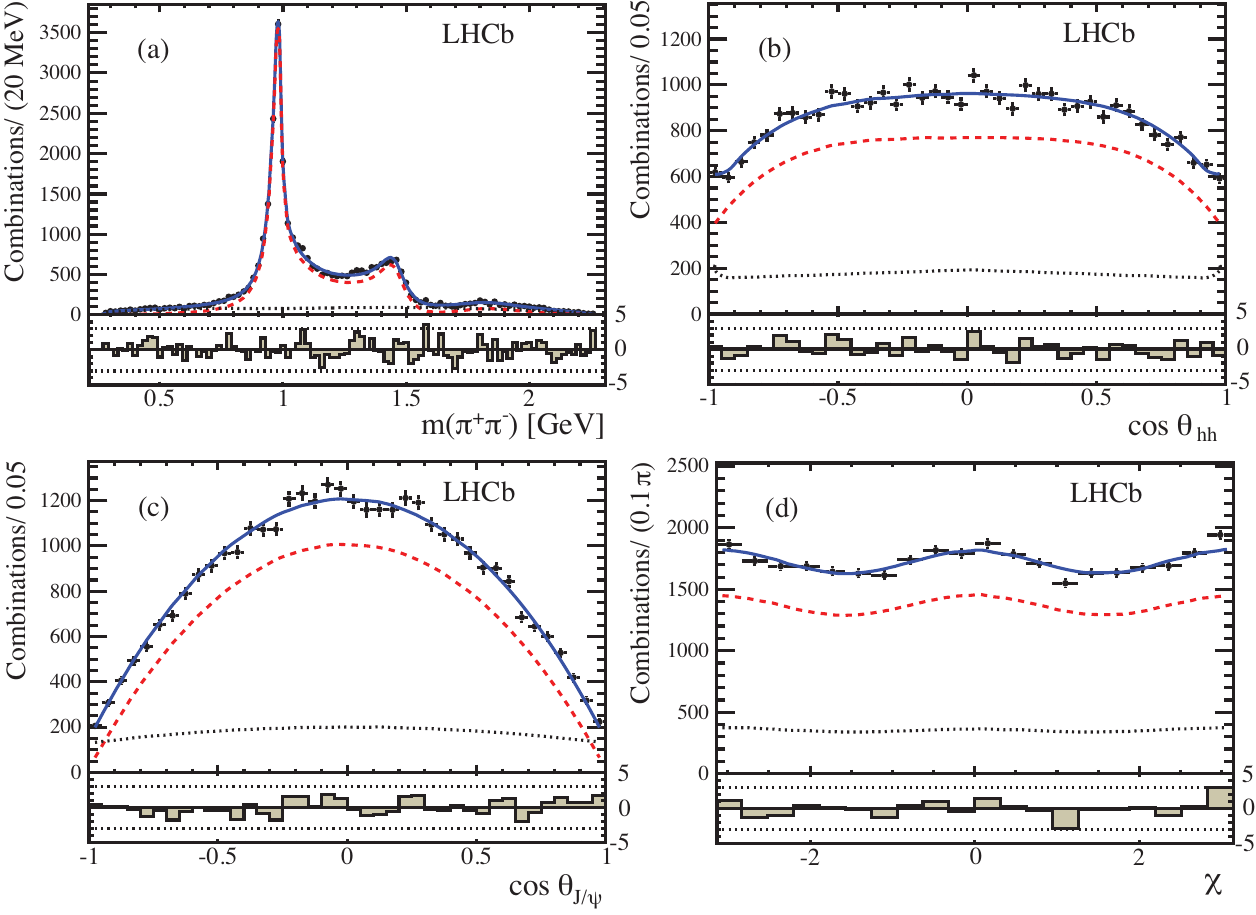}
 \caption{Projections of (a) $m(\pip\pim)$, (b) $\cos{\theta _{\pip\pim}}$, (c) $\cos{\theta_{\jpsi}}$ , (d) $\chi$ for the solution with the five resonance discussed in the text.
 The points with error bars represent data. The (red) dashed line represents the signal, the (black) dotted line represent the background, and the (blue) solid line represents the total fit. This figure is taken from Ref.~\cite{Aaij:2014emv}.}  \label{5R-1NR}
\end{figure}

\begin{table*}[t]
  \begin{center}
    \begin{tabular}{c c c  l}
      \hline
      Experiment & Mode & $\phi_s$ (rad)  & Ref. \\
      \hline
      CDF    & $\jpsi\phi$
       & $[-0.60,\, 0.12]$, 68\% CL
       & \cite{Aaltonen:2012ie} \\

\dzero & $\jpsi\phi$
       & $-0.55^{+0.38}_{-0.36}$  \
       & \cite{Abazov:2011ry} \\
ATLAS  & $\jpsi\phi$
       & $+0.12\pm 0.25 \pm 0.05$
       & \cite{Aad:2014cqa} \\
 ATLAS & $\jpsi\phi$
       & $-0.123\pm 0.089 \pm 0.041$
       & \cite{Aad:2016tdj} \\
CMS    & $\jpsi\phi$
       & $-0.075 \pm 0.097 \pm 0.031$
       & \cite{Khachatryan:2015nza}\\
LHCb   & $\jpsi K^+K^-$
       & $-0.058\pm0.049\pm0.006$
       & \cite{Aaij:2014zsa} \\
LHCb   & $\jpsi\pi^+\pi^-$
       & $+0.070 \pm0.068 \pm 0.008$         & \cite{Aaij:2014dka} \\
LHCb   & $\jpsi h^+h^-$
       & $-0.010\pm0.039(\rm tot)$
       & \cite{Aaij:2014zsa}$^a$ \\
LHCb   & $D_s^+D_s^-$
       & $+0.02 \pm0.17 \pm 0.02$
       & \cite{Aaij:2014ywt} \\
\hline
\multicolumn{2}{c}{All combined (HFAG 2016)} & $-0.033 \pm 0.033$  & ~ \\
\hline
\multicolumn{4}{l}{$^a$ {\footnotesize LHCb combination of $\jpsi K^+K^-$~\cite{Aaij:2014zsa} and $\jpsi\pi^+\pi^-$~\cite{Aaij:2014dka}.}}\\[-0.8ex]
\end{tabular}
        \caption{Measurements of the mixing phase $\phi_s$ in different $b \to c\bar{c} s$ channels, like $\Bs \to \jpsi \phi$, 
                  $\Bs \to \jpsi K^+ K^-$, $\Bs \to \jpsi \pi^+ \pi^-$, $\Bs \to \jpsi h+ h^-$  and  $\Bs \to D_s^+ D_s^-$. The Standard Model expectation (neglecting penguins)
                 for the phase $\phi_s$ reads $-0.0366 \pm 0.0020$.
    }
    \label{tabphi}
  \end{center}
\end{table*}

The combination of all the $\phi_s$ measurements performed by HFAG \cite{Amhis:2014hma} in the Spring 2016 gives
\begin{eqnarray}
\phi_s & = & (-0.033 \pm 0.033) \, {\rm rad} \, ,
       \\
       & = & (-1.89 \pm 1.89)^\circ
\,.
\end{eqnarray}
Individual experimental results are summarised in Table \ref{tabphi} and 
in Fig. \ref{phis_hfag2016}. 
The  current experimental uncertainty in $\phi_s$ is commensurate with the central value
of the SM prediction given in Eqs.(\ref{phisbetas},\ref{betas}).

\begin{figure}
\centering{
\includegraphics[width=0.45\textwidth]{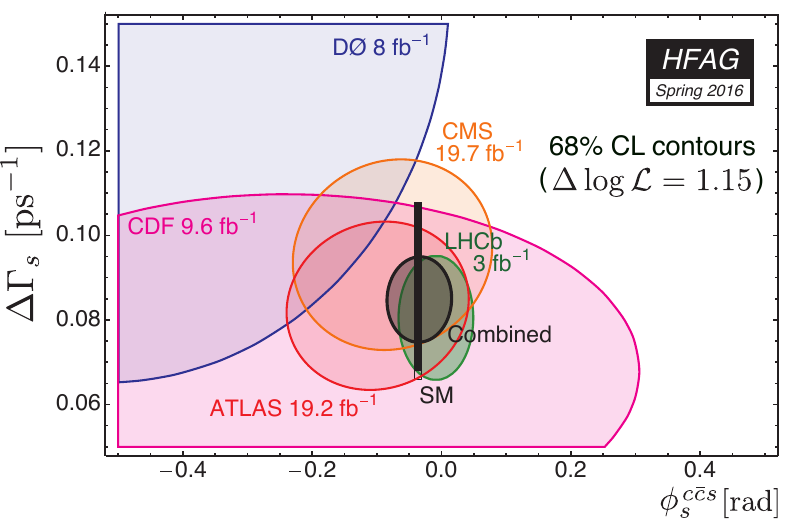}
}
\caption{
68\% CL regions in $\Bs$ width difference $\Delta \Gamma_s$ and weak phase $\phi_s$
obtained from individual and combined CDF, D0, ATLAS, CMS, and LHCb 
likelihoods of $\Bs \to \jpsi \phi$, $\Bs \to \jpsi K^+ K^-$, and $\Bs \to \jpsi \pi^+ \pi^-$
The expectation within the
Standard Model is shown as the black rectangle. The figure is taken from Ref. \cite{Amhis:2014hma}.
}

\label{phis_hfag2016}
\end{figure}

The average value of $\phi _s$ is consistent with the Standard Model, but subtle effects 
produced by diagrams mediated by new particles may yet be uncovered. The level of precision required to improve upon current status 
requires the consideration of effects neglected so far, such as the penguin contributions described above. Thus experiments have 
started to investigate decays that may constrain such contributions. The first such measurement is the study of the decay $\Bd\to\jpsi\pip\pim$.
This mode has both penguin and tree diagrams shown in Fig.~\ref{FeyPenguin}. Theoretical models predict that in this case the penguin 
diagram is greatly enhanced with respect to the decay $\Bd\to\jpsi K_S$.
%
The two decays $\Bd\to\jpsi\rho$ and $\Bs\to\jpsi\phi$ are related by SU(3) symmetry  if we
also assume that the difference between the $\phi$ being mostly a singlet state, and the $\rho$ an
octet state causes negligible breaking. If we assume equality between the penguin amplitudes and the strong phases in the two decay
and neglecting higher order diagrams (see \cite{Aaij:2014vda}), LHCb finds the penguin phase to be $\delta^{\rm Peng} = (0.05\pm 0.56)^\circ=0.9\pm 9.8$ mrad.
At 95\% CL, the penguin contribution in the $\Bs\to\jpsi\phi$ decay is within the interval (-1.05,+1.18). Relaxing these assumptions changes the limits on the possible penguin
induced shift. Figure~\ref{pen-con} shows how $\delta^{\rm Peng}$ varies as a function of the difference in strong phases  between the two decays, $\theta - \theta ^\prime$, indicating that the
95\% CL limit on penguin pollution can increase to at most 1.2$^\circ$. The phase $\delta^{\rm Peng}$ is
proportional to the ratio between penguin amplitudes $a/a^\prime$. As this ratio varies over the interval 0.5 to 1.5, the limit on
the penguin shift spans the range ($\pm 0.9,\pm 1.8$), even allowing for maximal
breaking between $\theta$ and $\theta ^\prime$.
\begin{figure}[hbt]
  \begin{center}
     \includegraphics[width=3.5in]{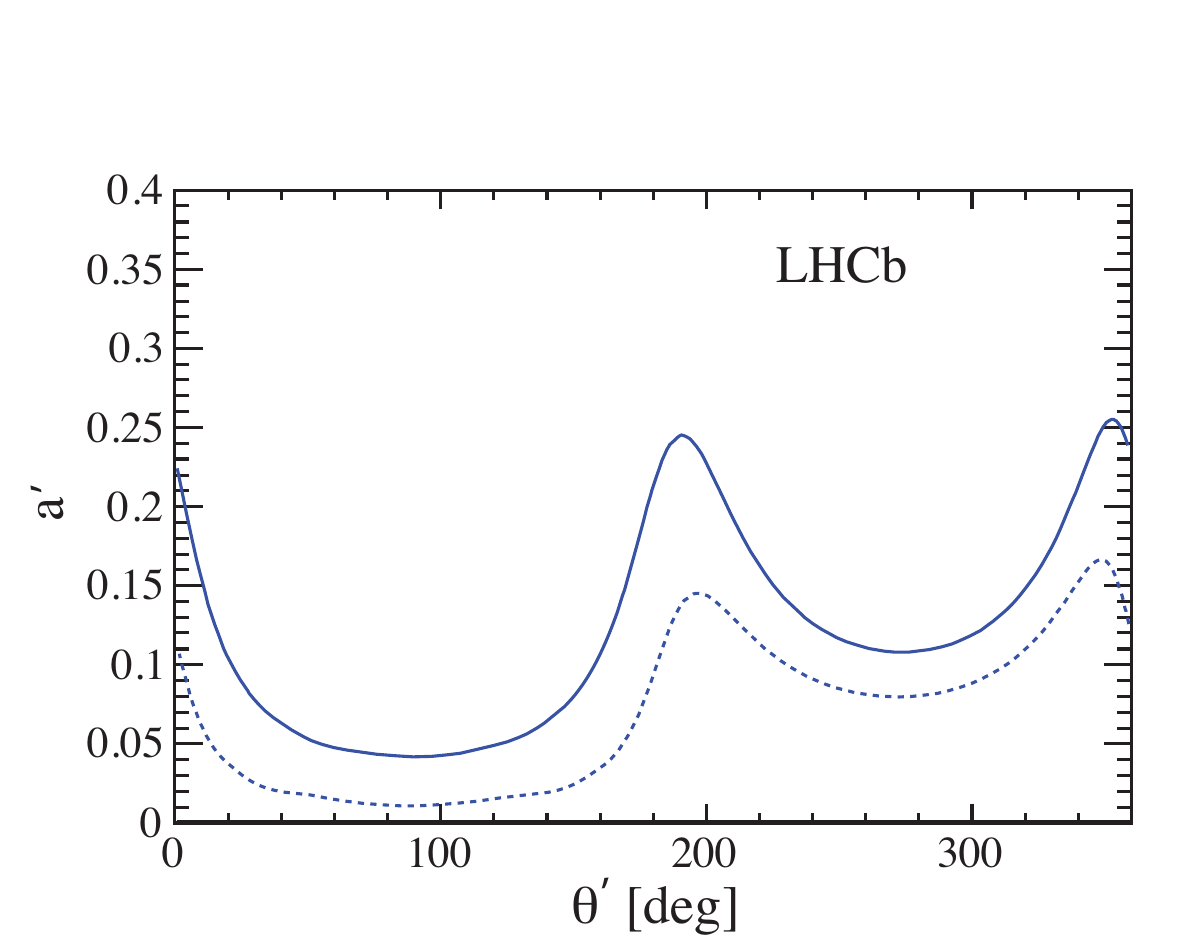}
     \vspace{-2mm}
    \caption{Contours corresponding to 68\% (dashed) and 95\% (solid) confidence levels for ndf of two, respectively, for the penguin amplitude parameters $a^\prime$ and $\theta^\prime$. The figure is taken from Ref.~\cite{Aaij:2014vda}.} \label{pen-con}
  \end{center}
\end{figure}
\begin{figure}[hbt]
\center
\includegraphics[width=2.7in]{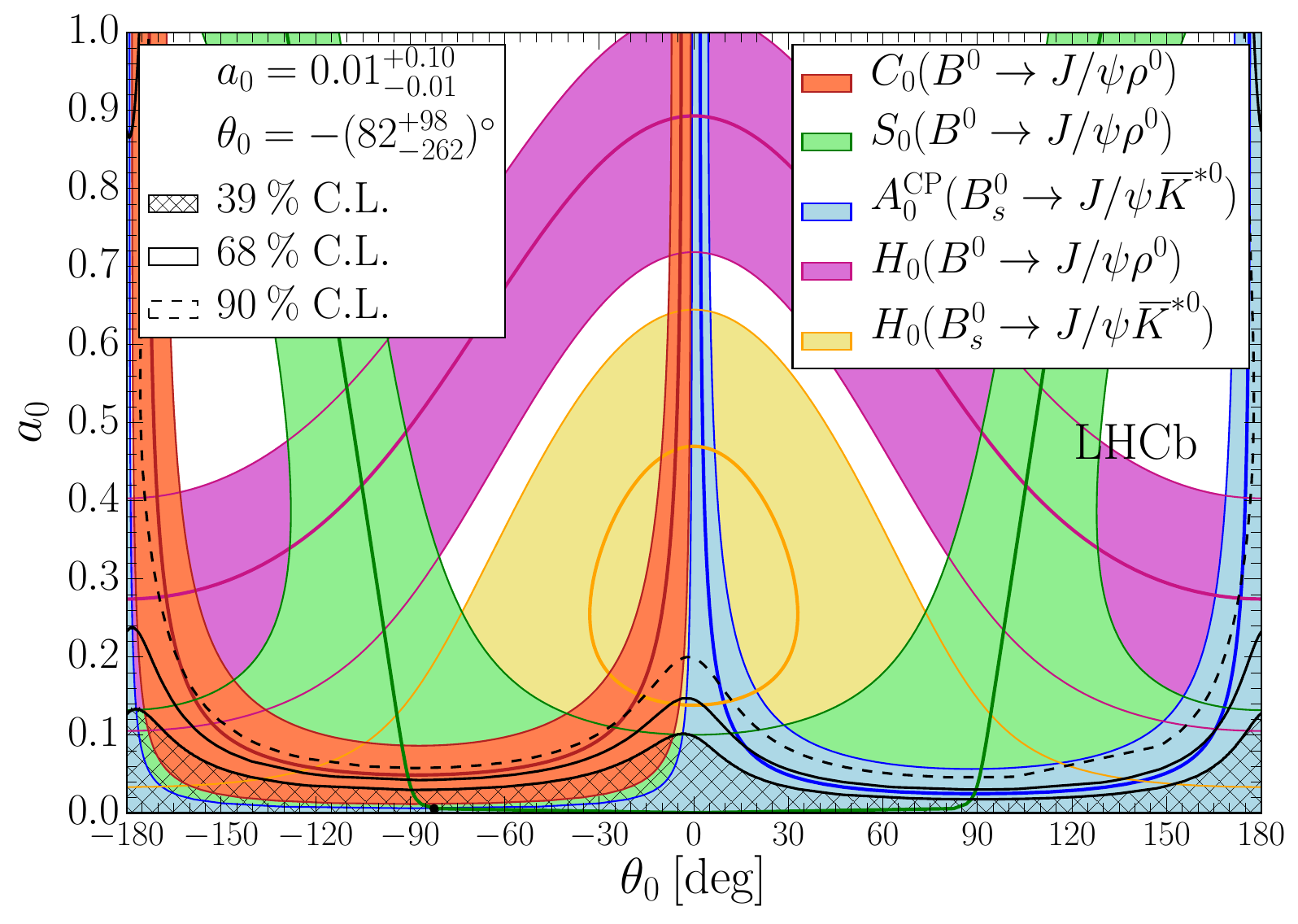}
\includegraphics[width=2.7in]{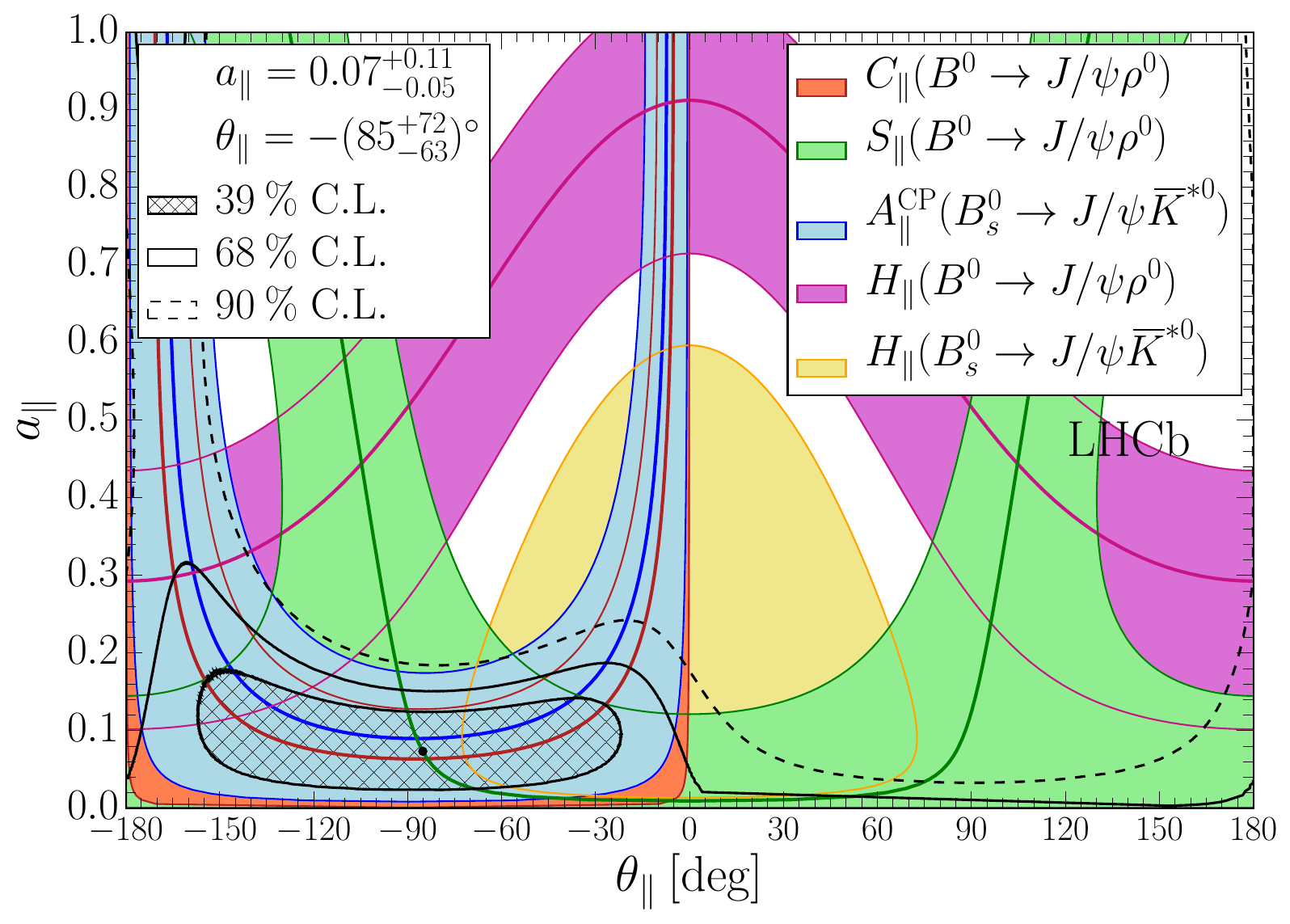}
\includegraphics[width=2.7in]{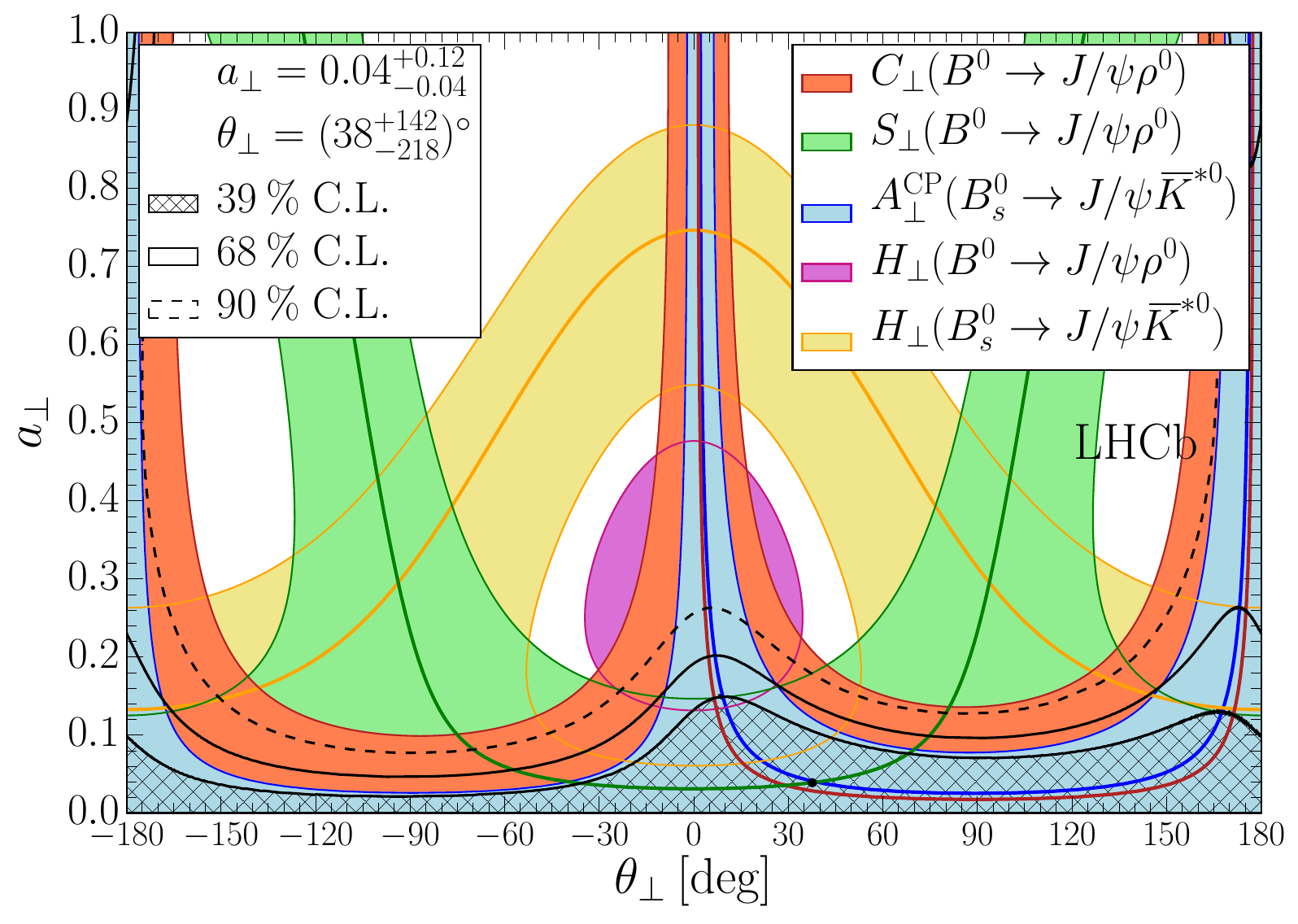}
\caption{Limits on the penguin parameters $a_i$ and $\theta_i$ obtained from intersecting contours derived from the \CP\  asymmetries and branching fraction information in $\Bs\to
\jpsi\bar{K}^{\star 0}$ and $\Bd\to\jpsi\rho^0$.
Superimposed are the confidence level contours obtained from a $\chi^2$ fit to the data.
The longitudinal (top), parallel (middle) and perpendicular (bottom) polarisations are shown. This figure is taken from Ref. \cite{Aaij:2015mea}.}
\label{fig:penguin}
\end{figure}

 A complementary approach is based on the study of the polarisation-dependent decay amplitudes of the decay $\Bs\to
\jpsi\bar{K}^{\star 0}$ \cite{Aaij:2015mea}.The results of  Ref.~\cite{Aaij:2014vda} and Ref.~\cite{Aaij:2015mea} are combined to produce the limits on penguin pollution shown in Fig.~\ref{fig:penguin}.

Finally, the decay $\Bs\to\phi\phi$ is analogous to $\Bs\to\jpsi\phi$, but is forbidden at tree level. It proceeds mostly via the $b\to s s\bar{s}$ penguin, 
thus providing an excellent probe for the manifestation  of interference of new physics particles with the penguin loop.  CP 
violation in this decay has been studied by LHCb \cite{Aaij:2013qha}. They perform an unbinned maximum likelihood fit to $d\Gamma/(dtd\cos{\theta _1}d\cos{\theta _2}d\Phi)$, 
where $t$ is the decay time and  $\theta _{1,2}$ is the angle between the $\Kp$ track momentum in the $\phi_{1,2}$ meson rest frame and the $\phi _{1,2}$ 
meson parent momentum in the $\Bs$ rest frame, and $\Phi$ is the angle between the two $\phi$ decay planes. The background is taken into account by 
assigning a weight to each candidate derived with an sPlot technique (see \cite{Pivk:2004ty}), using the invariant mass of the four $K$ system as a discriminating variable. 
The resulting fit
projections are shown
in Fig.~\ref{fig:cpvtwophi}. The CP-violating phase is found to be in the interval $[-2.46,-0.76]$ rad at 68\% confidence level. 
The $p$-value of the SM prediction is 16\%. The precision of the $\phi _s$ determination is dominated by the statistical uncertainty and is expected 
to improve with more data. The current results are based on a sample of 1 fb$^{-1}$.

\begin{figure}[ht]
\centering{
  \includegraphics[width=0.28\textwidth]{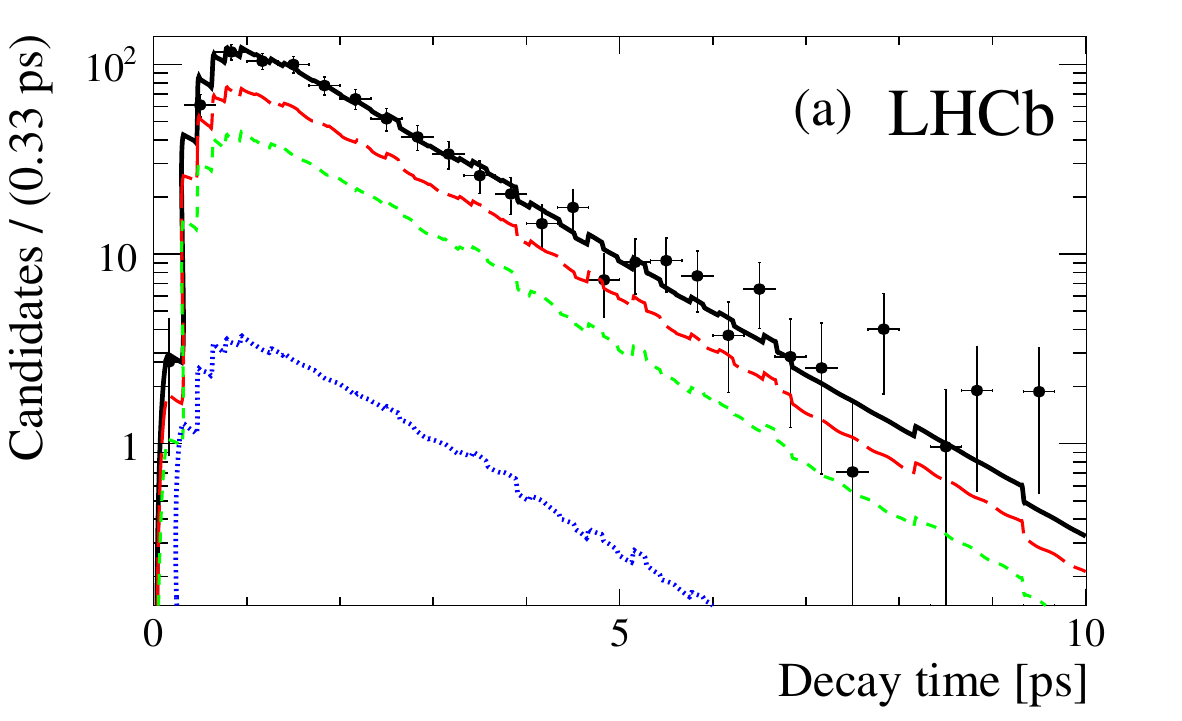}
  \includegraphics[width=0.28\textwidth]{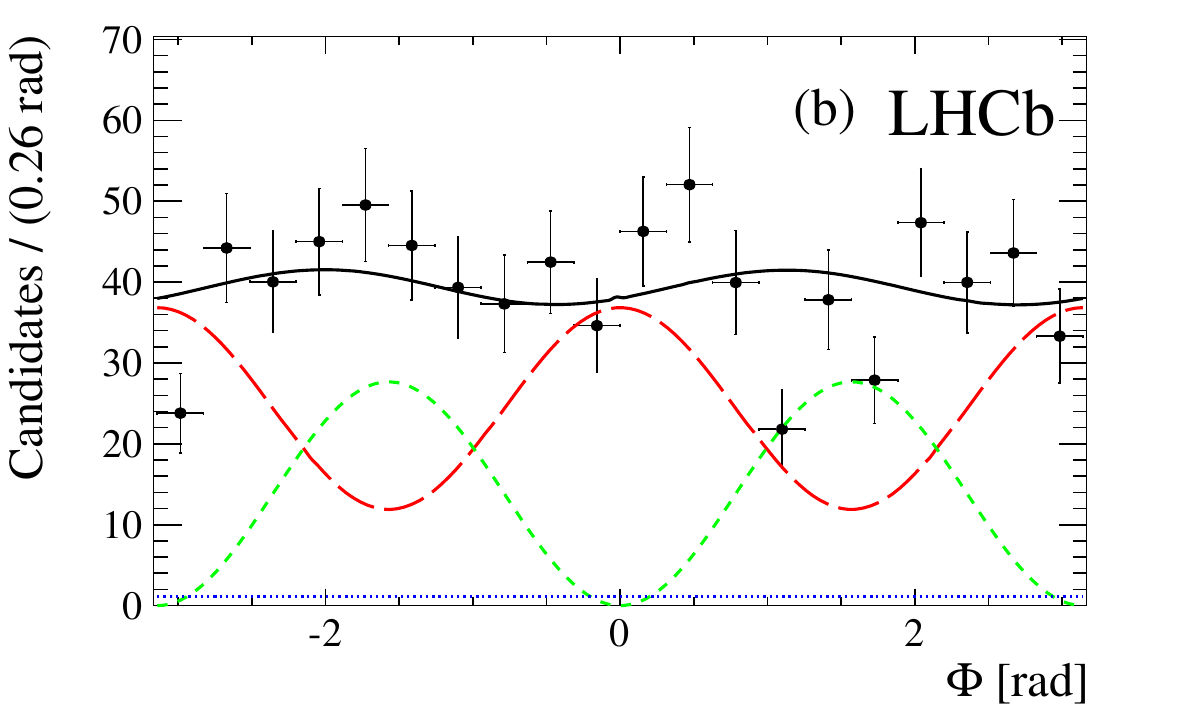}\\
  \includegraphics[width=0.28\textwidth]{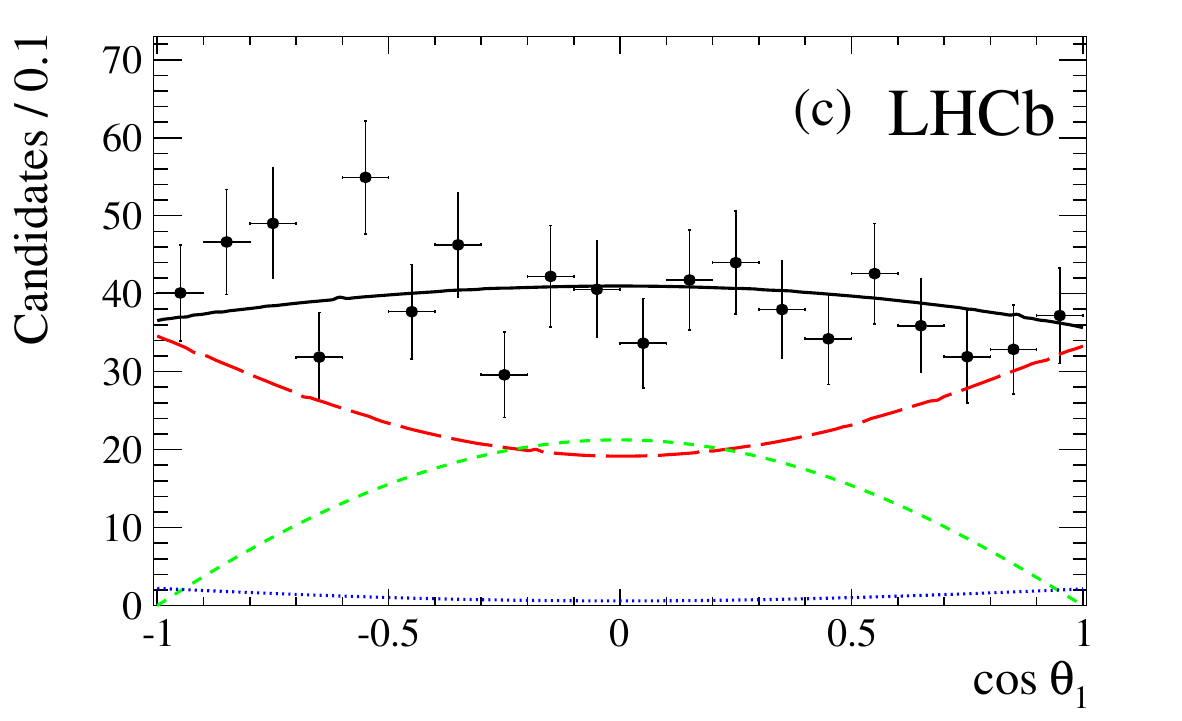}
  \includegraphics[width=0.28\textwidth]{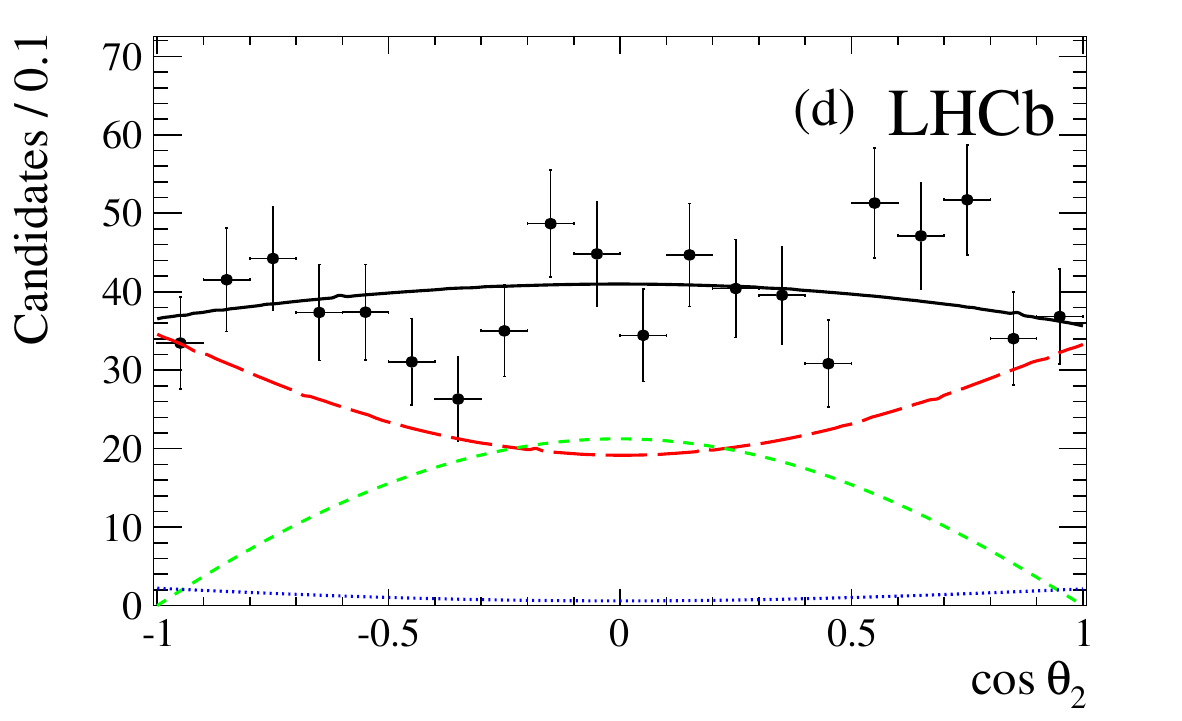}
}
 \caption{One-dimensional projections of the $\Bs\to\phi\phi$ fit for:
(a) decay-time, (b) helicity angle $\phi$, (c) and (d) cosines of the helicity-angles $\theta _1$ and $\theta _2$ respectively. The data are represented as points, with the one-dimensional fit projections overlaid.
The solid blue line shows the total signal contribution, which is composed
of CP-even (long-dashed red), CP-odd (short-dashed green) and S-wave (dotted-blue) contributions. The figure is taken from Ref.~\cite{Aaij:2013qha}.
}
\label{fig:cpvtwophi}
\end{figure}

\section{CP violation in decays and direct measurements of $\gamma$}
\label{CPVdecay}

\subsection{Theory}
CP violation in decays, also called {\it direct CP violation},  can arise
if we have $|{\cal{A}}_{f}| \neq |\bar{\cal A }_{\bar{f}}|$.
In that case we expect  the following CP asymmetry
\begin{equation}
A_{{\rm dir.CP},f}(t) = 
\frac{\Gamma \left( \barBs(t) \to \bar{f} \right) 
      - 
      \Gamma \left( \Bs(t) \to f \right)}
     {\Gamma \left( \barBs(t) \to \bar{f} \right) 
      + 
      \Gamma \left( \Bs(t) \to f \right)}
\; ,
\label{CPasydir}
\end{equation}
to give a non-vanishing value. Inserting the time evolution for the 
decay rates from Eq.(\ref{GammaBf}) and Eq.(\ref{GammabarBbarf}), we get 
a complicated expression, that vanishes, however for
$|{\cal A}_{f}|     = |\bar{\cal A}_{\bar{f}}|$, 
$|\bar{\cal A}_{f}| = |{\cal A}_{\bar{f}}|$ and neglecting terms 
of order $a_{\rm fs}^s$.
Neglecting mixing in a first step, 
i.e. setting $\Delta M_s$ and $\Delta \Gamma_s$ equal to 
zero, we get the simplified expression
\begin{equation}
A_{{\rm dirCP},f}(t) = 
\frac{\left|\bar{\cal A}_{\bar{f}}\right|^2 - 
      \left|  {  \cal A}_{f}      \right|^2}
     {\left|\bar{\cal A}_{\bar{f}}\right|^2 +
      \left|  {  \cal A}_{f}      \right|^2} 
\; .
\label{CPasydir2}
\end{equation}
Using the definitions in Eq.(\ref{amplitude_2})
and in  Eqs.(\ref{amplitude_3},\ref{amplitude_4}) we can write the 
two amplitudes as
\begin{eqnarray}
{\cal A}_f & = & 
\left|{{\cal A}}^{\rm Tree}_f\right| 
e^{i \left[\phi^{\rm QCD}_{\rm Tree} + \arg(\lambda_c)\right]} 
+
\left|{{\cal A}}^{\rm Peng}_f\right| 
e^{i \left[\phi^{\rm QCD}_{\rm Peng} + \arg(\lambda_u) \right]} 
 \; ,
\nonumber
\\
\\
\bar{\cal A}_{\bar{f}} & = & 
\left|{{\cal A}}^{\rm Tree}_f\right| 
e^{i \left[\phi^{\rm QCD}_{\rm Tree} - \arg(\lambda_c)\right]} 
+ 
\left|{{\cal A}}^{\rm Peng}_f\right| 
e^{i \left[\phi^{\rm QCD}_{\rm Peng} - \arg(\lambda_u) \right]} 
 \; 
\nonumber
\\
\end{eqnarray}
and we find for $A_{{\rm dirCP},f}(t)$ the following expression
\begin{widetext}
\begin{eqnarray}
A_{{\rm dirCP},f}(t) & = &  
\frac{2 \left|{{\cal A}}^{\rm Tree}_f\right| 
        \left|{{\cal A}}^{\rm Peng}_f\right| 
        \sin \left( \phi^{\rm QCD}_{\rm Peng}-\phi^{\rm QCD}_{\rm Tree}\right)
        \sin \left[ \arg(\lambda_u) - \arg(\lambda_c)\right]
     }
     {\left|{{\cal A}}^{\rm Tree}_f\right|^2+
      \left|{{\cal A}}^{\rm Peng}_f\right|^2+ 
       2 \left|{{\cal A}}^{\rm Tree}_f\right| 
        \left|{{\cal A}}^{\rm Peng}_f\right| 
        \cos \left( \phi^{\rm QCD}_{\rm Peng}-\phi^{\rm QCD}_{\rm Tree}\right)
        \cos \left[ \arg(\lambda_u) - \arg(\lambda_c)\right]
     }
\nonumber
\\
& = &
\frac{2 |r|
        \sin \left( \phi^{\rm QCD}_{\rm Peng}-\phi^{\rm QCD}_{\rm Tree}\right)
        \sin \left[ \arg(\lambda_u) - \arg(\lambda_c)\right]
     }
     {1 +|r|^2 + 2 |r| 
        \cos \left( \phi^{\rm QCD}_{\rm Peng}-\phi^{\rm QCD}_{\rm Tree}\right)
        \cos \left[ \arg(\lambda_u) - \arg(\lambda_c)\right]
     }\; ,
\end{eqnarray}
\end{widetext}
where $|r|$ gives the modulus of the ratios of the penguin amplitude
and the tree amplitude, analogous to Eq.(\ref{r}).
This simplified formula, that holds only in the absence of mixing, shows that
we can have a direct CP violation in decay only, if we have at least 
two different
CKM contributions with different weak and different strong phases.
The size of the CP asymmetry is also proportional to the modulus of the 
penguin contributions normalised to the tree contributions.
Thus such a asymmetry could in principle, e.g. arise in the decays 
$\Bs \to K^- \pi^+$ and $\barBs \to K^+ \pi^-$ (see Fig.~\ref{penguincont}), 
where we expected large penguins.
Using the definition of the CKM angle $\gamma$
\begin{equation}
\gamma = \arg \left( - \frac{V_{ud} V_{ub}^*}{V_{cd} V_{cb}^*} \right) \;
,
\label{gamma_def}
\end{equation}
we can write to a very good approximation
\begin{eqnarray}
A_{{\rm dirCP},f}(t) & = &  
\frac{2 |r|
        \sin \left( \phi^{\rm QCD}_{\rm Peng}-\phi^{\rm QCD}_{\rm Tree}\right)
        \sin  \gamma 
     }
     {1 +|r|^2 - 2 |r| 
        \cos \left( \phi^{\rm QCD}_{\rm Peng}-\phi^{\rm QCD}_{\rm Tree}\right)
        \cos \gamma
     }\; .
\nonumber
\\
\end{eqnarray}
If $|r|$ and the strong phases were known,  this direct CP asymmetry
could be used to determine the CKM angle $\gamma$. We already pointed out 
several times the difficulty of  determining these hadronic parameters from
a first principle calculation. Further strategies to determine $\gamma$ will
be discussed below.
On the other hand, using a measured value of $\gamma$,  the direct 
CP asymmetry can give indications about the size of hadronic parameters, 
which is a very useful input in the investigation of penguin 
pollution.
\\
Another possibility in the search for direct CP violation
is the investigation of final states, that are common to $\Bs$ and $\barBs$, 
as e.g. in $\Bs \to \jpsi \phi$ or $\Bs \to K^+ K^-$. 
According to the definition of the asymmetry in Eq.(\ref{CPasyminter}) 
the coefficient of $\cos(\Delta M_s t)$ will be proportional to 
$A_{\rm CP}^{\rm dir}$, which describes direct CP violation and which is 
non-zero if $|\lambda_f| \neq 1$. Here again the ratio $r$ will be the crucial
parameter.
\\
It is worth also mentioning that $\Bs$ decays provide also 
information about the CKM phase
$\gamma$, which was defined in Eq.~\ref{gamma_def}. 
This phase is directly proportional to the amount of CP violation
in the SM. Thus any measurement of $\gamma$ is a measurement of CP violation.
\\
In the case of the tree-level decay $\Bs \to D_s^\pm K^\mp$ the 
extraction of $\gamma$ 
was e.g. discussed in 
\cite{Dunietz:1987bv,Aleksan:1991nh,Fleischer:2003yb,Fleischer:2011ne,Gligorov:2011id,DeBruyn:2012jp}.
$\Bs \to D_s^+ K^-$ proceeds via a colour-allowed $\bar{b} \to \bar{u} c \bar{s}$ transition
and
$\Bs \to D_s^- K^+$ proceeds via a colour-allowed $\bar{b} \to \bar{c} u \bar{s}$ transition, see 
Fig.~{\ref{BstoDsK}.
\begin{figure*}
\includegraphics[width=0.45 \textwidth,angle=0]{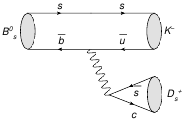}
\hfill
\includegraphics[width=0.45 \textwidth,angle=0]{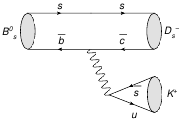}
\caption{\label{BstoDsK} Tree-level contribution to the decays $\Bs \to D_s^+ K^-$ and
                         $\Bs \to D_s^- K^+$.
                         Both diagrams are colour-allowed and their CKM structure is similar
                         in size, although the difference of the CKM phases is given by the CKM angle
                         $\gamma$.}
\end{figure*}
Doing a naive counting of powers of the Wolfenstein parameter $\lambda$ one expects that both amplitudes
have a similar size, while the phase difference is given by the CKM angle $\gamma$, which is 
more or less the phase of the CKM element $V_{ub}$.
From the diagrams in Fig.~{\ref{BstoDsK} one sees that both the $\Bs$ and $\barBs$-meson can decay
into the same final state. Thus an interference between mixing and decay can arise, and in the end
the value of  $\phi_s+\gamma$ can be extracted from measuring CP asymmetries.
Such an extraction of $\gamma$ became very popular, using $\Bm\to D\Km$, because 
tree-level decays are supposed not to be  affected by new physics effects. 
In view of the increasing experimental precision this assumption should, however,  be challenged.
A recent study \cite{Brod:2014bfa} found  that current experimental bounds on different
flavour observables that are dominated by tree-level effects,
allow beyond SM effects to be as large as about 
$\pm 0.1$ in the tree-level Wilson coefficients $C_1$ and $C_2$. 
A new physics contribution to the imaginary part of $C_1$ of about $0.1$ would modify the 
measurement of $\gamma$ from tree-level decays by about $4^\circ$ \cite{Brod:2014bfa}, 
which is smaller than the current experimental uncertainty 
of $\gamma$ (about $7^\circ$ according to Eq.(\ref{gamma_exp})), but larger
than the expected future uncertainty of about $1^\circ$ \cite{LHCb:2011dta,Abe:2010gxa}. 
Here clearly more studies are necessary in order to constrain the possible space for new physics effects 
in tree level decays.
Currently $\Bs \to D_s^\pm K^\mp$ decays lead to value of $\gamma = {115^{+28}_{-43}}^\circ$
(see \cite{Aaij:2014fba}), which is not competitive.
An extraction of this angle from  $\Bd \to \pi^+ \pi^-$, $\Bs \to K^+ K^-$ and 
$B_{d,s} \to \pi^\pm K^\mp$ decays, which have also loop contributions 
was e.g. discussed in 
\cite{Ciuchini:2012gd,Fleischer:2007hj,Fleischer:2010ib,Fleischer:1999pa}.
Assuming the SM value for $\beta_s$ and neglecting Standard Model penguins one gets a very precise
value of  $ \gamma = {63.5^{+7.2}_{-6.7}}^\circ$ (see \cite{Aaij:2013tna}).
For this decay the usual argument about the theoretical cleanliness of the extraction
does, however, not hold.
Finally \cite{Bhattacharya:2015uua} discussed also the extraction of the CKM angle $\gamma$ 
from three-body decays $\Bd, \Bs \to K_S h^+ h^-$ with
$(h= K, \pi)$. 

\subsection{Experiment}

The discovery of CP violation in charmless two-body decays of $\Bd$ and 
$B^+$ mesons by the BaBar and Belle experiments provide very interesting 
data, whose impact is difficult to ascertain in view of the challenges in 
determining precisely the hadronic matrix element relating the observed 
asymmetries with fundamental phases. The first observables of interests 
are the direct CP asymmetries. So far flavor SU(3) symmetry has been used 
to provide at least a theoretical framework to related such asymmetries 
measured in different decays. First principle calculations of the hadronic 
matrix elements involved will enable to fully exploit these measurements to 
test SM predictions. 
The study of  direct CP asymmetries  in  $\Bs$ decays provides valuable additional constraints.

The LHCb collaboration has measured CP violation asymmetries in $\Bs\to\Km\pip$ (\cite{Aaij:2012qe}) and $\Bs\to\Kp\Km$ (\cite{Aaij:2013tna}). 
These measurements share the same level of complexity as the measurements of asymmetries mediated by the interference between 
$\Bs-\barBs$ mixing and CP violation in direct decays: they require a determination of the flavor of the decaying $\Bs$, a time-dependent 
analysis to disentangle $A_{CP}$ from the $\Bs$ production asymmetry, in addition to a careful determination of all the instrumental asymmetries
 discussed before.  An important advantage that enables the LHCb experiment to perform these measurements with high precision is the excellent 
hadron identification efficiency and purity provided by the ring imaging Cherenkov (RICH) detectors (see \cite{Adinolfi:2012qfa}). As an illustration, 
Fig.~\ref{fig:invmass} shows the invariant mass spectra for different species of $B\to hh$ final states. There is excellent separation between different particle species.
\begin{figure}[tbh]
  \includegraphics[width=0.45 \textwidth,angle=0]{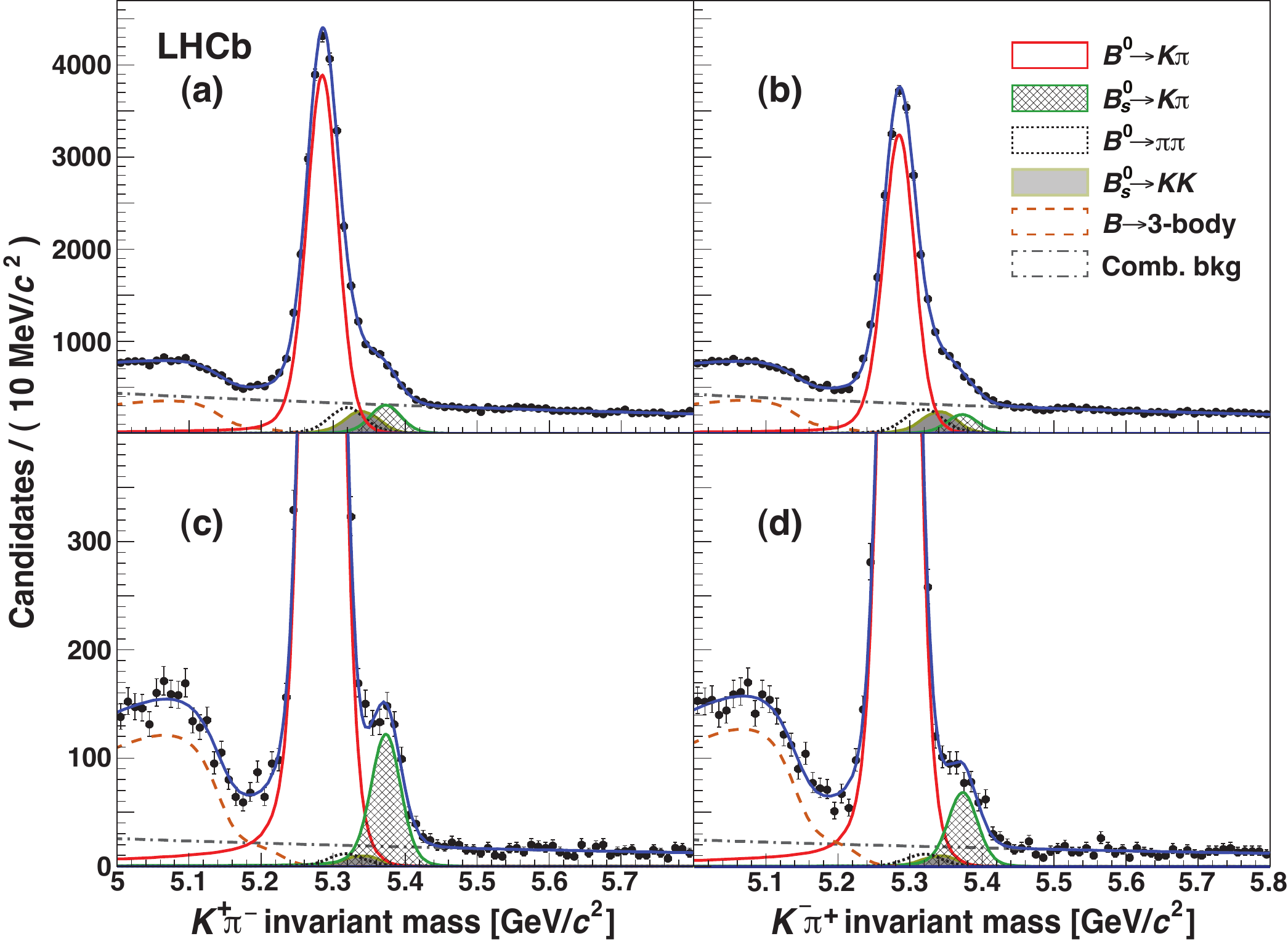}
  \caption{Panels
(a) and (b) show the invariant mass spectra obtained using the event selection adopted for the best sensitivity to $A_{CP}(\Bd\to\Kp\pim$; panels (c) and (d) 
show the  invariant mass spectra obtained using the event selection adopted for the best sensitivity to $A_{CP}(\Bs\to\Kp\pim$. Panels (a) and (c) show the $\Kp\pim$ invariant mass, while panels (b) and (d) the $\Km\pip$ invariant mass. The plot is taken from \cite{Aaij:2013iua}.
  }
  \label{fig:invmass}
\end{figure}

Using the formalism of Ref.~\cite{Aaij:2013iua}, the CP asymmetry is related to the raw asymmetry through
\begin{equation}
A_{CP}=A_{raw}-A_\Delta
\end{equation}
with
\begin{equation}
A_\Delta(\Bs\to \Kp\pim)=-A_D(\Kp\pim)+\kappa _s A_P(\Bs)
\end{equation}
where $A_D$ represents the detection efficiency asymmetry, that is derived from raw asymmetries measured for decays with known $A_{CP}$,  $\kappa _s=-0.033\pm 0.003$\cite{Aaij:2012qe}, 
and $A_P$ is the $\Bs-\barBs$ production asymmetry, derived from a fit to the time-dependent measured asymmetry.  The parameter $\kappa _s$ accounts for the dilution of the effect of the production asymmetry due to the fast $\Bs$ oscillations and is given by
\begin{equation}
\kappa _s= \frac{\int_{0}^{\infty}e^{-\Gamma _s t}\cos{(\Delta m_s t)}\epsilon(\Bs\to K\pi,t)dt}{\int_{0}^{\infty}e^{-\Gamma _s t}\cosh{(0.5 \Delta \Gamma_s t)}\epsilon(\Bs\to K\pi,t)dt}
\end{equation}
$A_P$ introduced an oscillatory component that makes 
it possible to measure the production asymmetry unambiguously.   Note that $A_P$ has a very marginal effect on $A_{CP}$ 
because the fast flavor oscillations greatly diminish the correlation between the flavor at  decay time with the flavor at production time. The final result is 
$A_{CP}(\Bs\to\Km\pip)=0.27\pm 0.04\pm 0.01$ where the first error is statistical and the second systematic. 

The study of the CP asymmetry and branching fraction of the decay $\Bs \to \Kp\Km$, combined with the knowledge of the corresponding observables in $\Bd\to\pip\pim$ 
can in principle be used to determine the CKM angle $\gamma$, defined in Eq.~(\ref{gamma_def}), or $-2\beta _s$, 
defined in Eq.~(\ref{betas}),
if U-spin be a valid symmetry of the strong interaction. The LHCb collaboration, using their 
measurements of $CPV$ observables in $\Bs\to\Kp\Km$, performs two analyses to determine either $\gamma$ or $\beta _s$ \cite{Aaij:2014xba}. Here we quote the 
first analysis, that uses the measured value of $\beta _s$ (and neglecting Standard Model penguins) to derive
\begin{equation}
\gamma=(63.5^{+7.2}_{-6.7})^\circ .
\label{gamma_exp}
\end{equation}
This value is consistent with the $\gamma$ value derived from tree-level decays. Further understanding of U-spin symmetry breaking as well as penguin pollution 
is needed to assess the impact of this measurement.

The decay $\Bs\to \Ds\Km$ is sensitive to the angle $\gamma$ of the Cabibbo-Kobayashi-Maskawa matrix. This is an example of a determination of $\gamma$ from a tree-level process, and thus, in  principle, not sensitive to effects induced by most new-physics models currently considered. 
Other such determinations of $\gamma$ from tree-level mediated processes have been performed at the $B$-factories and LHCb, 
through the study of $\Bd\to\Dm\pip$, and $\Bd\to\Dm\Kp$ decays. In these decays, 
the ratio $r_{D^{(\star)}}\equiv {\cal A}(\Bd\to D^{(\star)-}\pip)/\Bd\to D^{(\star)+}\pim/{\cal A}(\Bd\to D^{(\star)+}\pim)$ is small,
 $r_{D^{(\star)}}\approx 0.02$, while in the case of 
$\Bs\to\Dsp\Km$
the interfering amplitudes are of the same order of magnitude. Moreover, the decay width difference in the $\Bs$ system,
$\dg _s$, is non-zero,
which allows a determination of  $\gamma -2\beta _s$
from the sinusoidal and hyperbolic terms
in the decay time evolution, up to a two-fold ambiguity.

The measurement is sensitive to the combination $\gamma-2\betas$, and, 
as we have seen that $\beta _s$ is now measured with great precision from the study of $\Bs\to \jpsi h^+h^-$, if Standard Model penguins are neglected, 
it can be directly translated into a measurement of $\gamma$. This decay has been studied by the LHCb collaboration using 1 fb$^{-1}$ of data and the
 measurement requires a fit to the decay-time distribution of the selected $\Bs \to \Dsm\Kp$ candidates. 
It is a very challenging measurement because it requires the determination of the time-dependent efficiency, 
as well as the determination of the flavor of the decaying $B_s$. The kinematically similar mode $\Bs \to \Dsm\pip$ helps 
in constraining the time-dependent efficiency and the flavor tagging performance. In order to derive the CP-violating parameters, 
two different approaches have been pursued: the first, labelled $s$fit (see \cite{Xie:2009rka}), consists of a statistical method to subtract background in maximum likelihood fit, 
without relying on any separate sideband or simulation for background modelling, whereas the second, labelled $c$fit separates signal 
from background by fitting for these two components with separate PDFs. Fig.~\ref{fig:dsk} 
shows the results of the time-dependent fits, and Tab.~\ref{tab:dsk} shows the fitted values of the CP observables in this decay.

\begin{figure}[hbt]
 \centering
 \includegraphics[width=2in]{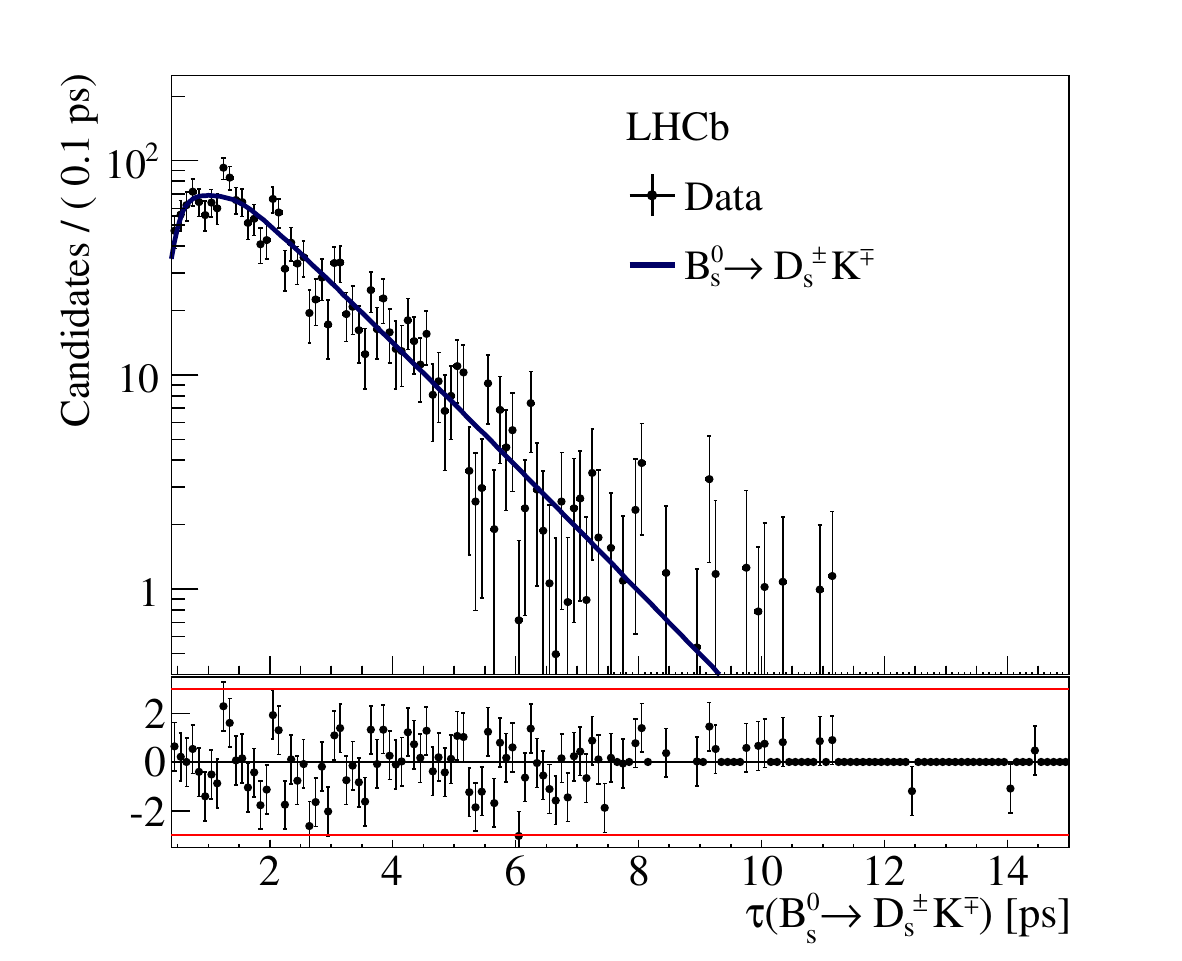}
 \includegraphics[width=2in]{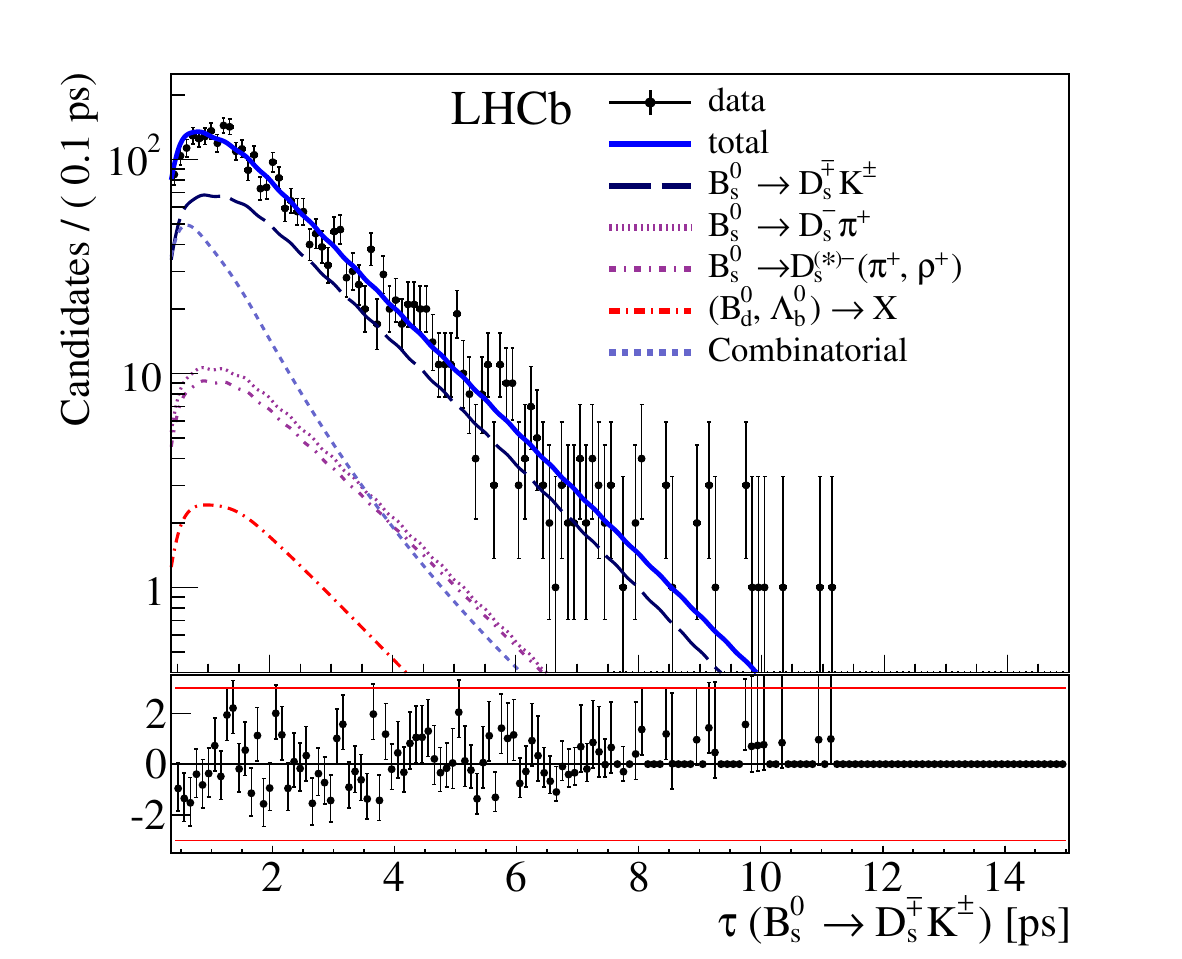}
\caption{Result of the decay-time (top) $s$fit and (bottom) $c$fit to
     the $\Bs\to \Ds K$ candidates. This figure is taken from Ref.~\cite{Aaij:2014fba}.}
 \label{fig:dsk}
\end{figure}

\begin{table}[b]
\centering
\caption{Fitted values of the CP observables to the $\Bs\to\Ds K$ time distribution
for (left) $s$fit and (right) $c$fit, where the first uncertainty is statistical,
the second is systematic.  All parameters other than the CP observables are
constrained in the fit. The results are taken from Ref.~\cite{Aaij:2014fba}.}
\label{tab:dsk}
\begin{tabular}{lcc}
  \hline
  Parameter     & $s$fit fitted value & $c$fit fitted value\\
  \hline
  \Cpar 	& $\phantom{-}0.52 \pm 0.25 \pm 0.04$  & $\phantom{-}0.53 \pm 0.25 \pm 0.04$ \\  
  \Dpar		& $\phantom{-}0.29 \pm 0.42 \pm 0.17$  & $\phantom{-}0.37 \pm 0.42 \pm 0.20$ \\  
  \Dbpar	& $\phantom{-}0.14 \pm 0.41 \pm 0.18$  & $\phantom{-}0.20 \pm 0.41 \pm 0.20$ \\
  \Spar 	& $          -0.90 \pm 0.31 \pm 0.06$  & $          -1.09 \pm 0.33 \pm 0.08$ \\  
  \Sbpar       	& $          -0.36 \pm 0.34 \pm 0.06$  & $          -0.36 \pm 0.34 \pm 0.08$ \\  
  \hline
\end{tabular}
\vspace{-3mm}
\end{table}

The study of $\Bs$ decays into two vector particles ($\Bs\to V_1V_2$), with the  vector particles  decaying 
into two pseudo-scalar mesons, has three helicity states that are allowed by angular momentum conservation, 
with amplitudes identified as $H_{+1}$,  $H_{-1}$ and $H_0$. It is convenient to map these amplitudes in terms 
of three transversity states to be considered, identified as  ``longitudinal" (0), ``perpendicular" ($\perp$), 
and ``parallel" ($\parallel$). 
They are related as
\begin{align}
A_0=H_0\\\nonumber
 A_\perp=\frac{H_{+1}-H_{=1}}{\sqrt{2}}\\\nonumber
 A_\parallel=\frac{H_{+1}+H_{=1}}{\sqrt{2}}\\\nonumber
 \end{align}
 Two such decays have been studied at LHCb: $\Bs\to\phi\phi$ (discussed in the previous section), and $\Bs\to\Kstz\Kstzb$.

 The study of the CP asymmetries and polarisation fractions in $\Bs\to\Kstz\Kstzb$ (see \cite{Aaij:2015kba}) takes a somewhat different approach. 
 In view of the limited statistics, rather than trying to implement a flavor tagged time-dependent analysis, a study of the triple product and direct 
 CP violation asymmetries is performed with a time-integrated analysis of $\Bs\to\Kstz\Kstzb$ , without determining the flavor of the decaying $\Bs$. 
 In $B$ mesons decays there are two possible triple products
 \begin{align}
 T_1=(\hat{n}_{V_1}\times\hat{n}_{V_2})\cdot \hat{p}_{V_1}=\sin{\phi}\\\nonumber
 T_2=2(\hat{n}_{V_1}\cdot{n}_{V_2})(\hat{n}_{V_1}\times {n}_{V_2})\cdot\hat{p}_{V_1}=\sin{2\phi}\\\nonumber
 \end{align}

They are found to be compatible with the Standard Model.

\section{Model independent constraints on new physics}
\label{model}
Indirect searches for new physics effects can be performed by assuming certain
extensions of the SM and calculating then the contribution of this
model to different flavour observables, e.g. $M_{12}^{s, \rm NP}$. 
Combining these calculations with the SM contributions 
(e.g. $M_{12}^{s, \rm SM}$) one gets a theory prediction for flavour observables 
that depends on unknown parameters $x, y,...$ of the considered new physics model.
Currently a comparison of experimental numbers and these new theory predictions
enables to bound the parameter space of new physics models, e.g.
\begin{equation}
\Delta M_s^{\rm Exp} = 2 \left|
M_{12}^{s, \rm NP} (x,y,...) + M_{12}^{s, \rm SM}
\right| \; .
\end{equation}
In future, this program could lead to a discovery of new physics effects, provided there is
a sufficient control over the theoretical uncertainties.
But also if physics beyond the SM will be first found by direct
detection of new particles, the above discussed 
investigations will be crucial in order to determine the flavour couplings
of the new model. 
There is a huge literature determining contributions of specific new physics models
to the observables discussed in this review,
in particular $\Bs$ mixing. 
We present here some examples,  not an exhaustive list:
super-symmetric contributions were discussed e.g. by
\cite{Hayakawa:2012ua,Altmannshofer:2011iv,Crivellin:2011sj,Ishimori:2011nv,Kifune:2007fj,Kawashima:2009jv,Kubo:2010mh,Kaburaki:2010xc,Wang:2011ax,Girrbach:2011an,Endo:2010yt,Buras:2010pm,Endo:2010fk,Wang:2010vv};
contributions of 2 Higgs-double models were discussed e.g. by
\cite{Chang:2015rva,Dutta:2011kg,Urban:1997gw};
extra dimensions were discussed in 
\cite{Goertz:2011nx,Datta:2010yq};
L-R symmetric models were discussed e.g. in
\cite{Bertolini:2014sua,Lee:2011kn};
extended gauge sectors were discussed e.g. in
\cite{Chang:2013hba,Sahoo:2013vuf,Kim:2012rpa,Li:2012xc,Chang:2011zza,Sahoo:2011zz,Fox:2011qd,Alok:2010ij};
additional fermions\footnote{A sequential, chiral, perturbative fourth generation
of fermions is already excluded by experiment, see e.g. 
\cite{Eberhardt:2012gv,Eberhardt:2012sb,Djouadi:2012ae,Eberhardt:2012ck,Kuflik:2012ai,Buchkremer:2012yy}. 
This exclusion holds, however, not for e.g. vector-like quarks or a
combination of a fourth chiral family with an additional modification of the SM,
see e.g. \cite{Lenz:2013iha}.}
which were discussed e.g. in
\cite{Alok:2012xm,Botella:2012ju}.
\\
In order to minimise the risk of betting on the wrong model, we
will discuss here a little more in detail the model-independent 
approach, where one tries to identify new physics effects without assuming
a specific model.
To start, it seems to be reasonable to assume that new physics only acts in mixing, in particular 
in $M_{12}$, but not in tree-level decays. 
For simplicity we also assume no penguin contributions. Later on we will
soften these restrictions.
Thus we postulate a general modification of $M_{12}^s$ by an a priori arbitrary
complex parameter  $\Delta_s$, while $\Gamma_{12}^s$ is just given by the 
SM prediction.
\begin{eqnarray}
M_{12}^s     & = & M_{12}^{s, \rm SM} |\Delta_s| e^{i \phi_s^\Delta} \; ,
\\
\Gamma_{12}^s & = & \Gamma_{12}^{s, \rm SM} \; .
\end{eqnarray}
Such a modification changes the mixing observables in the following 
way \footnote{The correction factor  $1/8 |\Gamma_{12}^{s, \rm SM}/M_{12}^{s, \rm SM}|^2
\left|1/ \Delta_s \right|^2
\sin \phi_{12}^s$ in Eq.(\ref{DMdef}) and Eq.(\ref{DGdef})
stays still small.
}
\begin{eqnarray}
\Delta M_s^{\rm Exp} & = & 2 \left|M_{12}^{s, \rm SM}\right| \cdot      |\Delta_s| \; ,
\label{DeltaMsNP}
\\
\Delta \Gamma_s^{\rm Exp} & = & 2 \left|\Gamma_{12}^{s, \rm SM} \right|
                      \cos \left( \phi_{12}^{s, \rm SM} + \phi_s^\Delta \right) \; ,
\\
a_{sl}^{s,\rm Exp} & = &  \frac{ \left|\Gamma_{12}^{s, \rm SM}\right|}{\left|M_{12}^{s, \rm SM}\right|}
                       \cdot
                \frac{\sin \left( \phi_{12}^{s, \rm SM} + \phi_s^\Delta  \right)}
                     { \left|{\Delta}_s\right| } \; .
\end{eqnarray}
Also the phases $\phi_{12}^s$  and $\phi_s$ will get new
contributions 
\begin{eqnarray}
\phi_{12}^{s, \rm Exp}       & = &  \phi_{12}^{s, \rm SM} + \phi_s^\Delta \; ,
\\
\phi_s^{\rm Exp}            & = & - 2 \beta_s + \phi_s^\Delta 
                                                     \; .
\label{phisNP}
\end{eqnarray}
Now a comparison of experimental numbers and SM predictions
can be used to obtain the bounds on the complex parameter $\Delta_s$.
If there is no new physics present, the comparison should result in 
$\Delta_s= 1 + 0 \times i$.
For a specific new physics model the parameter $\Delta_s$ can also be explicitly
calculated in dependence on the new physics parameters $x,y,...$. One gets
\begin{equation}
\Delta_s = \frac{M_{12}^{s, \rm NP} (x,y,...) + M_{12}^{s, \rm SM}}{M_{12}^{s, \rm SM}} \; . 
\end{equation}
General model-independent investigations, using the above introduced notation,  were done e.g. in 
\cite{Lenz:2006hd,Lenz:2010gu,Lenz:2012az} and \cite{Charles:2013aka,Charles:2015gya}.
Below we discuss different approaches.
Early investigations actually pointed towards large deviations from 
the SM. Unfortunately more data brought the extracted
value for $\Delta_s$ in perfect agreement with the SM.
The most recent result of such a investigation is
depicted in Fig.~\ref{NPmixing}\footnote{These plots
are taken from the CKMfitter  web-page (Summer 2014 - see \cite{Charles:2004jd}).}. 
For completeness we show also the result for the $B_d^0$-system.
\begin{figure*}
\includegraphics[width=0.45\textwidth]{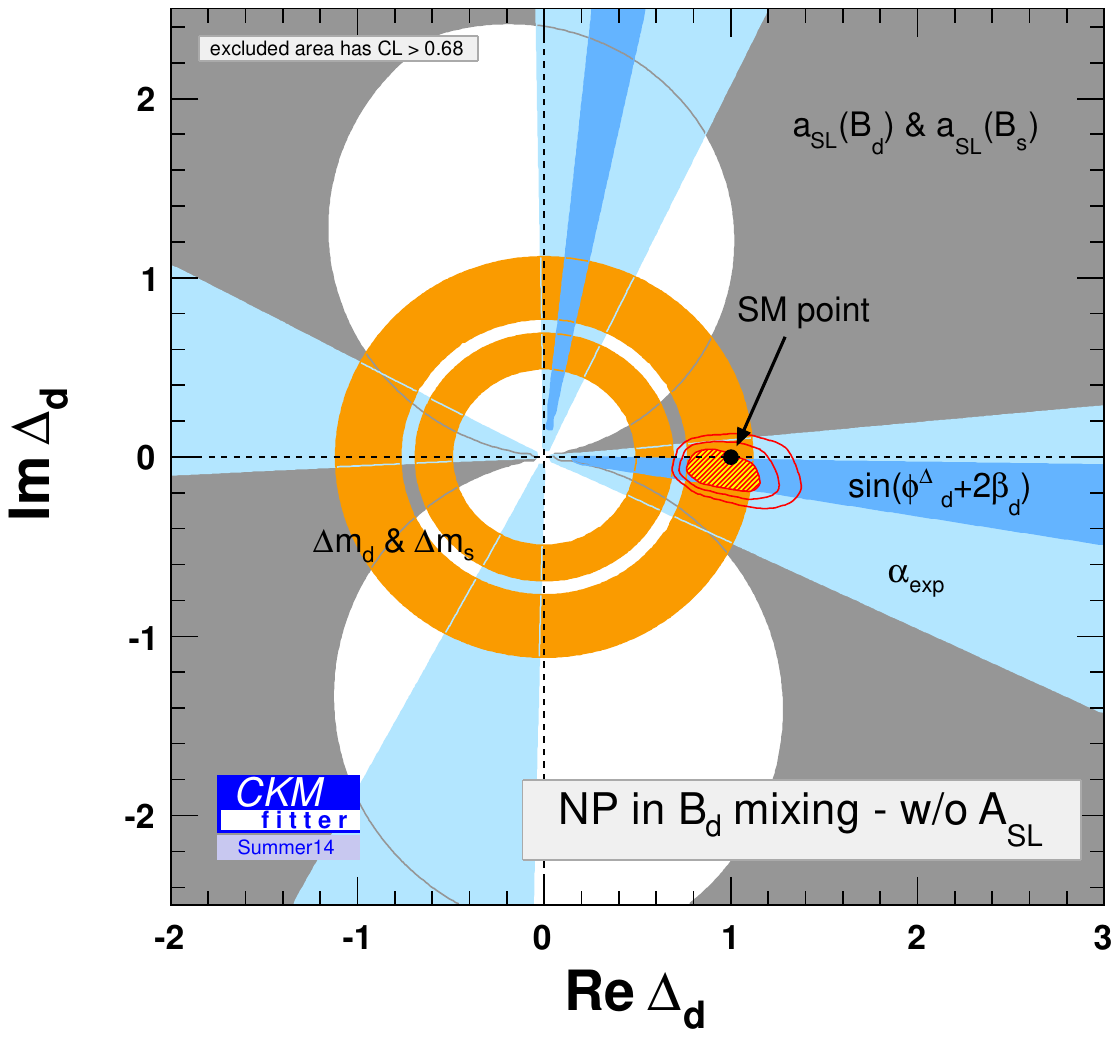}
\hfill
\includegraphics[width=0.45\textwidth]{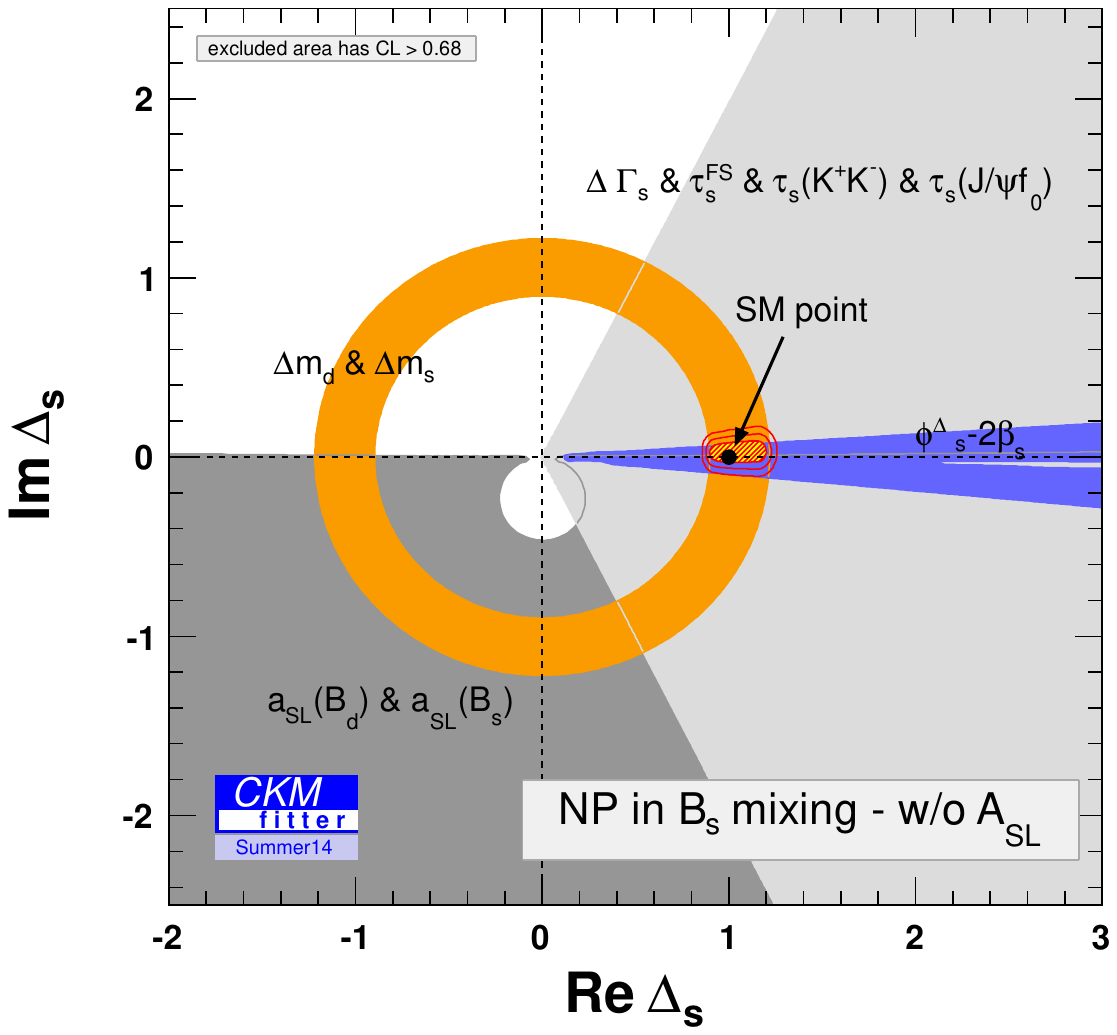}
\caption{\label{NPmixing} Current bounds (summer 2014)
on the complex parameters $\Delta_d$ (left) and
$\Delta_s$ (right) from different mixing observables. 
The point $\Delta_q = 1 + 0i$ corresponds to
the SM - no deviation from the SM is visible. 
Plots are taken from the CKMfitter 
web-page (see \cite{Charles:2004jd}).}
\end{figure*}
The constraint from the mass difference, Eq.(\ref{DeltaMsNP}), is denoted by the orange 
ring. The finite size of the ring is mostly due to the theory uncertainty of $\Delta M_q$.
In the case of $\Bd$ mesons we have two rings, due to two different values for the CKM
parameters $\rho$ and $\eta$ in the CKM-fit.
The purple region stems from the measurement of the phase $\phi_s$. According to
Eq.(\ref{phisNP}) this constrains also $\phi_s^\Delta$. Here one has to keep in mind that 
this assumes no sizable SM penguins and also no new physics penguins. 
The dark-grey area stems from the semileptonic asymmetries. Here we are currently 
limited by the experimental precision.
The overlap region of all experimental bounds is plotted in red. 
All in all we find in both mixing systems a perfect agreement with the SM, 
but there is still some sizable space (of the order of $10 \%$ in $|\Delta_q|$ and
several degrees in the phase $\phi_s^\Delta$) for new physics effects 
in $B_d^0$ and $\Bs$ mixing. 
It is entertaining and may be instructive - in the view of the currently discussed
deviations of experiment and SM - to show the corresponding plots from
2010  \cite{Lenz:2010gu} in Fig.~\ref{NPmixing2}.
\begin{figure*}
\includegraphics[width=0.45\textwidth]{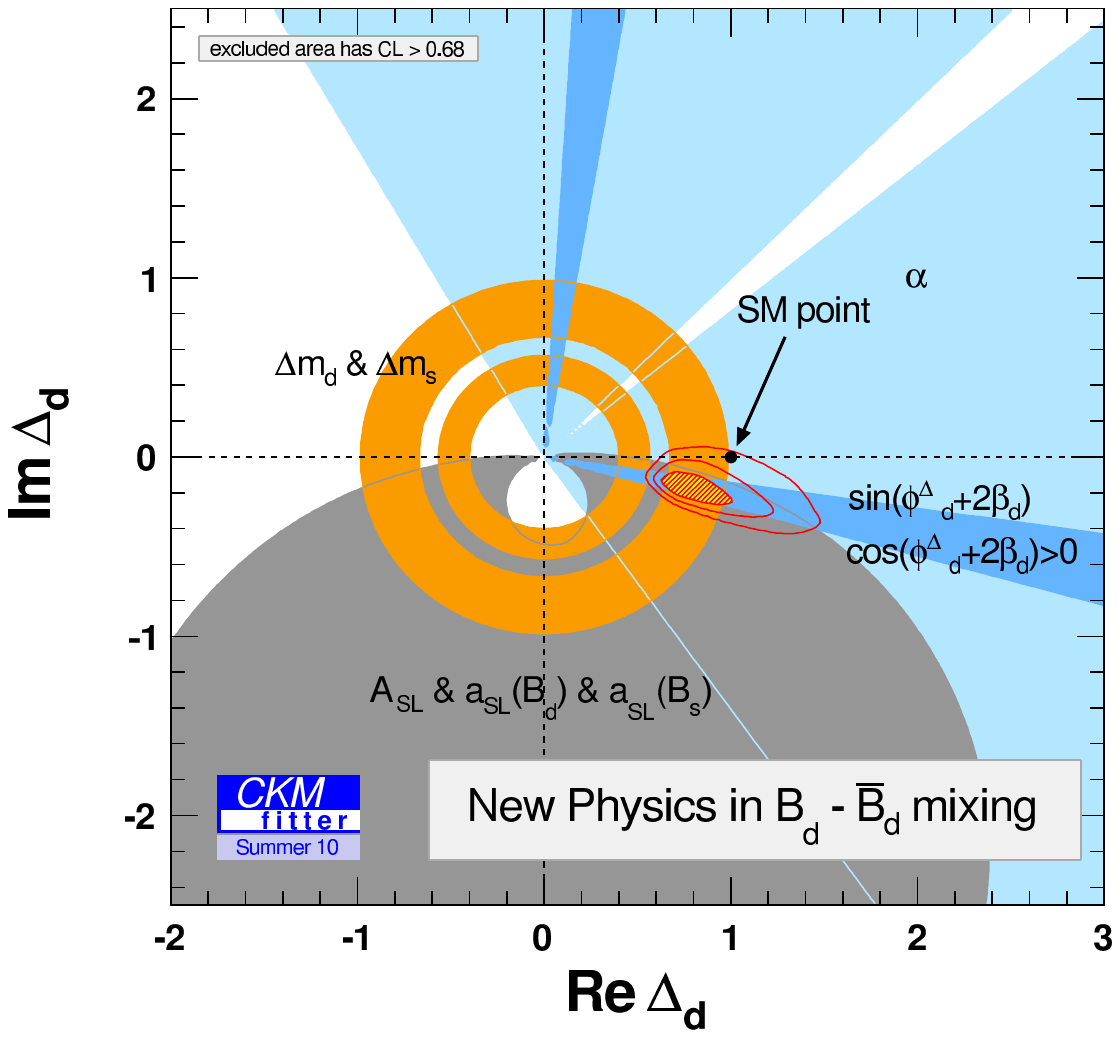}
\hfill
\includegraphics[width=0.45\textwidth]{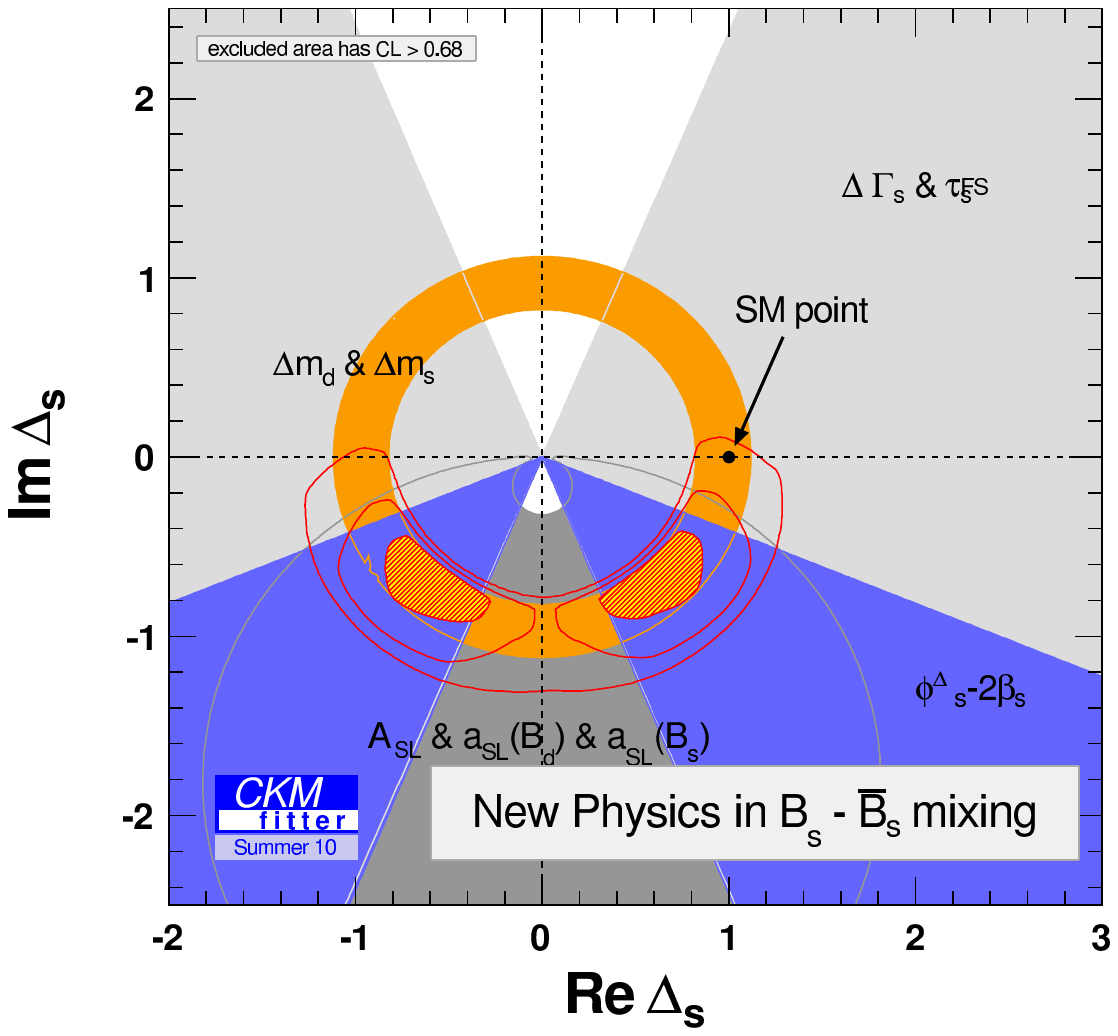}
\caption{
\label{NPmixing2} 
Bounds on the complex parameters $\Delta_d$ (left) and
$\Delta_s$ (right) from different mixing observables with data available till 2010. 
The point $\Delta_q = 1 + 0i$ corresponds to
the SM. Plots are taken from \cite{Lenz:2010gu}. Unfortunately this 
quite clear hint for new physics effects has completely vanished by now.}
\end{figure*}
Here a quite clear hint for new physics effects can be seen, actually in both
mixing systems, which unfortunately vanished completely in the last years.
\\
Similar investigations had been performed by  \cite{Fox:2007in}  
and the UTfit group (see e.g. the web-update of
\cite{Bevan:2013kaa,Bona:2006sa,Bona:2007vi}. In their notation one has
\begin{eqnarray}
C_{\Bs} e^{2 i \phi_{\Bs}} & = & \Delta_s  \; ,
\\
C_{\Bs} & = & |\Delta_s| \; ,
\\
\phi_{\Bs} & = & \frac12 \phi^\Delta_s \; .
\end{eqnarray}
Having only two parameters $C_{\Bs}$ and $ \phi_{\Bs}$ for parameterising
new physics effects in $\Bs$ mixing corresponds to making the same assumptions
as above: no new physics effects in $\Gamma_{12}^s$ and neglecting penguins.
Investigating all available mixing observables UFfit finds the following 
preferred parameter ranges
\begin{eqnarray}
C_{\Bs} & = & 1.052 \pm 0.084 \; ,
\\
\phi_{\Bs} & = & 0.72^\circ \pm 2.06^\circ \; . 
\end{eqnarray}
Again, everything seems to be perfectly consistent with the SM, while
leaving room for sizable new physics effects, i.e. of the order of $10 \%$ in  $C_{\Bs}$
and of the order of a factor of 10 in the phase $ \phi_{\Bs}$.
The corresponding allowed parameter regions for the $\Bd$-system read
\begin{eqnarray}
C_{\Bd} & = & 1.07 \pm 0.17 \; ,
\\
\phi_{\Bd} & = & -2.0^\circ \pm 3.2^\circ \; ,
\end{eqnarray}
yielding similar conclusions as in the $\Bs$-system.
\\
Sometimes these bounds are transferred into bounds on a hypothetical new physics scale.
\cite{Charles:2013aka} use the following notation for a deviation of $M_{12}^s$ from its SM value
\begin{eqnarray}
M_{12}^s & = & M_{12}^{s, \rm SM} 
\left( 
1 + h_s e^{2 i \sigma_s} 
\right) \; ,
\nonumber
\\
1 + h_s e^{2 i \sigma_s} & = & |\Delta_s| e^{i \phi_s^\Delta} \; .
\end{eqnarray} 
Assuming further the operator
\begin{equation}
\frac{C_{ij}^2}{\Lambda^2} \left( \bar{q}_{i,L} \gamma^\mu q_{j,L} \right)^2
\end{equation}
to describe the new physics contribution to $\Bs$ mixing (i.e. $i=s$ and $j=b$), they find
\begin{eqnarray}
h_s  & \approx &  \frac{C_{sb}^2}{\lambda_{sb}^2} \left( \frac{4.5 \; \rm TeV}{\Lambda} \right)^2 \; ,
\\
\sigma_s   & = & \arg \left( C_{sb} \lambda_{sb}^* \right) \; .
\end{eqnarray}
Here $C_{ij}$ denotes the size of the new physics couplings and $\Lambda$ is the mass 
scale of new physics. Both of these parameters are a priori unknown, because new physics 
has not been detected yet and an investigation of current experimental bounds on $\Bs$ mixing 
gives only information about the ratio $C_{ij}^2/ \Lambda^2$, but not about the individual values
of the couplings and of the scale. $\lambda_{sb} = V_{ts}^* V_{tb}$ denotes the CKM structure of the
SM contribution to $\Bs$ mixing.
\\
To make some statements about the new physics scale additional assumptions have to be made. Assuming that
the coefficients $C_{sb}$ have the same size as the CKM couplings, i.e. $C_{sb}=\lambda_{sb}$ 
\cite{Charles:2013aka} got a new physics scale $\lambda$ of about $19 \; \rm TeV$. Assuming instead
$C_{sb}=1$ the new physics scale increases to roughly $ 500 \; \rm TeV$. In particular the second scale
is far above the direct reach of LHC and thus $\Bs$ mixing could in principle probe new physics scales that 
are far from being accessible via direct measurements. On the other hand one should not  forget that
the assumption about the size of the coupling is in principle arbitrary. If the new physics couplings are very 
small then also the new physics scale that can be probed is very low.
\\
In order to fufill our final goal of unambiguously disentangling hypothetical new effects
from mixing observables  a strict control over uncertainties is 
mandatory. Also the assumptions made above, have to be challenged.
First of all we have to include penguins, both from the SM as well as from new
sources, this will modify the phase $\phi_s$ to 
\begin{eqnarray}
\phi_s            & = & - 2 \beta_s + \phi_s^\Delta + \delta^{\rm Peng, SM}
                                                    + \delta^{\rm Peng, NP} \; .
\label{phisNP2}
\end{eqnarray}
SM penguins are expected to contribute at most up to $1^\circ$, while
new physics penguins are less constrained.
General new physics effects in $M_{12}^s$ will be treated as above
\begin{eqnarray}
M_{12}^s     & = & M_{12}^{s, \rm SM}      |\Delta_s|         e^{i \phi_s^\Delta} \; .
\label{M12NP2}
\end{eqnarray}
In addition we will also allow new effects in $\Gamma_{12}^s$, encoded by the parameter
$\tilde{\Delta}_s$
 \begin{eqnarray}
 \Gamma_{12}^s & = & \Gamma_{12}^{s, \rm SM} |\tilde{\Delta}_s| e^{- i \tilde{\phi}_s^\Delta} \; ,
\label{Gamma12NP2}
 \end{eqnarray}
resulting in a modified mixing phase $\phi_{12}^s$ 
\begin{eqnarray}
\phi_{12}^s       & = &  \phi_{12}^{s, \rm SM} + \phi_s^\Delta + \tilde{\phi}_s^\Delta \; .
\\
\end{eqnarray}
New contributions to $\Gamma_{12}^s$ can be due to new penguins and/or 
new contributions to tree-level decays. For a long time new physics effects
in tree-level decays were considered to be negligible. Due to the dramatically
improved experimental precision, this possibility has, however, to be considered.
Taking only experimental constraints into account and no bias from model building,
first studies performed by \cite{Bobeth:2014rda}, \cite{Bobeth:2014rra} 
and \cite{Brod:2014bfa} find that the tree-level Wilson coefficients 
$C_1$ and $C_2$ can easily be affected by new effects of the order of $0.1$.
Such a deviation can sometimes have dramatic effects, e.g. a modification of the
imaginary part of $C_1$ by about $0.1$ would modify the extracted value of the CKM angle 
$\gamma$ by about $4^\circ$, see \cite{Brod:2014bfa}, which is larger than the expected
future experimental uncertainty. Thus these possibilities should be taken into account for
quantitative studies about new physics effects in mixing.
For a future disentangling of new effects in mixing, penguins and tree-level decays
clearly more theoretical work has to be done.
\\
The above modification of $M_{12}^s$ (see Eq.(\ref{M12NP2})) 
and $\Gamma_{12}^s$ (see Eq.(\ref{Gamma12NP2}))  changes the mixing 
observables in the following way\footnote{Again, the correction factor  $1/8 |\Gamma_{12}^{s, \rm SM}/M_{12}^{s, \rm SM}|^2
\left|\tilde{\Delta}_s / \Delta_s \right|^2
\sin \phi_{12}^s$ stays small.
}.
\begin{eqnarray}
\Delta M_s & = & 2 \left|M_{12}^{s, \rm SM}\right| \cdot      |\Delta_s| \; ,
\\
\Delta \Gamma_s & = & 2 \left|\Gamma_{12}^{s, \rm SM} \right| \left|\tilde{\Delta}_s\right|
                      \cos \left( \phi_{12}^{s, \rm SM} + \phi_s^\Delta + \tilde{\phi}_s^\Delta\right) \; ,
\\
a_{sl}^s & = &  \frac{ \left|\Gamma_{12}^{s, \rm SM}\right|}{\left|M_{12}^{s, \rm SM}\right|}
                       \cdot
                \frac{\left|\tilde{\Delta}_s\right|}{ \left|\tilde{\Delta}_s\right| }
                      \sin \left( \phi_{12}^{s, \rm SM} + \phi_s^\Delta + \tilde{\phi}_s^\Delta\right) \; ,
\end{eqnarray}
First steps in that direction haven been done in the analysis of \cite{Lenz:2012az}, where in 
Scenario IV new physics in $\Gamma_{12}^s$ was introduced by the parameter $\delta_s$.
\begin{equation}
\delta_q = 
\frac{\frac{\Gamma_{12}^s}{M_{12}^s}}{\Re \left(\frac{\Gamma_{12}^{s, \rm SM}}{M_{12}^{s, \rm SM}} \right)} \; .
\end{equation}
This parameter is related to mixing observables in the following way:
\begin{eqnarray}
\Re \left( \delta_s \right)  =  \frac{\frac{\Delta \Gamma_s}{\Delta M_s}}{\frac{\Delta \Gamma_s^{\rm SM}}{\Delta M_s^{\rm SM}}} \; , \; \; \; \; \; \; 
&&
\Im \left( \delta_s \right)  =  \frac{-a_{\rm sl}^s}{\frac{\Delta \Gamma_s^{\rm SM}}{\Delta M_s^{\rm SM}}}
\; .
\end{eqnarray}
In 2012 the fit of \cite{Lenz:2012az} seemed to prefer some deviations of $\Im \left( \delta_s \right)$
and $\Im \left( \delta_s \right)$, which were mostly triggered by an 
interpretation of the dimuon asymmetry, which was commonly accepted  at that time commonly,
but turned out to be incomplete.
\\
In future these model independent investigations should include new 
physics effects in $M_{12}^s$, in $\Gamma_{12}^s$  and in penguins. Doing the latter might also 
include a combination of $\Delta B = 2$ and $\Delta B =1$ observables.

\section{Conclusion/Outlook}
\label{conclusion}

The study of CP violation phenomena in the $\Bs$ system has been the focus of experimental and theoretical efforts.
%
It was started by the Tevatron experiments CDF and D0, who made the first measurements in
this system.
Among their main achievements are the measurement of the $\Bs$ meson
mass difference $\Delta M_s$ \cite{Abulencia:2006ze}
and the study of the semi leptonic charge asymmetry of the $\Bs$ meson
$a_{\rm sl}^s$ \cite{Abazov:2012zz,Abazov:2012hha,Abazov:2013uma}.
The measured value of $\asls$  based on the study of $\Bs\to\Dsp\mun\neumb$ is still contributing to
the average with the LHCb result, based on one-third of the run 1 data.
 The Tevatron experiments also initiated the
studies of other CP-violating phenomena, such as the mixing phase $\phi_s$ in the
$\Bs \to J/\psi \phi$ decay, albeit with large uncertainties.
\\
The  pioneering work of the Tevatron experiment
is continued and refined at the LHC, with a new
level of precision allowed by high statistics, improved detector performance, and
new analysis techniques.
%
In particular, the LHCb experiment has performed the most
precise measurement of all types of CP violation
(see \cite{Aaij:2013gta,Aaij:2014zsa,Aaij:2012qe}),
as well as that of $\Delta M_s$ and $\Delta \Gamma_s$ \cite{Aaij:2013mpa}. They measure
the CKM angle $\gamma$ not only in $\Bd$ decays previously studied by the $\epm$ $B$-factories, but also in $\Bs$ decays
both in tree-mediated processes, and in loop-mediated processes. Finally, they observe direct CP violation  in several $\Bs$ channels.
\\
 The current data do not confirm CP violation in 
 the $\Bs$ system in excess of the SM prediction, as it was originally hoped for.
Still, some room
for new physics manifestations remains.
In CP violation in the interference of decays and mixing quantified by the angle $\phi_s$
the experimental uncertainty is getting very close
to the SM central value.
In this respect, the emphasis on understanding small corrections such as penguin pollution 
is a field of active investigation in
the theoretical and experimental community.
The theory prediction for  CP violation in mixing is still orders
of magnitude smaller than the experimental uncertainty.
%
%
The level of understanding of the SM expectations for mixing observables and
CP violating phenomena in the $\Bs$ system is now very advanced.
Experimental studies have not only proven the CKM-mechanism to be the primary source of
quark-mixing and CP violation, but they have also confirmed the validity of theoretical approaches
such as the HQE to an unprecedented accuracy.
\\
The uncertainty on the theory prediction for the mass difference $\Delta M_s$ is about
$\pm 15\%$, thus allowing for new effects of the same order in this observable.
To improve the accuracy in $\Delta M_s$ further,  more precise lattice evaluations
of bag parameters and decay constants are mandatory. In this respect, an uncertainty of
about $\pm 5\%$ seems to be achievable in the next years\footnote{Here we assume an accuracy of  
  lattice values for dimension six operators considerably below $5\%$.}.
The calculation of the width difference according the HQE
seems on less solid theoretical grounds. The assumption of quark hadron duality was
questioned many times, see e.g. \cite{Ligeti:2010ia} or the discussion by \cite{Lenz:2011zz},
and deviations of more than $100 \%$ were discussed.
Such a failure of the HQE is now clearly ruled out.
The measurement of the width
difference $\Delta \Gamma_s$ has shown that the HQE works also in the most
challenging channel - $b \to c \bar{c} s$ - with an accuracy of at least $20 \%$\footnote{For very recent
estimates of the possible size of duality violating effects, see \cite{Jubb:2016mvq}.}.
For further independent tests of the precision of the HQE, lattice determinations of the matrix
elements that arise in lifetime difference of different $b$ hadrons, like
$\tau (B^+) / \tau (\Bd)$,
$\tau (\Bs) / \tau (\Bd)$
and
$\tau (\Lambda_b) / \tau (\Bd)$
are urgently needed, see the detailed discussion in \cite{Lenz:2014jha}. Here it might also
be insightful to study the charm sector, in particular the ratio $\tau (D^+) / \tau (D_0)$ and
$\tau (D_s^+) / \tau (D_0)$.
To reduce the uncertainty on the theory prediction of $\Delta \Gamma_s$ a first non-perturbative determination
of dimension 7 matrix elements is needed, i.e. the bag parameters $B_{R_0}$, $B_{R_2}$,
 $B_{R_3}$, $B_{\tilde{R}_2}$ and $B_{\tilde{R}_3}$. Currently, these parameters contribute 
the biggest individual uncertainty. Next, more precise lattice values of the complete SUSY-basis of
$\Delta B =2$ four quark operators are 
needed\footnote{While preparing this paper a new study of the Fermilab Lattice and MILC Collaborations 
was made public \cite{Bazavov:2016nty}.}. 
In parallel to these non-perturbative improvements, NNLO-QCD corrections\footnote{See \cite{Asatrian:2012tp} 
for a first step 
in that direction.}
have to be calculated (i.e. $\Gamma_3^{s,(2)}$ and  $\Gamma_4^{s,(1)}$ in our notation).
Having all these improvements
at hand, a final accuracy of about $5 \%$ for the $\Delta \Gamma_s$ prediction might also be feasible in the 
next years\footnote{Here we assume an accuracy of
 lattice values for dimension six operators considerably below $5\%$, an accuracy of about $10\%$ for the bag
 parameters $B_R$ of the dimension seven operators and a reduction of the renormalisation scale dependence by at least 
 a factor of two due to NNLO-QCD corrections.}.
\\
The current experimental uncertainty on $a_{sl}^s$ is still about a factor of 130 larger than the
tiny central value of the Standard Model expectations, thus still allowing plenty of room for new physics effects.
Turning to indirect CP violation, we find that the current experimental precision is coming close to
SM central value and also to the intrinsic theoretical uncertainties due to penguin contributions.
In principle the weak phase $\phi_s$ measured e.g. in $\Bs \to \jpsi \phi$ is a  null test similar to
the semileptonic asymmetries. In practice the theory prediction of the latter one is much more robust
than the one for $\phi_s$.
To fully exploit the  improving experimental precision  extended studies of penguin effects
and a quantification of them are mandatory.
\\
All LHC experiments expect to continue  data taking at least up to 2030.
The LHCb collaboration is currently engaging in a detector upgrade that should increase its sensitivity by a factor
of 10, with a combination of operating at higher instantaneous luminosity, and the implementation of  a 
purely software based trigger system, which will have to process the full 30 MHz of inelastic 
collisions delivered by the LHC. 
The physics opportunities offered by such an upgrade have been quantified by \cite{LHCb-PUB-2014-040} assuming a total integrated luminosity of 50 fb$^{-1}$. Several key measurements have been studied. Table~\ref{tab:lhcb-upgrade} summarises the prospects for some of the observables described in this paper.
\begin{table}
\begin{center}
\begin{tabular}{cccc}
\hline
Observable & LHCb 2018 & Upgrade & Theory Uncertainty \\
\hline
$\phi_s                      $ ($\Bs\to\jpsi\phi$) &0.025       & 0.009       & $\approx$ 0.002 \\
$a_{sl}^s                    $ ($10^{-3}$)         & 1.4        & 0. 5        & 0.003           \\
$\phi_s^{eff}(\Bs\to\phi\phi)$                     & 0.10       & 0.018       & 0.02            \\
\hline
$\gamma(\Bs\to\Ds K)         $                     & $11^\circ$ & 2.0$^\circ$ & negligible\\
\hline
\end{tabular}
\end{center}
\caption{Statistical sensitivity of the LHCb upgrade to key observables discussed in this paper. 
For each observables the projected sensitivity at the end of Run II and with a 
luminosity of 50 fb$^{-1}$ (phase I upgrade) are given. For a comparison we show 
also the current theory uncertainty of the Standard Model predictions, given in Eqs.~(\ref{phisbetas},\ref{betas}) and Eq.~(\ref{asls_sm}). 
The theory error in $\phi_s$ holds only for neglecting penguins.}
\label{tab:lhcb-upgrade}
\end{table}

 The plans
of other LHC collaborations are less ambitious. For example, the ATLAS
experiment projects to measure the value of $\phi_s$ with the precision of
0.022 by 2030 \cite{ATLAS-PHYS-PUB-2013-010}.
The precision of LHC measurements will allow to achieve the SM level in this quantity
and to perform unprecedented tests of the contribution of new models beyond the SM.
The huge statistics, which will become available during the next ten years, will also
allow to measure the CP violating phenomena in other channels like $\Bs \to J/\psi \eta$.
Advancement in theory, in particular in lattice QCD and other approaches to constrain the hadronic matrix elements needed to access fundamental quantities,
are expected to follow a similar path.
Thus, a new exciting era of  $\Bs$ meson studies is ahead of us.

\section{Acknowledgements}
We would like to thank Ikaros Bigi for many helpful comments on the manuscript.
M.A.  would like to thank the US National Science Foundation for their support, and S. Stone and P. Koppenburg for useful discussions.
A.L. would like to thank
Christine Davies, Tomomi Ishikawa, Andreas J{\"u}ttner and James Simone 
for helpful information about lattice inputs,
Gilberto Tetlalmatzi-Xolocotzi for creating the diagrams
in Fig. \ref{box}, Fig.\ref{penguincont} and Fig.{\ref{BstoDsK},
checking some of the numerics and proof-reading,
Matthew Kirk for providing Fig. \ref{lifetime}, spotting a tricky sign error and proof-reading,
Lucy Budge, Jonathan Cullen and Xiu Liu for checking some of
the numerical updates.

\appendix

\section{Numerical input for theory predictions}
\label{app:input}
In this appendix we list all input parameters that were used for our numerical updates
of several Standard Model predictions.
We start with listing some very well-known parameters in Table \ref{parameter1} that are mostly taken
from the PDG \cite{Agashe:2014kda}.
%
%
\begin{table}
\begin{tabular}{|l|l|l|}
\hline
Parameter & Value & Reference
\\
\hline
\hline
$M_W$   & $80.385(15)  \; \mbox{GeV}$ & $\mbox{PDG 2015}$
\\
\hline
$M_Z $  & $ 91.1876(21)  \; \mbox{GeV}$ & $\mbox{PDG 2015}$
\\
\hline
$G_F $  & $1.1663787(6)   10^{-5}   \; \mbox{GeV}^{-2} $&$\mbox{PDG 2015}$
\\
\hline
$\hbar$ & $6.58211928(15) 10^{-25} \; \mbox{GeV s}$ & $\mbox{PDG 2015}$
\\
\hline
$M_{\Bs} $& $  5.3667(4) \ \mbox{GeV} $& $\mbox{PDG 2015}$
\\
\hline
$\bar{m}_b  (\bar{m}_b) $  & $  4.18(3) \; \mbox{GeV} $& $\mbox{PDG 2015}$
\\
\hline
$\bar{m}_c  (\bar{m}_c) $  &  $ 1.275(25)  \; \mbox{GeV}$ & $\mbox{PDG 2015}$
\\
\hline
$\bar{m}_s  (2 \, \mbox{GeV})  $ & $  0.0935(25)  \; \mbox{GeV} $& $\mbox{PDG 2015}$
\\
\hline
$\alpha_s(M_Z) $ & $0.1185(6)  $& $ \mbox{PDG 2015}$
\\
\hline
\end{tabular}
\caption{List of precisely known input parameters needed for an update of the theory
prediction of different mixing observables.}
\label{parameter1}
\end{table}
Next we list in Table \ref{parameter2} some not so well determined input parameters. 
For lattice values our standard reference is FLAG \cite{Aoki:2013ldr}. 
In the case of $\tilde{B}_S/B$ FLAG did not provide an average, so we took 
the values from \cite{Becirevic:2001xt}, \cite{Bouchard:2011xj}, \cite{Carrasco:2013zta}
and \cite{Dowdall:2014qka} and did our own naive average.
For $B_{R_0}$ we took the preliminary value that can be read off the plots given
by \cite{Dowdall:2014qka}.
$B_{R_1}$ and  $B_{\tilde{R}_1}$ can be deduced from
\cite{Becirevic:2001xt}, \cite{Bouchard:2011xj}, \cite{Carrasco:2013zta}
and \cite{Dowdall:2014qka}.
The operators $R_1$ and $\tilde{R}_1$ are denoted by $O_4$ and $O_5$ in the lattice literature
\begin{equation}
R_1         \equiv \frac{m_s}{m_b} O_4
\; ,
\tilde{R}_1 \equiv \frac{m_s}{m_b} O_5
\; .
\end{equation}
The Fermilab-MILC Collaboration \cite{Bouchard:2011xj} uses again an additional factor 4
\begin{equation}
R_1         \equiv \frac{m_s}{m_b} 4 O_4
\; ,
\tilde{R}_1 \equiv \frac{m_s}{m_b} 4 O_5
\; .
\end{equation}
Moreover one has to be aware of different normalisation factors
used in the definition of the matrix elements.
In e.g. \cite{Beneke:1996gn} and \cite{Lenz:2006hd}
\begin{eqnarray}
\langle R_1 \rangle         & = & \frac73 \frac{m_s}{m_b} M_{\Bs}^2 f_{B_s}^2 B_{R_1} \; ,
\\
\langle \tilde{R}_1 \rangle & = & \frac53 \frac{m_s}{m_b} M_{\Bs}^2 f_{B_s}^2 B_{\tilde{R}_1} \; 
\end{eqnarray}
was used. This definition ensures that in vacuum insertion approximation the bag parameters
$B_{R_1}$ and $B_{\tilde{R}_1}$ have the value one.
In the lattice literature different normalisation factors, compared to $\frac73$ and $\frac53$, 
are used.
\cite{Becirevic:2001xt} and \cite{Carrasco:2013zta}
have
\begin{eqnarray}
\langle R_1 \rangle         & = & 2       \frac{m_s}{m_b} M_{\Bs}^2 f_{B_s}^2 B_4' \; ,
\\
\langle \tilde{R}_1 \rangle & = & \frac23 \frac{m_s}{m_b} M_{\Bs}^2 f_{B_s}^2 B_5' \; , 
\end{eqnarray}
 while
\cite{Bouchard:2011xj} and
and \cite{Dowdall:2014qka} use
\begin{eqnarray}
\langle R_1 \rangle         & = & 2       \frac{m_s}{m_b} M_{\Bs}^2 f_{B_s}^2 B_4 \; ,
\\
\langle \tilde{R}_1 \rangle & = & \frac23 \frac{m_s}{m_b} M_{\Bs}^2 f_{B_s}^2 B_5 \; . 
\end{eqnarray}
For the top-quark mass
we did not take the PDG value, but a first combination of TeVatron and LHC results,
presented by \cite{ATLAS:2014wva}. $\Lambda_{QCD}^{(5)}$ we derived from the NLO running
of $\alpha_s$ using $\alpha_s(M_Z)$ and $M_Z$ given above as an input. The values of
the CKM-elements were taken from the web-update of the CKMfitter group
\cite{Charles:2004jd}, similar results are given by UTfit \cite{Bona:2006ah}.
Here the value of  $V_{ub}$ is taken from the fit and not from either an inclusive or an exclusive 
determination.
\begin{table}
\begin{tabular}{|l|l|l|}
\hline
Parameter & Value & Reference
\\
\hline
\hline
$f_{B_s} \sqrt{B} $&  $216(15) \; \mbox{MeV} $& $\mbox{FLAG}$
\\
\hline
$f_{B_d} \sqrt{B} $&  $175(12) \; \mbox{MeV} $& $\mbox{FLAG}$
\\
\hline
$\tilde{B}_S/B $   &  $ 1.07(6)     $      & $\mbox{own average}$
\\
\hline
$\tilde{B}_S/B (B_d)$   &  $ 1.04(12)     $      & $\mbox{own average}$
\\
\hline
$\bar{z}$ &       $0.0543964(229532) $ & $ \mbox{own evaluation}$
\\
\hline
$m_t $&$173.34(76) \, \mbox{GeV} $& $\mbox{arXiv:1403.4427}$
\\
\hline
$\bar{m}_t(\bar{m}_t)$ &$165.696(73)  \, \mbox{GeV}$ &$ \mbox{own evaluation}$
\\
\hline
$\Lambda_{QCD}^{(5)}$ & $233(8) \, \mbox{MeV} $& $\mbox{derived from NLO } \alpha_s$
\\
\hline
$V_{us} $& $0.22548^{+0.00068}_{-0.00034}$ & $\mbox{CKMfitter}$
\\
\hline
$V_{cb} $& $0.04117^{+0.00090}_{-0.00114} $& $\mbox{CKMfitter}$
\\
\hline
$V_{ub}/V_{cb}$ &$ 0.0862278 \pm 0.00442474 $& $\mbox{CKMfitter}$
\\
\hline
$\gamma $ & $ 1.17077^{+0.0169297}_{ - 0.0378736}$ & $\mbox{CKMfitter}$
\\
\hline
$B_{R_0}/B $& $ 1 \pm 0.3 $& $\mbox{HPQCD preliminary}$
\\
\hline
$B_{R_1}/B$ & $ 1.71 \pm 0.26$ & $\mbox{own average}$
\\
\hline
$B_{R_2} $&  $1 \pm 0.5 $& $\mbox{VIA assumption}$
\\
\hline
$B_{R_3} $& $ 1 \pm 0.5$ &$ \mbox{VIA assumption}$
\\
\hline
$B_{\tilde{R}_1}/B $& $ 1.27 \pm 0.16 $& $\mbox{own average}$
\\
\hline
$B_{\tilde{R}_3} $&  $1 \pm 0.5 $& $\mbox{VIA assumption}$
\\
\hline
\end{tabular}
\caption{List of less precisely known input parameters needed for an update of the theory
prediction of different mixing observables.}
\label{parameter2}
\end{table}
Finally we also present  in Table \ref{parameter3} a list of additional lattice determinations
for $f_{B_s} \sqrt{B}$ and $\tilde{B}_S/B$, given by
HQPCD (LATTICE 2014 update by \cite{Dowdall:2014qka}),
ETMC \cite{Carrasco:2013zta},
the LATTICE 2015 update from the Fermilab-MILC Collaboration \cite{Bouchard:2011xj}
and the LATTICE 2015 update from RBC-UKQCD of \cite{Aoki:2014nga}.
\begin{table}
\begin{tabular}{|l|l|l|}
\hline
Parameter & Value & Collaboration
\\
\hline
\hline
$f_{B_s} \sqrt{B} $&  $200(5-10) \; \mbox{MeV} $& $\mbox{HPQCD}$
\\
\hline
$\tilde{B}_S/B $   &  $ 1.03  $         &
\\
\hline
$B_{R_1}/B$        &  $ 1.98 $          &
\\
\hline
$B_{\tilde{R}_1}/B $ &  $ 1.48 $          &
\\
\hline
\hline
$f_{B_s} \sqrt{B} $&  $211(8) \; \mbox{MeV}$ & $\mbox{ETMC}$
\\
\hline
$\tilde{B}_S/B $  &  $ 1.03 $          &
\\
\hline
$B_{R_1}/B $       &  $ 1.46 $          &
\\
\hline
$B_{\tilde{R}_1}/B $ & $  1.15 $          &
\\
\hline
\hline
$f_{B_s} \sqrt{B} $& $ 227(7) \; \mbox{MeV}$ & $\mbox{Fermi-MILC}$
\\
\hline
$\tilde{B}_S/B $   & $  1.15 $          &
\\
\hline
$B_{R_1}/B$        &  $ 1.60 $          &
\\
\hline
$B_{\tilde{R}_1}/B $ & $  1.17$           &
\\
\hline
\hline
$f_{B_s} \sqrt{B} $&  $262(?) \; \mbox{MeV} $ & $\mbox{RBC-UKQCD}$
\\
\hline
\end{tabular}
\caption{List of additional and mostly preliminary 
determinations of lattice parameters needed for an update of the theory
prediction of different mixing observables. Some of the values given here were simply read
off plots provided by the different collaborations. The error of the RBC-UK evaluation  cannot be estimated currently, 
because of missing  $1/m_b$ corrections.}
\label{parameter3}
\end{table}

\section{Error budget of the theory predictions}
\label{app:error}
In this appendix we compare the error budget or our new Standard Model predictions
with the ones given in 2011 by \cite{Lenz:2011ti}
and the ones given in 2006 by \cite{Lenz:2006hd}.
\\
The error budget for the updated Standard Model prediction of $\Delta M_s^{\rm SM}$ is given in
Table \ref{error1}.
\begin{table}
\begin{tabular}{|c||c|c|c|}
\hline
$\Delta M_s^{\rm SM} $   &   $\mbox{This work}  $&  $\mbox{LN 2011}  $      &  $ {\mbox{LN 2006}} $
\\
\hline
\hline
 $\mbox{Central Value} $   &   $18.3 \, \mbox{ps}^{-1} $ &  $ 17.3 \, \mbox{ps}^{-1 } $ &  $ 19.3 \, \mbox{ps}^{-1} $
\\
\hline
 $\delta (f_{B_s} \sqrt{B}) $ &  $ 13.9\%  $          & $  13.5 \%   $              & $  34.1 \% $
\\
\hline
 $\delta (V_{cb})    $     & $  4.9 \%   $         &  $   3.4 \%   $             &  $ 4.9 \% $
\\
\hline
 $\delta (m_t)     $       & $ 0.7 \%   $          &  $  1.1 \%    $             & $  1.8 \% $
\\ 
\hline 
 $\delta (\alpha_s)  $     & $ 0.1 \%    $        &  $  0.4 \%      $           &  $ 2.0 \% $
\\
\hline
 $\delta (\gamma)   $      &  $0.1 \%  $           &  $  0.3 \%      $           &  $ 1.0 \% $
\\
\hline
 $\delta (|V_{ub}/V_{cb}|) $ &  $0.1 \%   $        &  $  0.2 \%    $             &   $0.5 \% $
\\
\hline
 $\delta (\ov m_b)  $      &  $<0.1 \%    $        &  $  0.1 \%   $              &  $ --- $
\\
\hline
\hline
 $\sum \delta  $           &  $ 14.8 \%   $        &  $14.0 \%    $             &  $34.6 \% $
\\
\hline
\end{tabular}
\caption{List of the individual contributions to the theoretical error of the mass difference 
$\Delta M_s$ within the Standard Model and comparison with the values obtained in \cite{Lenz:2011ti}
and \cite{Lenz:2006hd}.}
\label{error1}
\end{table}
For the mass difference we observe no improvement in accuracy compared to the 2011
prediction, because the by far
dominant uncertainty (close to $14 \%$) stems from $f_{B_s} \sqrt{B}$ and
here the inputs are more or less  unchanged. This will change as soon as new
lattice values will be available.
The next important uncertainty is the accuracy of the CKM element $V_{cb}$, which
contributes about $5 \%$ to the error budget. If one gives up the assumption of the
unitarity of the 3 times 3 CKM matrix, the uncertainty can go up considerably.
The uncertainties due to the remaining parameters, play no important role.
All in all we are left with an overall uncertainty of close to $15 \%$, which
has to be compared to the experimental uncertainty of about 1 per mille.
This situation leaves currently some space for new physics contributions to the
mass difference $\Delta M_s$. With future improvements on the non-perturbative
parameters a theoretical uncertainty in the range of $5 \%$ till $10 \%$ is feasible.
\\
Next we study the error budget of the decay rate difference $\Delta \Gamma_s$ in Table \ref{error2}.
\begin{table}
\begin{tabular}{|c||c|c|c|}
\hline
$\Delta \Gamma_s^{\rm SM} $&$  \mbox{this work} $&$\mbox{LN 2011}  $&${\mbox{LN 2006}}$
\\
\hline
\hline
$\mbox{Central Value}   $&$  0.088 \, \mbox{ps}^{-1} $&$0.087 \, \mbox{ps}^{-1} $&$ 0.096 \, \mbox{ps}^{-1}$
\\
\hline
$\delta (B_{\widetilde R_2})$&$ 14.8 \% $&$17.2 \% $&$15.7 \%$
\\
\hline
$\delta (f_{B_s} \sqrt{B})$ &$ 13.9 \% $&$13.5 \% $&$34.0 \%$
\\
\hline
$\delta (\mu)            $  &$  8.4 \% $&$ 7.8 \% $&$13.7 \%$
\\
\hline
$\delta (V_{cb})         $  &$  4.9 \% $&$ 3.4 \% $&$ 4.9 \%$
\\
\hline
$\delta (\widetilde{B}_S)$  &$  2.1 \% $&$  4.8 \% $&$ 3.1 \%$
\\
\hline
$\delta (B_{R_0})       $   &$  2.1 \% $&$ 3.4 \% $&$ 3.0 \%$
\\
\hline
$\delta (\bar z )       $   &$  1.1 \% $&$ 1.5 \% $&$ 1.9 \%$
\\
\hline
$\delta (m_b )           $  &$  0.8 \% $&$ 0.1 \% $&$  1.0 \%$
\\
\hline
$\delta (B_{\tilde{R}_1})$  &$  0.7\%  $&$ 1.9 \% $&$  ---$
\\
\hline
$\delta (B_{\tilde{R}_3}) $  &$  0.6 \% $&$ 0.5 \% $&$  ----$
\\
\hline
$\delta (B_{R_1})          $&$ 0.5 \% $&$ 0.8 \% $&$  ---$
\\
\hline
$\delta (B_{R_3})           $&$ 0.2 \% $&$ 0.2 \% $&$ ---$
\\
\hline
$\delta (m_s )             $&$ 0.1 \% $&$ 1.0 \%  $&$  1.0 \%$
\\
\hline
$\delta (\gamma)           $&$ 0.1 \% $&$ 0.3 \%  $&$ 1.0 \%$
\\
\hline
$\delta (\alpha_s)         $&$ 0.1 \% $&$ 0.4 \%  $&$ 0.1 \%$
\\
\hline
$\delta (|V_{ub}/V_{cb}|)  $&$ 0.1 \% $&$  0.2 \% $&$ 0.5 \%$
\\
\hline
$\delta (\bar{m}_t(\bar{m}_t) $&$ 0.0 \% $&$ 0.0 \%  $&$ 0.0 \%$
\\
\hline
\hline
$\sum \delta               $&$22.8 \% $&$ 24.5 \%                $&$40.5 \%$
\\
\hline
\end{tabular}
\caption{List of the individual contributions to the theoretical error of the decay rate difference 
$\Delta \Gamma_s$ within the Standard Model and comparison with the values obtained in \cite{Lenz:2011ti}
and \cite{Lenz:2006hd}.}
\label{error2}
\end{table}
The uncertainty in the decay rate difference also did not change considerably
compared to 2011. The dominant uncertainty is still the unknown bag parameter of
the power suppressed operator $R_2$. This input did not improve since 2011.
Here and  in \cite{Lenz:2011ti} and
\cite{Lenz:2006hd} we took the very conservative assumption of $B_{R_{2,3}} = 1 \pm 0.5$. If in
future these parameters could be determined with an uncertainty of about $\pm 10 \%$, then
an overall uncertainty of less than $\pm 10 \%$ in $\Delta \Gamma_s$ would become feasible.
First steps in the direction of a non-perturbative determination of $B_{R_2}$
within the framework of QCD sum rules have been done by
\cite{Mannel:2007am,Mannel:2011zza}. There, however, only sub-leading contributions
were determined. Thus a calculation of the leading (three-loop) contribution
would be very desirable. 
The second largest uncertainty stems from $f_{B_s} \sqrt{B}$, whose value did 
also not improve since 2011. There are, however, several new (mostly preliminary) results on market
- HQPCD (LATTICE 2014 update by \cite{Dowdall:2014qka}), ETMC \cite{Carrasco:2013zta},
the LATTICE 2015 update from the Fermilab-MILC Collaboration \cite{Bouchard:2011xj}
and the LATTICE 2015 update from RBC-UKQCD of \cite{Aoki:2014nga} - that seem to indicate
that $f_{B_s} \sqrt{B}$ can be determined with an uncertainty as low as $5 \%$ in the near future.
In most of these works not only the matrix element of $Q$, but also the full $\Delta B = 2$
operator basis is studied. This will provide improved values for the bag parameters
$B_S$, $\tilde{B}_S$, $B_{R_1}$, $B_{\tilde{R}_1}$ and $B_{R_0}$, 
via Eq.(\ref{relation}).
Number three in the error budget is the dependence on the renormalisation scale, here
a calculation of NNLO-QCD corrections would be necessary to further reduce the error.
First steps of such an endeavour were done by \cite{Asatrian:2012tp}.
The next important dependence is the CKM element $V_{cb}$, which leads currently
to an uncertainty of about $5 \%$.
\\
In the ratio $\Delta  \Gamma_s^{\rm SM} / \Delta M_s^{\rm SM}$ one of the dominant
uncertainties, the dependence on $f_{B_s}^2 B$ is cancelling and we get for the
error budget the values given in Table \ref{error4}.
\begin{table}
\begin{tabular}{|c||c|c|c|}
\hline
$\Delta  \Gamma_s^{\rm SM} / \Delta M_s^{\rm SM}  $&$  \mbox{this work} $&$ \mbox{LN 2011}
        $&$ {\mbox{LN 2006}}$
\\
\hline
\hline
$\mbox{Central Value}    $&$48.1 \cdot 10^{-4}$&$50.4 \cdot 10^{-4}$&$  49.7 \cdot 10^{-4}$
\\
\hline
$\delta (B_{R_2})        $&$ 14.8 \%          $&$   17.2 \%        $&$  15.7 \%$
\\
\hline
$\delta (\mu)            $&$  8.4 \%          $&$   7.8 \%         $&$   9.1 \%$
\\
\hline
$\delta (\widetilde{B}_S)$&$  2.1 \%          $&$   4.8 \%         $&$  3.1 \%$
\\
\hline
$\delta (B_{R_0})        $&$  2.1 \%          $&$   3.4 \%         $&$   3.0 \%$
\\
\hline
$\delta (\bar z)         $&$  1.1 \%          $&$   1.5 \%         $&$   1.9 \%$
\\
\hline
$\delta (m_b)            $&$   0.8 \%         $&$  1.4 \%          $&$   1.0 \%$
\\
\hline
$\delta (m_t)            $&$   0.7 \%         $&$  1.1 \%          $&$   1.8 \%$
\\
\hline
$\delta (B_{\tilde{R}_1})$&$   0.7 \%          $&$   1.9 \%         $&$   ---$
\\
\hline
$\delta (B_{\tilde{R}_3})$&$   0.6 \%         $&$  0.5 \%          $&$   ----$
\\
\hline
$\delta (B_{R_1})        $&$   0.5 \%         $&$  0.8 \%          $&$   ---$
\\
\hline
$\delta (B_{R_3})        $&$   0.2 \%         $&$  0.2 \%          $&$  ---$
\\
\hline
$\delta (\alpha_s)       $&$   0.2 \%         $&$  0.8 \%          $&$  0.1 \%$
\\
\hline
$\delta (m_s)            $&$   0.1 \%         $&$ 1.0 \%           $&$   0.1 \%$
\\
\hline
$\delta (\gamma)         $&$   0.0 \%         $&$ 0.0 \%           $&$  0.1 \%$
\\
\hline
$\delta (|V_{ub}/V_{cb}|)$&$   0.0 \%         $&$ 0.0 \%           $&$  0.1 \%$
\\
\hline
$\delta (V_{cb})         $&$   0.0 \%         $&$  0.0 \%          $&$ 0.0 \%$
\\
\hline
\hline
$\sum \delta             $&$  17.3 \%         $&$ 20.1 \%          $&$ 18.9 \%$
\\
\hline
\end{tabular}
\caption{List of the individual contributions to the theoretical error of the ratio
$\Delta \Gamma_s$/$\Delta M_s$ within the Standard Model and comparison with the values obtained in \cite{Lenz:2011ti}
and \cite{Lenz:2006hd}.}
\label{error4}
\end{table}
For $\dg_s / \dm_s$ we see a tiny improvement in the theoretical precision compared to 2011.
The dominant uncertainty is  given by the unknown matrix element of the dimension 7 operator $R_2$,
followed by the uncertainty due to the renormalisation scale dependence. The overall uncertainty
is currently $17.3 \%$, which is also the final theoretical uncertainty that can currently be achieved
for $\Delta \Gamma_s$. Future investigations, i.e. non-perturbative determinations of the matrix 
element of $R_2$ and NNLO-QCD corrections might bring down this uncertainty to maybe $5 \%$.
\\
The error budget for the semileptonic CP asymmetries is finally listed in
Table \ref{error5}.
\begin{table}
\begin{tabular}{|c||c|c|c|}
\hline
 $a_{\rm fs}^{s,{\rm SM}} $&$ \mbox{this work}   $&$  \mbox{LN 2011}     $&$  {\mbox{LN 2006}} $
\\
\hline
\hline
 $\mbox{Central Value}    $&$2.22 \cdot 10^{-5}  $&$  1.90\cdot 10^{-5} $&$  2.06 \cdot 10^{-5} $
\\
\hline
 $\delta (\mu)            $&$9.5  \%             $&$    8.9 \%           $&$  12.7 \% $
\\
\hline
 $\delta (|V_{ub}/V_{cb}|)$&$5.0 \%              $&$  11.6 \%            $&$  19.5 \% $
\\
\hline
 $\delta (\bar z)         $&$4.6 \%              $&$   7.9 \%            $&$  9.3 \% $
\\
\hline
 $\delta (B_{\tilde{R}_3})$&$2.6 \%              $&$  2.8 \%             $&$   2.5 \% $
\\
\hline
 $\delta (\gamma)        $&$ 1.3 \%              $&$   3.1 \%            $&$  11.3 \% $
\\
\hline
 $\delta (B_{R_3})       $&$ 1.1 \%              $&$   1.2 \%            $&$  1.1 \% $
\\
\hline
 $\delta (m_b)           $&$ 1.0 \%              $&$   2.0 \%            $&$  3.7 \% $
\\
\hline
 $\delta (m_t)           $&$ 0.7 \%              $&$   1.1 \%            $&$  1.8 \% $
\\
\hline
 $\delta (\alpha_s)      $&$ 0.5 \%              $&$   1.8 \%            $&$  0.7 \% $
\\
\hline
 $\delta (B_{\tilde{R}_1})$&$0.5 \%              $&$  0.2 \%             $&$   --- $
\\
 \hline
 $\delta (\widetilde{B}_{S})$&$ 0.3 \%       $&$   0.6 \%                   $&$  0.4 \% $
\\
\hline 
 $\delta ({\cal B}_{R_0}) $&$ 0.2 \%       $&$    0.3 \%                  $&$   --- $
\\
\hline
 $\delta ({\cal B}_{R_2}) $&$ 0.1 \%       $&$   0.1 \%                   $&$  --- $
\\
\hline
 $\delta (m_s)           $&$ 0.1 \%        $&$   0.1 \%                   $&$  0.1 \% $
\\
\hline
 $\delta ({\cal B}_{R_1}) $&$ <0.1 \%       $&$    0.0 \%                  $&$   --- $
\\
\hline
 $\delta (V_{cb})        $&$  0.0 \%       $&$   0.0 \%                  $&$ 0.0 \% $
\\
\hline
\hline
 $\sum \delta            $&$  12.2 \%      $&$  17.3 \%                   $&$ 27.9 \% $
\\
\hline
\end{tabular}
\caption{List of the individual contributions to the theoretical error of the semileptonic CP 
asymmetries
$a_{sl}^s$ within the Standard Model and comparison with the values obtained in \cite{Lenz:2011ti}
and \cite{Lenz:2006hd}.}
\label{error5}
\end{table}
Here we witness some sizable reduction of the theory error. This quantity does not depend
on $f_{B_s} \sqrt{B}$ and has only a weak dependence on $R_2$, thus the two least known
parameters in the mixing sector do not affect the semileptonic asymmetries.
The increase in precision stems mostly from better known CKM elements in particular of $V_{ub}$,
in comparison to 2011.
Currently the dominant uncertainty stems from the renormalisation scale dependence followed by
the dependence on $V_{ub}$. For a reduction of the overall theoretical uncertainty considerably below 
$10 \%$ a NNLO-QCD calculation is mandatory.
\\
Finally we present in Table \ref{error6} 
also the theory errors for the observables in the $\Bd$-sector.
\begin{table}
\begin{tabular}{|c||c|c|c|}
\hline
                  &$  \Delta M_d^{\rm SM}  $&$ \Delta \Gamma_d^{\rm SM}  $&$a_{\rm sl}^{d, \rm SM}$
\\
\hline
\hline
$\mbox{Central Value}$&$  0.528 \, \mbox{ps}^{-1} $&$2.61 \cdot 10^{-3} \, \mbox{ps}^{-1} $&$ -4.7 \cdot 10^{-4}  $
\\
\hline
$\delta(B_{\widetilde R_2}) $&$ ---  $&$14.4\% $&$0.1 \%$
\\
\hline
$\delta(f_{B_d} \sqrt{B})   $&$13.7\%$&$13.7\% $&$ ---  $
\\
\hline
$\delta(\mu)                $&$ ---  $&$ 7.9\% $&$ 9.4 \%$
\\
\hline
$\delta(V_{cb})             $&$ 4.9\%$&$ 4.9\% $&$ 0.0 \%$
\\
\hline
$\delta(\widetilde{B}_S)    $&$  --- $&$ 4.0\% $&$ 0.6 \%$
\\
\hline
$\delta(B_{R_0})            $&$  --- $&$ 2.5\% $&$ 0.2 \%$
\\
\hline
$\delta(\bar z )            $&$  --- $&$ 1.1\% $&$ 4.9 \%$
\\
\hline
$\delta(m_b )               $&$ 0.0\%$&$ 0.8\% $&$  1.3 \%$
\\
\hline
$\delta(B_{\tilde{R}_1})    $&$  --- $&$ 0\%   $&$  ---$
\\
\hline
$\delta(B_{\tilde{R}_3})    $&$  --- $&$ 0.5\% $&$  2.7 \%$
\\
\hline
$\delta(B_{R_1})            $&$  --- $&$ 0\%   $&$  ---$
\\
\hline
$\delta(B_{R_3})            $&$  --- $&$ 0.2\% $&$ 1.2 \%$
\\
\hline
$\delta(m_s )               $&$  --- $&$ ---  $&$  --- $
\\
\hline
$\delta(\gamma)             $&$ 0.2\%$&$ 0.2 \%  $&$ 1.1 \%$
\\
\hline
$\delta(\alpha_s)           $&$ 0.0\%$&$ 0.1 \%  $&$ 0.5 \%$
\\
\hline
$\delta(|V_{ub}/V_{cb}|)    $&$ 0.0\%$&$  0.1 \% $&$ 5.2 \%$
\\
\hline
$\delta(\bar{m}_t(\bar{m}_t)$&$ 0.1\%$&$ 0.0 \%  $&$ 0.7 \%$
\\
\hline
\hline
$\sum \delta               $&$14.8 \%$&$22.7\%$&$12.3 \%$
\\
\hline
\end{tabular}
\caption{List of the individual contributions to the theoretical 
error of the mixing quantities $\Delta M_d$, $\Delta \Gamma_d$ and $a_{\rm sl}^{d, \rm SM}$ in the $B^0$-sector.}
\label{error6}
\end{table}

\bibliography{CPVBsR1}

\begin{thebibliography}{247}%
\makeatletter
\providecommand \@ifxundefined [1]{%
 \@ifx{#1\undefined}
}%
\providecommand \@ifnum [1]{%
 \ifnum #1\expandafter \@firstoftwo
 \else \expandafter \@secondoftwo
 \fi
}%
\providecommand \@ifx [1]{%
 \ifx #1\expandafter \@firstoftwo
 \else \expandafter \@secondoftwo
 \fi
}%
\providecommand \natexlab [1]{#1}%
\providecommand \enquote  [1]{``#1''}%
\providecommand \bibnamefont  [1]{#1}%
\providecommand \bibfnamefont [1]{#1}%
\providecommand \citenamefont [1]{#1}%
\providecommand \href@noop [0]{\@secondoftwo}%
\providecommand \href [0]{\begingroup \@sanitize@url \@href}%
\providecommand \@href[1]{\@@startlink{#1}\@@href}%
\providecommand \@@href[1]{\endgroup#1\@@endlink}%
\providecommand \@sanitize@url [0]{\catcode `\\12\catcode `\$12\catcode
  `\&12\catcode `\#12\catcode `\^12\catcode `\_12\catcode `\%12\relax}%
\providecommand \@@startlink[1]{}%
\providecommand \@@endlink[0]{}%
\providecommand \url  [0]{\begingroup\@sanitize@url \@url }%
\providecommand \@url [1]{\endgroup\@href {#1}{\urlprefix }}%
\providecommand \urlprefix  [0]{URL }%
\providecommand \Eprint [0]{\href }%
\providecommand \doibase [0]{http://dx.doi.org/}%
\providecommand \selectlanguage [0]{\@gobble}%
\providecommand \bibinfo  [0]{\@secondoftwo}%
\providecommand \bibfield  [0]{\@secondoftwo}%
\providecommand \translation [1]{[#1]}%
\providecommand \BibitemOpen [0]{}%
\providecommand \bibitemStop [0]{}%
\providecommand \bibitemNoStop [0]{.\EOS\space}%
\providecommand \EOS [0]{\spacefactor3000\relax}%
\providecommand \BibitemShut  [1]{\csname bibitem#1\endcsname}%
\let\auto@bib@innerbib\@empty
\bibitem [{\citenamefont {Aad}\ \emph {et~al.}(2014)\citenamefont {Aad} \emph
  {et~al.}}]{Aad:2014cqa}%
  \BibitemOpen
  \bibfield  {author} {\bibinfo {author} {\bibnamefont {Aad}, \bibfnamefont
  {Georges}},  \emph {et~al.} (\bibinfo {collaboration} {ATLAS})} (\bibinfo
  {year} {2014}),\ \bibfield  {title} {\enquote {\bibinfo {title} {{Flavor
  tagged time-dependent angular analysis of the $B_s \rightarrow J/\psi \phi$
  decay and extraction of $\Delta\Gamma$s and the weak phase $\phi_s$ in
  ATLAS}},}\ }\href {\doibase 10.1103/PhysRevD.90.052007} {\bibfield  {journal}
  {\bibinfo  {journal} {Phys. Rev.}\ }\textbf {\bibinfo {volume}
  {D90}}~(\bibinfo {number} {5}),\ \bibinfo {pages} {052007}},\ \Eprint
  {http://arxiv.org/abs/1407.1796} {arXiv:1407.1796 [hep-ex]} \BibitemShut
  {NoStop}%
\bibitem [{\citenamefont {Aad}\ \emph {et~al.}(2016)\citenamefont {Aad} \emph
  {et~al.}}]{Aad:2016tdj}%
  \BibitemOpen
  \bibfield  {author} {\bibinfo {author} {\bibnamefont {Aad}, \bibfnamefont
  {Georges}},  \emph {et~al.} (\bibinfo {collaboration} {ATLAS})} (\bibinfo
  {year} {2016}),\ \bibfield  {title} {\enquote {\bibinfo {title} {{Measurement
  of the CP-violating phase $\phi_s$ and the $B^0_s$ meson decay width
  difference with $B^0_s \to J/\psi\phi$ decays in ATLAS}},}\ }\href@noop {} {\
  }\Eprint {http://arxiv.org/abs/1601.03297} {arXiv:1601.03297 [hep-ex]}
  \BibitemShut {NoStop}%
\bibitem [{\citenamefont {Aaij}\ \emph
  {et~al.}(2012{\natexlab{a}})\citenamefont {Aaij} \emph
  {et~al.}}]{LHCb:2012ae}%
  \BibitemOpen
  \bibfield  {author} {\bibinfo {author} {\bibnamefont {Aaij}, \bibfnamefont
  {R}},  \emph {et~al.} (\bibinfo {collaboration} {LHCb})} (\bibinfo {year}
  {2012}{\natexlab{a}}),\ \bibfield  {title} {\enquote {\bibinfo {title}
  {{Analysis of the resonant components in $\Bs \to \jpsi\pi^+\pi^-$}},}\
  }\href {\doibase 10.1103/PhysRevD.86.052006} {\bibfield  {journal} {\bibinfo
  {journal} {Phys. Rev.}\ }\textbf {\bibinfo {volume} {D86}},\ \bibinfo {pages}
  {052006}},\ \Eprint {http://arxiv.org/abs/1204.5643} {arXiv:1204.5643
  [hep-ex]} \BibitemShut {NoStop}%
\bibitem [{\citenamefont {Aaij}\ \emph
  {et~al.}(2012{\natexlab{b}})\citenamefont {Aaij} \emph
  {et~al.}}]{Aaij:2012qe}%
  \BibitemOpen
  \bibfield  {author} {\bibinfo {author} {\bibnamefont {Aaij}, \bibfnamefont
  {R}},  \emph {et~al.} (\bibinfo {collaboration} {LHCb})} (\bibinfo {year}
  {2012}{\natexlab{b}}),\ \bibfield  {title} {\enquote {\bibinfo {title}
  {{First evidence of direct $C\!P$ violation in charmless two-body decays of
  $B^0_s$ mesons}},}\ }\href {\doibase 10.1103/PhysRevLett.108.201601}
  {\bibfield  {journal} {\bibinfo  {journal} {Phys. Rev. Lett.}\ }\textbf
  {\bibinfo {volume} {108}},\ \bibinfo {pages} {201601}},\ \Eprint
  {http://arxiv.org/abs/1202.6251} {arXiv:1202.6251 [hep-ex]} \BibitemShut
  {NoStop}%
\bibitem [{\citenamefont {Aaij}\ \emph
  {et~al.}(2012{\natexlab{c}})\citenamefont {Aaij} \emph
  {et~al.}}]{Aaij:2012nh}%
  \BibitemOpen
  \bibfield  {author} {\bibinfo {author} {\bibnamefont {Aaij}, \bibfnamefont
  {R}},  \emph {et~al.} (\bibinfo {collaboration} {LHCb})} (\bibinfo {year}
  {2012}{\natexlab{c}}),\ \bibfield  {title} {\enquote {\bibinfo {title}
  {{Measurement of the $B^0_s \rightarrow J/\psi \bar{K}^{*0}$ branching
  fraction and angular amplitudes}},}\ }\href {\doibase
  10.1103/PhysRevD.86.071102} {\bibfield  {journal} {\bibinfo  {journal} {Phys.
  Rev.}\ }\textbf {\bibinfo {volume} {D86}},\ \bibinfo {pages} {071102}},\
  \Eprint {http://arxiv.org/abs/1208.0738} {arXiv:1208.0738 [hep-ex]}
  \BibitemShut {NoStop}%
\bibitem [{\citenamefont {Aaij}\ \emph
  {et~al.}(2012{\natexlab{d}})\citenamefont {Aaij} \emph
  {et~al.}}]{Aaij:2012nta}%
  \BibitemOpen
  \bibfield  {author} {\bibinfo {author} {\bibnamefont {Aaij}, \bibfnamefont
  {R}},  \emph {et~al.} (\bibinfo {collaboration} {LHCb})} (\bibinfo {year}
  {2012}{\natexlab{d}}),\ \bibfield  {title} {\enquote {\bibinfo {title}
  {{Measurement of the $B_s$ effective lifetime in the $J/\psi f_0(980)$ final
  state}},}\ }\href {\doibase 10.1103/PhysRevLett.109.152002} {\bibfield
  {journal} {\bibinfo  {journal} {Phys. Rev. Lett.}\ }\textbf {\bibinfo
  {volume} {109}},\ \bibinfo {pages} {152002}},\ \Eprint
  {http://arxiv.org/abs/1207.0878} {arXiv:1207.0878 [hep-ex]} \BibitemShut
  {NoStop}%
\bibitem [{\citenamefont {Aaij}\ \emph
  {et~al.}(2012{\natexlab{e}})\citenamefont {Aaij} \emph
  {et~al.}}]{Aaij:2012di}%
  \BibitemOpen
  \bibfield  {author} {\bibinfo {author} {\bibnamefont {Aaij}, \bibfnamefont
  {R}},  \emph {et~al.} (\bibinfo {collaboration} {LHCb})} (\bibinfo {year}
  {2012}{\natexlab{e}}),\ \bibfield  {title} {\enquote {\bibinfo {title}
  {{Measurement of the $B_s^0\to J/\psi K_S^0$ branching fraction}},}\ }\href
  {\doibase 10.1016/j.physletb.2012.05.062} {\bibfield  {journal} {\bibinfo
  {journal} {Phys. Lett.}\ }\textbf {\bibinfo {volume} {B713}},\ \bibinfo
  {pages} {172--179}},\ \Eprint {http://arxiv.org/abs/1205.0934}
  {arXiv:1205.0934 [hep-ex]} \BibitemShut {NoStop}%
\bibitem [{\citenamefont {Aaij}\ \emph
  {et~al.}(2012{\natexlab{f}})\citenamefont {Aaij} \emph
  {et~al.}}]{Aaij:2012jw}%
  \BibitemOpen
  \bibfield  {author} {\bibinfo {author} {\bibnamefont {Aaij}, \bibfnamefont
  {R}},  \emph {et~al.} (\bibinfo {collaboration} {LHCb})} (\bibinfo {year}
  {2012}{\natexlab{f}}),\ \bibfield  {title} {\enquote {\bibinfo {title}
  {{Measurements of the branching fractions and CP asymmetries of $B^{\pm} \to
  J/\psi\, \pi^{\pm}$ and $B^{\pm} \to \psi(2S) \pi^{\pm}$ decays}},}\ }\href
  {\doibase 10.1103/PhysRevD.85.091105} {\bibfield  {journal} {\bibinfo
  {journal} {Phys. Rev.}\ }\textbf {\bibinfo {volume} {D85}},\ \bibinfo {pages}
  {091105}},\ \Eprint {http://arxiv.org/abs/1203.3592} {arXiv:1203.3592
  [hep-ex]} \BibitemShut {NoStop}%
\bibitem [{\citenamefont {Aaij}\ \emph
  {et~al.}(2012{\natexlab{g}})\citenamefont {Aaij} \emph
  {et~al.}}]{Aaij:2012mu}%
  \BibitemOpen
  \bibfield  {author} {\bibinfo {author} {\bibnamefont {Aaij}, \bibfnamefont
  {R}},  \emph {et~al.} (\bibinfo {collaboration} {LHCb})} (\bibinfo {year}
  {2012}{\natexlab{g}}),\ \bibfield  {title} {\enquote {\bibinfo {title}
  {{Opposite-side flavour tagging of B mesons at the LHCb experiment}},}\
  }\href {\doibase 10.1140/epjc/s10052-012-2022-1} {\bibfield  {journal}
  {\bibinfo  {journal} {Eur. Phys. J.}\ }\textbf {\bibinfo {volume} {C72}},\
  \bibinfo {pages} {2022}},\ \Eprint {http://arxiv.org/abs/1202.4979}
  {arXiv:1202.4979 [hep-ex]} \BibitemShut {NoStop}%
\bibitem [{\citenamefont {Aaij}\ \emph
  {et~al.}(2013{\natexlab{a}})\citenamefont {Aaij} \emph
  {et~al.}}]{Aaij:2013qha}%
  \BibitemOpen
  \bibfield  {author} {\bibinfo {author} {\bibnamefont {Aaij}, \bibfnamefont
  {R}},  \emph {et~al.} (\bibinfo {collaboration} {LHCb})} (\bibinfo {year}
  {2013}{\natexlab{a}}),\ \bibfield  {title} {\enquote {\bibinfo {title}
  {{First measurement of the CP-violating phase in $B_s^0 \to \phi \phi$
  decays}},}\ }\href {\doibase 10.1103/PhysRevLett.110.241802} {\bibfield
  {journal} {\bibinfo  {journal} {Phys. Rev. Lett.}\ }\textbf {\bibinfo
  {volume} {110}}~(\bibinfo {number} {24}),\ \bibinfo {pages} {241802}},\
  \Eprint {http://arxiv.org/abs/1303.7125} {arXiv:1303.7125 [hep-ex]}
  \BibitemShut {NoStop}%
\bibitem [{\citenamefont {Aaij}\ \emph
  {et~al.}(2013{\natexlab{b}})\citenamefont {Aaij} \emph
  {et~al.}}]{Aaij:2013tna}%
  \BibitemOpen
  \bibfield  {author} {\bibinfo {author} {\bibnamefont {Aaij}, \bibfnamefont
  {R}},  \emph {et~al.} (\bibinfo {collaboration} {LHCb})} (\bibinfo {year}
  {2013}{\natexlab{b}}),\ \bibfield  {title} {\enquote {\bibinfo {title}
  {{First measurement of time-dependent $C\!P$ violation in $B^0_s \to K^+K^-$
  decays}},}\ }\href {\doibase 10.1007/JHEP10(2013)183} {\bibfield  {journal}
  {\bibinfo  {journal} {JHEP}\ }\textbf {\bibinfo {volume} {10}},\ \bibinfo
  {pages} {183}},\ \Eprint {http://arxiv.org/abs/1308.1428} {arXiv:1308.1428
  [hep-ex]} \BibitemShut {NoStop}%
\bibitem [{\citenamefont {Aaij}\ \emph
  {et~al.}(2013{\natexlab{c}})\citenamefont {Aaij} \emph
  {et~al.}}]{Aaij:2013mtm}%
  \BibitemOpen
  \bibfield  {author} {\bibinfo {author} {\bibnamefont {Aaij}, \bibfnamefont
  {R}},  \emph {et~al.} (\bibinfo {collaboration} {LHCb})} (\bibinfo {year}
  {2013}{\natexlab{c}}),\ \bibfield  {title} {\enquote {\bibinfo {title}
  {{First observation of $\bar B^0 \to J/\psi K^+K^-$ and search for $\bar B^0
  \to J/\psi\phi$ decays}},}\ }\href {\doibase 10.1103/PhysRevD.88.072005}
  {\bibfield  {journal} {\bibinfo  {journal} {Phys. Rev.}\ }\textbf {\bibinfo
  {volume} {D88}}~(\bibinfo {number} {7}),\ \bibinfo {pages} {072005}},\
  \Eprint {http://arxiv.org/abs/1308.5916} {arXiv:1308.5916 [hep-ex]}
  \BibitemShut {NoStop}%
\bibitem [{\citenamefont {Aaij}\ \emph
  {et~al.}(2013{\natexlab{d}})\citenamefont {Aaij} \emph
  {et~al.}}]{Aaij:2013iua}%
  \BibitemOpen
  \bibfield  {author} {\bibinfo {author} {\bibnamefont {Aaij}, \bibfnamefont
  {R}},  \emph {et~al.} (\bibinfo {collaboration} {LHCb})} (\bibinfo {year}
  {2013}{\natexlab{d}}),\ \bibfield  {title} {\enquote {\bibinfo {title}
  {{First observation of $CP$ violation in the decays of $B^0_s$ mesons}},}\
  }\href {\doibase 10.1103/PhysRevLett.110.221601} {\bibfield  {journal}
  {\bibinfo  {journal} {Phys. Rev. Lett.}\ }\textbf {\bibinfo {volume}
  {110}}~(\bibinfo {number} {22}),\ \bibinfo {pages} {221601}},\ \Eprint
  {http://arxiv.org/abs/1304.6173} {arXiv:1304.6173 [hep-ex]} \BibitemShut
  {NoStop}%
\bibitem [{\citenamefont {Aaij}\ \emph
  {et~al.}(2013{\natexlab{e}})\citenamefont {Aaij} \emph
  {et~al.}}]{Bediaga:2012py}%
  \BibitemOpen
  \bibfield  {author} {\bibinfo {author} {\bibnamefont {Aaij}, \bibfnamefont
  {R}},  \emph {et~al.} (\bibinfo {collaboration} {LHCb})} (\bibinfo {year}
  {2013}{\natexlab{e}}),\ \bibfield  {title} {\enquote {\bibinfo {title}
  {{Implications of LHCb measurements and future prospects}},}\ }\href
  {\doibase 10.1140/epjc/s10052-013-2373-2} {\bibfield  {journal} {\bibinfo
  {journal} {Eur. Phys. J.}\ }\textbf {\bibinfo {volume} {C73}}~(\bibinfo
  {number} {4}),\ \bibinfo {pages} {2373}},\ \Eprint
  {http://arxiv.org/abs/1208.3355} {arXiv:1208.3355 [hep-ex]} \BibitemShut
  {NoStop}%
\bibitem [{\citenamefont {Aaij}\ \emph
  {et~al.}(2013{\natexlab{f}})\citenamefont {Aaij} \emph
  {et~al.}}]{Aaij:2013oba}%
  \BibitemOpen
  \bibfield  {author} {\bibinfo {author} {\bibnamefont {Aaij}, \bibfnamefont
  {R}},  \emph {et~al.} (\bibinfo {collaboration} {LHCb})} (\bibinfo {year}
  {2013}{\natexlab{f}}),\ \bibfield  {title} {\enquote {\bibinfo {title}
  {{Measurement of CP violation and the $B^0_s$ meson decay width difference
  with $B^0_s \to J/\psi K^+K^-$ and $B^0_s \to J/\psi \pi^+ \pi^-$ decays}},}\
  }\href {\doibase 10.1103/PhysRevD.87.112010} {\bibfield  {journal} {\bibinfo
  {journal} {Phys. Rev.}\ }\textbf {\bibinfo {volume} {D87}}~(\bibinfo {number}
  {11}),\ \bibinfo {pages} {112010}},\ \Eprint {http://arxiv.org/abs/1304.2600}
  {arXiv:1304.2600 [hep-ex]} \BibitemShut {NoStop}%
\bibitem [{\citenamefont {Aaij}\ \emph
  {et~al.}(2013{\natexlab{g}})\citenamefont {Aaij} \emph
  {et~al.}}]{Aaij:2013eia}%
  \BibitemOpen
  \bibfield  {author} {\bibinfo {author} {\bibnamefont {Aaij}, \bibfnamefont
  {R}},  \emph {et~al.} (\bibinfo {collaboration} {LHCb})} (\bibinfo {year}
  {2013}{\natexlab{g}}),\ \bibfield  {title} {\enquote {\bibinfo {title}
  {{Measurement of the effective $B_s^0 \to J/{\psi} K_S^0$ lifetime}},}\
  }\href {\doibase 10.1016/j.nuclphysb.2013.04.021} {\bibfield  {journal}
  {\bibinfo  {journal} {Nucl.Phys.}\ }\textbf {\bibinfo {volume} {B873}},\
  \bibinfo {pages} {275--292}},\ \Eprint {http://arxiv.org/abs/1304.4500}
  {arXiv:1304.4500 [hep-ex]} \BibitemShut {NoStop}%
\bibitem [{\citenamefont {Aaij}\ \emph
  {et~al.}(2013{\natexlab{h}})\citenamefont {Aaij} \emph
  {et~al.}}]{Aaij:2013mpa}%
  \BibitemOpen
  \bibfield  {author} {\bibinfo {author} {\bibnamefont {Aaij}, \bibfnamefont
  {R}},  \emph {et~al.} (\bibinfo {collaboration} {LHCb})} (\bibinfo {year}
  {2013}{\natexlab{h}}),\ \bibfield  {title} {\enquote {\bibinfo {title}
  {{Precision measurement of the $B^{0}_{s}$-$\bar{B}^{0}_{s}$ oscillation
  frequency with the decay $B^{0}_{s}\rightarrow D^{-}_{s}\pi^{+}$}},}\ }\href
  {\doibase 10.1088/1367-2630/15/5/053021} {\bibfield  {journal} {\bibinfo
  {journal} {New J.Phys.}\ }\textbf {\bibinfo {volume} {15}},\ \bibinfo {pages}
  {053021}},\ \Eprint {http://arxiv.org/abs/1304.4741} {arXiv:1304.4741
  [hep-ex]} \BibitemShut {NoStop}%
\bibitem [{\citenamefont {Aaij}\ \emph
  {et~al.}(2013{\natexlab{i}})\citenamefont {Aaij} \emph
  {et~al.}}]{Aaij:2013oha}%
  \BibitemOpen
  \bibfield  {author} {\bibinfo {author} {\bibnamefont {Aaij}, \bibfnamefont
  {R}},  \emph {et~al.} (\bibinfo {collaboration} {LHCb})} (\bibinfo {year}
  {2013}{\natexlab{i}}),\ \bibfield  {title} {\enquote {\bibinfo {title}
  {{Precision measurement of the $\Lambda_b^0$ baryon lifetime}},}\ }\href
  {\doibase 10.1103/PhysRevLett.111.102003} {\bibfield  {journal} {\bibinfo
  {journal} {Phys.Rev.Lett.}\ }\textbf {\bibinfo {volume} {111}},\ \bibinfo
  {pages} {102003}},\ \Eprint {http://arxiv.org/abs/1307.2476} {arXiv:1307.2476
  [hep-ex]} \BibitemShut {NoStop}%
\bibitem [{\citenamefont {Aaij}\ \emph
  {et~al.}(2014{\natexlab{a}})\citenamefont {Aaij} \emph
  {et~al.}}]{Aaij:2014fia}%
  \BibitemOpen
  \bibfield  {author} {\bibinfo {author} {\bibnamefont {Aaij}, \bibfnamefont
  {R}},  \emph {et~al.} (\bibinfo {collaboration} {LHCb})} (\bibinfo {year}
  {2014}{\natexlab{a}}),\ \bibfield  {title} {\enquote {\bibinfo {title}
  {{Effective lifetime measurements in the $B_s^0 \to K^+K^- , B^0 \to K^+
  \pi^-$ and $B_s^0 \to \pi^+ K^-$ decays}},}\ }\href {\doibase
  10.1016/j.physletb.2014.07.051} {\bibfield  {journal} {\bibinfo  {journal}
  {Phys. Lett.}\ }\textbf {\bibinfo {volume} {B736}},\ \bibinfo {pages}
  {446--454}},\ \Eprint {http://arxiv.org/abs/1406.7204} {arXiv:1406.7204
  [hep-ex]} \BibitemShut {NoStop}%
\bibitem [{\citenamefont {Aaij}\ \emph
  {et~al.}(2014{\natexlab{b}})\citenamefont {Aaij} \emph
  {et~al.}}]{Aaij:2013bvd}%
  \BibitemOpen
  \bibfield  {author} {\bibinfo {author} {\bibnamefont {Aaij}, \bibfnamefont
  {R}},  \emph {et~al.} (\bibinfo {collaboration} {LHCb})} (\bibinfo {year}
  {2014}{\natexlab{b}}),\ \bibfield  {title} {\enquote {\bibinfo {title}
  {{Measurement of the $\bar{B}_s^0\to D_s^-D_s^+$ and $\bar{B}_s^0\to
  D^-D_s^+$ effective lifetimes}},}\ }\href {\doibase
  10.1103/PhysRevLett.112.111802} {\bibfield  {journal} {\bibinfo  {journal}
  {Phys. Rev. Lett.}\ }\textbf {\bibinfo {volume} {112}}~(\bibinfo {number}
  {11}),\ \bibinfo {pages} {111802}},\ \Eprint {http://arxiv.org/abs/1312.1217}
  {arXiv:1312.1217 [hep-ex]} \BibitemShut {NoStop}%
\bibitem [{\citenamefont {Aaij}\ \emph
  {et~al.}(2014{\natexlab{c}})\citenamefont {Aaij} \emph
  {et~al.}}]{Aaij:2013gta}%
  \BibitemOpen
  \bibfield  {author} {\bibinfo {author} {\bibnamefont {Aaij}, \bibfnamefont
  {R}},  \emph {et~al.} (\bibinfo {collaboration} {LHCb})} (\bibinfo {year}
  {2014}{\natexlab{c}}),\ \bibfield  {title} {\enquote {\bibinfo {title}
  {{Measurement of the flavour-specific CP-violating asymmetry $a_{sl}^s$ in
  $B_s^0$ decays}},}\ }\href {\doibase 10.1016/j.physletb.2013.12.030}
  {\bibfield  {journal} {\bibinfo  {journal} {Phys. Lett.}\ }\textbf {\bibinfo
  {volume} {B728}},\ \bibinfo {pages} {607--615}},\ \Eprint
  {http://arxiv.org/abs/1308.1048} {arXiv:1308.1048 [hep-ex]} \BibitemShut
  {NoStop}%
\bibitem [{\citenamefont {Aaij}\ \emph
  {et~al.}(2014{\natexlab{d}})\citenamefont {Aaij} \emph
  {et~al.}}]{Aaij:2014fba}%
  \BibitemOpen
  \bibfield  {author} {\bibinfo {author} {\bibnamefont {Aaij}, \bibfnamefont
  {Roel}},  \emph {et~al.} (\bibinfo {collaboration} {LHCb})} (\bibinfo {year}
  {2014}{\natexlab{d}}),\ \bibfield  {title} {\enquote {\bibinfo {title}
  {{Measurement of CP asymmetry in $B^0_s \rightarrow D^{\mp}_s K^{\pm}$
  decays}},}\ }\href {\doibase 10.1007/JHEP11(2014)060} {\bibfield  {journal}
  {\bibinfo  {journal} {JHEP}\ }\textbf {\bibinfo {volume} {11}},\ \bibinfo
  {pages} {060}},\ \Eprint {http://arxiv.org/abs/1407.6127} {arXiv:1407.6127
  [hep-ex]} \BibitemShut {NoStop}%
\bibitem [{\citenamefont {Aaij}\ \emph
  {et~al.}(2014{\natexlab{e}})\citenamefont {Aaij} \emph
  {et~al.}}]{Aaij:2014kxa}%
  \BibitemOpen
  \bibfield  {author} {\bibinfo {author} {\bibnamefont {Aaij}, \bibfnamefont
  {Roel}},  \emph {et~al.} (\bibinfo {collaboration} {LHCb})} (\bibinfo {year}
  {2014}{\natexlab{e}}),\ \bibfield  {title} {\enquote {\bibinfo {title}
  {{Measurement of CP violation in $B_s^0 \to \phi \phi$ decays}},}\ }\href
  {\doibase 10.1103/PhysRevD.90.052011} {\bibfield  {journal} {\bibinfo
  {journal} {Phys. Rev.}\ }\textbf {\bibinfo {volume} {D90}}~(\bibinfo {number}
  {5}),\ \bibinfo {pages} {052011}},\ \Eprint {http://arxiv.org/abs/1407.2222}
  {arXiv:1407.2222 [hep-ex]} \BibitemShut {NoStop}%
\bibitem [{\citenamefont {Aaij}\ \emph
  {et~al.}(2014{\natexlab{f}})\citenamefont {Aaij} \emph
  {et~al.}}]{Aaij:2014emv}%
  \BibitemOpen
  \bibfield  {author} {\bibinfo {author} {\bibnamefont {Aaij}, \bibfnamefont
  {Roel}},  \emph {et~al.} (\bibinfo {collaboration} {LHCb})} (\bibinfo {year}
  {2014}{\natexlab{f}}),\ \bibfield  {title} {\enquote {\bibinfo {title}
  {{Measurement of resonant and CP components in $\bar{B}_s^0 \to
  J/\psi\pi^+\pi^-$ decays}},}\ }\href {\doibase 10.1103/PhysRevD.89.092006}
  {\bibfield  {journal} {\bibinfo  {journal} {Phys. Rev.}\ }\textbf {\bibinfo
  {volume} {D89}}~(\bibinfo {number} {9}),\ \bibinfo {pages} {092006}},\
  \Eprint {http://arxiv.org/abs/1402.6248} {arXiv:1402.6248 [hep-ex]}
  \BibitemShut {NoStop}%
\bibitem [{\citenamefont {Aaij}\ \emph
  {et~al.}(2014{\natexlab{g}})\citenamefont {Aaij} \emph
  {et~al.}}]{Aaij:2014ywt}%
  \BibitemOpen
  \bibfield  {author} {\bibinfo {author} {\bibnamefont {Aaij}, \bibfnamefont
  {Roel}},  \emph {et~al.} (\bibinfo {collaboration} {LHCb})} (\bibinfo {year}
  {2014}{\natexlab{g}}),\ \bibfield  {title} {\enquote {\bibinfo {title}
  {{Measurement of the $CP$-violating phase $\phi_s$ in $\bar{B}^{0}_{s}\to
  D_{s}^{+}D_{s}^{-}$ decays}},}\ }\href {\doibase
  10.1103/PhysRevLett.113.211801} {\bibfield  {journal} {\bibinfo  {journal}
  {Phys. Rev. Lett.}\ }\textbf {\bibinfo {volume} {113}}~(\bibinfo {number}
  {21}),\ \bibinfo {pages} {211801}},\ \Eprint {http://arxiv.org/abs/1409.4619}
  {arXiv:1409.4619 [hep-ex]} \BibitemShut {NoStop}%
\bibitem [{\citenamefont {Aaij}\ \emph
  {et~al.}(2014{\natexlab{h}})\citenamefont {Aaij} \emph
  {et~al.}}]{Aaij:2014dka}%
  \BibitemOpen
  \bibfield  {author} {\bibinfo {author} {\bibnamefont {Aaij}, \bibfnamefont
  {Roel}},  \emph {et~al.} (\bibinfo {collaboration} {LHCb})} (\bibinfo {year}
  {2014}{\natexlab{h}}),\ \bibfield  {title} {\enquote {\bibinfo {title}
  {{Measurement of the CP-violating phase $\phi_s$ in
  $\overline{B}^0_s\rightarrow J/\psi \pi^+\pi^-$ decays}},}\ }\href {\doibase
  10.1016/j.physletb.2014.06.079} {\bibfield  {journal} {\bibinfo  {journal}
  {Phys. Lett.}\ }\textbf {\bibinfo {volume} {B736}},\ \bibinfo {pages}
  {186}},\ \Eprint {http://arxiv.org/abs/1405.4140} {arXiv:1405.4140 [hep-ex]}
  \BibitemShut {NoStop}%
\bibitem [{\citenamefont {Aaij}\ \emph
  {et~al.}(2014{\natexlab{i}})\citenamefont {Aaij} \emph
  {et~al.}}]{Aaij:2014siy}%
  \BibitemOpen
  \bibfield  {author} {\bibinfo {author} {\bibnamefont {Aaij}, \bibfnamefont
  {Roel}},  \emph {et~al.} (\bibinfo {collaboration} {LHCb})} (\bibinfo {year}
  {2014}{\natexlab{i}}),\ \bibfield  {title} {\enquote {\bibinfo {title}
  {{Measurement of the resonant and CP components in $\overline{B}^0\to J/\psi
  \pi^+\pi^-$ decays}},}\ }\href {\doibase 10.1103/PhysRevD.90.012003}
  {\bibfield  {journal} {\bibinfo  {journal} {Phys. Rev.}\ }\textbf {\bibinfo
  {volume} {D90}}~(\bibinfo {number} {1}),\ \bibinfo {pages} {012003}},\
  \Eprint {http://arxiv.org/abs/1404.5673} {arXiv:1404.5673 [hep-ex]}
  \BibitemShut {NoStop}%
\bibitem [{\citenamefont {Aaij}\ \emph
  {et~al.}(2014{\natexlab{j}})\citenamefont {Aaij} \emph
  {et~al.}}]{Aaij:2014owa}%
  \BibitemOpen
  \bibfield  {author} {\bibinfo {author} {\bibnamefont {Aaij}, \bibfnamefont
  {Roel}},  \emph {et~al.} (\bibinfo {collaboration} {LHCb})} (\bibinfo {year}
  {2014}{\natexlab{j}}),\ \bibfield  {title} {\enquote {\bibinfo {title}
  {{Measurements of the $B^+, B^0, B^0_s$ meson and $\Lambda^0_b$ baryon
  lifetimes}},}\ }\href {\doibase 10.1007/JHEP04(2014)114} {\bibfield
  {journal} {\bibinfo  {journal} {JHEP}\ }\textbf {\bibinfo {volume} {04}},\
  \bibinfo {pages} {114}},\ \Eprint {http://arxiv.org/abs/1402.2554}
  {arXiv:1402.2554 [hep-ex]} \BibitemShut {NoStop}%
\bibitem [{\citenamefont {Aaij}\ \emph
  {et~al.}(2014{\natexlab{k}})\citenamefont {Aaij} \emph
  {et~al.}}]{Aaij:2014zyy}%
  \BibitemOpen
  \bibfield  {author} {\bibinfo {author} {\bibnamefont {Aaij}, \bibfnamefont
  {Roel}},  \emph {et~al.} (\bibinfo {collaboration} {LHCb})} (\bibinfo {year}
  {2014}{\natexlab{k}}),\ \bibfield  {title} {\enquote {\bibinfo {title}
  {{Precision measurement of the ratio of the $\Lambda^0_b$ to $\overline{B}^0$
  lifetimes}},}\ }\href {\doibase 10.1016/j.physletb.2014.05.021} {\bibfield
  {journal} {\bibinfo  {journal} {Phys. Lett.}\ }\textbf {\bibinfo {volume}
  {B734}},\ \bibinfo {pages} {122--130}},\ \Eprint
  {http://arxiv.org/abs/1402.6242} {arXiv:1402.6242 [hep-ex]} \BibitemShut
  {NoStop}%
\bibitem [{\citenamefont {Aaij}\ \emph
  {et~al.}(2015{\natexlab{a}})\citenamefont {Aaij} \emph
  {et~al.}}]{Aaij:2015pha}%
  \BibitemOpen
  \bibfield  {author} {\bibinfo {author} {\bibnamefont {Aaij}, \bibfnamefont
  {Roel}},  \emph {et~al.} (\bibinfo {collaboration} {LHCb})} (\bibinfo {year}
  {2015}{\natexlab{a}}),\ \bibfield  {title} {\enquote {\bibinfo {title} {{$B$
  flavour tagging using charm decays at the LHCb experiment}},}\ }\href@noop {}
  {\ }\Eprint {http://arxiv.org/abs/1507.07892} {arXiv:1507.07892 [hep-ex]}
  \BibitemShut {NoStop}%
\bibitem [{\citenamefont {Aaij}\ \emph
  {et~al.}(2015{\natexlab{b}})\citenamefont {Aaij} \emph
  {et~al.}}]{Aaij:2014xba}%
  \BibitemOpen
  \bibfield  {author} {\bibinfo {author} {\bibnamefont {Aaij}, \bibfnamefont
  {Roel}},  \emph {et~al.} (\bibinfo {collaboration} {LHCb})} (\bibinfo {year}
  {2015}{\natexlab{b}}),\ \bibfield  {title} {\enquote {\bibinfo {title}
  {{Determination of $\gamma$ and 2$\beta_s$ from charmless two-body decays of
  beauty mesons}},}\ }\href {\doibase 10.1016/j.physletb.2014.12.015}
  {\bibfield  {journal} {\bibinfo  {journal} {Phys. Lett.}\ }\textbf {\bibinfo
  {volume} {B741}},\ \bibinfo {pages} {1--11}},\ \Eprint
  {http://arxiv.org/abs/1408.4368} {arXiv:1408.4368 [hep-ex]} \BibitemShut
  {NoStop}%
\bibitem [{\citenamefont {Aaij}\ \emph
  {et~al.}(2015{\natexlab{c}})\citenamefont {Aaij} \emph
  {et~al.}}]{Aaij:2015kba}%
  \BibitemOpen
  \bibfield  {author} {\bibinfo {author} {\bibnamefont {Aaij}, \bibfnamefont
  {Roel}},  \emph {et~al.} (\bibinfo {collaboration} {LHCb})} (\bibinfo {year}
  {2015}{\natexlab{c}}),\ \bibfield  {title} {\enquote {\bibinfo {title}
  {{Measurement of $CP$ asymmetries and polarisation fractions in $B_s^0
  \rightarrow K^{*0}\bar{K}{}^{*0}$ decays}},}\ }\href {\doibase
  10.1007/JHEP07(2015)166} {\bibfield  {journal} {\bibinfo  {journal} {JHEP}\
  }\textbf {\bibinfo {volume} {07}},\ \bibinfo {pages} {166}},\ \Eprint
  {http://arxiv.org/abs/1503.05362} {arXiv:1503.05362 [hep-ex]} \BibitemShut
  {NoStop}%
\bibitem [{\citenamefont {Aaij}\ \emph
  {et~al.}(2015{\natexlab{d}})\citenamefont {Aaij} \emph
  {et~al.}}]{Aaij:2015mea}%
  \BibitemOpen
  \bibfield  {author} {\bibinfo {author} {\bibnamefont {Aaij}, \bibfnamefont
  {Roel}},  \emph {et~al.} (\bibinfo {collaboration} {LHCb})} (\bibinfo {year}
  {2015}{\natexlab{d}}),\ \bibfield  {title} {\enquote {\bibinfo {title}
  {{Measurement of ${C\!P}$ violation parameters and polarisation fractions in
  ${B_s^0\to J/\psi \overline{K}^{*0}}$ decays}},}\ }\href@noop {} {\ }\Eprint
  {http://arxiv.org/abs/1509.00400} {arXiv:1509.00400 [hep-ex]} \BibitemShut
  {NoStop}%
\bibitem [{\citenamefont {Aaij}\ \emph
  {et~al.}(2015{\natexlab{e}})\citenamefont {Aaij} \emph
  {et~al.}}]{Aaij:2014vda}%
  \BibitemOpen
  \bibfield  {author} {\bibinfo {author} {\bibnamefont {Aaij}, \bibfnamefont
  {Roel}},  \emph {et~al.} (\bibinfo {collaboration} {LHCb})} (\bibinfo {year}
  {2015}{\natexlab{e}}),\ \bibfield  {title} {\enquote {\bibinfo {title}
  {{Measurement of the CP-violating phase $\beta$ in $B^0\rightarrow J/\psi
  \pi^+\pi^-$ decays and limits on penguin effects}},}\ }\href {\doibase
  10.1016/j.physletb.2015.01.008} {\bibfield  {journal} {\bibinfo  {journal}
  {Phys. Lett.}\ }\textbf {\bibinfo {volume} {B742}},\ \bibinfo {pages}
  {38--49}},\ \Eprint {http://arxiv.org/abs/1411.1634} {arXiv:1411.1634
  [hep-ex]} \BibitemShut {NoStop}%
\bibitem [{\citenamefont {Aaij}\ \emph
  {et~al.}(2015{\natexlab{f}})\citenamefont {Aaij} \emph
  {et~al.}}]{Aaij:2014nxa}%
  \BibitemOpen
  \bibfield  {author} {\bibinfo {author} {\bibnamefont {Aaij}, \bibfnamefont
  {Roel}},  \emph {et~al.} (\bibinfo {collaboration} {LHCb})} (\bibinfo {year}
  {2015}{\natexlab{f}}),\ \bibfield  {title} {\enquote {\bibinfo {title}
  {{Measurement of the semileptonic $CP$ asymmetry in $B^0-\overline{B}{}^0$
  mixing}},}\ }\href {\doibase 10.1103/PhysRevLett.114.041601} {\bibfield
  {journal} {\bibinfo  {journal} {Phys. Rev. Lett.}\ }\textbf {\bibinfo
  {volume} {114}},\ \bibinfo {pages} {041601}},\ \Eprint
  {http://arxiv.org/abs/1409.8586} {arXiv:1409.8586 [hep-ex]} \BibitemShut
  {NoStop}%
\bibitem [{\citenamefont {Aaij}\ \emph
  {et~al.}(2015{\natexlab{g}})\citenamefont {Aaij} \emph
  {et~al.}}]{Aaij:2015tza}%
  \BibitemOpen
  \bibfield  {author} {\bibinfo {author} {\bibnamefont {Aaij}, \bibfnamefont
  {Roel}},  \emph {et~al.} (\bibinfo {collaboration} {LHCb})} (\bibinfo {year}
  {2015}{\natexlab{g}}),\ \bibfield  {title} {\enquote {\bibinfo {title}
  {{Measurement of the time-dependent CP asymmetries in $B_s^0\rightarrow
  J/\psi K_{\rm S}^0$}},}\ }\href {\doibase 10.1007/JHEP06(2015)131} {\bibfield
   {journal} {\bibinfo  {journal} {JHEP}\ }\textbf {\bibinfo {volume} {06}},\
  \bibinfo {pages} {131}},\ \Eprint {http://arxiv.org/abs/1503.07055}
  {arXiv:1503.07055 [hep-ex]} \BibitemShut {NoStop}%
\bibitem [{\citenamefont {Aaij}\ \emph
  {et~al.}(2015{\natexlab{h}})\citenamefont {Aaij} \emph
  {et~al.}}]{Aaij:2014zsa}%
  \BibitemOpen
  \bibfield  {author} {\bibinfo {author} {\bibnamefont {Aaij}, \bibfnamefont
  {Roel}},  \emph {et~al.} (\bibinfo {collaboration} {LHCb})} (\bibinfo {year}
  {2015}{\natexlab{h}}),\ \bibfield  {title} {\enquote {\bibinfo {title}
  {{Precision measurement of $CP$ violation in $B_s^0 \to J/\psi K^+K^-$
  decays}},}\ }\href {\doibase 10.1103/PhysRevLett.114.041801} {\bibfield
  {journal} {\bibinfo  {journal} {Phys. Rev. Lett.}\ }\textbf {\bibinfo
  {volume} {114}}~(\bibinfo {number} {4}),\ \bibinfo {pages} {041801}},\
  \Eprint {http://arxiv.org/abs/1411.3104} {arXiv:1411.3104 [hep-ex]}
  \BibitemShut {NoStop}%
\bibitem [{\citenamefont {Aaij}\ \emph {et~al.}(2016)\citenamefont {Aaij} \emph
  {et~al.}}]{Aaij:2016yze}%
  \BibitemOpen
  \bibfield  {author} {\bibinfo {author} {\bibnamefont {Aaij}, \bibfnamefont
  {Roel}},  \emph {et~al.} (\bibinfo {collaboration} {LHCb})} (\bibinfo {year}
  {2016}),\ \bibfield  {title} {\enquote {\bibinfo {title} {{Measurement of the
  $CP$ asymmetry in $B_s^0-\overline{B}{}_s^0$ mixing}},}\ }\href@noop {} {\
  }\Eprint {http://arxiv.org/abs/1605.09768} {arXiv:1605.09768 [hep-ex]}
  \BibitemShut {NoStop}%
\bibitem [{\citenamefont {Aaltonen}\ \emph {et~al.}(2011)\citenamefont
  {Aaltonen} \emph {et~al.}}]{Aaltonen:2011nk}%
  \BibitemOpen
  \bibfield  {author} {\bibinfo {author} {\bibnamefont {Aaltonen},
  \bibfnamefont {T}},  \emph {et~al.} (\bibinfo {collaboration} {CDF})}
  (\bibinfo {year} {2011}),\ \bibfield  {title} {\enquote {\bibinfo {title}
  {{Measurement of branching ratio and $B_s^0$ lifetime in the decay $B_s^0
  \rightarrow J/\psi f_0(980)$ at CDF}},}\ }\href {\doibase
  10.1103/PhysRevD.84.052012} {\bibfield  {journal} {\bibinfo  {journal} {Phys.
  Rev.}\ }\textbf {\bibinfo {volume} {D84}},\ \bibinfo {pages} {052012}},\
  \Eprint {http://arxiv.org/abs/1106.3682} {arXiv:1106.3682 [hep-ex]}
  \BibitemShut {NoStop}%
\bibitem [{\citenamefont {Aaltonen}\ \emph {et~al.}(2012)\citenamefont
  {Aaltonen} \emph {et~al.}}]{Aaltonen:2012ie}%
  \BibitemOpen
  \bibfield  {author} {\bibinfo {author} {\bibnamefont {Aaltonen},
  \bibfnamefont {T}},  \emph {et~al.} (\bibinfo {collaboration} {CDF})}
  (\bibinfo {year} {2012}),\ \bibfield  {title} {\enquote {\bibinfo {title}
  {{Measurement of the Bottom-Strange Meson Mixing Phase in the Full CDF Data
  Set}},}\ }\href {\doibase 10.1103/PhysRevLett.109.171802} {\bibfield
  {journal} {\bibinfo  {journal} {Phys. Rev. Lett.}\ }\textbf {\bibinfo
  {volume} {109}},\ \bibinfo {pages} {171802}},\ \Eprint
  {http://arxiv.org/abs/1208.2967} {arXiv:1208.2967 [hep-ex]} \BibitemShut
  {NoStop}%
\bibitem [{\citenamefont {Aaltonen}\ \emph {et~al.}(2014)\citenamefont
  {Aaltonen} \emph {et~al.}}]{Aaltonen:2014wfa}%
  \BibitemOpen
  \bibfield  {author} {\bibinfo {author} {\bibnamefont {Aaltonen},
  \bibfnamefont {Timo~Antero}},  \emph {et~al.} (\bibinfo {collaboration}
  {CDF})} (\bibinfo {year} {2014}),\ \bibfield  {title} {\enquote {\bibinfo
  {title} {{Mass and lifetime measurements of bottom and charm baryons in
  $p\bar p$ collisions at $\sqrt{s}= 1.96$ TeV}},}\ }\href {\doibase
  10.1103/PhysRevD.89.072014} {\bibfield  {journal} {\bibinfo  {journal} {Phys.
  Rev.}\ }\textbf {\bibinfo {volume} {D89}}~(\bibinfo {number} {7}),\ \bibinfo
  {pages} {072014}},\ \Eprint {http://arxiv.org/abs/1403.8126} {arXiv:1403.8126
  [hep-ex]} \BibitemShut {NoStop}%
\bibitem [{\citenamefont {Abazov}\ \emph
  {et~al.}(2006{\natexlab{a}})\citenamefont {Abazov} \emph
  {et~al.}}]{Abazov:2006dm}%
  \BibitemOpen
  \bibfield  {author} {\bibinfo {author} {\bibnamefont {Abazov}, \bibfnamefont
  {V~M}},  \emph {et~al.} (\bibinfo {collaboration} {D0})} (\bibinfo {year}
  {2006}{\natexlab{a}}),\ \bibfield  {title} {\enquote {\bibinfo {title}
  {{First direct two-sided bound on the $B^0_{s}$ oscillation frequency}},}\
  }\href {\doibase 10.1103/PhysRevLett.97.021802} {\bibfield  {journal}
  {\bibinfo  {journal} {Phys. Rev. Lett.}\ }\textbf {\bibinfo {volume} {97}},\
  \bibinfo {pages} {021802}},\ \Eprint {http://arxiv.org/abs/hep-ex/0603029}
  {arXiv:hep-ex/0603029 [hep-ex]} \BibitemShut {NoStop}%
\bibitem [{\citenamefont {Abazov}\ \emph
  {et~al.}(2006{\natexlab{b}})\citenamefont {Abazov} \emph
  {et~al.}}]{Abazov:2006qp}%
  \BibitemOpen
  \bibfield  {author} {\bibinfo {author} {\bibnamefont {Abazov}, \bibfnamefont
  {V~M}},  \emph {et~al.} (\bibinfo {collaboration} {D0})} (\bibinfo {year}
  {2006}{\natexlab{b}}),\ \bibfield  {title} {\enquote {\bibinfo {title}
  {{Measurement of $B_d$ mixing using opposite-side flavor tagging}},}\ }\href
  {\doibase 10.1103/PhysRevD.74.112002} {\bibfield  {journal} {\bibinfo
  {journal} {Phys. Rev.}\ }\textbf {\bibinfo {volume} {D74}},\ \bibinfo {pages}
  {112002}},\ \Eprint {http://arxiv.org/abs/hep-ex/0609034}
  {arXiv:hep-ex/0609034 [hep-ex]} \BibitemShut {NoStop}%
\bibitem [{\citenamefont {Abazov}\ \emph
  {et~al.}(2006{\natexlab{c}})\citenamefont {Abazov} \emph
  {et~al.}}]{Abazov:2005pn}%
  \BibitemOpen
  \bibfield  {author} {\bibinfo {author} {\bibnamefont {Abazov}, \bibfnamefont
  {V~M}},  \emph {et~al.} (\bibinfo {collaboration} {D0})} (\bibinfo {year}
  {2006}{\natexlab{c}}),\ \bibfield  {title} {\enquote {\bibinfo {title} {{The
  Upgraded D0 detector}},}\ }\href {\doibase 10.1016/j.nima.2006.05.248}
  {\bibfield  {journal} {\bibinfo  {journal} {Nucl. Instrum. Meth.}\ }\textbf
  {\bibinfo {volume} {A565}},\ \bibinfo {pages} {463--537}},\ \Eprint
  {http://arxiv.org/abs/physics/0507191} {arXiv:physics/0507191
  [physics.ins-det]} \BibitemShut {NoStop}%
\bibitem [{\citenamefont {Abazov}\ \emph
  {et~al.}(2010{\natexlab{a}})\citenamefont {Abazov} \emph
  {et~al.}}]{Abazov:2010hj}%
  \BibitemOpen
  \bibfield  {author} {\bibinfo {author} {\bibnamefont {Abazov}, \bibfnamefont
  {V~M}},  \emph {et~al.} (\bibinfo {collaboration} {D0})} (\bibinfo {year}
  {2010}{\natexlab{a}}),\ \bibfield  {title} {\enquote {\bibinfo {title}
  {{Evidence for an anomalous like-sign dimuon charge asymmetry}},}\ }\href
  {\doibase 10.1103/PhysRevLett.105.081801} {\bibfield  {journal} {\bibinfo
  {journal} {Phys. Rev. Lett.}\ }\textbf {\bibinfo {volume} {105}},\ \bibinfo
  {pages} {081801}},\ \Eprint {http://arxiv.org/abs/1007.0395} {arXiv:1007.0395
  [hep-ex]} \BibitemShut {NoStop}%
\bibitem [{\citenamefont {Abazov}\ \emph
  {et~al.}(2010{\natexlab{b}})\citenamefont {Abazov} \emph
  {et~al.}}]{Abazov:2010hv}%
  \BibitemOpen
  \bibfield  {author} {\bibinfo {author} {\bibnamefont {Abazov}, \bibfnamefont
  {V~M}},  \emph {et~al.} (\bibinfo {collaboration} {D0})} (\bibinfo {year}
  {2010}{\natexlab{b}}),\ \bibfield  {title} {\enquote {\bibinfo {title}
  {{Evidence for an anomalous like-sign dimuon charge asymmetry}},}\ }\href
  {\doibase 10.1103/PhysRevD.82.032001} {\bibfield  {journal} {\bibinfo
  {journal} {Phys. Rev.}\ }\textbf {\bibinfo {volume} {D82}},\ \bibinfo {pages}
  {032001}},\ \Eprint {http://arxiv.org/abs/1005.2757} {arXiv:1005.2757
  [hep-ex]} \BibitemShut {NoStop}%
\bibitem [{\citenamefont {Abazov}\ \emph {et~al.}(2011)\citenamefont {Abazov}
  \emph {et~al.}}]{Abazov:2011yk}%
  \BibitemOpen
  \bibfield  {author} {\bibinfo {author} {\bibnamefont {Abazov}, \bibfnamefont
  {V~M}},  \emph {et~al.} (\bibinfo {collaboration} {D0})} (\bibinfo {year}
  {2011}),\ \bibfield  {title} {\enquote {\bibinfo {title} {{Measurement of the
  anomalous like-sign dimuon charge asymmetry with 9 fb${^-1}$ of $p\bar{p}$
  collisions}},}\ }\href {\doibase 10.1103/PhysRevD.84.052007} {\bibfield
  {journal} {\bibinfo  {journal} {Phys. Rev.}\ }\textbf {\bibinfo {volume}
  {D84}},\ \bibinfo {pages} {052007}},\ \Eprint
  {http://arxiv.org/abs/1106.6308} {arXiv:1106.6308 [hep-ex]} \BibitemShut
  {NoStop}%
\bibitem [{\citenamefont {Abazov}\ \emph
  {et~al.}(2012{\natexlab{a}})\citenamefont {Abazov} \emph
  {et~al.}}]{Abazov:2011ry}%
  \BibitemOpen
  \bibfield  {author} {\bibinfo {author} {\bibnamefont {Abazov}, \bibfnamefont
  {V~M}},  \emph {et~al.} (\bibinfo {collaboration} {D0})} (\bibinfo {year}
  {2012}{\natexlab{a}}),\ \bibfield  {title} {\enquote {\bibinfo {title}
  {{Measurement of the CP-violating phase $\phi_s^{J/\psi \phi}$ using the
  flavor-tagged decay $B_s^0 \rightarrow J/\psi \phi$ in 8 fb$^{-1}$ of $p \bar
  p$ collisions}},}\ }\href {\doibase 10.1103/PhysRevD.85.032006} {\bibfield
  {journal} {\bibinfo  {journal} {Phys. Rev.}\ }\textbf {\bibinfo {volume}
  {D85}},\ \bibinfo {pages} {032006}},\ \Eprint
  {http://arxiv.org/abs/1109.3166} {arXiv:1109.3166 [hep-ex]} \BibitemShut
  {NoStop}%
\bibitem [{\citenamefont {Abazov}\ \emph
  {et~al.}(2012{\natexlab{b}})\citenamefont {Abazov} \emph
  {et~al.}}]{Abazov:2012hha}%
  \BibitemOpen
  \bibfield  {author} {\bibinfo {author} {\bibnamefont {Abazov}, \bibfnamefont
  {V~M}},  \emph {et~al.} (\bibinfo {collaboration} {D0})} (\bibinfo {year}
  {2012}{\natexlab{b}}),\ \bibfield  {title} {\enquote {\bibinfo {title}
  {{Measurement of the semileptonic charge asymmetry in $B^0$ meson mixing with
  the D0 detector}},}\ }\href {\doibase 10.1103/PhysRevD.86.072009} {\bibfield
  {journal} {\bibinfo  {journal} {Phys. Rev.}\ }\textbf {\bibinfo {volume}
  {D86}},\ \bibinfo {pages} {072009}},\ \Eprint
  {http://arxiv.org/abs/1208.5813} {arXiv:1208.5813 [hep-ex]} \BibitemShut
  {NoStop}%
\bibitem [{\citenamefont {Abazov}\ \emph {et~al.}(2013)\citenamefont {Abazov}
  \emph {et~al.}}]{Abazov:2012zz}%
  \BibitemOpen
  \bibfield  {author} {\bibinfo {author} {\bibnamefont {Abazov}, \bibfnamefont
  {V~M}},  \emph {et~al.} (\bibinfo {collaboration} {D0})} (\bibinfo {year}
  {2013}),\ \bibfield  {title} {\enquote {\bibinfo {title} {{Measurement of the
  Semileptonic Charge Asymmetry using $B_s^0 \to D_s \mu X$ Decays}},}\ }\href
  {\doibase 10.1103/PhysRevLett.110.011801} {\bibfield  {journal} {\bibinfo
  {journal} {Phys. Rev. Lett.}\ }\textbf {\bibinfo {volume} {110}},\ \bibinfo
  {pages} {011801}},\ \Eprint {http://arxiv.org/abs/1207.1769} {arXiv:1207.1769
  [hep-ex]} \BibitemShut {NoStop}%
\bibitem [{\citenamefont {Abazov}\ \emph {et~al.}(2014)\citenamefont {Abazov}
  \emph {et~al.}}]{Abazov:2013uma}%
  \BibitemOpen
  \bibfield  {author} {\bibinfo {author} {\bibnamefont {Abazov}, \bibfnamefont
  {V~M}},  \emph {et~al.} (\bibinfo {collaboration} {D0})} (\bibinfo {year}
  {2014}),\ \bibfield  {title} {\enquote {\bibinfo {title} {{Study of CP
  -violating charge asymmetries of single muons and like-sign dimuons in $p
  \bar{p}$ collisions}},}\ }\href {\doibase 10.1103/PhysRevD.89.012002}
  {\bibfield  {journal} {\bibinfo  {journal} {Phys. Rev.}\ }\textbf {\bibinfo
  {volume} {D89}}~(\bibinfo {number} {1}),\ \bibinfo {pages} {012002}},\
  \Eprint {http://arxiv.org/abs/1310.0447} {arXiv:1310.0447 [hep-ex]}
  \BibitemShut {NoStop}%
\bibitem [{\citenamefont {Abe}\ \emph {et~al.}(2010)\citenamefont {Abe} \emph
  {et~al.}}]{Abe:2010gxa}%
  \BibitemOpen
  \bibfield  {author} {\bibinfo {author} {\bibnamefont {Abe}, \bibfnamefont
  {T}},  \emph {et~al.} (\bibinfo {collaboration} {Belle-II})} (\bibinfo {year}
  {2010}),\ \bibfield  {title} {\enquote {\bibinfo {title} {{Belle II Technical
  Design Report}},}\ }\href@noop {} {\ }\Eprint
  {http://arxiv.org/abs/1011.0352} {arXiv:1011.0352 [physics.ins-det]}
  \BibitemShut {NoStop}%
\bibitem [{\citenamefont {Abulencia}\ \emph {et~al.}(2006)\citenamefont
  {Abulencia} \emph {et~al.}}]{Abulencia:2006ze}%
  \BibitemOpen
  \bibfield  {author} {\bibinfo {author} {\bibnamefont {Abulencia},
  \bibfnamefont {A}},  \emph {et~al.} (\bibinfo {collaboration} {CDF})}
  (\bibinfo {year} {2006}),\ \bibfield  {title} {\enquote {\bibinfo {title}
  {{Observation of $B^0_s - \bar{B}^0_s$ Oscillations}},}\ }\href {\doibase
  10.1103/PhysRevLett.97.242003} {\bibfield  {journal} {\bibinfo  {journal}
  {Phys. Rev. Lett.}\ }\textbf {\bibinfo {volume} {97}},\ \bibinfo {pages}
  {242003}},\ \Eprint {http://arxiv.org/abs/hep-ex/0609040}
  {arXiv:hep-ex/0609040 [hep-ex]} \BibitemShut {NoStop}%
\bibitem [{\citenamefont {Adinolfi}\ \emph {et~al.}(2013)\citenamefont
  {Adinolfi} \emph {et~al.}}]{Adinolfi:2012qfa}%
  \BibitemOpen
  \bibfield  {author} {\bibinfo {author} {\bibnamefont {Adinolfi},
  \bibfnamefont {M}},  \emph {et~al.} (\bibinfo {collaboration} {LHCb RICH
  Group})} (\bibinfo {year} {2013}),\ \bibfield  {title} {\enquote {\bibinfo
  {title} {{Performance of the LHCb RICH detector at the LHC}},}\ }\href
  {\doibase 10.1140/epjc/s10052-013-2431-9} {\bibfield  {journal} {\bibinfo
  {journal} {Eur. Phys. J.}\ }\textbf {\bibinfo {volume} {C73}},\ \bibinfo
  {pages} {2431}},\ \Eprint {http://arxiv.org/abs/1211.6759} {arXiv:1211.6759
  [physics.ins-det]} \BibitemShut {NoStop}%
\bibitem [{\citenamefont {Aleksan}\ \emph {et~al.}(1992)\citenamefont
  {Aleksan}, \citenamefont {Dunietz},\ and\ \citenamefont
  {Kayser}}]{Aleksan:1991nh}%
  \BibitemOpen
  \bibfield  {author} {\bibinfo {author} {\bibnamefont {Aleksan}, \bibfnamefont
  {Roy}}, \bibinfo {author} {\bibfnamefont {Isard}\ \bibnamefont {Dunietz}}, \
  and\ \bibinfo {author} {\bibfnamefont {Boris}\ \bibnamefont {Kayser}}}
  (\bibinfo {year} {1992}),\ \bibfield  {title} {\enquote {\bibinfo {title}
  {{Determining the CP violating phase gamma}},}\ }\href {\doibase
  10.1007/BF01559494} {\bibfield  {journal} {\bibinfo  {journal} {Z. Phys.}\
  }\textbf {\bibinfo {volume} {C54}},\ \bibinfo {pages} {653--660}}\BibitemShut
  {NoStop}%
\bibitem [{\citenamefont {Alok}\ \emph {et~al.}(2011)\citenamefont {Alok},
  \citenamefont {Baek},\ and\ \citenamefont {London}}]{Alok:2010ij}%
  \BibitemOpen
  \bibfield  {author} {\bibinfo {author} {\bibnamefont {Alok}, \bibfnamefont
  {Ashutosh~Kumar}}, \bibinfo {author} {\bibfnamefont {Seungwon}\ \bibnamefont
  {Baek}}, \ and\ \bibinfo {author} {\bibfnamefont {David}\ \bibnamefont
  {London}}} (\bibinfo {year} {2011}),\ \bibfield  {title} {\enquote {\bibinfo
  {title} {{Neutral Gauge Boson Contributions to the Dimuon Charge Asymmetry in
  B Decays}},}\ }\href {\doibase 10.1007/JHEP07(2011)111} {\bibfield  {journal}
  {\bibinfo  {journal} {JHEP}\ }\textbf {\bibinfo {volume} {07}},\ \bibinfo
  {pages} {111}},\ \Eprint {http://arxiv.org/abs/1010.1333} {arXiv:1010.1333
  [hep-ph]} \BibitemShut {NoStop}%
\bibitem [{\citenamefont {Alok}\ and\ \citenamefont
  {Gangal}(2012)}]{Alok:2012xm}%
  \BibitemOpen
  \bibfield  {author} {\bibinfo {author} {\bibnamefont {Alok}, \bibfnamefont
  {Ashutosh~Kumar}}, \ and\ \bibinfo {author} {\bibfnamefont {Shireen}\
  \bibnamefont {Gangal}}} (\bibinfo {year} {2012}),\ \bibfield  {title}
  {\enquote {\bibinfo {title} {{$b \to s$ Decays in a model with Z-mediated
  flavor changing neutral current}},}\ }\href {\doibase
  10.1103/PhysRevD.86.114009} {\bibfield  {journal} {\bibinfo  {journal} {Phys.
  Rev.}\ }\textbf {\bibinfo {volume} {D86}},\ \bibinfo {pages} {114009}},\
  \Eprint {http://arxiv.org/abs/1209.1987} {arXiv:1209.1987 [hep-ph]}
  \BibitemShut {NoStop}%
\bibitem [{\citenamefont {Altmannshofer}\ and\ \citenamefont
  {Carena}(2012)}]{Altmannshofer:2011iv}%
  \BibitemOpen
  \bibfield  {author} {\bibinfo {author} {\bibnamefont {Altmannshofer},
  \bibfnamefont {Wolfgang}}, \ and\ \bibinfo {author} {\bibfnamefont {Marcela}\
  \bibnamefont {Carena}}} (\bibinfo {year} {2012}),\ \bibfield  {title}
  {\enquote {\bibinfo {title} {{B Meson Mixing in Effective Theories of
  Supersymmetric Higgs Bosons}},}\ }\href {\doibase 10.1103/PhysRevD.85.075006}
  {\bibfield  {journal} {\bibinfo  {journal} {Phys. Rev.}\ }\textbf {\bibinfo
  {volume} {D85}},\ \bibinfo {pages} {075006}},\ \Eprint
  {http://arxiv.org/abs/1110.0843} {arXiv:1110.0843 [hep-ph]} \BibitemShut
  {NoStop}%
\bibitem [{\citenamefont {Amhis}\ \emph {et~al.}(2014)\citenamefont {Amhis}
  \emph {et~al.}}]{Amhis:2014hma}%
  \BibitemOpen
  \bibfield  {author} {\bibinfo {author} {\bibnamefont {Amhis}, \bibfnamefont
  {Y}},  \emph {et~al.} (\bibinfo {collaboration} {Heavy Flavor Averaging Group
  (HFAG)})} (\bibinfo {year} {2014}),\ \bibfield  {title} {\enquote {\bibinfo
  {title} {{Averages of $b$-hadron, $c$-hadron, and $\tau$-lepton properties as
  of summer 2014}},}\ }\href@noop {} {\ }\Eprint
  {http://arxiv.org/abs/1412.7515} {arXiv:1412.7515 [hep-ex]} \BibitemShut
  {NoStop}%
\bibitem [{\citenamefont {Anikeev}\ \emph {et~al.}(2001)\citenamefont {Anikeev}
  \emph {et~al.}}]{Anikeev:2001rk}%
  \BibitemOpen
  \bibfield  {author} {\bibinfo {author} {\bibnamefont {Anikeev}, \bibfnamefont
  {K}},  \emph {et~al.}} (\bibinfo {year} {2001}),\ \bibfield  {title}
  {\enquote {\bibinfo {title} {{$B$ physics at the Tevatron: Run II and
  beyond}},}\ }in\ \href {http://lss.fnal.gov/cgi-bin/find_paper.pl?pub-01-197}
  {\emph {\bibinfo {booktitle} {{Workshop on B Physics at the Tevatron: Run II
  and Beyond Batavia, Illinois, September 23-25, 1999}}}},\ \Eprint
  {http://arxiv.org/abs/hep-ph/0201071} {arXiv:hep-ph/0201071 [hep-ph]}
  \BibitemShut {NoStop}%
\bibitem [{\citenamefont {Aoki}\ \emph {et~al.}(2014)\citenamefont {Aoki} \emph
  {et~al.}}]{Aoki:2013ldr}%
  \BibitemOpen
  \bibfield  {author} {\bibinfo {author} {\bibnamefont {Aoki}, \bibfnamefont
  {Sinya}},  \emph {et~al.}} (\bibinfo {year} {2014}),\ \bibfield  {title}
  {\enquote {\bibinfo {title} {{Review of lattice results concerning low-energy
  particle physics}},}\ }\href {\doibase 10.1140/epjc/s10052-014-2890-7}
  {\bibfield  {journal} {\bibinfo  {journal} {Eur. Phys. J.}\ }\textbf
  {\bibinfo {volume} {C74}},\ \bibinfo {pages} {2890}},\ \Eprint
  {http://arxiv.org/abs/1310.8555} {arXiv:1310.8555 [hep-lat]} \BibitemShut
  {NoStop}%
\bibitem [{\citenamefont {Aoki}\ \emph {et~al.}(2015)\citenamefont {Aoki},
  \citenamefont {Ishikawa}, \citenamefont {Izubuchi}, \citenamefont {Lehner},\
  and\ \citenamefont {Soni}}]{Aoki:2014nga}%
  \BibitemOpen
  \bibfield  {author} {\bibinfo {author} {\bibnamefont {Aoki}, \bibfnamefont
  {Yasumichi}}, \bibinfo {author} {\bibfnamefont {Tomomi}\ \bibnamefont
  {Ishikawa}}, \bibinfo {author} {\bibfnamefont {Taku}\ \bibnamefont
  {Izubuchi}}, \bibinfo {author} {\bibfnamefont {Christoph}\ \bibnamefont
  {Lehner}}, \ and\ \bibinfo {author} {\bibfnamefont {Amarjit}\ \bibnamefont
  {Soni}}} (\bibinfo {year} {2015}),\ \bibfield  {title} {\enquote {\bibinfo
  {title} {{Neutral $B$ meson mixings and $B$ meson decay constants with static
  heavy and domain-wall light quarks}},}\ }\href {\doibase
  10.1103/PhysRevD.91.114505} {\bibfield  {journal} {\bibinfo  {journal} {Phys.
  Rev.}\ }\textbf {\bibinfo {volume} {D91}}~(\bibinfo {number} {11}),\ \bibinfo
  {pages} {114505}},\ \Eprint {http://arxiv.org/abs/1406.6192} {arXiv:1406.6192
  [hep-lat]} \BibitemShut {NoStop}%
\bibitem [{\citenamefont {Asatrian}\ \emph {et~al.}(2012)\citenamefont
  {Asatrian}, \citenamefont {Hovhannisyan},\ and\ \citenamefont
  {Yeghiazaryan}}]{Asatrian:2012tp}%
  \BibitemOpen
  \bibfield  {author} {\bibinfo {author} {\bibnamefont {Asatrian},
  \bibfnamefont {H~M}}, \bibinfo {author} {\bibfnamefont {A.}~\bibnamefont
  {Hovhannisyan}}, \ and\ \bibinfo {author} {\bibfnamefont {A.}~\bibnamefont
  {Yeghiazaryan}}} (\bibinfo {year} {2012}),\ \bibfield  {title} {\enquote
  {\bibinfo {title} {{The phase space analysis for three and four massive
  particles in final states}},}\ }\href {\doibase 10.1103/PhysRevD.86.114023}
  {\bibfield  {journal} {\bibinfo  {journal} {Phys. Rev.}\ }\textbf {\bibinfo
  {volume} {D86}},\ \bibinfo {pages} {114023}},\ \Eprint
  {http://arxiv.org/abs/1210.7939} {arXiv:1210.7939 [hep-ph]} \BibitemShut
  {NoStop}%
\bibitem [{\citenamefont {ATLAS}\ and\ \citenamefont
  {Collaborations}(2014)}]{ATLAS:2014wva}%
  \BibitemOpen
  \bibfield  {author} {\bibinfo {author} {\bibnamefont {ATLAS}, \bibfnamefont
  {CDF, CMS}}, \ and\ \bibinfo {author} {\bibfnamefont {D0}~\bibnamefont
  {Collaborations}}} (\bibinfo {year} {2014}),\ \bibfield  {title} {\enquote
  {\bibinfo {title} {{First combination of Tevatron and LHC measurements of the
  top-quark mass}},}\ }\href@noop {} {\ }\Eprint
  {http://arxiv.org/abs/1403.4427} {arXiv:1403.4427 [hep-ex]} \BibitemShut
  {NoStop}%
\bibitem [{\citenamefont {ATLAS}(2013)}]{ATLAS-PHYS-PUB-2013-010}%
  \BibitemOpen
  \bibfield  {author} {\bibinfo {author} {\bibnamefont {ATLAS}, \bibfnamefont
  {Collaboration}}} (\bibinfo {year} {2013}),\ \href
  {http://cds.cern.ch/record/1604429} {\emph {\bibinfo {title} {{ATLAS
  B-physics studies at increased LHC luminosity, potential for CP-violation
  measurement in the $\Bs \to J/\psi \phi$ decay}}}},\ \bibinfo {type} {Tech.
  Rep.}\ \bibinfo {number} {ATL-PHYS-PUB-2013-010}\ (\bibinfo  {institution}
  {CERN},\ \bibinfo {address} {Geneva})\BibitemShut {NoStop}%
\bibitem [{\citenamefont {Badin}\ \emph {et~al.}(2007)\citenamefont {Badin},
  \citenamefont {Gabbiani},\ and\ \citenamefont {Petrov}}]{Badin:2007bv}%
  \BibitemOpen
  \bibfield  {author} {\bibinfo {author} {\bibnamefont {Badin}, \bibfnamefont
  {Andriy}}, \bibinfo {author} {\bibfnamefont {Fabrizio}\ \bibnamefont
  {Gabbiani}}, \ and\ \bibinfo {author} {\bibfnamefont {Alexey~A.}\
  \bibnamefont {Petrov}}} (\bibinfo {year} {2007}),\ \bibfield  {title}
  {\enquote {\bibinfo {title} {{Lifetime difference in $B_s$ mixing: Standard
  model and beyond}},}\ }\href {\doibase 10.1016/j.physletb.2007.07.049}
  {\bibfield  {journal} {\bibinfo  {journal} {Phys. Lett.}\ }\textbf {\bibinfo
  {volume} {B653}},\ \bibinfo {pages} {230--240}},\ \Eprint
  {http://arxiv.org/abs/0707.0294} {arXiv:0707.0294 [hep-ph]} \BibitemShut
  {NoStop}%
\bibitem [{\citenamefont {Bagan}\ \emph {et~al.}(1995)\citenamefont {Bagan},
  \citenamefont {Ball}, \citenamefont {Fiol},\ and\ \citenamefont
  {Gosdzinsky}}]{Bagan:1995yf}%
  \BibitemOpen
  \bibfield  {author} {\bibinfo {author} {\bibnamefont {Bagan}, \bibfnamefont
  {E}}, \bibinfo {author} {\bibfnamefont {Patricia}\ \bibnamefont {Ball}},
  \bibinfo {author} {\bibfnamefont {B.}~\bibnamefont {Fiol}}, \ and\ \bibinfo
  {author} {\bibfnamefont {P.}~\bibnamefont {Gosdzinsky}}} (\bibinfo {year}
  {1995}),\ \bibfield  {title} {\enquote {\bibinfo {title} {{Next-to-leading
  order radiative corrections to the decay $ b \to c \bar{c} s$}},}\ }\href
  {\doibase 10.1016/0370-2693(95)00437-P} {\bibfield  {journal} {\bibinfo
  {journal} {Phys. Lett.}\ }\textbf {\bibinfo {volume} {B351}},\ \bibinfo
  {pages} {546--554}},\ \Eprint {http://arxiv.org/abs/hep-ph/9502338}
  {arXiv:hep-ph/9502338 [hep-ph]} \BibitemShut {NoStop}%
\bibitem [{\citenamefont {Balitsky}\ \emph {et~al.}(1989)\citenamefont
  {Balitsky}, \citenamefont {Braun},\ and\ \citenamefont
  {Kolesnichenko}}]{Balitsky:1989ry}%
  \BibitemOpen
  \bibfield  {author} {\bibinfo {author} {\bibnamefont {Balitsky},
  \bibfnamefont {I~I}}, \bibinfo {author} {\bibfnamefont {Vladimir~M.}\
  \bibnamefont {Braun}}, \ and\ \bibinfo {author} {\bibfnamefont {A.~V.}\
  \bibnamefont {Kolesnichenko}}} (\bibinfo {year} {1989}),\ \bibfield  {title}
  {\enquote {\bibinfo {title} {{Radiative Decay $\Sigma^+ \to p \gamma$ in
  Quantum Chromodynamics}},}\ }\href {\doibase 10.1016/0550-3213(89)90570-1}
  {\bibfield  {journal} {\bibinfo  {journal} {Nucl. Phys.}\ }\textbf {\bibinfo
  {volume} {B312}},\ \bibinfo {pages} {509--550}}\BibitemShut {NoStop}%
\bibitem [{\citenamefont {Bambi}\ and\ \citenamefont
  {Dolgov}(2015)}]{Bambi:2015mba}%
  \BibitemOpen
  \bibfield  {author} {\bibinfo {author} {\bibnamefont {Bambi}, \bibfnamefont
  {Cosimo}}, \ and\ \bibinfo {author} {\bibfnamefont {Alexandre~D.}\
  \bibnamefont {Dolgov}}} (\bibinfo {year} {2015}),\ \href {\doibase
  10.1007/978-3-662-48078-6} {\emph {\bibinfo {title} {{Introduction to
  Particle Cosmology}}}},\ UNITEXT for Physics\ (\bibinfo  {publisher}
  {Springer})\BibitemShut {NoStop}%
\bibitem [{\citenamefont {Bander}\ \emph {et~al.}(1979)\citenamefont {Bander},
  \citenamefont {Silverman},\ and\ \citenamefont {Soni}}]{Bander:1979px}%
  \BibitemOpen
  \bibfield  {author} {\bibinfo {author} {\bibnamefont {Bander}, \bibfnamefont
  {Myron}}, \bibinfo {author} {\bibfnamefont {D.}~\bibnamefont {Silverman}}, \
  and\ \bibinfo {author} {\bibfnamefont {A.}~\bibnamefont {Soni}}} (\bibinfo
  {year} {1979}),\ \bibfield  {title} {\enquote {\bibinfo {title} {{CP
  Noninvariance in the Decays of Heavy Charged Quark Systems}},}\ }\href
  {\doibase 10.1103/PhysRevLett.43.242} {\bibfield  {journal} {\bibinfo
  {journal} {Phys. Rev. Lett.}\ }\textbf {\bibinfo {volume} {43}},\ \bibinfo
  {pages} {242}}\BibitemShut {NoStop}%
\bibitem [{\citenamefont {Bardeen}\ \emph {et~al.}(1978)\citenamefont
  {Bardeen}, \citenamefont {Buras}, \citenamefont {Duke},\ and\ \citenamefont
  {Muta}}]{Bardeen:1978yd}%
  \BibitemOpen
  \bibfield  {author} {\bibinfo {author} {\bibnamefont {Bardeen}, \bibfnamefont
  {William~A}}, \bibinfo {author} {\bibfnamefont {A.~J.}\ \bibnamefont
  {Buras}}, \bibinfo {author} {\bibfnamefont {D.~W.}\ \bibnamefont {Duke}}, \
  and\ \bibinfo {author} {\bibfnamefont {T.}~\bibnamefont {Muta}}} (\bibinfo
  {year} {1978}),\ \bibfield  {title} {\enquote {\bibinfo {title} {{Deep
  Inelastic Scattering Beyond the Leading Order in Asymptotically Free Gauge
  Theories}},}\ }\href {\doibase 10.1103/PhysRevD.18.3998} {\bibfield
  {journal} {\bibinfo  {journal} {Phys. Rev.}\ }\textbf {\bibinfo {volume}
  {D18}},\ \bibinfo {pages} {3998}}\BibitemShut {NoStop}%
\bibitem [{\citenamefont {Bartsch}\ \emph {et~al.}(2008)\citenamefont
  {Bartsch}, \citenamefont {Buchalla},\ and\ \citenamefont
  {Kraus}}]{Bartsch:2008ps}%
  \BibitemOpen
  \bibfield  {author} {\bibinfo {author} {\bibnamefont {Bartsch}, \bibfnamefont
  {Matthaus}}, \bibinfo {author} {\bibfnamefont {Gerhard}\ \bibnamefont
  {Buchalla}}, \ and\ \bibinfo {author} {\bibfnamefont {Christina}\
  \bibnamefont {Kraus}}} (\bibinfo {year} {2008}),\ \bibfield  {title}
  {\enquote {\bibinfo {title} {{$B \to V(L) V(L)$ Decays at Next-to-Leading
  Order in QCD}},}\ }\href@noop {} {\ }\Eprint {http://arxiv.org/abs/0810.0249}
  {arXiv:0810.0249 [hep-ph]} \BibitemShut {NoStop}%
\bibitem [{\citenamefont {Bauer}\ \emph {et~al.}(2001)\citenamefont {Bauer},
  \citenamefont {Fleming}, \citenamefont {Pirjol},\ and\ \citenamefont
  {Stewart}}]{Bauer:2000yr}%
  \BibitemOpen
  \bibfield  {author} {\bibinfo {author} {\bibnamefont {Bauer}, \bibfnamefont
  {Christian~W}}, \bibinfo {author} {\bibfnamefont {Sean}\ \bibnamefont
  {Fleming}}, \bibinfo {author} {\bibfnamefont {Dan}\ \bibnamefont {Pirjol}}, \
  and\ \bibinfo {author} {\bibfnamefont {Iain~W.}\ \bibnamefont {Stewart}}}
  (\bibinfo {year} {2001}),\ \bibfield  {title} {\enquote {\bibinfo {title}
  {{An Effective field theory for collinear and soft gluons: Heavy to light
  decays}},}\ }\href {\doibase 10.1103/PhysRevD.63.114020} {\bibfield
  {journal} {\bibinfo  {journal} {Phys. Rev.}\ }\textbf {\bibinfo {volume}
  {D63}},\ \bibinfo {pages} {114020}},\ \Eprint
  {http://arxiv.org/abs/hep-ph/0011336} {arXiv:hep-ph/0011336 [hep-ph]}
  \BibitemShut {NoStop}%
\bibitem [{\citenamefont {Bauer}\ \emph {et~al.}(2004)\citenamefont {Bauer},
  \citenamefont {Pirjol}, \citenamefont {Rothstein},\ and\ \citenamefont
  {Stewart}}]{Bauer:2004tj}%
  \BibitemOpen
  \bibfield  {author} {\bibinfo {author} {\bibnamefont {Bauer}, \bibfnamefont
  {Christian~W}}, \bibinfo {author} {\bibfnamefont {Dan}\ \bibnamefont
  {Pirjol}}, \bibinfo {author} {\bibfnamefont {Ira~Z.}\ \bibnamefont
  {Rothstein}}, \ and\ \bibinfo {author} {\bibfnamefont {Iain~W.}\ \bibnamefont
  {Stewart}}} (\bibinfo {year} {2004}),\ \bibfield  {title} {\enquote {\bibinfo
  {title} {{$B \to M(1) M(2)$: Factorization, charming penguins, strong phases,
  and polarization}},}\ }\href {\doibase 10.1103/PhysRevD.70.054015} {\bibfield
   {journal} {\bibinfo  {journal} {Phys. Rev.}\ }\textbf {\bibinfo {volume}
  {D70}},\ \bibinfo {pages} {054015}},\ \Eprint
  {http://arxiv.org/abs/hep-ph/0401188} {arXiv:hep-ph/0401188 [hep-ph]}
  \BibitemShut {NoStop}%
\bibitem [{\citenamefont {Bauer}\ \emph {et~al.}(2002)\citenamefont {Bauer},
  \citenamefont {Pirjol},\ and\ \citenamefont {Stewart}}]{Bauer:2001yt}%
  \BibitemOpen
  \bibfield  {author} {\bibinfo {author} {\bibnamefont {Bauer}, \bibfnamefont
  {Christian~W}}, \bibinfo {author} {\bibfnamefont {Dan}\ \bibnamefont
  {Pirjol}}, \ and\ \bibinfo {author} {\bibfnamefont {Iain~W.}\ \bibnamefont
  {Stewart}}} (\bibinfo {year} {2002}),\ \bibfield  {title} {\enquote {\bibinfo
  {title} {{Soft collinear factorization in effective field theory}},}\ }\href
  {\doibase 10.1103/PhysRevD.65.054022} {\bibfield  {journal} {\bibinfo
  {journal} {Phys. Rev.}\ }\textbf {\bibinfo {volume} {D65}},\ \bibinfo {pages}
  {054022}},\ \Eprint {http://arxiv.org/abs/hep-ph/0109045}
  {arXiv:hep-ph/0109045 [hep-ph]} \BibitemShut {NoStop}%
\bibitem [{\citenamefont {Bazavov}\ \emph {et~al.}(2016)\citenamefont {Bazavov}
  \emph {et~al.}}]{Bazavov:2016nty}%
  \BibitemOpen
  \bibfield  {author} {\bibinfo {author} {\bibnamefont {Bazavov}, \bibfnamefont
  {A}},  \emph {et~al.} (\bibinfo {collaboration} {Fermilab Lattice, MILC})}
  (\bibinfo {year} {2016}),\ \bibfield  {title} {\enquote {\bibinfo {title}
  {{$B^0_{(s)}$-Mixing Matrix Elements from Lattice QCD for the Standard Model
  and Beyond}},}\ }\href@noop {} {\ }\Eprint {http://arxiv.org/abs/1602.03560}
  {arXiv:1602.03560 [hep-lat]} \BibitemShut {NoStop}%
\bibitem [{\citenamefont {Becirevic}\ \emph {et~al.}(2002)\citenamefont
  {Becirevic}, \citenamefont {Gimenez}, \citenamefont {Martinelli},
  \citenamefont {Papinutto},\ and\ \citenamefont {Reyes}}]{Becirevic:2001xt}%
  \BibitemOpen
  \bibfield  {author} {\bibinfo {author} {\bibnamefont {Becirevic},
  \bibfnamefont {D}}, \bibinfo {author} {\bibfnamefont {V.}~\bibnamefont
  {Gimenez}}, \bibinfo {author} {\bibfnamefont {G.}~\bibnamefont {Martinelli}},
  \bibinfo {author} {\bibfnamefont {M.}~\bibnamefont {Papinutto}}, \ and\
  \bibinfo {author} {\bibfnamefont {J.}~\bibnamefont {Reyes}}} (\bibinfo {year}
  {2002}),\ \bibfield  {title} {\enquote {\bibinfo {title} {{B parameters of
  the complete set of matrix elements of $\Delta B = 2$ operators from the
  lattice}},}\ }\href {\doibase 10.1088/1126-6708/2002/04/025} {\bibfield
  {journal} {\bibinfo  {journal} {JHEP}\ }\textbf {\bibinfo {volume} {04}},\
  \bibinfo {pages} {025}},\ \Eprint {http://arxiv.org/abs/hep-lat/0110091}
  {arXiv:hep-lat/0110091 [hep-lat]} \BibitemShut {NoStop}%
\bibitem [{\citenamefont {Becirevic}(2001)}]{Becirevic:2001fy}%
  \BibitemOpen
  \bibfield  {author} {\bibinfo {author} {\bibnamefont {Becirevic},
  \bibfnamefont {Damir}}} (\bibinfo {year} {2001}),\ \bibfield  {title}
  {\enquote {\bibinfo {title} {{Theoretical progress in describing the B meson
  lifetimes}},}\ }\bibfield  {booktitle} {\emph {\bibinfo {booktitle}
  {{Proceedings, 2001 Europhysics Conference on High Energy Physics (EPS-HEP
  2001)}}},\ }\href@noop {} {\bibfield  {journal} {\bibinfo  {journal} {PoS}\
  }\textbf {\bibinfo {volume} {HEP2001}},\ \bibinfo {pages} {098}},\ \Eprint
  {http://arxiv.org/abs/hep-ph/0110124} {arXiv:hep-ph/0110124 [hep-ph]}
  \BibitemShut {NoStop}%
\bibitem [{\citenamefont {Beneke}\ \emph {et~al.}(1996)\citenamefont {Beneke},
  \citenamefont {Buchalla},\ and\ \citenamefont {Dunietz}}]{Beneke:1996gn}%
  \BibitemOpen
  \bibfield  {author} {\bibinfo {author} {\bibnamefont {Beneke}, \bibfnamefont
  {M}}, \bibinfo {author} {\bibfnamefont {G.}~\bibnamefont {Buchalla}}, \ and\
  \bibinfo {author} {\bibfnamefont {I.}~\bibnamefont {Dunietz}}} (\bibinfo
  {year} {1996}),\ \bibfield  {title} {\enquote {\bibinfo {title} {{Width
  Difference in the $B_s-\bar{B_s}$ System}},}\ }\href {\doibase
  10.1103/PhysRevD.54.4419, 10.1103/PhysRevD.83.119902} {\bibfield  {journal}
  {\bibinfo  {journal} {Phys. Rev.}\ }\textbf {\bibinfo {volume} {D54}},\
  \bibinfo {pages} {4419--4431}},\ \bibinfo {note} {[Erratum: Phys.
  Rev.D83,119902(2011)]},\ \Eprint {http://arxiv.org/abs/hep-ph/9605259}
  {arXiv:hep-ph/9605259 [hep-ph]} \BibitemShut {NoStop}%
\bibitem [{\citenamefont {Beneke}\ \emph
  {et~al.}(1999{\natexlab{a}})\citenamefont {Beneke}, \citenamefont {Buchalla},
  \citenamefont {Greub}, \citenamefont {Lenz},\ and\ \citenamefont
  {Nierste}}]{Beneke:1998sy}%
  \BibitemOpen
  \bibfield  {author} {\bibinfo {author} {\bibnamefont {Beneke}, \bibfnamefont
  {M}}, \bibinfo {author} {\bibfnamefont {G.}~\bibnamefont {Buchalla}},
  \bibinfo {author} {\bibfnamefont {C.}~\bibnamefont {Greub}}, \bibinfo
  {author} {\bibfnamefont {A.}~\bibnamefont {Lenz}}, \ and\ \bibinfo {author}
  {\bibfnamefont {U.}~\bibnamefont {Nierste}}} (\bibinfo {year}
  {1999}{\natexlab{a}}),\ \bibfield  {title} {\enquote {\bibinfo {title}
  {{Next-to-leading order QCD corrections to the lifetime difference of $\Bs$
  mesons}},}\ }\href {\doibase 10.1016/S0370-2693(99)00684-X} {\bibfield
  {journal} {\bibinfo  {journal} {Phys. Lett.}\ }\textbf {\bibinfo {volume}
  {B459}},\ \bibinfo {pages} {631--640}},\ \Eprint
  {http://arxiv.org/abs/hep-ph/9808385} {arXiv:hep-ph/9808385 [hep-ph]}
  \BibitemShut {NoStop}%
\bibitem [{\citenamefont {Beneke}\ \emph
  {et~al.}(1999{\natexlab{b}})\citenamefont {Beneke}, \citenamefont {Buchalla},
  \citenamefont {Neubert},\ and\ \citenamefont {Sachrajda}}]{Beneke:1999br}%
  \BibitemOpen
  \bibfield  {author} {\bibinfo {author} {\bibnamefont {Beneke}, \bibfnamefont
  {M}}, \bibinfo {author} {\bibfnamefont {G.}~\bibnamefont {Buchalla}},
  \bibinfo {author} {\bibfnamefont {M.}~\bibnamefont {Neubert}}, \ and\
  \bibinfo {author} {\bibfnamefont {Christopher~T.}\ \bibnamefont {Sachrajda}}}
  (\bibinfo {year} {1999}{\natexlab{b}}),\ \bibfield  {title} {\enquote
  {\bibinfo {title} {{QCD factorization for $B \to \pi \pi$ decays: Strong
  phases and CP violation in the heavy quark limit}},}\ }\href {\doibase
  10.1103/PhysRevLett.83.1914} {\bibfield  {journal} {\bibinfo  {journal}
  {Phys.Rev.Lett.}\ }\textbf {\bibinfo {volume} {83}},\ \bibinfo {pages}
  {1914--1917}},\ \Eprint {http://arxiv.org/abs/hep-ph/9905312}
  {arXiv:hep-ph/9905312 [hep-ph]} \BibitemShut {NoStop}%
\bibitem [{\citenamefont {Beneke}\ \emph {et~al.}(2000)\citenamefont {Beneke},
  \citenamefont {Buchalla}, \citenamefont {Neubert},\ and\ \citenamefont
  {Sachrajda}}]{Beneke:2000ry}%
  \BibitemOpen
  \bibfield  {author} {\bibinfo {author} {\bibnamefont {Beneke}, \bibfnamefont
  {M}}, \bibinfo {author} {\bibfnamefont {G.}~\bibnamefont {Buchalla}},
  \bibinfo {author} {\bibfnamefont {M.}~\bibnamefont {Neubert}}, \ and\
  \bibinfo {author} {\bibfnamefont {Christopher~T.}\ \bibnamefont {Sachrajda}}}
  (\bibinfo {year} {2000}),\ \bibfield  {title} {\enquote {\bibinfo {title}
  {{QCD factorization for exclusive, nonleptonic B meson decays: General
  arguments and the case of heavy light final states}},}\ }\href {\doibase
  10.1016/S0550-3213(00)00559-9} {\bibfield  {journal} {\bibinfo  {journal}
  {Nucl.Phys.}\ }\textbf {\bibinfo {volume} {B591}},\ \bibinfo {pages}
  {313--418}},\ \Eprint {http://arxiv.org/abs/hep-ph/0006124}
  {arXiv:hep-ph/0006124 [hep-ph]} \BibitemShut {NoStop}%
\bibitem [{\citenamefont {Beneke}\ \emph {et~al.}(2001)\citenamefont {Beneke},
  \citenamefont {Buchalla}, \citenamefont {Neubert},\ and\ \citenamefont
  {Sachrajda}}]{Beneke:2001ev}%
  \BibitemOpen
  \bibfield  {author} {\bibinfo {author} {\bibnamefont {Beneke}, \bibfnamefont
  {M}}, \bibinfo {author} {\bibfnamefont {G.}~\bibnamefont {Buchalla}},
  \bibinfo {author} {\bibfnamefont {M.}~\bibnamefont {Neubert}}, \ and\
  \bibinfo {author} {\bibfnamefont {Christopher~T.}\ \bibnamefont {Sachrajda}}}
  (\bibinfo {year} {2001}),\ \bibfield  {title} {\enquote {\bibinfo {title}
  {{QCD factorization in $ B \to \pi K, \pi \pi$ decays and extraction of
  Wolfenstein parameters}},}\ }\href {\doibase 10.1016/S0550-3213(01)00251-6}
  {\bibfield  {journal} {\bibinfo  {journal} {Nucl.Phys.}\ }\textbf {\bibinfo
  {volume} {B606}},\ \bibinfo {pages} {245--321}},\ \Eprint
  {http://arxiv.org/abs/hep-ph/0104110} {arXiv:hep-ph/0104110 [hep-ph]}
  \BibitemShut {NoStop}%
\bibitem [{\citenamefont {Beneke}\ \emph {et~al.}(2002)\citenamefont {Beneke},
  \citenamefont {Buchalla}, \citenamefont {Greub}, \citenamefont {Lenz},\ and\
  \citenamefont {Nierste}}]{Beneke:2002rj}%
  \BibitemOpen
  \bibfield  {author} {\bibinfo {author} {\bibnamefont {Beneke}, \bibfnamefont
  {Martin}}, \bibinfo {author} {\bibfnamefont {Gerhard}\ \bibnamefont
  {Buchalla}}, \bibinfo {author} {\bibfnamefont {Christoph}\ \bibnamefont
  {Greub}}, \bibinfo {author} {\bibfnamefont {Alexander}\ \bibnamefont {Lenz}},
  \ and\ \bibinfo {author} {\bibfnamefont {Ulrich}\ \bibnamefont {Nierste}}}
  (\bibinfo {year} {2002}),\ \bibfield  {title} {\enquote {\bibinfo {title}
  {{The $B^+ - B^0_d$ lifetime difference beyond leading logarithms}},}\ }\href
  {\doibase 10.1016/S0550-3213(02)00561-8} {\bibfield  {journal} {\bibinfo
  {journal} {Nucl. Phys.}\ }\textbf {\bibinfo {volume} {B639}},\ \bibinfo
  {pages} {389--407}},\ \Eprint {http://arxiv.org/abs/hep-ph/0202106}
  {arXiv:hep-ph/0202106 [hep-ph]} \BibitemShut {NoStop}%
\bibitem [{\citenamefont {Beneke}\ \emph {et~al.}(2003)\citenamefont {Beneke},
  \citenamefont {Buchalla}, \citenamefont {Lenz},\ and\ \citenamefont
  {Nierste}}]{Beneke:2003az}%
  \BibitemOpen
  \bibfield  {author} {\bibinfo {author} {\bibnamefont {Beneke}, \bibfnamefont
  {Martin}}, \bibinfo {author} {\bibfnamefont {Gerhard}\ \bibnamefont
  {Buchalla}}, \bibinfo {author} {\bibfnamefont {Alexander}\ \bibnamefont
  {Lenz}}, \ and\ \bibinfo {author} {\bibfnamefont {Ulrich}\ \bibnamefont
  {Nierste}}} (\bibinfo {year} {2003}),\ \bibfield  {title} {\enquote {\bibinfo
  {title} {{CP asymmetry in flavor specific $B$ decays beyond leading
  logarithms}},}\ }\href {\doibase 10.1016/j.physletb.2003.09.089} {\bibfield
  {journal} {\bibinfo  {journal} {Phys. Lett.}\ }\textbf {\bibinfo {volume}
  {B576}},\ \bibinfo {pages} {173--183}},\ \Eprint
  {http://arxiv.org/abs/hep-ph/0307344} {arXiv:hep-ph/0307344 [hep-ph]}
  \BibitemShut {NoStop}%
\bibitem [{\citenamefont {Beneke}\ and\ \citenamefont
  {Neubert}(2003)}]{Beneke:2003zv}%
  \BibitemOpen
  \bibfield  {author} {\bibinfo {author} {\bibnamefont {Beneke}, \bibfnamefont
  {Martin}}, \ and\ \bibinfo {author} {\bibfnamefont {Matthias}\ \bibnamefont
  {Neubert}}} (\bibinfo {year} {2003}),\ \bibfield  {title} {\enquote {\bibinfo
  {title} {{QCD factorization for $B \to PP$ and $B \to PV $ decays}},}\ }\href
  {\doibase 10.1016/j.nuclphysb.2003.09.026} {\bibfield  {journal} {\bibinfo
  {journal} {Nucl. Phys.}\ }\textbf {\bibinfo {volume} {B675}},\ \bibinfo
  {pages} {333--415}},\ \Eprint {http://arxiv.org/abs/hep-ph/0308039}
  {arXiv:hep-ph/0308039 [hep-ph]} \BibitemShut {NoStop}%
\bibitem [{\citenamefont {Beneke}\ \emph {et~al.}(2007)\citenamefont {Beneke},
  \citenamefont {Rohrer},\ and\ \citenamefont {Yang}}]{Beneke:2006hg}%
  \BibitemOpen
  \bibfield  {author} {\bibinfo {author} {\bibnamefont {Beneke}, \bibfnamefont
  {Martin}}, \bibinfo {author} {\bibfnamefont {Johannes}\ \bibnamefont
  {Rohrer}}, \ and\ \bibinfo {author} {\bibfnamefont {Deshan}\ \bibnamefont
  {Yang}}} (\bibinfo {year} {2007}),\ \bibfield  {title} {\enquote {\bibinfo
  {title} {{Branching fractions, polarisation and asymmetries of $B \to VV$
  decays}},}\ }\href {\doibase 10.1016/j.nuclphysb.2007.03.020} {\bibfield
  {journal} {\bibinfo  {journal} {Nucl. Phys.}\ }\textbf {\bibinfo {volume}
  {B774}},\ \bibinfo {pages} {64--101}},\ \Eprint
  {http://arxiv.org/abs/hep-ph/0612290} {arXiv:hep-ph/0612290 [hep-ph]}
  \BibitemShut {NoStop}%
\bibitem [{\citenamefont {Bertolini}\ \emph {et~al.}(2014)\citenamefont
  {Bertolini}, \citenamefont {Maiezza},\ and\ \citenamefont
  {Nesti}}]{Bertolini:2014sua}%
  \BibitemOpen
  \bibfield  {author} {\bibinfo {author} {\bibnamefont {Bertolini},
  \bibfnamefont {Stefano}}, \bibinfo {author} {\bibfnamefont {Alessio}\
  \bibnamefont {Maiezza}}, \ and\ \bibinfo {author} {\bibfnamefont {Fabrizio}\
  \bibnamefont {Nesti}}} (\bibinfo {year} {2014}),\ \bibfield  {title}
  {\enquote {\bibinfo {title} {{Present and Future K and B Meson Mixing
  Constraints on TeV Scale Left-Right Symmetry}},}\ }\href {\doibase
  10.1103/PhysRevD.89.095028} {\bibfield  {journal} {\bibinfo  {journal} {Phys.
  Rev.}\ }\textbf {\bibinfo {volume} {D89}}~(\bibinfo {number} {9}),\ \bibinfo
  {pages} {095028}},\ \Eprint {http://arxiv.org/abs/1403.7112} {arXiv:1403.7112
  [hep-ph]} \BibitemShut {NoStop}%
\bibitem [{\citenamefont {Bevan}\ \emph {et~al.}(2013)\citenamefont {Bevan}
  \emph {et~al.}}]{Bevan:2013kaa}%
  \BibitemOpen
  \bibfield  {author} {\bibinfo {author} {\bibnamefont {Bevan}, \bibfnamefont
  {A}},  \emph {et~al.}} (\bibinfo {year} {2013}),\ \bibfield  {title}
  {\enquote {\bibinfo {title} {{Standard Model updates and new physics analysis
  with the Unitarity Triangle fit}},}\ }\bibfield  {booktitle} {\emph {\bibinfo
  {booktitle} {{Proceedings, 4th Workshop on Theory, Phenomenology and
  Experiments in Heavy Flavour Physics}}},\ }\href {\doibase
  10.1016/j.nuclphysbps.2013.06.015} {\bibfield  {journal} {\bibinfo  {journal}
  {Nucl. Phys. Proc. Suppl.}\ }\textbf {\bibinfo {volume} {241-242}},\ \bibinfo
  {pages} {89--94}}\BibitemShut {NoStop}%
\bibitem [{\citenamefont {Bevan}\ \emph {et~al.}(2014)\citenamefont {Bevan}
  \emph {et~al.}}]{Bevan:2014iga}%
  \BibitemOpen
  \bibfield  {author} {\bibinfo {author} {\bibnamefont {Bevan}, \bibfnamefont
  {A~J}},  \emph {et~al.} (\bibinfo {collaboration} {Belle, BaBar})} (\bibinfo
  {year} {2014}),\ \bibfield  {title} {\enquote {\bibinfo {title} {{The Physics
  of the $B$ Factories}},}\ }\href {\doibase 10.1140/epjc/s10052-014-3026-9}
  {\bibfield  {journal} {\bibinfo  {journal} {Eur. Phys. J.}\ }\textbf
  {\bibinfo {volume} {C74}},\ \bibinfo {pages} {3026}},\ \Eprint
  {http://arxiv.org/abs/1406.6311} {arXiv:1406.6311 [hep-ex]} \BibitemShut
  {NoStop}%
\bibitem [{\citenamefont {Bhattacharya}\ \emph {et~al.}(2013)\citenamefont
  {Bhattacharya}, \citenamefont {Datta},\ and\ \citenamefont
  {London}}]{Bhattacharya:2012ph}%
  \BibitemOpen
  \bibfield  {author} {\bibinfo {author} {\bibnamefont {Bhattacharya},
  \bibfnamefont {Bhubanjyoti}}, \bibinfo {author} {\bibfnamefont {Alakabha}\
  \bibnamefont {Datta}}, \ and\ \bibinfo {author} {\bibfnamefont {David}\
  \bibnamefont {London}}} (\bibinfo {year} {2013}),\ \bibfield  {title}
  {\enquote {\bibinfo {title} {{Reducing Penguin Pollution}},}\ }\href
  {\doibase 10.1142/S0217751X13500632} {\bibfield  {journal} {\bibinfo
  {journal} {Int. J. Mod. Phys.}\ }\textbf {\bibinfo {volume} {A28}},\ \bibinfo
  {pages} {1350063}},\ \Eprint {http://arxiv.org/abs/1209.1413}
  {arXiv:1209.1413 [hep-ph]} \BibitemShut {NoStop}%
\bibitem [{\citenamefont {Bhattacharya}\ and\ \citenamefont
  {London}(2015)}]{Bhattacharya:2015uua}%
  \BibitemOpen
  \bibfield  {author} {\bibinfo {author} {\bibnamefont {Bhattacharya},
  \bibfnamefont {Bhubanjyoti}}, \ and\ \bibinfo {author} {\bibfnamefont
  {David}\ \bibnamefont {London}}} (\bibinfo {year} {2015}),\ \bibfield
  {title} {\enquote {\bibinfo {title} {{Using U spin to extract $\gamma$ from
  charmless $B \to PPP$ decays}},}\ }\href {\doibase 10.1007/JHEP04(2015)154}
  {\bibfield  {journal} {\bibinfo  {journal} {JHEP}\ }\textbf {\bibinfo
  {volume} {04}},\ \bibinfo {pages} {154}},\ \Eprint
  {http://arxiv.org/abs/1503.00737} {arXiv:1503.00737 [hep-ph]} \BibitemShut
  {NoStop}%
\bibitem [{\citenamefont {Bigi}\ \emph {et~al.}(1989)\citenamefont {Bigi},
  \citenamefont {Khoze}, \citenamefont {Uraltsev},\ and\ \citenamefont
  {Sanda}}]{Bigi:1987in}%
  \BibitemOpen
  \bibfield  {author} {\bibinfo {author} {\bibnamefont {Bigi}, \bibfnamefont
  {Ikaros I~Y}}, \bibinfo {author} {\bibfnamefont {Valery~A.}\ \bibnamefont
  {Khoze}}, \bibinfo {author} {\bibfnamefont {N.~G.}\ \bibnamefont {Uraltsev}},
  \ and\ \bibinfo {author} {\bibfnamefont {A.~I.}\ \bibnamefont {Sanda}}}
  (\bibinfo {year} {1989}),\ \bibfield  {title} {\enquote {\bibinfo {title}
  {{The Question of {CP} Noninvariance - as Seen Through the Eyes of Neutral
  Beauty}},}\ }\href {\doibase 10.1142/9789814503280_0004} {\bibfield
  {journal} {\bibinfo  {journal} {Adv. Ser. Direct. High Energy Phys.}\
  }\textbf {\bibinfo {volume} {3}},\ \bibinfo {pages} {175--248}}\BibitemShut
  {NoStop}%
\bibitem [{\citenamefont {Bigi}\ \emph {et~al.}(1997)\citenamefont {Bigi},
  \citenamefont {Shifman},\ and\ \citenamefont {Uraltsev}}]{Bigi:1997fj}%
  \BibitemOpen
  \bibfield  {author} {\bibinfo {author} {\bibnamefont {Bigi}, \bibfnamefont
  {Ikaros I~Y}}, \bibinfo {author} {\bibfnamefont {Mikhail~A.}\ \bibnamefont
  {Shifman}}, \ and\ \bibinfo {author} {\bibfnamefont {N.}~\bibnamefont
  {Uraltsev}}} (\bibinfo {year} {1997}),\ \bibfield  {title} {\enquote
  {\bibinfo {title} {{Aspects of heavy quark theory}},}\ }\href {\doibase
  10.1146/annurev.nucl.47.1.591} {\bibfield  {journal} {\bibinfo  {journal}
  {Ann. Rev. Nucl. Part. Sci.}\ }\textbf {\bibinfo {volume} {47}},\ \bibinfo
  {pages} {591--661}},\ \Eprint {http://arxiv.org/abs/hep-ph/9703290}
  {arXiv:hep-ph/9703290 [hep-ph]} \BibitemShut {NoStop}%
\bibitem [{\citenamefont {Bigi}\ and\ \citenamefont
  {Uraltsev}(1992)}]{Bigi:1991ir}%
  \BibitemOpen
  \bibfield  {author} {\bibinfo {author} {\bibnamefont {Bigi}, \bibfnamefont
  {Ikaros I~Y}}, \ and\ \bibinfo {author} {\bibfnamefont {N.~G.}\ \bibnamefont
  {Uraltsev}}} (\bibinfo {year} {1992}),\ \bibfield  {title} {\enquote
  {\bibinfo {title} {{Gluonic enhancements in non-spectator beauty decays: An
  Inclusive mirage though an exclusive possibility}},}\ }\href {\doibase
  10.1016/0370-2693(92)90066-D} {\bibfield  {journal} {\bibinfo  {journal}
  {Phys. Lett.}\ }\textbf {\bibinfo {volume} {B280}},\ \bibinfo {pages}
  {271--280}}\BibitemShut {NoStop}%
\bibitem [{\citenamefont {Bigi}\ \emph {et~al.}(1992)\citenamefont {Bigi},
  \citenamefont {Uraltsev},\ and\ \citenamefont {Vainshtein}}]{Bigi:1992su}%
  \BibitemOpen
  \bibfield  {author} {\bibinfo {author} {\bibnamefont {Bigi}, \bibfnamefont
  {Ikaros I~Y}}, \bibinfo {author} {\bibfnamefont {N.~G.}\ \bibnamefont
  {Uraltsev}}, \ and\ \bibinfo {author} {\bibfnamefont {A.~I.}\ \bibnamefont
  {Vainshtein}}} (\bibinfo {year} {1992}),\ \bibfield  {title} {\enquote
  {\bibinfo {title} {{Nonperturbative corrections to inclusive beauty and charm
  decays: QCD versus phenomenological models}},}\ }\href {\doibase
  10.1016/0370-2693(92)90908-M} {\bibfield  {journal} {\bibinfo  {journal}
  {Phys. Lett.}\ }\textbf {\bibinfo {volume} {B293}},\ \bibinfo {pages}
  {430--436}},\ \bibinfo {note} {[Erratum: Phys. Lett.B297,477(1993)]},\
  \Eprint {http://arxiv.org/abs/hep-ph/9207214} {arXiv:hep-ph/9207214 [hep-ph]}
  \BibitemShut {NoStop}%
\bibitem [{\citenamefont {Bigi}\ and\ \citenamefont
  {Sanda}(1981)}]{Bigi:1981qs}%
  \BibitemOpen
  \bibfield  {author} {\bibinfo {author} {\bibnamefont {Bigi}, \bibfnamefont
  {Ikaros~IY}}, \ and\ \bibinfo {author} {\bibfnamefont {A.I.}\ \bibnamefont
  {Sanda}}} (\bibinfo {year} {1981}),\ \bibfield  {title} {\enquote {\bibinfo
  {title} {{Notes on the Observability of CP Violations in B Decays}},}\ }\href
  {\doibase 10.1016/0550-3213(81)90519-8} {\bibfield  {journal} {\bibinfo
  {journal} {Nucl.Phys.}\ }\textbf {\bibinfo {volume} {B193}},\ \bibinfo
  {pages} {85}}\BibitemShut {NoStop}%
\bibitem [{\citenamefont {Blok}\ and\ \citenamefont
  {Shifman}(1993{\natexlab{a}})}]{Blok:1992hw}%
  \BibitemOpen
  \bibfield  {author} {\bibinfo {author} {\bibnamefont {Blok}, \bibfnamefont
  {B}}, \ and\ \bibinfo {author} {\bibfnamefont {Mikhail~A.}\ \bibnamefont
  {Shifman}}} (\bibinfo {year} {1993}{\natexlab{a}}),\ \bibfield  {title}
  {\enquote {\bibinfo {title} {{The Rule of discarding $1/N_c$ in inclusive
  weak decays. 1.}}}\ }\href {\doibase 10.1016/0550-3213(93)90504-I} {\bibfield
   {journal} {\bibinfo  {journal} {Nucl. Phys.}\ }\textbf {\bibinfo {volume}
  {B399}},\ \bibinfo {pages} {441--458}},\ \Eprint
  {http://arxiv.org/abs/hep-ph/9207236} {arXiv:hep-ph/9207236 [hep-ph]}
  \BibitemShut {NoStop}%
\bibitem [{\citenamefont {Blok}\ and\ \citenamefont
  {Shifman}(1993{\natexlab{b}})}]{Blok:1992he}%
  \BibitemOpen
  \bibfield  {author} {\bibinfo {author} {\bibnamefont {Blok}, \bibfnamefont
  {B}}, \ and\ \bibinfo {author} {\bibfnamefont {Mikhail~A.}\ \bibnamefont
  {Shifman}}} (\bibinfo {year} {1993}{\natexlab{b}}),\ \bibfield  {title}
  {\enquote {\bibinfo {title} {{The Rule of discarding $1/N_c$ in inclusive
  weak decays. 2.}}}\ }\href {\doibase 10.1016/0550-3213(93)90505-J} {\bibfield
   {journal} {\bibinfo  {journal} {Nucl. Phys.}\ }\textbf {\bibinfo {volume}
  {B399}},\ \bibinfo {pages} {459--476}},\ \Eprint
  {http://arxiv.org/abs/hep-ph/9209289} {arXiv:hep-ph/9209289 [hep-ph]}
  \BibitemShut {NoStop}%
\bibitem [{\citenamefont {Bobeth}\ \emph {et~al.}(2015)\citenamefont {Bobeth},
  \citenamefont {Gorbahn},\ and\ \citenamefont {Vickers}}]{Bobeth:2014rra}%
  \BibitemOpen
  \bibfield  {author} {\bibinfo {author} {\bibnamefont {Bobeth}, \bibfnamefont
  {Christoph}}, \bibinfo {author} {\bibfnamefont {Martin}\ \bibnamefont
  {Gorbahn}}, \ and\ \bibinfo {author} {\bibfnamefont {Stefan}\ \bibnamefont
  {Vickers}}} (\bibinfo {year} {2015}),\ \bibfield  {title} {\enquote {\bibinfo
  {title} {{Weak annihilation and new physics in charmless $B \to M M$
  decays}},}\ }\href {\doibase 10.1140/epjc/s10052-015-3535-1} {\bibfield
  {journal} {\bibinfo  {journal} {Eur. Phys. J.}\ }\textbf {\bibinfo {volume}
  {C75}}~(\bibinfo {number} {7}),\ \bibinfo {pages} {340}},\ \Eprint
  {http://arxiv.org/abs/1409.3252} {arXiv:1409.3252 [hep-ph]} \BibitemShut
  {NoStop}%
\bibitem [{\citenamefont {Bobeth}\ \emph {et~al.}(2014)\citenamefont {Bobeth},
  \citenamefont {Haisch}, \citenamefont {Lenz}, \citenamefont {Pecjak},\ and\
  \citenamefont {Tetlalmatzi-Xolocotzi}}]{Bobeth:2014rda}%
  \BibitemOpen
  \bibfield  {author} {\bibinfo {author} {\bibnamefont {Bobeth}, \bibfnamefont
  {Christoph}}, \bibinfo {author} {\bibfnamefont {Ulrich}\ \bibnamefont
  {Haisch}}, \bibinfo {author} {\bibfnamefont {Alexander}\ \bibnamefont
  {Lenz}}, \bibinfo {author} {\bibfnamefont {Ben}\ \bibnamefont {Pecjak}}, \
  and\ \bibinfo {author} {\bibfnamefont {Gilberto}\ \bibnamefont
  {Tetlalmatzi-Xolocotzi}}} (\bibinfo {year} {2014}),\ \bibfield  {title}
  {\enquote {\bibinfo {title} {{On new physics in $\Delta\Gamma_{d}$}},}\
  }\href {\doibase 10.1007/JHEP06(2014)040} {\bibfield  {journal} {\bibinfo
  {journal} {JHEP}\ }\textbf {\bibinfo {volume} {06}},\ \bibinfo {pages}
  {040}},\ \Eprint {http://arxiv.org/abs/1404.2531} {arXiv:1404.2531 [hep-ph]}
  \BibitemShut {NoStop}%
\bibitem [{\citenamefont {Bobrowski}\ \emph {et~al.}(2010)\citenamefont
  {Bobrowski}, \citenamefont {Lenz}, \citenamefont {Riedl},\ and\ \citenamefont
  {Rohrwild}}]{Bobrowski:2010xg}%
  \BibitemOpen
  \bibfield  {author} {\bibinfo {author} {\bibnamefont {Bobrowski},
  \bibfnamefont {M}}, \bibinfo {author} {\bibfnamefont {A.}~\bibnamefont
  {Lenz}}, \bibinfo {author} {\bibfnamefont {J.}~\bibnamefont {Riedl}}, \ and\
  \bibinfo {author} {\bibfnamefont {J.}~\bibnamefont {Rohrwild}}} (\bibinfo
  {year} {2010}),\ \bibfield  {title} {\enquote {\bibinfo {title} {{How Large
  Can the SM Contribution to CP Violation in $D^0-\bar D^0$ Mixing Be?}}}\
  }\href {\doibase 10.1007/JHEP03(2010)009} {\bibfield  {journal} {\bibinfo
  {journal} {JHEP}\ }\textbf {\bibinfo {volume} {03}},\ \bibinfo {pages}
  {009}},\ \Eprint {http://arxiv.org/abs/1002.4794} {arXiv:1002.4794 [hep-ph]}
  \BibitemShut {NoStop}%
\bibitem [{\citenamefont {Bona}\ \emph
  {et~al.}(2006{\natexlab{a}})\citenamefont {Bona} \emph
  {et~al.}}]{Bona:2006sa}%
  \BibitemOpen
  \bibfield  {author} {\bibinfo {author} {\bibnamefont {Bona}, \bibfnamefont
  {M}},  \emph {et~al.} (\bibinfo {collaboration} {UTfit})} (\bibinfo {year}
  {2006}{\natexlab{a}}),\ \bibfield  {title} {\enquote {\bibinfo {title}
  {{Constraints on new physics from the quark mixing unitarity triangle}},}\
  }\href {\doibase 10.1103/PhysRevLett.97.151803} {\bibfield  {journal}
  {\bibinfo  {journal} {Phys. Rev. Lett.}\ }\textbf {\bibinfo {volume} {97}},\
  \bibinfo {pages} {151803}},\ \Eprint {http://arxiv.org/abs/hep-ph/0605213}
  {arXiv:hep-ph/0605213 [hep-ph]} \BibitemShut {NoStop}%
\bibitem [{\citenamefont {Bona}\ \emph
  {et~al.}(2006{\natexlab{b}})\citenamefont {Bona} \emph
  {et~al.}}]{Bona:2006ah}%
  \BibitemOpen
  \bibfield  {author} {\bibinfo {author} {\bibnamefont {Bona}, \bibfnamefont
  {M}},  \emph {et~al.} (\bibinfo {collaboration} {UTfit})} (\bibinfo {year}
  {2006}{\natexlab{b}}),\ \bibfield  {title} {\enquote {\bibinfo {title} {{The
  Unitarity Triangle Fit in the Standard Model and Hadronic Parameters from
  Lattice QCD: A Reappraisal after the Measurements of $\Delta M_s$ and $BR(B
  \to \tau \nu_\tau)$}},}\ }\href {\doibase 10.1088/1126-6708/2006/10/081}
  {\bibfield  {journal} {\bibinfo  {journal} {JHEP}\ }\textbf {\bibinfo
  {volume} {10}},\ \bibinfo {pages} {081}},\ \Eprint
  {http://arxiv.org/abs/hep-ph/0606167} {arXiv:hep-ph/0606167 [hep-ph]}
  \BibitemShut {NoStop}%
\bibitem [{\citenamefont {Bona}\ \emph {et~al.}(2008)\citenamefont {Bona} \emph
  {et~al.}}]{Bona:2007vi}%
  \BibitemOpen
  \bibfield  {author} {\bibinfo {author} {\bibnamefont {Bona}, \bibfnamefont
  {M}},  \emph {et~al.} (\bibinfo {collaboration} {UTfit})} (\bibinfo {year}
  {2008}),\ \bibfield  {title} {\enquote {\bibinfo {title} {{Model-independent
  constraints on $\Delta F=2$ operators and the scale of new physics}},}\
  }\href {\doibase 10.1088/1126-6708/2008/03/049} {\bibfield  {journal}
  {\bibinfo  {journal} {JHEP}\ }\textbf {\bibinfo {volume} {03}},\ \bibinfo
  {pages} {049}},\ \Eprint {http://arxiv.org/abs/0707.0636} {arXiv:0707.0636
  [hep-ph]} \BibitemShut {NoStop}%
\bibitem [{\citenamefont {Boos}\ \emph {et~al.}(2004)\citenamefont {Boos},
  \citenamefont {Mannel},\ and\ \citenamefont {Reuter}}]{Boos:2004xp}%
  \BibitemOpen
  \bibfield  {author} {\bibinfo {author} {\bibnamefont {Boos}, \bibfnamefont
  {Heike}}, \bibinfo {author} {\bibfnamefont {Thomas}\ \bibnamefont {Mannel}},
  \ and\ \bibinfo {author} {\bibfnamefont {Jurgen}\ \bibnamefont {Reuter}}}
  (\bibinfo {year} {2004}),\ \bibfield  {title} {\enquote {\bibinfo {title}
  {{The Gold plated mode revisited: $\sin(2 \beta)$ and $\Bd \to \jpsi K_S$ in
  the standard model}},}\ }\href {\doibase 10.1103/PhysRevD.70.036006}
  {\bibfield  {journal} {\bibinfo  {journal} {Phys. Rev.}\ }\textbf {\bibinfo
  {volume} {D70}},\ \bibinfo {pages} {036006}},\ \Eprint
  {http://arxiv.org/abs/hep-ph/0403085} {arXiv:hep-ph/0403085 [hep-ph]}
  \BibitemShut {NoStop}%
\bibitem [{\citenamefont {Borissov}\ and\ \citenamefont
  {Hoeneisen}(2013)}]{Borissov:2013wwa}%
  \BibitemOpen
  \bibfield  {author} {\bibinfo {author} {\bibnamefont {Borissov},
  \bibfnamefont {G}}, \ and\ \bibinfo {author} {\bibfnamefont {B.}~\bibnamefont
  {Hoeneisen}}} (\bibinfo {year} {2013}),\ \bibfield  {title} {\enquote
  {\bibinfo {title} {{Understanding the like-sign dimuon charge asymmetry in $p
  \bar{p}$ collisions}},}\ }\href {\doibase 10.1103/PhysRevD.87.074020}
  {\bibfield  {journal} {\bibinfo  {journal} {Phys. Rev.}\ }\textbf {\bibinfo
  {volume} {D87}}~(\bibinfo {number} {7}),\ \bibinfo {pages} {074020}},\
  \Eprint {http://arxiv.org/abs/1303.0175} {arXiv:1303.0175 [hep-ex]}
  \BibitemShut {NoStop}%
\bibitem [{\citenamefont {Botella}\ \emph {et~al.}(2012)\citenamefont
  {Botella}, \citenamefont {Branco},\ and\ \citenamefont
  {Nebot}}]{Botella:2012ju}%
  \BibitemOpen
  \bibfield  {author} {\bibinfo {author} {\bibnamefont {Botella}, \bibfnamefont
  {F~J}}, \bibinfo {author} {\bibfnamefont {G.~C.}\ \bibnamefont {Branco}}, \
  and\ \bibinfo {author} {\bibfnamefont {M.}~\bibnamefont {Nebot}}} (\bibinfo
  {year} {2012}),\ \bibfield  {title} {\enquote {\bibinfo {title} {{The Hunt
  for New Physics in the Flavour Sector with up vector-like quarks}},}\ }\href
  {\doibase 10.1007/JHEP12(2012)040} {\bibfield  {journal} {\bibinfo  {journal}
  {JHEP}\ }\textbf {\bibinfo {volume} {12}},\ \bibinfo {pages} {040}},\ \Eprint
  {http://arxiv.org/abs/1207.4440} {arXiv:1207.4440 [hep-ph]} \BibitemShut
  {NoStop}%
\bibitem [{\citenamefont {Bouchard}\ \emph {et~al.}(2011)\citenamefont
  {Bouchard}, \citenamefont {Freeland}, \citenamefont {Bernard}, \citenamefont
  {El-Khadra}, \citenamefont {Gamiz}, \citenamefont {Kronfeld}, \citenamefont
  {Laiho},\ and\ \citenamefont {Van~de Water}}]{Bouchard:2011xj}%
  \BibitemOpen
  \bibfield  {author} {\bibinfo {author} {\bibnamefont {Bouchard},
  \bibfnamefont {C~M}}, \bibinfo {author} {\bibfnamefont {E.~D.}\ \bibnamefont
  {Freeland}}, \bibinfo {author} {\bibfnamefont {C.}~\bibnamefont {Bernard}},
  \bibinfo {author} {\bibfnamefont {A.~X.}\ \bibnamefont {El-Khadra}}, \bibinfo
  {author} {\bibfnamefont {E.}~\bibnamefont {Gamiz}}, \bibinfo {author}
  {\bibfnamefont {A.~S.}\ \bibnamefont {Kronfeld}}, \bibinfo {author}
  {\bibfnamefont {J.}~\bibnamefont {Laiho}}, \ and\ \bibinfo {author}
  {\bibfnamefont {R.~S.}\ \bibnamefont {Van~de Water}}} (\bibinfo {year}
  {2011}),\ \bibfield  {title} {\enquote {\bibinfo {title} {{Neutral $B$ mixing
  from $2+1$ flavor lattice-QCD: the Standard Model and beyond}},}\ }\bibfield
  {booktitle} {\emph {\bibinfo {booktitle} {{Proceedings, 29th International
  Symposium on Lattice field theory (Lattice 2011)}}},\ }\href@noop {}
  {\bibfield  {journal} {\bibinfo  {journal} {PoS}\ }\textbf {\bibinfo {volume}
  {LATTICE2011}},\ \bibinfo {pages} {274}},\ \Eprint
  {http://arxiv.org/abs/1112.5642} {arXiv:1112.5642 [hep-lat]} \BibitemShut
  {NoStop}%
\bibitem [{\citenamefont {Brod}\ \emph {et~al.}(2015)\citenamefont {Brod},
  \citenamefont {Lenz}, \citenamefont {Tetlalmatzi-Xolocotzi},\ and\
  \citenamefont {Wiebusch}}]{Brod:2014bfa}%
  \BibitemOpen
  \bibfield  {author} {\bibinfo {author} {\bibnamefont {Brod}, \bibfnamefont
  {Joachim}}, \bibinfo {author} {\bibfnamefont {Alexander}\ \bibnamefont
  {Lenz}}, \bibinfo {author} {\bibfnamefont {Gilberto}\ \bibnamefont
  {Tetlalmatzi-Xolocotzi}}, \ and\ \bibinfo {author} {\bibfnamefont {Martin}\
  \bibnamefont {Wiebusch}}} (\bibinfo {year} {2015}),\ \bibfield  {title}
  {\enquote {\bibinfo {title} {{New physics effects in tree-level decays and
  the precision in the determination of the quark mixing angle $\gamma$ }},}\
  }\href {\doibase 10.1103/PhysRevD.92.033002} {\bibfield  {journal} {\bibinfo
  {journal} {Phys. Rev.}\ }\textbf {\bibinfo {volume} {D92}}~(\bibinfo {number}
  {3}),\ \bibinfo {pages} {033002}},\ \Eprint {http://arxiv.org/abs/1412.1446}
  {arXiv:1412.1446 [hep-ph]} \BibitemShut {NoStop}%
\bibitem [{\citenamefont {Buchalla}\ \emph {et~al.}(1996)\citenamefont
  {Buchalla}, \citenamefont {Buras},\ and\ \citenamefont
  {Lautenbacher}}]{Buchalla:1995vs}%
  \BibitemOpen
  \bibfield  {author} {\bibinfo {author} {\bibnamefont {Buchalla},
  \bibfnamefont {Gerhard}}, \bibinfo {author} {\bibfnamefont {Andrzej~J.}\
  \bibnamefont {Buras}}, \ and\ \bibinfo {author} {\bibfnamefont {Markus~E.}\
  \bibnamefont {Lautenbacher}}} (\bibinfo {year} {1996}),\ \bibfield  {title}
  {\enquote {\bibinfo {title} {{Weak decays beyond leading logarithms}},}\
  }\href {\doibase 10.1103/RevModPhys.68.1125} {\bibfield  {journal} {\bibinfo
  {journal} {Rev. Mod. Phys.}\ }\textbf {\bibinfo {volume} {68}},\ \bibinfo
  {pages} {1125--1144}},\ \Eprint {http://arxiv.org/abs/hep-ph/9512380}
  {arXiv:hep-ph/9512380 [hep-ph]} \BibitemShut {NoStop}%
\bibitem [{\citenamefont {Buchkremer}\ \emph {et~al.}(2012)\citenamefont
  {Buchkremer}, \citenamefont {Gerard},\ and\ \citenamefont
  {Maltoni}}]{Buchkremer:2012yy}%
  \BibitemOpen
  \bibfield  {author} {\bibinfo {author} {\bibnamefont {Buchkremer},
  \bibfnamefont {Mathieu}}, \bibinfo {author} {\bibfnamefont {Jean-Marc}\
  \bibnamefont {Gerard}}, \ and\ \bibinfo {author} {\bibfnamefont {Fabio}\
  \bibnamefont {Maltoni}}} (\bibinfo {year} {2012}),\ \bibfield  {title}
  {\enquote {\bibinfo {title} {{Closing in on a perturbative fourth
  generation}},}\ }\href {\doibase 10.1007/JHEP06(2012)135} {\bibfield
  {journal} {\bibinfo  {journal} {JHEP}\ }\textbf {\bibinfo {volume} {06}},\
  \bibinfo {pages} {135}},\ \Eprint {http://arxiv.org/abs/1204.5403}
  {arXiv:1204.5403 [hep-ph]} \BibitemShut {NoStop}%
\bibitem [{\citenamefont {Buras}\ \emph {et~al.}(1984)\citenamefont {Buras},
  \citenamefont {Slominski},\ and\ \citenamefont {Steger}}]{Buras:1984pq}%
  \BibitemOpen
  \bibfield  {author} {\bibinfo {author} {\bibnamefont {Buras}, \bibfnamefont
  {A~J}}, \bibinfo {author} {\bibfnamefont {W.}~\bibnamefont {Slominski}}, \
  and\ \bibinfo {author} {\bibfnamefont {H.}~\bibnamefont {Steger}}} (\bibinfo
  {year} {1984}),\ \bibfield  {title} {\enquote {\bibinfo {title} {{$B^0-
  \bar{B}^0$ Mixing, CP Violation and the B Meson Decay}},}\ }\href {\doibase
  10.1016/0550-3213(84)90437-1} {\bibfield  {journal} {\bibinfo  {journal}
  {Nucl. Phys.}\ }\textbf {\bibinfo {volume} {B245}},\ \bibinfo {pages}
  {369}}\BibitemShut {NoStop}%
\bibitem [{\citenamefont {Buras}(1998)}]{Buras:1998raa}%
  \BibitemOpen
  \bibfield  {author} {\bibinfo {author} {\bibnamefont {Buras}, \bibfnamefont
  {Andrzej~J}}} (\bibinfo {year} {1998}),\ \bibfield  {title} {\enquote
  {\bibinfo {title} {{Weak Hamiltonian, CP violation and rare decays}},}\ }in\
  \href {http://alice.cern.ch/format/showfull?sysnb=0282793} {\emph {\bibinfo
  {booktitle} {{Probing the standard model of particle interactions.
  Proceedings, Summer School in Theoretical Physics, NATO Advanced Study
  Institute, 68th session, Les Houches, France, July 28-September 5, 1997. Pt.
  1, 2}}}},\ pp.\ \bibinfo {pages} {281--539},\ \Eprint
  {http://arxiv.org/abs/hep-ph/9806471} {arXiv:hep-ph/9806471 [hep-ph]}
  \BibitemShut {NoStop}%
\bibitem [{\citenamefont {Buras}\ \emph {et~al.}(1990)\citenamefont {Buras},
  \citenamefont {Jamin},\ and\ \citenamefont {Weisz}}]{Buras:1990fn}%
  \BibitemOpen
  \bibfield  {author} {\bibinfo {author} {\bibnamefont {Buras}, \bibfnamefont
  {Andrzej~J}}, \bibinfo {author} {\bibfnamefont {Matthias}\ \bibnamefont
  {Jamin}}, \ and\ \bibinfo {author} {\bibfnamefont {Peter~H.}\ \bibnamefont
  {Weisz}}} (\bibinfo {year} {1990}),\ \bibfield  {title} {\enquote {\bibinfo
  {title} {{Leading and Next-to-leading {QCD} Corrections to $\epsilon$
  Parameter and $B^0 - \bar{B}^0$ Mixing in the Presence of a Heavy Top
  Quark}},}\ }\href {\doibase 10.1016/0550-3213(90)90373-L} {\bibfield
  {journal} {\bibinfo  {journal} {Nucl.Phys.}\ }\textbf {\bibinfo {volume}
  {B347}},\ \bibinfo {pages} {491--536}}\BibitemShut {NoStop}%
\bibitem [{\citenamefont {Buras}\ \emph {et~al.}(2011)\citenamefont {Buras},
  \citenamefont {Nagai},\ and\ \citenamefont {Paradisi}}]{Buras:2010pm}%
  \BibitemOpen
  \bibfield  {author} {\bibinfo {author} {\bibnamefont {Buras}, \bibfnamefont
  {Andrzej~J}}, \bibinfo {author} {\bibfnamefont {Minoru}\ \bibnamefont
  {Nagai}}, \ and\ \bibinfo {author} {\bibfnamefont {Paride}\ \bibnamefont
  {Paradisi}}} (\bibinfo {year} {2011}),\ \bibfield  {title} {\enquote
  {\bibinfo {title} {{Footprints of SUSY GUTs in Flavour Physics}},}\ }\href
  {\doibase 10.1007/JHEP05(2011)005} {\bibfield  {journal} {\bibinfo  {journal}
  {JHEP}\ }\textbf {\bibinfo {volume} {05}},\ \bibinfo {pages} {005}},\ \Eprint
  {http://arxiv.org/abs/1011.4853} {arXiv:1011.4853 [hep-ph]} \BibitemShut
  {NoStop}%
\bibitem [{\citenamefont {Cabibbo}(1963)}]{Cabibbo:1963yz}%
  \BibitemOpen
  \bibfield  {author} {\bibinfo {author} {\bibnamefont {Cabibbo}, \bibfnamefont
  {Nicola}}} (\bibinfo {year} {1963}),\ \bibfield  {title} {\enquote {\bibinfo
  {title} {{Unitary Symmetry and Leptonic Decays}},}\ }\href {\doibase
  10.1103/PhysRevLett.10.531} {\bibfield  {journal} {\bibinfo  {journal} {Phys.
  Rev. Lett.}\ }\textbf {\bibinfo {volume} {10}},\ \bibinfo {pages}
  {531--533}}\BibitemShut {NoStop}%
\bibitem [{\citenamefont {Carrasco}\ \emph {et~al.}(2014)\citenamefont
  {Carrasco} \emph {et~al.}}]{Carrasco:2013zta}%
  \BibitemOpen
  \bibfield  {author} {\bibinfo {author} {\bibnamefont {Carrasco},
  \bibfnamefont {N}},  \emph {et~al.} (\bibinfo {collaboration} {ETM})}
  (\bibinfo {year} {2014}),\ \bibfield  {title} {\enquote {\bibinfo {title}
  {{B-physics from $N_f$ = 2 tmQCD: the Standard Model and beyond}},}\ }\href
  {\doibase 10.1007/JHEP03(2014)016} {\bibfield  {journal} {\bibinfo  {journal}
  {JHEP}\ }\textbf {\bibinfo {volume} {1403}},\ \bibinfo {pages} {016}},\
  \Eprint {http://arxiv.org/abs/1308.1851} {arXiv:1308.1851 [hep-lat]}
  \BibitemShut {NoStop}%
\bibitem [{\citenamefont {Carter}\ and\ \citenamefont
  {Sanda}(1981)}]{Carter:1980tk}%
  \BibitemOpen
  \bibfield  {author} {\bibinfo {author} {\bibnamefont {Carter}, \bibfnamefont
  {Ashton~B}}, \ and\ \bibinfo {author} {\bibfnamefont {A.I.}\ \bibnamefont
  {Sanda}}} (\bibinfo {year} {1981}),\ \bibfield  {title} {\enquote {\bibinfo
  {title} {{CP Violation in B Meson Decays}},}\ }\href {\doibase
  10.1103/PhysRevD.23.1567} {\bibfield  {journal} {\bibinfo  {journal}
  {Phys.Rev.}\ }\textbf {\bibinfo {volume} {D23}},\ \bibinfo {pages}
  {1567}}\BibitemShut {NoStop}%
\bibitem [{\citenamefont {Chang}\ \emph {et~al.}(2015)\citenamefont {Chang},
  \citenamefont {Li},\ and\ \citenamefont {Li}}]{Chang:2015rva}%
  \BibitemOpen
  \bibfield  {author} {\bibinfo {author} {\bibnamefont {Chang}, \bibfnamefont
  {Qin}}, \bibinfo {author} {\bibfnamefont {Pei-Fu}\ \bibnamefont {Li}}, \ and\
  \bibinfo {author} {\bibfnamefont {Xin-Qiang}\ \bibnamefont {Li}}} (\bibinfo
  {year} {2015}),\ \bibfield  {title} {\enquote {\bibinfo {title}
  {{$B_s^0-\bar{B}_s^0$ mixing within minimal flavor-violating
  two-Higgs-doublet models}},}\ }\href@noop {} {\ }\Eprint
  {http://arxiv.org/abs/1505.03650} {arXiv:1505.03650 [hep-ph]} \BibitemShut
  {NoStop}%
\bibitem [{\citenamefont {Chang}\ \emph {et~al.}(2014)\citenamefont {Chang},
  \citenamefont {Li},\ and\ \citenamefont {Yang}}]{Chang:2013hba}%
  \BibitemOpen
  \bibfield  {author} {\bibinfo {author} {\bibnamefont {Chang}, \bibfnamefont
  {Qin}}, \bibinfo {author} {\bibfnamefont {Xin-Qiang}\ \bibnamefont {Li}}, \
  and\ \bibinfo {author} {\bibfnamefont {Ya-Dong}\ \bibnamefont {Yang}}}
  (\bibinfo {year} {2014}),\ \bibfield  {title} {\enquote {\bibinfo {title} {{A
  comprehensive analysis of hadronic b $\to$ s transitions in a family
  non-universal $Z'$ model}},}\ }\href {\doibase
  10.1088/0954-3899/41/10/105002} {\bibfield  {journal} {\bibinfo  {journal}
  {J. Phys.}\ }\textbf {\bibinfo {volume} {G41}},\ \bibinfo {pages} {105002}},\
  \Eprint {http://arxiv.org/abs/1312.1302} {arXiv:1312.1302 [hep-ph]}
  \BibitemShut {NoStop}%
\bibitem [{\citenamefont {Chang}\ \emph {et~al.}(2011)\citenamefont {Chang},
  \citenamefont {Wang}, \citenamefont {Xu},\ and\ \citenamefont
  {Cui}}]{Chang:2011zza}%
  \BibitemOpen
  \bibfield  {author} {\bibinfo {author} {\bibnamefont {Chang}, \bibfnamefont
  {Qin}}, \bibinfo {author} {\bibfnamefont {Ru-Min}\ \bibnamefont {Wang}},
  \bibinfo {author} {\bibfnamefont {Yuan-Guo}\ \bibnamefont {Xu}}, \ and\
  \bibinfo {author} {\bibfnamefont {Xiao-Wei}\ \bibnamefont {Cui}}} (\bibinfo
  {year} {2011}),\ \bibfield  {title} {\enquote {\bibinfo {title} {{Large
  dimuon asymmetry and a non-universal $Z'$ boson in the $\Bs -
  \barBs$-system}},}\ }\href {\doibase 10.1088/0256-307X/28/8/081301}
  {\bibfield  {journal} {\bibinfo  {journal} {Chin. Phys. Lett.}\ }\textbf
  {\bibinfo {volume} {28}},\ \bibinfo {pages} {081301}}\BibitemShut {NoStop}%
\bibitem [{\citenamefont {Charles}\ \emph {et~al.}(2005)\citenamefont
  {Charles}, \citenamefont {Hocker}, \citenamefont {Lacker}, \citenamefont
  {Laplace}, \citenamefont {Le~Diberder}, \citenamefont {Malcles},
  \citenamefont {Ocariz}, \citenamefont {Pivk},\ and\ \citenamefont
  {Roos}}]{Charles:2004jd}%
  \BibitemOpen
  \bibfield  {author} {\bibinfo {author} {\bibnamefont {Charles}, \bibfnamefont
  {J}}, \bibinfo {author} {\bibfnamefont {Andreas}\ \bibnamefont {Hocker}},
  \bibinfo {author} {\bibfnamefont {H.}~\bibnamefont {Lacker}}, \bibinfo
  {author} {\bibfnamefont {S.}~\bibnamefont {Laplace}}, \bibinfo {author}
  {\bibfnamefont {F.~R.}\ \bibnamefont {Le~Diberder}}, \bibinfo {author}
  {\bibfnamefont {J.}~\bibnamefont {Malcles}}, \bibinfo {author} {\bibfnamefont
  {J.}~\bibnamefont {Ocariz}}, \bibinfo {author} {\bibfnamefont
  {M.}~\bibnamefont {Pivk}}, \ and\ \bibinfo {author} {\bibfnamefont
  {L.}~\bibnamefont {Roos}} (\bibinfo {collaboration} {CKMfitter Group})}
  (\bibinfo {year} {2005}),\ \bibfield  {title} {\enquote {\bibinfo {title}
  {{CP violation and the CKM matrix: Assessing the impact of the asymmetric $B$
  factories}},}\ }\href {\doibase 10.1140/epjc/s2005-02169-1} {\bibfield
  {journal} {\bibinfo  {journal} {Eur. Phys. J.}\ }\textbf {\bibinfo {volume}
  {C41}},\ \bibinfo {pages} {1--131}},\ \Eprint
  {http://arxiv.org/abs/hep-ph/0406184} {arXiv:hep-ph/0406184 [hep-ph]}
  \BibitemShut {NoStop}%
\bibitem [{\citenamefont {Charles}\ \emph {et~al.}(2015)\citenamefont {Charles}
  \emph {et~al.}}]{Charles:2015gya}%
  \BibitemOpen
  \bibfield  {author} {\bibinfo {author} {\bibnamefont {Charles}, \bibfnamefont
  {J}},  \emph {et~al.}} (\bibinfo {year} {2015}),\ \bibfield  {title}
  {\enquote {\bibinfo {title} {{Current status of the Standard Model CKM fit
  and constraints on $\Delta F=2$ New Physics}},}\ }\href {\doibase
  10.1103/PhysRevD.91.073007} {\bibfield  {journal} {\bibinfo  {journal} {Phys.
  Rev.}\ }\textbf {\bibinfo {volume} {D91}}~(\bibinfo {number} {7}),\ \bibinfo
  {pages} {073007}},\ \Eprint {http://arxiv.org/abs/1501.05013}
  {arXiv:1501.05013 [hep-ph]} \BibitemShut {NoStop}%
\bibitem [{\citenamefont {Charles}\ \emph {et~al.}(2014)\citenamefont
  {Charles}, \citenamefont {Descotes-Genon}, \citenamefont {Ligeti},
  \citenamefont {Monteil}, \citenamefont {Papucci} \emph
  {et~al.}}]{Charles:2013aka}%
  \BibitemOpen
  \bibfield  {author} {\bibinfo {author} {\bibnamefont {Charles}, \bibfnamefont
  {Jérôme}}, \bibinfo {author} {\bibfnamefont {Sebastien}\ \bibnamefont
  {Descotes-Genon}}, \bibinfo {author} {\bibfnamefont {Zoltan}\ \bibnamefont
  {Ligeti}}, \bibinfo {author} {\bibfnamefont {Stéphane}\ \bibnamefont
  {Monteil}}, \bibinfo {author} {\bibfnamefont {Michele}\ \bibnamefont
  {Papucci}},  \emph {et~al.}} (\bibinfo {year} {2014}),\ \bibfield  {title}
  {\enquote {\bibinfo {title} {{Future sensitivity to new physics in $B_d,
  B_s$, and K mixings}},}\ }\href {\doibase 10.1103/PhysRevD.89.033016}
  {\bibfield  {journal} {\bibinfo  {journal} {Phys.Rev.}\ }\textbf {\bibinfo
  {volume} {D89}}~(\bibinfo {number} {3}),\ \bibinfo {pages} {033016}},\
  \Eprint {http://arxiv.org/abs/1309.2293} {arXiv:1309.2293 [hep-ph]}
  \BibitemShut {NoStop}%
\bibitem [{\citenamefont {Chau}(1983)}]{Chau:1982da}%
  \BibitemOpen
  \bibfield  {author} {\bibinfo {author} {\bibnamefont {Chau}, \bibfnamefont
  {Ling-Lie}}} (\bibinfo {year} {1983}),\ \bibfield  {title} {\enquote
  {\bibinfo {title} {{Quark Mixing in Weak Interactions}},}\ }\bibfield
  {booktitle} {\emph {\bibinfo {booktitle} {{European Sympos.1982:297}}},\
  }\href {\doibase 10.1016/0370-1573(83)90043-1} {\bibfield  {journal}
  {\bibinfo  {journal} {Phys. Rept.}\ }\textbf {\bibinfo {volume} {95}},\
  \bibinfo {pages} {1--94}}\BibitemShut {NoStop}%
\bibitem [{\citenamefont {Cheng}\ and\ \citenamefont
  {Chua}(2009)}]{Cheng:2009mu}%
  \BibitemOpen
  \bibfield  {author} {\bibinfo {author} {\bibnamefont {Cheng}, \bibfnamefont
  {Hai-Yang}}, \ and\ \bibinfo {author} {\bibfnamefont {Chun-Khiang}\
  \bibnamefont {Chua}}} (\bibinfo {year} {2009}),\ \bibfield  {title} {\enquote
  {\bibinfo {title} {{{QCD} Factorization for Charmless Hadronic $B_s$ Decays
  Revisited}},}\ }\href {\doibase 10.1103/PhysRevD.80.114026} {\bibfield
  {journal} {\bibinfo  {journal} {Phys. Rev.}\ }\textbf {\bibinfo {volume}
  {D80}},\ \bibinfo {pages} {114026}},\ \Eprint
  {http://arxiv.org/abs/0910.5237} {arXiv:0910.5237 [hep-ph]} \BibitemShut
  {NoStop}%
\bibitem [{\citenamefont {Christenson}\ \emph {et~al.}(1964)\citenamefont
  {Christenson}, \citenamefont {Cronin}, \citenamefont {Fitch},\ and\
  \citenamefont {Turlay}}]{Christenson:1964fg}%
  \BibitemOpen
  \bibfield  {author} {\bibinfo {author} {\bibnamefont {Christenson},
  \bibfnamefont {J~H}}, \bibinfo {author} {\bibfnamefont {J.~W.}\ \bibnamefont
  {Cronin}}, \bibinfo {author} {\bibfnamefont {V.~L.}\ \bibnamefont {Fitch}}, \
  and\ \bibinfo {author} {\bibfnamefont {R.}~\bibnamefont {Turlay}}} (\bibinfo
  {year} {1964}),\ \bibfield  {title} {\enquote {\bibinfo {title} {{Evidence
  for the $2 \pi$ Decay of the $K_2^0$ Meson}},}\ }\href {\doibase
  10.1103/PhysRevLett.13.138} {\bibfield  {journal} {\bibinfo  {journal} {Phys.
  Rev. Lett.}\ }\textbf {\bibinfo {volume} {13}},\ \bibinfo {pages}
  {138--140}}\BibitemShut {NoStop}%
\bibitem [{\citenamefont {Ciuchini}\ \emph {et~al.}(2003)\citenamefont
  {Ciuchini}, \citenamefont {Franco}, \citenamefont {Lubicz}, \citenamefont
  {Mescia},\ and\ \citenamefont {Tarantino}}]{Ciuchini:2003ww}%
  \BibitemOpen
  \bibfield  {author} {\bibinfo {author} {\bibnamefont {Ciuchini},
  \bibfnamefont {M}}, \bibinfo {author} {\bibfnamefont {E.}~\bibnamefont
  {Franco}}, \bibinfo {author} {\bibfnamefont {V.}~\bibnamefont {Lubicz}},
  \bibinfo {author} {\bibfnamefont {F.}~\bibnamefont {Mescia}}, \ and\ \bibinfo
  {author} {\bibfnamefont {C.}~\bibnamefont {Tarantino}}} (\bibinfo {year}
  {2003}),\ \bibfield  {title} {\enquote {\bibinfo {title} {{Lifetime
  differences and CP violation parameters of neutral $B$ mesons at the
  next-to-leading order in QCD}},}\ }\href {\doibase
  10.1088/1126-6708/2003/08/031} {\bibfield  {journal} {\bibinfo  {journal}
  {JHEP}\ }\textbf {\bibinfo {volume} {08}},\ \bibinfo {pages} {031}},\ \Eprint
  {http://arxiv.org/abs/hep-ph/0308029} {arXiv:hep-ph/0308029 [hep-ph]}
  \BibitemShut {NoStop}%
\bibitem [{\citenamefont {Ciuchini}\ \emph {et~al.}(2012)\citenamefont
  {Ciuchini}, \citenamefont {Franco}, \citenamefont {Mishima},\ and\
  \citenamefont {Silvestrini}}]{Ciuchini:2012gd}%
  \BibitemOpen
  \bibfield  {author} {\bibinfo {author} {\bibnamefont {Ciuchini},
  \bibfnamefont {M}}, \bibinfo {author} {\bibfnamefont {E.}~\bibnamefont
  {Franco}}, \bibinfo {author} {\bibfnamefont {S.}~\bibnamefont {Mishima}}, \
  and\ \bibinfo {author} {\bibfnamefont {L.}~\bibnamefont {Silvestrini}}}
  (\bibinfo {year} {2012}),\ \bibfield  {title} {\enquote {\bibinfo {title}
  {{Testing the Standard Model and Searching for New Physics with $B_d \to \pi
  \pi$ and $B_s \to K K$ Decays}},}\ }\href {\doibase 10.1007/JHEP10(2012)029}
  {\bibfield  {journal} {\bibinfo  {journal} {JHEP}\ }\textbf {\bibinfo
  {volume} {10}},\ \bibinfo {pages} {029}},\ \Eprint
  {http://arxiv.org/abs/1205.4948} {arXiv:1205.4948 [hep-ph]} \BibitemShut
  {NoStop}%
\bibitem [{\citenamefont {Ciuchini}\ \emph {et~al.}(2005)\citenamefont
  {Ciuchini}, \citenamefont {Pierini},\ and\ \citenamefont
  {Silvestrini}}]{Ciuchini:2005mg}%
  \BibitemOpen
  \bibfield  {author} {\bibinfo {author} {\bibnamefont {Ciuchini},
  \bibfnamefont {M}}, \bibinfo {author} {\bibfnamefont {M.}~\bibnamefont
  {Pierini}}, \ and\ \bibinfo {author} {\bibfnamefont {L.}~\bibnamefont
  {Silvestrini}}} (\bibinfo {year} {2005}),\ \bibfield  {title} {\enquote
  {\bibinfo {title} {{The Effect of penguins in the $\Bd \to \jpsi K^0$ CP
  asymmetry}},}\ }\href {\doibase 10.1103/PhysRevLett.95.221804} {\bibfield
  {journal} {\bibinfo  {journal} {Phys. Rev. Lett.}\ }\textbf {\bibinfo
  {volume} {95}},\ \bibinfo {pages} {221804}},\ \Eprint
  {http://arxiv.org/abs/hep-ph/0507290} {arXiv:hep-ph/0507290 [hep-ph]}
  \BibitemShut {NoStop}%
\bibitem [{\citenamefont {Ciuchini}\ \emph {et~al.}(2011)\citenamefont
  {Ciuchini}, \citenamefont {Pierini},\ and\ \citenamefont
  {Silvestrini}}]{Ciuchini:2011kd}%
  \BibitemOpen
  \bibfield  {author} {\bibinfo {author} {\bibnamefont {Ciuchini},
  \bibfnamefont {Marco}}, \bibinfo {author} {\bibfnamefont {Maurizio}\
  \bibnamefont {Pierini}}, \ and\ \bibinfo {author} {\bibfnamefont {Luca}\
  \bibnamefont {Silvestrini}}} (\bibinfo {year} {2011}),\ \bibfield  {title}
  {\enquote {\bibinfo {title} {{Theoretical uncertainty in sin $2\beta$: An
  Update}},}\ }in\ \href
  {https://inspirehep.net/record/886527/files/arXiv:1102.0392.pdf} {\emph
  {\bibinfo {booktitle} {{CKM unitarity triangle. Proceedings, 6th
  International Workshop, CKM 2010, Warwick, UK, September 6-10, 2010}}}},\
  \Eprint {http://arxiv.org/abs/1102.0392} {arXiv:1102.0392 [hep-ph]}
  \BibitemShut {NoStop}%
\bibitem [{\citenamefont {Colangelo}\ \emph {et~al.}(2011)\citenamefont
  {Colangelo}, \citenamefont {De~Fazio},\ and\ \citenamefont
  {Wang}}]{Colangelo:2010wg}%
  \BibitemOpen
  \bibfield  {author} {\bibinfo {author} {\bibnamefont {Colangelo},
  \bibfnamefont {Pietro}}, \bibinfo {author} {\bibfnamefont {Fulvia}\
  \bibnamefont {De~Fazio}}, \ and\ \bibinfo {author} {\bibfnamefont {Wei}\
  \bibnamefont {Wang}}} (\bibinfo {year} {2011}),\ \bibfield  {title} {\enquote
  {\bibinfo {title} {{Nonleptonic $B_s$ to charmonium decays: analyses in
  pursuit of determining the weak phase $\beta_s$}},}\ }\href {\doibase
  10.1103/PhysRevD.83.094027} {\bibfield  {journal} {\bibinfo  {journal} {Phys.
  Rev.}\ }\textbf {\bibinfo {volume} {D83}},\ \bibinfo {pages} {094027}},\
  \Eprint {http://arxiv.org/abs/1009.4612} {arXiv:1009.4612 [hep-ph]}
  \BibitemShut {NoStop}%
\bibitem [{\citenamefont {Collaboration}(2011)}]{LHCb:2011dta}%
  \BibitemOpen
  \bibfield  {author} {\bibinfo {author} {\bibnamefont {Collaboration},
  \bibfnamefont {The~LHCb}} (\bibinfo {collaboration} {LHCb})} (\bibinfo {year}
  {2011}),\ \bibfield  {title} {\enquote {\bibinfo {title} {{Letter of Intent
  for the LHCb Upgrade}},}\ }\href@noop {} {\ }\BibitemShut {NoStop}%
\bibitem [{\citenamefont {Crivellin}\ \emph {et~al.}(2011)\citenamefont
  {Crivellin}, \citenamefont {Hofer}, \citenamefont {Nierste},\ and\
  \citenamefont {Scherer}}]{Crivellin:2011sj}%
  \BibitemOpen
  \bibfield  {author} {\bibinfo {author} {\bibnamefont {Crivellin},
  \bibfnamefont {Andreas}}, \bibinfo {author} {\bibfnamefont {Lars}\
  \bibnamefont {Hofer}}, \bibinfo {author} {\bibfnamefont {Ulrich}\
  \bibnamefont {Nierste}}, \ and\ \bibinfo {author} {\bibfnamefont {Dominik}\
  \bibnamefont {Scherer}}} (\bibinfo {year} {2011}),\ \bibfield  {title}
  {\enquote {\bibinfo {title} {{Phenomenological consequences of radiative
  flavor violation in the MSSM}},}\ }\href {\doibase
  10.1103/PhysRevD.84.035030} {\bibfield  {journal} {\bibinfo  {journal} {Phys.
  Rev.}\ }\textbf {\bibinfo {volume} {D84}},\ \bibinfo {pages} {035030}},\
  \Eprint {http://arxiv.org/abs/1105.2818} {arXiv:1105.2818 [hep-ph]}
  \BibitemShut {NoStop}%
\bibitem [{\citenamefont {Datta}\ \emph {et~al.}(2011)\citenamefont {Datta},
  \citenamefont {Duraisamy},\ and\ \citenamefont {Khalil}}]{Datta:2010yq}%
  \BibitemOpen
  \bibfield  {author} {\bibinfo {author} {\bibnamefont {Datta}, \bibfnamefont
  {Alakabha}}, \bibinfo {author} {\bibfnamefont {Murugeswaran}\ \bibnamefont
  {Duraisamy}}, \ and\ \bibinfo {author} {\bibfnamefont {Shaaban}\ \bibnamefont
  {Khalil}}} (\bibinfo {year} {2011}),\ \bibfield  {title} {\enquote {\bibinfo
  {title} {{Like-sign dimuon charge asymmetry in Randall-Sundurm model}},}\
  }\href {\doibase 10.1103/PhysRevD.83.094501} {\bibfield  {journal} {\bibinfo
  {journal} {Phys. Rev.}\ }\textbf {\bibinfo {volume} {D83}},\ \bibinfo {pages}
  {094501}},\ \Eprint {http://arxiv.org/abs/1011.5979} {arXiv:1011.5979
  [hep-ph]} \BibitemShut {NoStop}%
\bibitem [{\citenamefont {Datta}\ \emph {et~al.}(2012)\citenamefont {Datta},
  \citenamefont {Duraisamy},\ and\ \citenamefont {London}}]{Datta:2012ky}%
  \BibitemOpen
  \bibfield  {author} {\bibinfo {author} {\bibnamefont {Datta}, \bibfnamefont
  {Alakabha}}, \bibinfo {author} {\bibfnamefont {Murugeswaran}\ \bibnamefont
  {Duraisamy}}, \ and\ \bibinfo {author} {\bibfnamefont {David}\ \bibnamefont
  {London}}} (\bibinfo {year} {2012}),\ \bibfield  {title} {\enquote {\bibinfo
  {title} {{New Physics in $b \to s$ Transitions and the $B_{d,s}^0 \to V_1
  V_2$ Angular Analysis}},}\ }\href {\doibase 10.1103/PhysRevD.86.076011}
  {\bibfield  {journal} {\bibinfo  {journal} {Phys. Rev.}\ }\textbf {\bibinfo
  {volume} {D86}},\ \bibinfo {pages} {076011}},\ \Eprint
  {http://arxiv.org/abs/1207.4495} {arXiv:1207.4495 [hep-ph]} \BibitemShut
  {NoStop}%
\bibitem [{\citenamefont {De~Bruyn}\ and\ \citenamefont
  {Fleischer}(2015)}]{DeBruyn:2014oga}%
  \BibitemOpen
  \bibfield  {author} {\bibinfo {author} {\bibnamefont {De~Bruyn},
  \bibfnamefont {Kristof}}, \ and\ \bibinfo {author} {\bibfnamefont {Robert}\
  \bibnamefont {Fleischer}}} (\bibinfo {year} {2015}),\ \bibfield  {title}
  {\enquote {\bibinfo {title} {{A Roadmap to Control Penguin Effects in
  $B^0_d\to J/\psi K_{\rm S}^0$ and $B^0_s\to J/\psi \phi$}},}\ }\href
  {\doibase 10.1007/JHEP03(2015)145} {\bibfield  {journal} {\bibinfo  {journal}
  {JHEP}\ }\textbf {\bibinfo {volume} {03}},\ \bibinfo {pages} {145}},\ \Eprint
  {http://arxiv.org/abs/1412.6834} {arXiv:1412.6834 [hep-ph]} \BibitemShut
  {NoStop}%
\bibitem [{\citenamefont {De~Bruyn}\ \emph {et~al.}(2013)\citenamefont
  {De~Bruyn}, \citenamefont {Fleischer}, \citenamefont {Knegjens},
  \citenamefont {Merk}, \citenamefont {Schiller},\ and\ \citenamefont
  {Tuning}}]{DeBruyn:2012jp}%
  \BibitemOpen
  \bibfield  {author} {\bibinfo {author} {\bibnamefont {De~Bruyn},
  \bibfnamefont {Kristof}}, \bibinfo {author} {\bibfnamefont {Robert}\
  \bibnamefont {Fleischer}}, \bibinfo {author} {\bibfnamefont {Robert}\
  \bibnamefont {Knegjens}}, \bibinfo {author} {\bibfnamefont {Marcel}\
  \bibnamefont {Merk}}, \bibinfo {author} {\bibfnamefont {Manuel}\ \bibnamefont
  {Schiller}}, \ and\ \bibinfo {author} {\bibfnamefont {Niels}\ \bibnamefont
  {Tuning}}} (\bibinfo {year} {2013}),\ \bibfield  {title} {\enquote {\bibinfo
  {title} {{Exploring $B_s \to D_s^{(*)\pm} K^\mp$ Decays in the Presence of a
  Sizable Width Difference $\Delta\Gamma_s$}},}\ }\href {\doibase
  10.1016/j.nuclphysb.2012.11.012} {\bibfield  {journal} {\bibinfo  {journal}
  {Nucl. Phys.}\ }\textbf {\bibinfo {volume} {B868}},\ \bibinfo {pages}
  {351--367}},\ \Eprint {http://arxiv.org/abs/1208.6463} {arXiv:1208.6463
  [hep-ph]} \BibitemShut {NoStop}%
\bibitem [{\citenamefont {Di~Donato}\ \emph {et~al.}(2012)\citenamefont
  {Di~Donato}, \citenamefont {Ricciardi},\ and\ \citenamefont
  {Bigi}}]{DiDonato:2011kr}%
  \BibitemOpen
  \bibfield  {author} {\bibinfo {author} {\bibnamefont {Di~Donato},
  \bibfnamefont {Camilla}}, \bibinfo {author} {\bibfnamefont {Giulia}\
  \bibnamefont {Ricciardi}}, \ and\ \bibinfo {author} {\bibfnamefont {Ikaros}\
  \bibnamefont {Bigi}}} (\bibinfo {year} {2012}),\ \bibfield  {title} {\enquote
  {\bibinfo {title} {{$\eta - \eta'$ Mixing - From electromagnetic transitions
  to weak decays of charm and beauty hadrons}},}\ }\href {\doibase
  10.1103/PhysRevD.85.013016} {\bibfield  {journal} {\bibinfo  {journal} {Phys.
  Rev.}\ }\textbf {\bibinfo {volume} {D85}},\ \bibinfo {pages} {013016}},\
  \Eprint {http://arxiv.org/abs/1105.3557} {arXiv:1105.3557 [hep-ph]}
  \BibitemShut {NoStop}%
\bibitem [{\citenamefont {Di~Pierro}\ \emph {et~al.}(1999)\citenamefont
  {Di~Pierro}, \citenamefont {Sachrajda},\ and\ \citenamefont
  {Michael}}]{DiPierro:1999tb}%
  \BibitemOpen
  \bibfield  {author} {\bibinfo {author} {\bibnamefont {Di~Pierro},
  \bibfnamefont {Massimo}}, \bibinfo {author} {\bibfnamefont {Christopher~T}\
  \bibnamefont {Sachrajda}}, \ and\ \bibinfo {author} {\bibfnamefont
  {Christopher}\ \bibnamefont {Michael}} (\bibinfo {collaboration} {UKQCD})}
  (\bibinfo {year} {1999}),\ \bibfield  {title} {\enquote {\bibinfo {title}
  {{An Exploratory lattice study of spectator effects in inclusive decays of
  the $\Lambda_b$-baryon}},}\ }\href {\doibase 10.1016/S0370-2693(99)01166-1}
  {\bibfield  {journal} {\bibinfo  {journal} {Phys. Lett.}\ }\textbf {\bibinfo
  {volume} {B468}},\ \bibinfo {pages} {143}},\ \Eprint
  {http://arxiv.org/abs/hep-lat/9906031} {arXiv:hep-lat/9906031 [hep-lat]}
  \BibitemShut {NoStop}%
\bibitem [{\citenamefont {Dighe}\ \emph {et~al.}(2002)\citenamefont {Dighe},
  \citenamefont {Hurth}, \citenamefont {Kim},\ and\ \citenamefont
  {Yoshikawa}}]{Dighe:2001gc}%
  \BibitemOpen
  \bibfield  {author} {\bibinfo {author} {\bibnamefont {Dighe}, \bibfnamefont
  {A~S}}, \bibinfo {author} {\bibfnamefont {T.}~\bibnamefont {Hurth}}, \bibinfo
  {author} {\bibfnamefont {C.~S.}\ \bibnamefont {Kim}}, \ and\ \bibinfo
  {author} {\bibfnamefont {T.}~\bibnamefont {Yoshikawa}}} (\bibinfo {year}
  {2002}),\ \bibfield  {title} {\enquote {\bibinfo {title} {{Measurement of the
  lifetime difference of $B_d$ mesons: Possible and worthwhile?}}}\ }\href
  {\doibase 10.1016/S0550-3213(01)00655-1} {\bibfield  {journal} {\bibinfo
  {journal} {Nucl. Phys.}\ }\textbf {\bibinfo {volume} {B624}},\ \bibinfo
  {pages} {377--404}},\ \Eprint {http://arxiv.org/abs/hep-ph/0109088}
  {arXiv:hep-ph/0109088 [hep-ph]} \BibitemShut {NoStop}%
\bibitem [{\citenamefont {Dighe}\ \emph {et~al.}(1999)\citenamefont {Dighe},
  \citenamefont {Dunietz},\ and\ \citenamefont {Fleischer}}]{Dighe:1998vk}%
  \BibitemOpen
  \bibfield  {author} {\bibinfo {author} {\bibnamefont {Dighe}, \bibfnamefont
  {Amol~S}}, \bibinfo {author} {\bibfnamefont {Isard}\ \bibnamefont {Dunietz}},
  \ and\ \bibinfo {author} {\bibfnamefont {Robert}\ \bibnamefont {Fleischer}}}
  (\bibinfo {year} {1999}),\ \bibfield  {title} {\enquote {\bibinfo {title}
  {{Extracting CKM phases and $B_s - \bar{B}_s$ mixing parameters from angular
  distributions of nonleptonic $B$ decays}},}\ }\href {\doibase
  10.1007/s100520050372} {\bibfield  {journal} {\bibinfo  {journal} {Eur. Phys.
  J.}\ }\textbf {\bibinfo {volume} {C6}},\ \bibinfo {pages} {647--662}},\
  \Eprint {http://arxiv.org/abs/hep-ph/9804253} {arXiv:hep-ph/9804253 [hep-ph]}
  \BibitemShut {NoStop}%
\bibitem [{\citenamefont {Dighe}\ \emph {et~al.}(1996)\citenamefont {Dighe},
  \citenamefont {Dunietz}, \citenamefont {Lipkin},\ and\ \citenamefont
  {Rosner}}]{Dighe:1995pd}%
  \BibitemOpen
  \bibfield  {author} {\bibinfo {author} {\bibnamefont {Dighe}, \bibfnamefont
  {Amol~S}}, \bibinfo {author} {\bibfnamefont {Isard}\ \bibnamefont {Dunietz}},
  \bibinfo {author} {\bibfnamefont {Harry~J.}\ \bibnamefont {Lipkin}}, \ and\
  \bibinfo {author} {\bibfnamefont {Jonathan~L.}\ \bibnamefont {Rosner}}}
  (\bibinfo {year} {1996}),\ \bibfield  {title} {\enquote {\bibinfo {title}
  {{Angular distributions and lifetime differences in $B_s \to J/\psi \phi$
  decays}},}\ }\href {\doibase 10.1016/0370-2693(95)01523-X} {\bibfield
  {journal} {\bibinfo  {journal} {Phys. Lett.}\ }\textbf {\bibinfo {volume}
  {B369}},\ \bibinfo {pages} {144--150}},\ \Eprint
  {http://arxiv.org/abs/hep-ph/9511363} {arXiv:hep-ph/9511363 [hep-ph]}
  \BibitemShut {NoStop}%
\bibitem [{\citenamefont {Djouadi}\ and\ \citenamefont
  {Lenz}(2012)}]{Djouadi:2012ae}%
  \BibitemOpen
  \bibfield  {author} {\bibinfo {author} {\bibnamefont {Djouadi}, \bibfnamefont
  {Abdelhak}}, \ and\ \bibinfo {author} {\bibfnamefont {Alexander}\
  \bibnamefont {Lenz}}} (\bibinfo {year} {2012}),\ \bibfield  {title} {\enquote
  {\bibinfo {title} {{Sealing the fate of a fourth generation of fermions}},}\
  }\href {\doibase 10.1016/j.physletb.2012.07.060} {\bibfield  {journal}
  {\bibinfo  {journal} {Phys. Lett.}\ }\textbf {\bibinfo {volume} {B715}},\
  \bibinfo {pages} {310--314}},\ \Eprint {http://arxiv.org/abs/1204.1252}
  {arXiv:1204.1252 [hep-ph]} \BibitemShut {NoStop}%
\bibitem [{\citenamefont {Dowdall}\ \emph {et~al.}(2014)\citenamefont
  {Dowdall}, \citenamefont {Davies}, \citenamefont {Horgan}, \citenamefont
  {Lepage}, \citenamefont {Monahan},\ and\ \citenamefont
  {Shigemitsu}}]{Dowdall:2014qka}%
  \BibitemOpen
  \bibfield  {author} {\bibinfo {author} {\bibnamefont {Dowdall}, \bibfnamefont
  {R~J}}, \bibinfo {author} {\bibfnamefont {C.~T.~H.}\ \bibnamefont {Davies}},
  \bibinfo {author} {\bibfnamefont {R.~R.}\ \bibnamefont {Horgan}}, \bibinfo
  {author} {\bibfnamefont {G.~Peter}\ \bibnamefont {Lepage}}, \bibinfo {author}
  {\bibfnamefont {C.~J.}\ \bibnamefont {Monahan}}, \ and\ \bibinfo {author}
  {\bibfnamefont {J.}~\bibnamefont {Shigemitsu}}} (\bibinfo {year} {2014}),\
  \bibfield  {title} {\enquote {\bibinfo {title} {{B-meson mixing from full
  lattice QCD with physical u, d, s and c quarks}},}\ }\href@noop {} {\
  }\Eprint {http://arxiv.org/abs/1411.6989} {arXiv:1411.6989 [hep-lat]}
  \BibitemShut {NoStop}%
\bibitem [{\citenamefont {Dunietz}(1995)}]{Dunietz:1995cp}%
  \BibitemOpen
  \bibfield  {author} {\bibinfo {author} {\bibnamefont {Dunietz}, \bibfnamefont
  {Isard}}} (\bibinfo {year} {1995}),\ \bibfield  {title} {\enquote {\bibinfo
  {title} {{$\Bs - \barBs$ mixing, CP violation and extraction of CKM phases
  from untagged $\Bs$ data samples}},}\ }\href {\doibase
  10.1103/PhysRevD.52.3048} {\bibfield  {journal} {\bibinfo  {journal} {Phys.
  Rev.}\ }\textbf {\bibinfo {volume} {D52}},\ \bibinfo {pages} {3048--3064}},\
  \Eprint {http://arxiv.org/abs/hep-ph/9501287} {arXiv:hep-ph/9501287 [hep-ph]}
  \BibitemShut {NoStop}%
\bibitem [{\citenamefont {Dunietz}\ \emph {et~al.}(2001)\citenamefont
  {Dunietz}, \citenamefont {Fleischer},\ and\ \citenamefont
  {Nierste}}]{Dunietz:2000cr}%
  \BibitemOpen
  \bibfield  {author} {\bibinfo {author} {\bibnamefont {Dunietz}, \bibfnamefont
  {Isard}}, \bibinfo {author} {\bibfnamefont {Robert}\ \bibnamefont
  {Fleischer}}, \ and\ \bibinfo {author} {\bibfnamefont {Ulrich}\ \bibnamefont
  {Nierste}}} (\bibinfo {year} {2001}),\ \bibfield  {title} {\enquote {\bibinfo
  {title} {{In pursuit of new physics with $B_s$ decays}},}\ }\href {\doibase
  10.1103/PhysRevD.63.114015} {\bibfield  {journal} {\bibinfo  {journal} {Phys.
  Rev.}\ }\textbf {\bibinfo {volume} {D63}},\ \bibinfo {pages} {114015}},\
  \Eprint {http://arxiv.org/abs/hep-ph/0012219} {arXiv:hep-ph/0012219 [hep-ph]}
  \BibitemShut {NoStop}%
\bibitem [{\citenamefont {Dunietz}\ and\ \citenamefont
  {Sachs}(1988)}]{Dunietz:1987bv}%
  \BibitemOpen
  \bibfield  {author} {\bibinfo {author} {\bibnamefont {Dunietz}, \bibfnamefont
  {Isard}}, \ and\ \bibinfo {author} {\bibfnamefont {Robert~G.}\ \bibnamefont
  {Sachs}}} (\bibinfo {year} {1988}),\ \bibfield  {title} {\enquote {\bibinfo
  {title} {{Asymmetry Between Inclusive Charmed and Anticharmed Modes in B0,
  Anti-b0 Decay as a Measure of {CP} Violation}},}\ }\href {\doibase
  10.1103/PhysRevD.37.3186, 10.1103/PhysRevD.39.3515} {\bibfield  {journal}
  {\bibinfo  {journal} {Phys. Rev.}\ }\textbf {\bibinfo {volume} {D37}},\
  \bibinfo {pages} {3186}},\ \bibinfo {note} {[Erratum: Phys.
  Rev.D39,3515(1989)]}\BibitemShut {NoStop}%
\bibitem [{\citenamefont {Dutta}\ \emph {et~al.}(2012)\citenamefont {Dutta},
  \citenamefont {Khalil}, \citenamefont {Mimura},\ and\ \citenamefont
  {Shafi}}]{Dutta:2011kg}%
  \BibitemOpen
  \bibfield  {author} {\bibinfo {author} {\bibnamefont {Dutta}, \bibfnamefont
  {Bhaskar}}, \bibinfo {author} {\bibfnamefont {Shaaban}\ \bibnamefont
  {Khalil}}, \bibinfo {author} {\bibfnamefont {Yukihiro}\ \bibnamefont
  {Mimura}}, \ and\ \bibinfo {author} {\bibfnamefont {Qaisar}\ \bibnamefont
  {Shafi}}} (\bibinfo {year} {2012}),\ \bibfield  {title} {\enquote {\bibinfo
  {title} {{Dimuon CP Asymmetry in B Decays and Wjj Excess in Two Higgs Doublet
  Models}},}\ }\href {\doibase 10.1007/JHEP05(2012)131} {\bibfield  {journal}
  {\bibinfo  {journal} {JHEP}\ }\textbf {\bibinfo {volume} {05}},\ \bibinfo
  {pages} {131}},\ \Eprint {http://arxiv.org/abs/1104.5209} {arXiv:1104.5209
  [hep-ph]} \BibitemShut {NoStop}%
\bibitem [{\citenamefont {Eberhardt}\ \emph
  {et~al.}(2012{\natexlab{a}})\citenamefont {Eberhardt}, \citenamefont
  {Herbert}, \citenamefont {Lacker}, \citenamefont {Lenz}, \citenamefont
  {Menzel}, \citenamefont {Nierste},\ and\ \citenamefont
  {Wiebusch}}]{Eberhardt:2012gv}%
  \BibitemOpen
  \bibfield  {author} {\bibinfo {author} {\bibnamefont {Eberhardt},
  \bibfnamefont {Otto}}, \bibinfo {author} {\bibfnamefont {Geoffrey}\
  \bibnamefont {Herbert}}, \bibinfo {author} {\bibfnamefont {Heiko}\
  \bibnamefont {Lacker}}, \bibinfo {author} {\bibfnamefont {Alexander}\
  \bibnamefont {Lenz}}, \bibinfo {author} {\bibfnamefont {Andreas}\
  \bibnamefont {Menzel}}, \bibinfo {author} {\bibfnamefont {Ulrich}\
  \bibnamefont {Nierste}}, \ and\ \bibinfo {author} {\bibfnamefont {Martin}\
  \bibnamefont {Wiebusch}}} (\bibinfo {year} {2012}{\natexlab{a}}),\ \bibfield
  {title} {\enquote {\bibinfo {title} {{Impact of a Higgs boson at a mass of
  126 GeV on the standard model with three and four fermion generations}},}\
  }\href {\doibase 10.1103/PhysRevLett.109.241802} {\bibfield  {journal}
  {\bibinfo  {journal} {Phys. Rev. Lett.}\ }\textbf {\bibinfo {volume} {109}},\
  \bibinfo {pages} {241802}},\ \Eprint {http://arxiv.org/abs/1209.1101}
  {arXiv:1209.1101 [hep-ph]} \BibitemShut {NoStop}%
\bibitem [{\citenamefont {Eberhardt}\ \emph
  {et~al.}(2012{\natexlab{b}})\citenamefont {Eberhardt}, \citenamefont
  {Herbert}, \citenamefont {Lacker}, \citenamefont {Lenz}, \citenamefont
  {Menzel}, \citenamefont {Nierste},\ and\ \citenamefont
  {Wiebusch}}]{Eberhardt:2012sb}%
  \BibitemOpen
  \bibfield  {author} {\bibinfo {author} {\bibnamefont {Eberhardt},
  \bibfnamefont {Otto}}, \bibinfo {author} {\bibfnamefont {Geoffrey}\
  \bibnamefont {Herbert}}, \bibinfo {author} {\bibfnamefont {Heiko}\
  \bibnamefont {Lacker}}, \bibinfo {author} {\bibfnamefont {Alexander}\
  \bibnamefont {Lenz}}, \bibinfo {author} {\bibfnamefont {Andreas}\
  \bibnamefont {Menzel}}, \bibinfo {author} {\bibfnamefont {Ulrich}\
  \bibnamefont {Nierste}}, \ and\ \bibinfo {author} {\bibfnamefont {Martin}\
  \bibnamefont {Wiebusch}}} (\bibinfo {year} {2012}{\natexlab{b}}),\ \bibfield
  {title} {\enquote {\bibinfo {title} {{Joint analysis of Higgs decays and
  electroweak precision observables in the Standard Model with a sequential
  fourth generation}},}\ }\href {\doibase 10.1103/PhysRevD.86.013011}
  {\bibfield  {journal} {\bibinfo  {journal} {Phys. Rev.}\ }\textbf {\bibinfo
  {volume} {D86}},\ \bibinfo {pages} {013011}},\ \Eprint
  {http://arxiv.org/abs/1204.3872} {arXiv:1204.3872 [hep-ph]} \BibitemShut
  {NoStop}%
\bibitem [{\citenamefont {Eberhardt}\ \emph
  {et~al.}(2012{\natexlab{c}})\citenamefont {Eberhardt}, \citenamefont {Lenz},
  \citenamefont {Menzel}, \citenamefont {Nierste},\ and\ \citenamefont
  {Wiebusch}}]{Eberhardt:2012ck}%
  \BibitemOpen
  \bibfield  {author} {\bibinfo {author} {\bibnamefont {Eberhardt},
  \bibfnamefont {Otto}}, \bibinfo {author} {\bibfnamefont {Alexander}\
  \bibnamefont {Lenz}}, \bibinfo {author} {\bibfnamefont {Andreas}\
  \bibnamefont {Menzel}}, \bibinfo {author} {\bibfnamefont {Ulrich}\
  \bibnamefont {Nierste}}, \ and\ \bibinfo {author} {\bibfnamefont {Martin}\
  \bibnamefont {Wiebusch}}} (\bibinfo {year} {2012}{\natexlab{c}}),\ \bibfield
  {title} {\enquote {\bibinfo {title} {{Status of the fourth fermion generation
  before ICHEP2012: Higgs data and electroweak precision observables}},}\
  }\href {\doibase 10.1103/PhysRevD.86.074014} {\bibfield  {journal} {\bibinfo
  {journal} {Phys. Rev.}\ }\textbf {\bibinfo {volume} {D86}},\ \bibinfo {pages}
  {074014}},\ \Eprint {http://arxiv.org/abs/1207.0438} {arXiv:1207.0438
  [hep-ph]} \BibitemShut {NoStop}%
\bibitem [{\citenamefont {Ellis}\ \emph {et~al.}(1977)\citenamefont {Ellis},
  \citenamefont {Gaillard}, \citenamefont {Nanopoulos},\ and\ \citenamefont
  {Rudaz}}]{Ellis:1977uk}%
  \BibitemOpen
  \bibfield  {author} {\bibinfo {author} {\bibnamefont {Ellis}, \bibfnamefont
  {John~R}}, \bibinfo {author} {\bibfnamefont {M.~K.}\ \bibnamefont
  {Gaillard}}, \bibinfo {author} {\bibfnamefont {Dimitri~V.}\ \bibnamefont
  {Nanopoulos}}, \ and\ \bibinfo {author} {\bibfnamefont {S.}~\bibnamefont
  {Rudaz}}} (\bibinfo {year} {1977}),\ \bibfield  {title} {\enquote {\bibinfo
  {title} {{The Phenomenology of the Next Left-Handed Quarks}},}\ }\href
  {\doibase 10.1016/0550-3213(77)90374-1} {\bibfield  {journal} {\bibinfo
  {journal} {Nucl. Phys.}\ }\textbf {\bibinfo {volume} {B131}},\ \bibinfo
  {pages} {285}},\ \bibinfo {note} {[Erratum: Nucl.
  Phys.B132,541(1978)]}\BibitemShut {NoStop}%
\bibitem [{\citenamefont {Endo}\ \emph {et~al.}(2011)\citenamefont {Endo},
  \citenamefont {Shirai},\ and\ \citenamefont {Yanagida}}]{Endo:2010fk}%
  \BibitemOpen
  \bibfield  {author} {\bibinfo {author} {\bibnamefont {Endo}, \bibfnamefont
  {Motoi}}, \bibinfo {author} {\bibfnamefont {Satoshi}\ \bibnamefont {Shirai}},
  \ and\ \bibinfo {author} {\bibfnamefont {Tsutomu~T.}\ \bibnamefont
  {Yanagida}}} (\bibinfo {year} {2011}),\ \bibfield  {title} {\enquote
  {\bibinfo {title} {{Split Generation in the SUSY Mass Spectrum and $B_s -
  \bar{B}_s$ Mixing}},}\ }\href {\doibase 10.1143/PTP.125.921} {\bibfield
  {journal} {\bibinfo  {journal} {Prog. Theor. Phys.}\ }\textbf {\bibinfo
  {volume} {125}},\ \bibinfo {pages} {921--932}},\ \Eprint
  {http://arxiv.org/abs/1009.3366} {arXiv:1009.3366 [hep-ph]} \BibitemShut
  {NoStop}%
\bibitem [{\citenamefont {Endo}\ and\ \citenamefont
  {Yokozaki}(2011)}]{Endo:2010yt}%
  \BibitemOpen
  \bibfield  {author} {\bibinfo {author} {\bibnamefont {Endo}, \bibfnamefont
  {Motoi}}, \ and\ \bibinfo {author} {\bibfnamefont {Norimi}\ \bibnamefont
  {Yokozaki}}} (\bibinfo {year} {2011}),\ \bibfield  {title} {\enquote
  {\bibinfo {title} {{Large CP Violation in $B_s$ Meson Mixing with EDM
  constraint in Supersymmetry}},}\ }\href {\doibase 10.1007/JHEP03(2011)130}
  {\bibfield  {journal} {\bibinfo  {journal} {JHEP}\ }\textbf {\bibinfo
  {volume} {03}},\ \bibinfo {pages} {130}},\ \Eprint
  {http://arxiv.org/abs/1012.5501} {arXiv:1012.5501 [hep-ph]} \BibitemShut
  {NoStop}%
\bibitem [{\citenamefont {Faller}\ \emph
  {et~al.}(2009{\natexlab{a}})\citenamefont {Faller}, \citenamefont
  {Fleischer},\ and\ \citenamefont {Mannel}}]{Faller:2008gt}%
  \BibitemOpen
  \bibfield  {author} {\bibinfo {author} {\bibnamefont {Faller}, \bibfnamefont
  {Sven}}, \bibinfo {author} {\bibfnamefont {Robert}\ \bibnamefont
  {Fleischer}}, \ and\ \bibinfo {author} {\bibfnamefont {Thomas}\ \bibnamefont
  {Mannel}}} (\bibinfo {year} {2009}{\natexlab{a}}),\ \bibfield  {title}
  {\enquote {\bibinfo {title} {{Precision Physics with $B^0_s \to J/\psi \phi$
  at the LHC: The Quest for New Physics}},}\ }\href {\doibase
  10.1103/PhysRevD.79.014005} {\bibfield  {journal} {\bibinfo  {journal} {Phys.
  Rev.}\ }\textbf {\bibinfo {volume} {D79}},\ \bibinfo {pages} {014005}},\
  \Eprint {http://arxiv.org/abs/0810.4248} {arXiv:0810.4248 [hep-ph]}
  \BibitemShut {NoStop}%
\bibitem [{\citenamefont {Faller}\ \emph
  {et~al.}(2009{\natexlab{b}})\citenamefont {Faller}, \citenamefont {Jung},
  \citenamefont {Fleischer},\ and\ \citenamefont {Mannel}}]{Faller:2008zc}%
  \BibitemOpen
  \bibfield  {author} {\bibinfo {author} {\bibnamefont {Faller}, \bibfnamefont
  {Sven}}, \bibinfo {author} {\bibfnamefont {Martin}\ \bibnamefont {Jung}},
  \bibinfo {author} {\bibfnamefont {Robert}\ \bibnamefont {Fleischer}}, \ and\
  \bibinfo {author} {\bibfnamefont {Thomas}\ \bibnamefont {Mannel}}} (\bibinfo
  {year} {2009}{\natexlab{b}}),\ \bibfield  {title} {\enquote {\bibinfo {title}
  {{The Golden Modes $\Bd \to \jpsi K_{(S,L)}$ in the Era of Precision Flavour
  Physics}},}\ }\href {\doibase 10.1103/PhysRevD.79.014030} {\bibfield
  {journal} {\bibinfo  {journal} {Phys. Rev.}\ }\textbf {\bibinfo {volume}
  {D79}},\ \bibinfo {pages} {014030}},\ \Eprint
  {http://arxiv.org/abs/0809.0842} {arXiv:0809.0842 [hep-ph]} \BibitemShut
  {NoStop}%
\bibitem [{\citenamefont {Fleischer}(1999{\natexlab{a}})}]{Fleischer:1999zi}%
  \BibitemOpen
  \bibfield  {author} {\bibinfo {author} {\bibnamefont {Fleischer},
  \bibfnamefont {Robert}}} (\bibinfo {year} {1999}{\natexlab{a}}),\ \bibfield
  {title} {\enquote {\bibinfo {title} {{Extracting CKM phases from angular
  distributions of B(d,s) decays into admixtures of CP eigenstates}},}\ }\href
  {\doibase 10.1103/PhysRevD.60.073008} {\bibfield  {journal} {\bibinfo
  {journal} {Phys. Rev.}\ }\textbf {\bibinfo {volume} {D60}},\ \bibinfo {pages}
  {073008}},\ \Eprint {http://arxiv.org/abs/hep-ph/9903540}
  {arXiv:hep-ph/9903540 [hep-ph]} \BibitemShut {NoStop}%
\bibitem [{\citenamefont {Fleischer}(1999{\natexlab{b}})}]{Fleischer:1999nz}%
  \BibitemOpen
  \bibfield  {author} {\bibinfo {author} {\bibnamefont {Fleischer},
  \bibfnamefont {Robert}}} (\bibinfo {year} {1999}{\natexlab{b}}),\ \bibfield
  {title} {\enquote {\bibinfo {title} {{Extracting $\gamma$ from $B(s/d) \to
  J/\psi K_{S}$ and $B(d/s) \to D^+(d/s) D^-(d/s)$}},}\ }\href {\doibase
  10.1007/s100529900099} {\bibfield  {journal} {\bibinfo  {journal} {Eur. Phys.
  J.}\ }\textbf {\bibinfo {volume} {C10}},\ \bibinfo {pages} {299--306}},\
  \Eprint {http://arxiv.org/abs/hep-ph/9903455} {arXiv:hep-ph/9903455 [hep-ph]}
  \BibitemShut {NoStop}%
\bibitem [{\citenamefont {Fleischer}(1999{\natexlab{c}})}]{Fleischer:1999pa}%
  \BibitemOpen
  \bibfield  {author} {\bibinfo {author} {\bibnamefont {Fleischer},
  \bibfnamefont {Robert}}} (\bibinfo {year} {1999}{\natexlab{c}}),\ \bibfield
  {title} {\enquote {\bibinfo {title} {{New strategies to extract $\eta$ and
  $\gamma$ from $\Bd \to \pi^+ \pi^-$ and $\Bs \to K^+ K^-$}},}\ }\href
  {\doibase 10.1016/S0370-2693(99)00640-1} {\bibfield  {journal} {\bibinfo
  {journal} {Phys. Lett.}\ }\textbf {\bibinfo {volume} {B459}},\ \bibinfo
  {pages} {306--320}},\ \Eprint {http://arxiv.org/abs/hep-ph/9903456}
  {arXiv:hep-ph/9903456 [hep-ph]} \BibitemShut {NoStop}%
\bibitem [{\citenamefont {Fleischer}(2003)}]{Fleischer:2003yb}%
  \BibitemOpen
  \bibfield  {author} {\bibinfo {author} {\bibnamefont {Fleischer},
  \bibfnamefont {Robert}}} (\bibinfo {year} {2003}),\ \bibfield  {title}
  {\enquote {\bibinfo {title} {{New strategies to obtain insights into CP
  violation through $\Bs \to D_s^\pm K^\mp$, $ D_s^{*\pm} K^\mp$, ... and $\Bd
  \to D^\pm \pi^\mp$, $ D^{*\pm} \pi^\mp$, ... decays}},}\ }\href {\doibase
  10.1016/j.nuclphysb.2003.08.010} {\bibfield  {journal} {\bibinfo  {journal}
  {Nucl. Phys.}\ }\textbf {\bibinfo {volume} {B671}},\ \bibinfo {pages}
  {459--482}},\ \Eprint {http://arxiv.org/abs/hep-ph/0304027}
  {arXiv:hep-ph/0304027 [hep-ph]} \BibitemShut {NoStop}%
\bibitem [{\citenamefont {Fleischer}(2007{\natexlab{a}})}]{Fleischer:2007hj}%
  \BibitemOpen
  \bibfield  {author} {\bibinfo {author} {\bibnamefont {Fleischer},
  \bibfnamefont {Robert}}} (\bibinfo {year} {2007}{\natexlab{a}}),\ \bibfield
  {title} {\enquote {\bibinfo {title} {{$B_{s,d} \to \pi \pi, \pi K, KK$:
  Status and Prospects}},}\ }\href {\doibase 10.1140/epjc/s10052-007-0391-7}
  {\bibfield  {journal} {\bibinfo  {journal} {Eur. Phys. J.}\ }\textbf
  {\bibinfo {volume} {C52}},\ \bibinfo {pages} {267--281}},\ \Eprint
  {http://arxiv.org/abs/0705.1121} {arXiv:0705.1121 [hep-ph]} \BibitemShut
  {NoStop}%
\bibitem [{\citenamefont {Fleischer}(2007{\natexlab{b}})}]{Fleischer:2007zn}%
  \BibitemOpen
  \bibfield  {author} {\bibinfo {author} {\bibnamefont {Fleischer},
  \bibfnamefont {Robert}}} (\bibinfo {year} {2007}{\natexlab{b}}),\ \bibfield
  {title} {\enquote {\bibinfo {title} {{Exploring CP violation and penguin
  effects through $B^0_{d} \to D^{+} D^{-}$ and $B^0_{s} \to D^+_{s}
  D^-_{s}$}},}\ }\href {\doibase 10.1140/epjc/s10052-007-0341-4} {\bibfield
  {journal} {\bibinfo  {journal} {Eur. Phys. J.}\ }\textbf {\bibinfo {volume}
  {C51}},\ \bibinfo {pages} {849--858}},\ \Eprint
  {http://arxiv.org/abs/0705.4421} {arXiv:0705.4421 [hep-ph]} \BibitemShut
  {NoStop}%
\bibitem [{\citenamefont {Fleischer}(2015)}]{Fleischer:2015mla}%
  \BibitemOpen
  \bibfield  {author} {\bibinfo {author} {\bibnamefont {Fleischer},
  \bibfnamefont {Robert}}} (\bibinfo {year} {2015}),\ \bibfield  {title}
  {\enquote {\bibinfo {title} {{Theoretical Prospects for B Physics}},}\ }in\
  \href {https://inspirehep.net/record/1391506/files/arXiv:1509.00601.pdf}
  {\emph {\bibinfo {booktitle} {{13th Conference on Flavor Physics and CP
  Violation (FPCP 2015) Nagoya, Japan, May 25-29, 2015}}}},\ \Eprint
  {http://arxiv.org/abs/1509.00601} {arXiv:1509.00601 [hep-ph]} \BibitemShut
  {NoStop}%
\bibitem [{\citenamefont {Fleischer}\ and\ \citenamefont
  {Knegjens}(2011{\natexlab{a}})}]{Fleischer:2011cw}%
  \BibitemOpen
  \bibfield  {author} {\bibinfo {author} {\bibnamefont {Fleischer},
  \bibfnamefont {Robert}}, \ and\ \bibinfo {author} {\bibfnamefont {Robert}\
  \bibnamefont {Knegjens}}} (\bibinfo {year} {2011}{\natexlab{a}}),\ \bibfield
  {title} {\enquote {\bibinfo {title} {{Effective Lifetimes of $B_s$ Decays and
  their Constraints on the $B_s^0$-$\bar B_s^0$ Mixing Parameters}},}\ }\href
  {\doibase 10.1140/epjc/s10052-011-1789-9} {\bibfield  {journal} {\bibinfo
  {journal} {Eur. Phys. J.}\ }\textbf {\bibinfo {volume} {C71}},\ \bibinfo
  {pages} {1789}},\ \Eprint {http://arxiv.org/abs/1109.5115} {arXiv:1109.5115
  [hep-ph]} \BibitemShut {NoStop}%
\bibitem [{\citenamefont {Fleischer}\ and\ \citenamefont
  {Knegjens}(2011{\natexlab{b}})}]{Fleischer:2010ib}%
  \BibitemOpen
  \bibfield  {author} {\bibinfo {author} {\bibnamefont {Fleischer},
  \bibfnamefont {Robert}}, \ and\ \bibinfo {author} {\bibfnamefont {Robert}\
  \bibnamefont {Knegjens}}} (\bibinfo {year} {2011}{\natexlab{b}}),\ \bibfield
  {title} {\enquote {\bibinfo {title} {{In Pursuit of New Physics With $B^0_s
  \to K^+K^-$}},}\ }\href {\doibase 10.1140/epjc/s10052-010-1532-y} {\bibfield
  {journal} {\bibinfo  {journal} {Eur. Phys. J.}\ }\textbf {\bibinfo {volume}
  {C71}},\ \bibinfo {pages} {1532}},\ \Eprint {http://arxiv.org/abs/1011.1096}
  {arXiv:1011.1096 [hep-ph]} \BibitemShut {NoStop}%
\bibitem [{\citenamefont {Fleischer}\ \emph
  {et~al.}(2011{\natexlab{a}})\citenamefont {Fleischer}, \citenamefont
  {Knegjens},\ and\ \citenamefont {Ricciardi}}]{Fleischer:2011au}%
  \BibitemOpen
  \bibfield  {author} {\bibinfo {author} {\bibnamefont {Fleischer},
  \bibfnamefont {Robert}}, \bibinfo {author} {\bibfnamefont {Robert}\
  \bibnamefont {Knegjens}}, \ and\ \bibinfo {author} {\bibfnamefont {Giulia}\
  \bibnamefont {Ricciardi}}} (\bibinfo {year} {2011}{\natexlab{a}}),\ \bibfield
   {title} {\enquote {\bibinfo {title} {{Anatomy of $B^0_{s,d} \to J/\psi
  f_0(980)$}},}\ }\href {\doibase 10.1140/epjc/s10052-011-1832-x} {\bibfield
  {journal} {\bibinfo  {journal} {Eur. Phys. J.}\ }\textbf {\bibinfo {volume}
  {C71}},\ \bibinfo {pages} {1832}},\ \Eprint {http://arxiv.org/abs/1109.1112}
  {arXiv:1109.1112 [hep-ph]} \BibitemShut {NoStop}%
\bibitem [{\citenamefont {Fleischer}\ \emph
  {et~al.}(2011{\natexlab{b}})\citenamefont {Fleischer}, \citenamefont
  {Knegjens},\ and\ \citenamefont {Ricciardi}}]{Fleischer:2011ib}%
  \BibitemOpen
  \bibfield  {author} {\bibinfo {author} {\bibnamefont {Fleischer},
  \bibfnamefont {Robert}}, \bibinfo {author} {\bibfnamefont {Robert}\
  \bibnamefont {Knegjens}}, \ and\ \bibinfo {author} {\bibfnamefont {Giulia}\
  \bibnamefont {Ricciardi}}} (\bibinfo {year} {2011}{\natexlab{b}}),\ \bibfield
   {title} {\enquote {\bibinfo {title} {{Exploring CP Violation and
  $\eta$-$\eta'$ Mixing with the $B^0_{s,d} \to J/\psi \eta^{(\prime)}$
  Systems}},}\ }\href {\doibase 10.1140/epjc/s10052-011-1798-8} {\bibfield
  {journal} {\bibinfo  {journal} {Eur. Phys. J.}\ }\textbf {\bibinfo {volume}
  {C71}},\ \bibinfo {pages} {1798}},\ \Eprint {http://arxiv.org/abs/1110.5490}
  {arXiv:1110.5490 [hep-ph]} \BibitemShut {NoStop}%
\bibitem [{\citenamefont {Fleischer}\ and\ \citenamefont
  {Ricciardi}(2011)}]{Fleischer:2011ne}%
  \BibitemOpen
  \bibfield  {author} {\bibinfo {author} {\bibnamefont {Fleischer},
  \bibfnamefont {Robert}}, \ and\ \bibinfo {author} {\bibfnamefont {Stefania}\
  \bibnamefont {Ricciardi}}} (\bibinfo {year} {2011}),\ \bibfield  {title}
  {\enquote {\bibinfo {title} {{Extraction of the weak angle $\gamma$ from B to
  charm decays}},}\ }in\ \href
  {https://inspirehep.net/record/896821/files/arXiv:1104.4029.pdf} {\emph
  {\bibinfo {booktitle} {{CKM unitarity triangle. Proceedings, 6th
  International Workshop, CKM 2010, Warwick, UK, September 6-10, 2010}}}},\
  \Eprint {http://arxiv.org/abs/1104.4029} {arXiv:1104.4029 [hep-ph]}
  \BibitemShut {NoStop}%
\bibitem [{\citenamefont {Fox}\ \emph {et~al.}(2008)\citenamefont {Fox},
  \citenamefont {Ligeti}, \citenamefont {Papucci}, \citenamefont {Perez},\ and\
  \citenamefont {Schwartz}}]{Fox:2007in}%
  \BibitemOpen
  \bibfield  {author} {\bibinfo {author} {\bibnamefont {Fox}, \bibfnamefont
  {Patrick~J}}, \bibinfo {author} {\bibfnamefont {Zoltan}\ \bibnamefont
  {Ligeti}}, \bibinfo {author} {\bibfnamefont {Michele}\ \bibnamefont
  {Papucci}}, \bibinfo {author} {\bibfnamefont {Gilad}\ \bibnamefont {Perez}},
  \ and\ \bibinfo {author} {\bibfnamefont {Matthew~D.}\ \bibnamefont
  {Schwartz}}} (\bibinfo {year} {2008}),\ \bibfield  {title} {\enquote
  {\bibinfo {title} {{Deciphering top flavor violation at the LHC with $B$
  factories}},}\ }\href {\doibase 10.1103/PhysRevD.78.054008} {\bibfield
  {journal} {\bibinfo  {journal} {Phys. Rev.}\ }\textbf {\bibinfo {volume}
  {D78}},\ \bibinfo {pages} {054008}},\ \Eprint
  {http://arxiv.org/abs/0704.1482} {arXiv:0704.1482 [hep-ph]} \BibitemShut
  {NoStop}%
\bibitem [{\citenamefont {Fox}\ \emph {et~al.}(2011)\citenamefont {Fox},
  \citenamefont {Liu}, \citenamefont {Tucker-Smith},\ and\ \citenamefont
  {Weiner}}]{Fox:2011qd}%
  \BibitemOpen
  \bibfield  {author} {\bibinfo {author} {\bibnamefont {Fox}, \bibfnamefont
  {Patrick~J}}, \bibinfo {author} {\bibfnamefont {Jia}\ \bibnamefont {Liu}},
  \bibinfo {author} {\bibfnamefont {David}\ \bibnamefont {Tucker-Smith}}, \
  and\ \bibinfo {author} {\bibfnamefont {Neal}\ \bibnamefont {Weiner}}}
  (\bibinfo {year} {2011}),\ \bibfield  {title} {\enquote {\bibinfo {title}
  {{An Effective Z'}},}\ }\href {\doibase 10.1103/PhysRevD.84.115006}
  {\bibfield  {journal} {\bibinfo  {journal} {Phys. Rev.}\ }\textbf {\bibinfo
  {volume} {D84}},\ \bibinfo {pages} {115006}},\ \Eprint
  {http://arxiv.org/abs/1104.4127} {arXiv:1104.4127 [hep-ph]} \BibitemShut
  {NoStop}%
\bibitem [{\citenamefont {Franco}\ \emph {et~al.}(1982)\citenamefont {Franco},
  \citenamefont {Lusignoli},\ and\ \citenamefont {Pugliese}}]{Franco:1981ea}%
  \BibitemOpen
  \bibfield  {author} {\bibinfo {author} {\bibnamefont {Franco}, \bibfnamefont
  {E}}, \bibinfo {author} {\bibfnamefont {Maurizio}\ \bibnamefont {Lusignoli}},
  \ and\ \bibinfo {author} {\bibfnamefont {A.}~\bibnamefont {Pugliese}}}
  (\bibinfo {year} {1982}),\ \bibfield  {title} {\enquote {\bibinfo {title}
  {{Strong Interaction Corrections to {CP} Violation in $B^0 -
  \bar{B}^0$-Mixing}},}\ }\href {\doibase 10.1016/0550-3213(82)90018-9}
  {\bibfield  {journal} {\bibinfo  {journal} {Nucl. Phys.}\ }\textbf {\bibinfo
  {volume} {B194}},\ \bibinfo {pages} {403}}\BibitemShut {NoStop}%
\bibitem [{\citenamefont {Frings}\ \emph {et~al.}(2015)\citenamefont {Frings},
  \citenamefont {Nierste},\ and\ \citenamefont {Wiebusch}}]{Frings:2015eva}%
  \BibitemOpen
  \bibfield  {author} {\bibinfo {author} {\bibnamefont {Frings}, \bibfnamefont
  {Philipp}}, \bibinfo {author} {\bibfnamefont {Ulrich}\ \bibnamefont
  {Nierste}}, \ and\ \bibinfo {author} {\bibfnamefont {Martin}\ \bibnamefont
  {Wiebusch}}} (\bibinfo {year} {2015}),\ \bibfield  {title} {\enquote
  {\bibinfo {title} {{Penguin contributions to CP phases in $B_{d,s}$ decays to
  charmonium}},}\ }\href {\doibase 10.1103/PhysRevLett.115.061802} {\bibfield
  {journal} {\bibinfo  {journal} {Phys. Rev. Lett.}\ }\textbf {\bibinfo
  {volume} {115}},\ \bibinfo {pages} {061802}},\ \Eprint
  {http://arxiv.org/abs/1503.00859} {arXiv:1503.00859 [hep-ph]} \BibitemShut
  {NoStop}%
\bibitem [{\citenamefont {Gamiz}\ \emph {et~al.}(2009)\citenamefont {Gamiz},
  \citenamefont {Davies}, \citenamefont {Lepage}, \citenamefont {Shigemitsu},\
  and\ \citenamefont {Wingate}}]{Gamiz:2009ku}%
  \BibitemOpen
  \bibfield  {author} {\bibinfo {author} {\bibnamefont {Gamiz}, \bibfnamefont
  {Elvira}}, \bibinfo {author} {\bibfnamefont {Christine T.~H.}\ \bibnamefont
  {Davies}}, \bibinfo {author} {\bibfnamefont {G.~Peter}\ \bibnamefont
  {Lepage}}, \bibinfo {author} {\bibfnamefont {Junko}\ \bibnamefont
  {Shigemitsu}}, \ and\ \bibinfo {author} {\bibfnamefont {Matthew}\
  \bibnamefont {Wingate}} (\bibinfo {collaboration} {HPQCD})} (\bibinfo {year}
  {2009}),\ \bibfield  {title} {\enquote {\bibinfo {title} {{Neutral $B$ Meson
  Mixing in Unquenched Lattice QCD}},}\ }\href {\doibase
  10.1103/PhysRevD.80.014503} {\bibfield  {journal} {\bibinfo  {journal} {Phys.
  Rev.}\ }\textbf {\bibinfo {volume} {D80}},\ \bibinfo {pages} {014503}},\
  \Eprint {http://arxiv.org/abs/0902.1815} {arXiv:0902.1815 [hep-lat]}
  \BibitemShut {NoStop}%
\bibitem [{\citenamefont {Girrbach}\ \emph {et~al.}(2011)\citenamefont
  {Girrbach}, \citenamefont {Jager}, \citenamefont {Knopf}, \citenamefont
  {Martens}, \citenamefont {Nierste}, \citenamefont {Scherrer},\ and\
  \citenamefont {Wiesenfeldt}}]{Girrbach:2011an}%
  \BibitemOpen
  \bibfield  {author} {\bibinfo {author} {\bibnamefont {Girrbach},
  \bibfnamefont {Jennifer}}, \bibinfo {author} {\bibfnamefont {Sebastian}\
  \bibnamefont {Jager}}, \bibinfo {author} {\bibfnamefont {Markus}\
  \bibnamefont {Knopf}}, \bibinfo {author} {\bibfnamefont {Waldemar}\
  \bibnamefont {Martens}}, \bibinfo {author} {\bibfnamefont {Ulrich}\
  \bibnamefont {Nierste}}, \bibinfo {author} {\bibfnamefont {Christian}\
  \bibnamefont {Scherrer}}, \ and\ \bibinfo {author} {\bibfnamefont {Soren}\
  \bibnamefont {Wiesenfeldt}}} (\bibinfo {year} {2011}),\ \bibfield  {title}
  {\enquote {\bibinfo {title} {{Flavor Physics in an SO(10) Grand Unified
  Model}},}\ }\href {\doibase 10.1007/JHEP06(2011)044, 10.1007/JHEP07(2011)001}
  {\bibfield  {journal} {\bibinfo  {journal} {JHEP}\ }\textbf {\bibinfo
  {volume} {06}},\ \bibinfo {pages} {044}},\ \bibinfo {note} {[Erratum:
  JHEP07,001(2011)]},\ \Eprint {http://arxiv.org/abs/1101.6047}
  {arXiv:1101.6047 [hep-ph]} \BibitemShut {NoStop}%
\bibitem [{\citenamefont {Glashow}(1961)}]{Glashow:1961tr}%
  \BibitemOpen
  \bibfield  {author} {\bibinfo {author} {\bibnamefont {Glashow}, \bibfnamefont
  {S~L}}} (\bibinfo {year} {1961}),\ \bibfield  {title} {\enquote {\bibinfo
  {title} {{Partial Symmetries of Weak Interactions}},}\ }\href {\doibase
  10.1016/0029-5582(61)90469-2} {\bibfield  {journal} {\bibinfo  {journal}
  {Nucl. Phys.}\ }\textbf {\bibinfo {volume} {22}},\ \bibinfo {pages}
  {579--588}}\BibitemShut {NoStop}%
\bibitem [{\citenamefont {Glashow}\ \emph {et~al.}(1970)\citenamefont
  {Glashow}, \citenamefont {Iliopoulos},\ and\ \citenamefont
  {Maiani}}]{Glashow:1970gm}%
  \BibitemOpen
  \bibfield  {author} {\bibinfo {author} {\bibnamefont {Glashow}, \bibfnamefont
  {S~L}}, \bibinfo {author} {\bibfnamefont {J.}~\bibnamefont {Iliopoulos}}, \
  and\ \bibinfo {author} {\bibfnamefont {L.}~\bibnamefont {Maiani}}} (\bibinfo
  {year} {1970}),\ \bibfield  {title} {\enquote {\bibinfo {title} {{Weak
  Interactions with Lepton-Hadron Symmetry}},}\ }\href {\doibase
  10.1103/PhysRevD.2.1285} {\bibfield  {journal} {\bibinfo  {journal} {Phys.
  Rev.}\ }\textbf {\bibinfo {volume} {D2}},\ \bibinfo {pages}
  {1285--1292}}\BibitemShut {NoStop}%
\bibitem [{\citenamefont {Gligorov}(2011)}]{Gligorov:2011id}%
  \BibitemOpen
  \bibfield  {author} {\bibinfo {author} {\bibnamefont {Gligorov},
  \bibfnamefont {Vladimir~Vava}} (\bibinfo {collaboration} {LHCb})} (\bibinfo
  {year} {2011}),\ \bibfield  {title} {\enquote {\bibinfo {title} {{Time
  dependent measurements of the CKM angle $\gamma$ at LHCb}},}\ }in\ \href
  {https://inspirehep.net/record/883508/files/arXiv:1101.1201.pdf} {\emph
  {\bibinfo {booktitle} {{CKM unitarity triangle. Proceedings, 6th
  International Workshop, CKM 2010, Warwick, UK, September 6-10, 2010}}}},\
  \Eprint {http://arxiv.org/abs/1101.1201} {arXiv:1101.1201 [hep-ex]}
  \BibitemShut {NoStop}%
\bibitem [{\citenamefont {Goertz}\ and\ \citenamefont
  {Pfoh}(2011)}]{Goertz:2011nx}%
  \BibitemOpen
  \bibfield  {author} {\bibinfo {author} {\bibnamefont {Goertz}, \bibfnamefont
  {Florian}}, \ and\ \bibinfo {author} {\bibfnamefont {Torsten}\ \bibnamefont
  {Pfoh}}} (\bibinfo {year} {2011}),\ \bibfield  {title} {\enquote {\bibinfo
  {title} {{Randall-Sundrum Corrections to the Width Difference and
  CP-Violating Phase in $B^0_s$-Meson Decays}},}\ }\href {\doibase
  10.1103/PhysRevD.84.095016} {\bibfield  {journal} {\bibinfo  {journal} {Phys.
  Rev.}\ }\textbf {\bibinfo {volume} {D84}},\ \bibinfo {pages} {095016}},\
  \Eprint {http://arxiv.org/abs/1105.1507} {arXiv:1105.1507 [hep-ph]}
  \BibitemShut {NoStop}%
\bibitem [{\citenamefont {Hagelin}(1981)}]{Hagelin:1981zk}%
  \BibitemOpen
  \bibfield  {author} {\bibinfo {author} {\bibnamefont {Hagelin}, \bibfnamefont
  {J~S}}} (\bibinfo {year} {1981}),\ \bibfield  {title} {\enquote {\bibinfo
  {title} {{Mass Mixing and CP Violation in the $B^0-\bar{B}^0$ system}},}\
  }\href {\doibase 10.1016/0550-3213(81)90521-6} {\bibfield  {journal}
  {\bibinfo  {journal} {Nucl. Phys.}\ }\textbf {\bibinfo {volume} {B193}},\
  \bibinfo {pages} {123--149}}\BibitemShut {NoStop}%
\bibitem [{\citenamefont {Hartkorn}\ and\ \citenamefont
  {Moser}(1999)}]{Hartkorn:1999ga}%
  \BibitemOpen
  \bibfield  {author} {\bibinfo {author} {\bibnamefont {Hartkorn},
  \bibfnamefont {K}}, \ and\ \bibinfo {author} {\bibfnamefont {H.~G.}\
  \bibnamefont {Moser}}} (\bibinfo {year} {1999}),\ \bibfield  {title}
  {\enquote {\bibinfo {title} {{A new method of measuring $\Delta
  \Gamma/\Gamma$ in the $\Bs - \barBs$- system}},}\ }\href {\doibase
  10.1007/s100520050472} {\bibfield  {journal} {\bibinfo  {journal} {Eur. Phys.
  J.}\ }\textbf {\bibinfo {volume} {C8}},\ \bibinfo {pages}
  {381--383}}\BibitemShut {NoStop}%
\bibitem [{\citenamefont {Hayakawa}\ \emph {et~al.}(2012)\citenamefont
  {Hayakawa}, \citenamefont {Shimizu}, \citenamefont {Tanimoto},\ and\
  \citenamefont {Yamamoto}}]{Hayakawa:2012ua}%
  \BibitemOpen
  \bibfield  {author} {\bibinfo {author} {\bibnamefont {Hayakawa},
  \bibfnamefont {Atsushi}}, \bibinfo {author} {\bibfnamefont {Yusuke}\
  \bibnamefont {Shimizu}}, \bibinfo {author} {\bibfnamefont {Morimitsu}\
  \bibnamefont {Tanimoto}}, \ and\ \bibinfo {author} {\bibfnamefont {Kei}\
  \bibnamefont {Yamamoto}}} (\bibinfo {year} {2012}),\ \bibfield  {title}
  {\enquote {\bibinfo {title} {{Squark flavor mixing and CP asymmetry of
  neutral $B$ mesons at LHCb}},}\ }\href {\doibase
  10.1016/j.physletb.2012.03.020} {\bibfield  {journal} {\bibinfo  {journal}
  {Phys. Lett.}\ }\textbf {\bibinfo {volume} {B710}},\ \bibinfo {pages}
  {446--453}},\ \Eprint {http://arxiv.org/abs/1202.0486} {arXiv:1202.0486
  [hep-ph]} \BibitemShut {NoStop}%
\bibitem [{\citenamefont {Inami}\ and\ \citenamefont
  {Lim}(1981)}]{Inami:1980fz}%
  \BibitemOpen
  \bibfield  {author} {\bibinfo {author} {\bibnamefont {Inami}, \bibfnamefont
  {T}}, \ and\ \bibinfo {author} {\bibfnamefont {C.S.}\ \bibnamefont {Lim}}}
  (\bibinfo {year} {1981}),\ \bibfield  {title} {\enquote {\bibinfo {title}
  {{Effects of Superheavy Quarks and Leptons in Low-Energy Weak Processes $K(L)
  \to \mu^+ \mu^-, K+ \to \pi^+ \nu \bar{\nu}$ and $K^0 \to \bar{K}^0$}},}\
  }\href {\doibase 10.1143/PTP.65.297} {\bibfield  {journal} {\bibinfo
  {journal} {Prog.Theor.Phys.}\ }\textbf {\bibinfo {volume} {65}},\ \bibinfo
  {pages} {297}}\BibitemShut {NoStop}%
\bibitem [{\citenamefont {Ishimori}\ \emph {et~al.}(2011)\citenamefont
  {Ishimori}, \citenamefont {Kajiyama}, \citenamefont {Shimizu},\ and\
  \citenamefont {Tanimoto}}]{Ishimori:2011nv}%
  \BibitemOpen
  \bibfield  {author} {\bibinfo {author} {\bibnamefont {Ishimori},
  \bibfnamefont {Hajime}}, \bibinfo {author} {\bibfnamefont {Yuji}\
  \bibnamefont {Kajiyama}}, \bibinfo {author} {\bibfnamefont {Yusuke}\
  \bibnamefont {Shimizu}}, \ and\ \bibinfo {author} {\bibfnamefont {Morimitsu}\
  \bibnamefont {Tanimoto}}} (\bibinfo {year} {2011}),\ \bibfield  {title}
  {\enquote {\bibinfo {title} {{Like-sign dimuon asymmetry of $B^0$ meson and
  LFV in SU(5) SUSY GUT with S4 flavor symmetry}},}\ }\href {\doibase
  10.1143/PTP.126.703} {\bibfield  {journal} {\bibinfo  {journal} {Prog. Theor.
  Phys.}\ }\textbf {\bibinfo {volume} {126}},\ \bibinfo {pages} {703--734}},\
  \Eprint {http://arxiv.org/abs/1103.5705} {arXiv:1103.5705 [hep-ph]}
  \BibitemShut {NoStop}%
\bibitem [{\citenamefont {Jubb}\ \emph {et~al.}(2016)\citenamefont {Jubb},
  \citenamefont {Kirk}, \citenamefont {Lenz},\ and\ \citenamefont
  {Tetlalmatzi-Xolocotzi}}]{Jubb:2016mvq}%
  \BibitemOpen
  \bibfield  {author} {\bibinfo {author} {\bibnamefont {Jubb}, \bibfnamefont
  {Thomas}}, \bibinfo {author} {\bibfnamefont {Matthew}\ \bibnamefont {Kirk}},
  \bibinfo {author} {\bibfnamefont {Alexander}\ \bibnamefont {Lenz}}, \ and\
  \bibinfo {author} {\bibfnamefont {Gilberto}\ \bibnamefont
  {Tetlalmatzi-Xolocotzi}}} (\bibinfo {year} {2016}),\ \bibfield  {title}
  {\enquote {\bibinfo {title} {{On the ultimate precision of meson mixing
  observables}},}\ }\href@noop {} {\ }\Eprint {http://arxiv.org/abs/1603.07770}
  {arXiv:1603.07770 [hep-ph]} \BibitemShut {NoStop}%
\bibitem [{\citenamefont {Jung}(2012)}]{Jung:2012mp}%
  \BibitemOpen
  \bibfield  {author} {\bibinfo {author} {\bibnamefont {Jung}, \bibfnamefont
  {Martin}}} (\bibinfo {year} {2012}),\ \bibfield  {title} {\enquote {\bibinfo
  {title} {{Determining weak phases from $B\to J/\psi P$ decays}},}\ }\href
  {\doibase 10.1103/PhysRevD.86.053008} {\bibfield  {journal} {\bibinfo
  {journal} {Phys. Rev.}\ }\textbf {\bibinfo {volume} {D86}},\ \bibinfo {pages}
  {053008}},\ \Eprint {http://arxiv.org/abs/1206.2050} {arXiv:1206.2050
  [hep-ph]} \BibitemShut {NoStop}%
\bibitem [{\citenamefont {Kaburaki}\ \emph {et~al.}(2011)\citenamefont
  {Kaburaki}, \citenamefont {Konya}, \citenamefont {Kubo},\ and\ \citenamefont
  {Lenz}}]{Kaburaki:2010xc}%
  \BibitemOpen
  \bibfield  {author} {\bibinfo {author} {\bibnamefont {Kaburaki},
  \bibfnamefont {Yoshiyuki}}, \bibinfo {author} {\bibfnamefont {Kazuhiro}\
  \bibnamefont {Konya}}, \bibinfo {author} {\bibfnamefont {Jisuke}\
  \bibnamefont {Kubo}}, \ and\ \bibinfo {author} {\bibfnamefont {Alexander}\
  \bibnamefont {Lenz}}} (\bibinfo {year} {2011}),\ \bibfield  {title} {\enquote
  {\bibinfo {title} {{Triangle Relation of Dark Matter, EDM and CP Violation in
  $B^0$ Mixing in a Supersymmetric Q6 Model}},}\ }\href {\doibase
  10.1103/PhysRevD.84.016007} {\bibfield  {journal} {\bibinfo  {journal} {Phys.
  Rev.}\ }\textbf {\bibinfo {volume} {D84}},\ \bibinfo {pages} {016007}},\
  \Eprint {http://arxiv.org/abs/1012.2435} {arXiv:1012.2435 [hep-ph]}
  \BibitemShut {NoStop}%
\bibitem [{\citenamefont {Kawashima}\ \emph {et~al.}(2009)\citenamefont
  {Kawashima}, \citenamefont {Kubo},\ and\ \citenamefont
  {Lenz}}]{Kawashima:2009jv}%
  \BibitemOpen
  \bibfield  {author} {\bibinfo {author} {\bibnamefont {Kawashima},
  \bibfnamefont {Kenji}}, \bibinfo {author} {\bibfnamefont {Jisuke}\
  \bibnamefont {Kubo}}, \ and\ \bibinfo {author} {\bibfnamefont {Alexander}\
  \bibnamefont {Lenz}}} (\bibinfo {year} {2009}),\ \bibfield  {title} {\enquote
  {\bibinfo {title} {{Testing the new CP phase in a Supersymmetric Model with
  $Q(6)$ Family Symmetry by $\Bs$ Mixing}},}\ }\href {\doibase
  10.1016/j.physletb.2009.09.064} {\bibfield  {journal} {\bibinfo  {journal}
  {Phys. Lett.}\ }\textbf {\bibinfo {volume} {B681}},\ \bibinfo {pages}
  {60--67}},\ \Eprint {http://arxiv.org/abs/0907.2302} {arXiv:0907.2302
  [hep-ph]} \BibitemShut {NoStop}%
\bibitem [{\citenamefont {Khachatryan}(2015)}]{Khachatryan:2015nza}%
  \BibitemOpen
  \bibfield  {author} {\bibinfo {author} {\bibnamefont {Khachatryan},
  \bibfnamefont {V}} (\bibinfo {collaboration} {CMS})} (\bibinfo {year}
  {2015}),\ \bibfield  {title} {\enquote {\bibinfo {title} {{Measurement of the
  CP-violating weak phase $\mathrm{ \phi_s }$ and the decay width difference $
  \Delta \Gamma_{ \mathrm{s} }$ using the $ \mathrm{B^0_s} \to \mathrm{J} /
  \psi \phi(1020) $ decay channel in pp collisions at $\sqrt{s} =$ 8 TeV}},}\
  }\href@noop {} {\ }\Eprint {http://arxiv.org/abs/1507.07527}
  {arXiv:1507.07527 [hep-ex]} \BibitemShut {NoStop}%
\bibitem [{\citenamefont {Khodjamirian}\ \emph {et~al.}(2003)\citenamefont
  {Khodjamirian}, \citenamefont {Mannel},\ and\ \citenamefont
  {Melic}}]{Khodjamirian:2003eq}%
  \BibitemOpen
  \bibfield  {author} {\bibinfo {author} {\bibnamefont {Khodjamirian},
  \bibfnamefont {A}}, \bibinfo {author} {\bibfnamefont {T.}~\bibnamefont
  {Mannel}}, \ and\ \bibinfo {author} {\bibfnamefont {B.}~\bibnamefont
  {Melic}}} (\bibinfo {year} {2003}),\ \bibfield  {title} {\enquote {\bibinfo
  {title} {{QCD light cone sum rule estimate of charming penguin contributions
  in $B \to \pi \pi$}},}\ }\href {\doibase 10.1016/j.physletb.2003.08.012}
  {\bibfield  {journal} {\bibinfo  {journal} {Phys. Lett.}\ }\textbf {\bibinfo
  {volume} {B571}},\ \bibinfo {pages} {75--84}},\ \bibinfo {note} {[Phys.
  Lett.B572,171(2003)]},\ \Eprint {http://arxiv.org/abs/hep-ph/0304179}
  {arXiv:hep-ph/0304179 [hep-ph]} \BibitemShut {NoStop}%
\bibitem [{\citenamefont {Khodjamirian}(2001)}]{Khodjamirian:2000mi}%
  \BibitemOpen
  \bibfield  {author} {\bibinfo {author} {\bibnamefont {Khodjamirian},
  \bibfnamefont {Alexander}}} (\bibinfo {year} {2001}),\ \bibfield  {title}
  {\enquote {\bibinfo {title} {{$B \to \pi \pi$ decay in QCD}},}\ }\href
  {\doibase 10.1016/S0550-3213(01)00194-8} {\bibfield  {journal} {\bibinfo
  {journal} {Nucl. Phys.}\ }\textbf {\bibinfo {volume} {B605}},\ \bibinfo
  {pages} {558--578}},\ \Eprint {http://arxiv.org/abs/hep-ph/0012271}
  {arXiv:hep-ph/0012271 [hep-ph]} \BibitemShut {NoStop}%
\bibitem [{\citenamefont {Khoze}\ and\ \citenamefont
  {Shifman}(1983)}]{Khoze:1983yp}%
  \BibitemOpen
  \bibfield  {author} {\bibinfo {author} {\bibnamefont {Khoze}, \bibfnamefont
  {Valery~A}}, \ and\ \bibinfo {author} {\bibfnamefont {Mikhail~A.}\
  \bibnamefont {Shifman}}} (\bibinfo {year} {1983}),\ \bibfield  {title}
  {\enquote {\bibinfo {title} {{HEAVY QUARKS}},}\ }\href {\doibase
  10.1070/PU1983v026n05ABEH004398} {\bibfield  {journal} {\bibinfo  {journal}
  {Sov. Phys. Usp.}\ }\textbf {\bibinfo {volume} {26}},\ \bibinfo {pages}
  {387}}\BibitemShut {NoStop}%
\bibitem [{\citenamefont {Khoze}\ \emph {et~al.}(1987)\citenamefont {Khoze},
  \citenamefont {Shifman}, \citenamefont {Uraltsev},\ and\ \citenamefont
  {Voloshin}}]{Khoze:1986fa}%
  \BibitemOpen
  \bibfield  {author} {\bibinfo {author} {\bibnamefont {Khoze}, \bibfnamefont
  {Valery~A}}, \bibinfo {author} {\bibfnamefont {Mikhail~A.}\ \bibnamefont
  {Shifman}}, \bibinfo {author} {\bibfnamefont {N.~G.}\ \bibnamefont
  {Uraltsev}}, \ and\ \bibinfo {author} {\bibfnamefont {M.~B.}\ \bibnamefont
  {Voloshin}}} (\bibinfo {year} {1987}),\ \bibfield  {title} {\enquote
  {\bibinfo {title} {{On Inclusive Hadronic Widths of Beautiful Particles}},}\
  }\href@noop {} {\bibfield  {journal} {\bibinfo  {journal} {Sov. J. Nucl.
  Phys.}\ }\textbf {\bibinfo {volume} {46}},\ \bibinfo {pages} {112}},\
  \bibinfo {note} {[Yad. Fiz.46,181(1987)]}\BibitemShut {NoStop}%
\bibitem [{\citenamefont {Kifune}\ \emph {et~al.}(2008)\citenamefont {Kifune},
  \citenamefont {Kubo},\ and\ \citenamefont {Lenz}}]{Kifune:2007fj}%
  \BibitemOpen
  \bibfield  {author} {\bibinfo {author} {\bibnamefont {Kifune}, \bibfnamefont
  {Naoko}}, \bibinfo {author} {\bibfnamefont {Jisuke}\ \bibnamefont {Kubo}}, \
  and\ \bibinfo {author} {\bibfnamefont {Alexander}\ \bibnamefont {Lenz}}}
  (\bibinfo {year} {2008}),\ \bibfield  {title} {\enquote {\bibinfo {title}
  {{Flavor Changing Neutral Higgs Bosons in a Supersymmetric Extension based on
  a $Q_{6}$ Family Symmetry}},}\ }\href {\doibase 10.1103/PhysRevD.77.076010}
  {\bibfield  {journal} {\bibinfo  {journal} {Phys. Rev.}\ }\textbf {\bibinfo
  {volume} {D77}},\ \bibinfo {pages} {076010}},\ \Eprint
  {http://arxiv.org/abs/0712.0503} {arXiv:0712.0503 [hep-ph]} \BibitemShut
  {NoStop}%
\bibitem [{\citenamefont {Kim}\ \emph {et~al.}(2013)\citenamefont {Kim},
  \citenamefont {Kim},\ and\ \citenamefont {Shin}}]{Kim:2012rpa}%
  \BibitemOpen
  \bibfield  {author} {\bibinfo {author} {\bibnamefont {Kim}, \bibfnamefont
  {Hyung~Do}}, \bibinfo {author} {\bibfnamefont {Sung-Gi}\ \bibnamefont {Kim}},
  \ and\ \bibinfo {author} {\bibfnamefont {Seodong}\ \bibnamefont {Shin}}}
  (\bibinfo {year} {2013}),\ \bibfield  {title} {\enquote {\bibinfo {title}
  {{D0 dimuon charge asymmetry from $B_s$ system with Z~ couplings and the
  recent LHCb result}},}\ }\href {\doibase 10.1103/PhysRevD.88.015005}
  {\bibfield  {journal} {\bibinfo  {journal} {Phys. Rev.}\ }\textbf {\bibinfo
  {volume} {D88}}~(\bibinfo {number} {1}),\ \bibinfo {pages} {015005}},\
  \Eprint {http://arxiv.org/abs/1205.6481} {arXiv:1205.6481 [hep-ph]}
  \BibitemShut {NoStop}%
\bibitem [{\citenamefont {Kobayashi}\ and\ \citenamefont
  {Maskawa}(1973)}]{Kobayashi:1973fv}%
  \BibitemOpen
  \bibfield  {author} {\bibinfo {author} {\bibnamefont {Kobayashi},
  \bibfnamefont {Makoto}}, \ and\ \bibinfo {author} {\bibfnamefont {Toshihide}\
  \bibnamefont {Maskawa}}} (\bibinfo {year} {1973}),\ \bibfield  {title}
  {\enquote {\bibinfo {title} {{CP Violation in the Renormalizable Theory of
  Weak Interaction}},}\ }\href {\doibase 10.1143/PTP.49.652} {\bibfield
  {journal} {\bibinfo  {journal} {Prog. Theor. Phys.}\ }\textbf {\bibinfo
  {volume} {49}},\ \bibinfo {pages} {652--657}}\BibitemShut {NoStop}%
\bibitem [{\citenamefont {Korner}\ \emph {et~al.}(2003)\citenamefont {Korner},
  \citenamefont {Onishchenko}, \citenamefont {Petrov},\ and\ \citenamefont
  {Pivovarov}}]{Korner:2003zk}%
  \BibitemOpen
  \bibfield  {author} {\bibinfo {author} {\bibnamefont {Korner}, \bibfnamefont
  {J~G}}, \bibinfo {author} {\bibfnamefont {A.~I.}\ \bibnamefont
  {Onishchenko}}, \bibinfo {author} {\bibfnamefont {Alexey~A.}\ \bibnamefont
  {Petrov}}, \ and\ \bibinfo {author} {\bibfnamefont {A.~A.}\ \bibnamefont
  {Pivovarov}}} (\bibinfo {year} {2003}),\ \bibfield  {title} {\enquote
  {\bibinfo {title} {{$B^0 - \bar{B}^0$ mixing beyond factorization}},}\ }\href
  {\doibase 10.1103/PhysRevLett.91.192002} {\bibfield  {journal} {\bibinfo
  {journal} {Phys. Rev. Lett.}\ }\textbf {\bibinfo {volume} {91}},\ \bibinfo
  {pages} {192002}},\ \Eprint {http://arxiv.org/abs/hep-ph/0306032}
  {arXiv:hep-ph/0306032 [hep-ph]} \BibitemShut {NoStop}%
\bibitem [{\citenamefont {Krinner}\ \emph {et~al.}(2013)\citenamefont
  {Krinner}, \citenamefont {Lenz},\ and\ \citenamefont
  {Rauh}}]{Krinner:2013cja}%
  \BibitemOpen
  \bibfield  {author} {\bibinfo {author} {\bibnamefont {Krinner}, \bibfnamefont
  {Fabian}}, \bibinfo {author} {\bibfnamefont {Alexander}\ \bibnamefont
  {Lenz}}, \ and\ \bibinfo {author} {\bibfnamefont {Thomas}\ \bibnamefont
  {Rauh}}} (\bibinfo {year} {2013}),\ \bibfield  {title} {\enquote {\bibinfo
  {title} {{The inclusive decay $b \to c\bar{c}s$ revisited}},}\ }\href
  {\doibase 10.1016/j.nuclphysb.2013.07.028} {\bibfield  {journal} {\bibinfo
  {journal} {Nucl.Phys.}\ }\textbf {\bibinfo {volume} {B876}},\ \bibinfo
  {pages} {31--54}},\ \Eprint {http://arxiv.org/abs/1305.5390} {arXiv:1305.5390
  [hep-ph]} \BibitemShut {NoStop}%
\bibitem [{\citenamefont {Kubo}\ and\ \citenamefont
  {Lenz}(2010)}]{Kubo:2010mh}%
  \BibitemOpen
  \bibfield  {author} {\bibinfo {author} {\bibnamefont {Kubo}, \bibfnamefont
  {Jisuke}}, \ and\ \bibinfo {author} {\bibfnamefont {Alexander}\ \bibnamefont
  {Lenz}}} (\bibinfo {year} {2010}),\ \bibfield  {title} {\enquote {\bibinfo
  {title} {{Large loop effects of extra {SUSY} Higgs doublets to {CP} violation
  in $B^0$ mixing}},}\ }\href {\doibase 10.1103/PhysRevD.82.075001} {\bibfield
  {journal} {\bibinfo  {journal} {Phys. Rev.}\ }\textbf {\bibinfo {volume}
  {D82}},\ \bibinfo {pages} {075001}},\ \Eprint
  {http://arxiv.org/abs/1007.0680} {arXiv:1007.0680 [hep-ph]} \BibitemShut
  {NoStop}%
\bibitem [{\citenamefont {Kuflik}\ \emph {et~al.}(2013)\citenamefont {Kuflik},
  \citenamefont {Nir},\ and\ \citenamefont {Volansky}}]{Kuflik:2012ai}%
  \BibitemOpen
  \bibfield  {author} {\bibinfo {author} {\bibnamefont {Kuflik}, \bibfnamefont
  {Eric}}, \bibinfo {author} {\bibfnamefont {Yosef}\ \bibnamefont {Nir}}, \
  and\ \bibinfo {author} {\bibfnamefont {Tomer}\ \bibnamefont {Volansky}}}
  (\bibinfo {year} {2013}),\ \bibfield  {title} {\enquote {\bibinfo {title}
  {{Implications of Higgs searches on the four generation standard model}},}\
  }\href {\doibase 10.1103/PhysRevLett.110.091801} {\bibfield  {journal}
  {\bibinfo  {journal} {Phys. Rev. Lett.}\ }\textbf {\bibinfo {volume}
  {110}}~(\bibinfo {number} {9}),\ \bibinfo {pages} {091801}},\ \Eprint
  {http://arxiv.org/abs/1204.1975} {arXiv:1204.1975 [hep-ph]} \BibitemShut
  {NoStop}%
\bibitem [{\citenamefont {Lee}\ and\ \citenamefont {Nam}(2012)}]{Lee:2011kn}%
  \BibitemOpen
  \bibfield  {author} {\bibinfo {author} {\bibnamefont {Lee}, \bibfnamefont
  {Kang~Young}}, \ and\ \bibinfo {author} {\bibfnamefont {Soo-hyeon}\
  \bibnamefont {Nam}}} (\bibinfo {year} {2012}),\ \bibfield  {title} {\enquote
  {\bibinfo {title} {{CP violating dimuon charge asymmetry in general
  left-right models}},}\ }\href {\doibase 10.1103/PhysRevD.85.035001}
  {\bibfield  {journal} {\bibinfo  {journal} {Phys. Rev.}\ }\textbf {\bibinfo
  {volume} {D85}},\ \bibinfo {pages} {035001}},\ \Eprint
  {http://arxiv.org/abs/1111.4666} {arXiv:1111.4666 [hep-ph]} \BibitemShut
  {NoStop}%
\bibitem [{\citenamefont {Lees}\ \emph {et~al.}(2013)\citenamefont {Lees} \emph
  {et~al.}}]{Lees:2013sua}%
  \BibitemOpen
  \bibfield  {author} {\bibinfo {author} {\bibnamefont {Lees}, \bibfnamefont
  {J~P}},  \emph {et~al.} (\bibinfo {collaboration} {BaBar})} (\bibinfo {year}
  {2013}),\ \bibfield  {title} {\enquote {\bibinfo {title} {{Search for $CP$
  Violation in $B^0$-$\bar{B}^0$ Mixing using Partial Reconstruction of $B^0
  \to D^{*-}X\ell^+ \nu_\ell$ and a Kaon Tag}},}\ }\href {\doibase
  10.1103/PhysRevLett.111.159901, 10.1103/PhysRevLett.111.101802} {\bibfield
  {journal} {\bibinfo  {journal} {Phys. Rev. Lett.}\ }\textbf {\bibinfo
  {volume} {111}}~(\bibinfo {number} {10}),\ \bibinfo {pages} {101802}},\
  \bibinfo {note} {[Addendum: Phys. Rev. Lett.111,no.15,159901(2013)]},\
  \Eprint {http://arxiv.org/abs/1305.1575} {arXiv:1305.1575 [hep-ex]}
  \BibitemShut {NoStop}%
\bibitem [{\citenamefont {Lees}\ \emph
  {et~al.}(2015{\natexlab{a}})\citenamefont {Lees} \emph
  {et~al.}}]{Lees:2014lra}%
  \BibitemOpen
  \bibfield  {author} {\bibinfo {author} {\bibnamefont {Lees}, \bibfnamefont
  {J~P}},  \emph {et~al.} (\bibinfo {collaboration} {BaBar})} (\bibinfo {year}
  {2015}{\natexlab{a}}),\ \bibfield  {title} {\enquote {\bibinfo {title}
  {{Study of $B^{\pm,0} \to J/\psi K^+ K^- K^{\pm,0}$ and search for $B^0 \to
  J/\psi\phi$ at BABAR}},}\ }\href {\doibase 10.1103/PhysRevD.91.012003}
  {\bibfield  {journal} {\bibinfo  {journal} {Phys. Rev.}\ }\textbf {\bibinfo
  {volume} {D91}}~(\bibinfo {number} {1}),\ \bibinfo {pages} {012003}},\
  \Eprint {http://arxiv.org/abs/1407.7244} {arXiv:1407.7244 [hep-ex]}
  \BibitemShut {NoStop}%
\bibitem [{\citenamefont {Lees}\ \emph
  {et~al.}(2015{\natexlab{b}})\citenamefont {Lees} \emph
  {et~al.}}]{Lees:2014kep}%
  \BibitemOpen
  \bibfield  {author} {\bibinfo {author} {\bibnamefont {Lees}, \bibfnamefont
  {J~P}},  \emph {et~al.} (\bibinfo {collaboration} {BaBar})} (\bibinfo {year}
  {2015}{\natexlab{b}}),\ \bibfield  {title} {\enquote {\bibinfo {title}
  {{Study of $CP$ Asymmetry in $B^0-\bar B^0$ Mixing with Inclusive Dilepton
  Events}},}\ }\href {\doibase 10.1103/PhysRevLett.114.081801} {\bibfield
  {journal} {\bibinfo  {journal} {Phys. Rev. Lett.}\ }\textbf {\bibinfo
  {volume} {114}}~(\bibinfo {number} {8}),\ \bibinfo {pages} {081801}},\
  \Eprint {http://arxiv.org/abs/1411.1842} {arXiv:1411.1842 [hep-ex]}
  \BibitemShut {NoStop}%
\bibitem [{\citenamefont {Lenz}\ \emph {et~al.}(2011)\citenamefont {Lenz},
  \citenamefont {Nierste}, \citenamefont {Charles}, \citenamefont
  {Descotes-Genon}, \citenamefont {Jantsch} \emph {et~al.}}]{Lenz:2010gu}%
  \BibitemOpen
  \bibfield  {author} {\bibinfo {author} {\bibnamefont {Lenz}, \bibfnamefont
  {A}}, \bibinfo {author} {\bibfnamefont {U.}~\bibnamefont {Nierste}}, \bibinfo
  {author} {\bibfnamefont {J.}~\bibnamefont {Charles}}, \bibinfo {author}
  {\bibfnamefont {S.}~\bibnamefont {Descotes-Genon}}, \bibinfo {author}
  {\bibfnamefont {A.}~\bibnamefont {Jantsch}},  \emph {et~al.}} (\bibinfo
  {year} {2011}),\ \bibfield  {title} {\enquote {\bibinfo {title} {{Anatomy of
  New Physics in $B - \bar{B}$ mixing}},}\ }\href {\doibase
  10.1103/PhysRevD.83.036004} {\bibfield  {journal} {\bibinfo  {journal}
  {Phys.Rev.}\ }\textbf {\bibinfo {volume} {D83}},\ \bibinfo {pages}
  {036004}},\ \Eprint {http://arxiv.org/abs/1008.1593} {arXiv:1008.1593
  [hep-ph]} \BibitemShut {NoStop}%
\bibitem [{\citenamefont {Lenz}\ \emph {et~al.}(2012)\citenamefont {Lenz},
  \citenamefont {Nierste}, \citenamefont {Charles}, \citenamefont
  {Descotes-Genon}, \citenamefont {Lacker} \emph {et~al.}}]{Lenz:2012az}%
  \BibitemOpen
  \bibfield  {author} {\bibinfo {author} {\bibnamefont {Lenz}, \bibfnamefont
  {A}}, \bibinfo {author} {\bibfnamefont {U.}~\bibnamefont {Nierste}}, \bibinfo
  {author} {\bibfnamefont {J.}~\bibnamefont {Charles}}, \bibinfo {author}
  {\bibfnamefont {S.}~\bibnamefont {Descotes-Genon}}, \bibinfo {author}
  {\bibfnamefont {H.}~\bibnamefont {Lacker}},  \emph {et~al.}} (\bibinfo {year}
  {2012}),\ \bibfield  {title} {\enquote {\bibinfo {title} {{Constraints on new
  physics in $B-\bar{B}$ mixing in the light of recent LHCb data}},}\ }\href
  {\doibase 10.1103/PhysRevD.86.033008} {\bibfield  {journal} {\bibinfo
  {journal} {Phys.Rev.}\ }\textbf {\bibinfo {volume} {D86}},\ \bibinfo {pages}
  {033008}},\ \Eprint {http://arxiv.org/abs/1203.0238} {arXiv:1203.0238
  [hep-ph]} \BibitemShut {NoStop}%
\bibitem [{\citenamefont {Lenz}(2000)}]{Lenz:2000kv}%
  \BibitemOpen
  \bibfield  {author} {\bibinfo {author} {\bibnamefont {Lenz}, \bibfnamefont
  {Alexander}}} (\bibinfo {year} {2000}),\ \bibfield  {title} {\enquote
  {\bibinfo {title} {{Some comments on the missing charm puzzle}},}\ }in\ \href
  {http://alice.cern.ch/format/showfull?sysnb=2231606} {\emph {\bibinfo
  {booktitle} {{Heavy flavours and CP violation. Proceedings, 8th UK
  Phenomenology Workshop, Durham, UK, September 19-24, 2000}}}},\ \Eprint
  {http://arxiv.org/abs/hep-ph/0011258} {arXiv:hep-ph/0011258 [hep-ph]}
  \BibitemShut {NoStop}%
\bibitem [{\citenamefont {Lenz}(2004)}]{Lenz:2004nx}%
  \BibitemOpen
  \bibfield  {author} {\bibinfo {author} {\bibnamefont {Lenz}, \bibfnamefont
  {Alexander}}} (\bibinfo {year} {2004}),\ \bibfield  {title} {\enquote
  {\bibinfo {title} {{Decay rate difference in the neutral B-system: $\Delta
  \Gamma (\Bs)$ and $\Delta \Gamma (B_d)$}},}\ }in\ \href@noop {} {\emph
  {\bibinfo {booktitle} {{3rd Conference on Flavor Physics and CP Violation
  (FPCP 2004) Daegu, Korea, October 4-9, 2004}}}},\ \Eprint
  {http://arxiv.org/abs/hep-ph/0412007} {arXiv:hep-ph/0412007 [hep-ph]}
  \BibitemShut {NoStop}%
\bibitem [{\citenamefont {Lenz}(2013)}]{Lenz:2013iha}%
  \BibitemOpen
  \bibfield  {author} {\bibinfo {author} {\bibnamefont {Lenz}, \bibfnamefont
  {Alexander}}} (\bibinfo {year} {2013}),\ \bibfield  {title} {\enquote
  {\bibinfo {title} {{Constraints on a fourth generation of fermions from Higgs
  Boson searches}},}\ }\href {\doibase 10.1155/2013/910275} {\bibfield
  {journal} {\bibinfo  {journal} {Adv. High Energy Phys.}\ }\textbf {\bibinfo
  {volume} {2013}},\ \bibinfo {pages} {910275}}\BibitemShut {NoStop}%
\bibitem [{\citenamefont {Lenz}(2014)}]{Lenz:2014jha}%
  \BibitemOpen
  \bibfield  {author} {\bibinfo {author} {\bibnamefont {Lenz}, \bibfnamefont
  {Alexander}}} (\bibinfo {year} {2014}),\ \bibfield  {title} {\enquote
  {\bibinfo {title} {{Lifetimes and HQE}},}\ }\href@noop {} {\ }\Eprint
  {http://arxiv.org/abs/1405.3601} {arXiv:1405.3601 [hep-ph]} \BibitemShut
  {NoStop}%
\bibitem [{\citenamefont {Lenz}\ and\ \citenamefont
  {Nierste}(2007)}]{Lenz:2006hd}%
  \BibitemOpen
  \bibfield  {author} {\bibinfo {author} {\bibnamefont {Lenz}, \bibfnamefont
  {Alexander}}, \ and\ \bibinfo {author} {\bibfnamefont {Ulrich}\ \bibnamefont
  {Nierste}}} (\bibinfo {year} {2007}),\ \bibfield  {title} {\enquote {\bibinfo
  {title} {{Theoretical update of $\Bs - \barBs$ mixing}},}\ }\href {\doibase
  10.1088/1126-6708/2007/06/072} {\bibfield  {journal} {\bibinfo  {journal}
  {JHEP}\ }\textbf {\bibinfo {volume} {06}},\ \bibinfo {pages} {072}},\ \Eprint
  {http://arxiv.org/abs/hep-ph/0612167} {arXiv:hep-ph/0612167 [hep-ph]}
  \BibitemShut {NoStop}%
\bibitem [{\citenamefont {Lenz}\ and\ \citenamefont
  {Nierste}(2011)}]{Lenz:2011ti}%
  \BibitemOpen
  \bibfield  {author} {\bibinfo {author} {\bibnamefont {Lenz}, \bibfnamefont
  {Alexander}}, \ and\ \bibinfo {author} {\bibfnamefont {Ulrich}\ \bibnamefont
  {Nierste}}} (\bibinfo {year} {2011}),\ \bibfield  {title} {\enquote {\bibinfo
  {title} {{Numerical Updates of Lifetimes and Mixing Parameters of B
  Mesons}},}\ }\href@noop {} {\ }\Eprint {http://arxiv.org/abs/1102.4274}
  {arXiv:1102.4274 [hep-ph]} \BibitemShut {NoStop}%
\bibitem [{\citenamefont {Lenz}\ \emph {et~al.}(1997)\citenamefont {Lenz},
  \citenamefont {Nierste},\ and\ \citenamefont {Ostermaier}}]{Lenz:1997aa}%
  \BibitemOpen
  \bibfield  {author} {\bibinfo {author} {\bibnamefont {Lenz}, \bibfnamefont
  {Alexander}}, \bibinfo {author} {\bibfnamefont {Ulrich}\ \bibnamefont
  {Nierste}}, \ and\ \bibinfo {author} {\bibfnamefont {Gaby}\ \bibnamefont
  {Ostermaier}}} (\bibinfo {year} {1997}),\ \bibfield  {title} {\enquote
  {\bibinfo {title} {{Penguin diagrams, charmless B decays and the missing
  charm puzzle}},}\ }\href {\doibase 10.1103/PhysRevD.56.7228} {\bibfield
  {journal} {\bibinfo  {journal} {Phys. Rev.}\ }\textbf {\bibinfo {volume}
  {D56}},\ \bibinfo {pages} {7228--7239}},\ \Eprint
  {http://arxiv.org/abs/hep-ph/9706501} {arXiv:hep-ph/9706501 [hep-ph]}
  \BibitemShut {NoStop}%
\bibitem [{\citenamefont {Lenz}\ and\ \citenamefont
  {Rauh}(2013)}]{Lenz:2013aua}%
  \BibitemOpen
  \bibfield  {author} {\bibinfo {author} {\bibnamefont {Lenz}, \bibfnamefont
  {Alexander}}, \ and\ \bibinfo {author} {\bibfnamefont {Thomas}\ \bibnamefont
  {Rauh}}} (\bibinfo {year} {2013}),\ \bibfield  {title} {\enquote {\bibinfo
  {title} {{D-meson lifetimes within the heavy quark expansion}},}\ }\href
  {\doibase 10.1103/PhysRevD.88.034004} {\bibfield  {journal} {\bibinfo
  {journal} {Phys.Rev.}\ }\textbf {\bibinfo {volume} {D88}},\ \bibinfo {pages}
  {034004}},\ \Eprint {http://arxiv.org/abs/1305.3588} {arXiv:1305.3588
  [hep-ph]} \BibitemShut {NoStop}%
\bibitem [{\citenamefont {Lenz}(2011)}]{Lenz:2011zz}%
  \BibitemOpen
  \bibfield  {author} {\bibinfo {author} {\bibnamefont {Lenz}, \bibfnamefont
  {Alexander~J}}} (\bibinfo {year} {2011}),\ \bibfield  {title} {\enquote
  {\bibinfo {title} {{A simple relation for $B_s$ mixing}},}\ }\href {\doibase
  10.1103/PhysRevD.84.031501} {\bibfield  {journal} {\bibinfo  {journal} {Phys.
  Rev.}\ }\textbf {\bibinfo {volume} {D84}},\ \bibinfo {pages} {031501}},\
  \Eprint {http://arxiv.org/abs/1106.3200} {arXiv:1106.3200 [hep-ph]}
  \BibitemShut {NoStop}%
\bibitem [{\citenamefont {LHCb}(2014)}]{LHCb-PUB-2014-040}%
  \BibitemOpen
  \bibfield  {author} {\bibinfo {author} {\bibnamefont {LHCb}, \bibfnamefont
  {Collaboration}}} (\bibinfo {year} {2014}),\ \href
  {http://cds.cern.ch/record/1748643} {\emph {\bibinfo {title} {{Impact of the
  LHCb upgrade detector design choices on physics and trigger performance}}}},\
  \bibinfo {type} {Tech. Rep.}\ \bibinfo {number} {LHCb-PUB-2014-040.
  CERN-LHCb-PUB-2014-040. LHCb-INT-2013-024}\ (\bibinfo  {institution} {CERN},\
  \bibinfo {address} {Geneva})\BibitemShut {NoStop}%
\bibitem [{\citenamefont {Li}\ and\ \citenamefont {Yu}(1996)}]{Li:1994iu}%
  \BibitemOpen
  \bibfield  {author} {\bibinfo {author} {\bibnamefont {Li}, \bibfnamefont
  {Hsiang-nan}}, \ and\ \bibinfo {author} {\bibfnamefont {Hoi-Lai}\
  \bibnamefont {Yu}}} (\bibinfo {year} {1996}),\ \bibfield  {title} {\enquote
  {\bibinfo {title} {{Perturbative QCD analysis of B meson decays}},}\ }\href
  {\doibase 10.1103/PhysRevD.53.2480} {\bibfield  {journal} {\bibinfo
  {journal} {Phys. Rev.}\ }\textbf {\bibinfo {volume} {D53}},\ \bibinfo {pages}
  {2480--2490}},\ \Eprint {http://arxiv.org/abs/hep-ph/9411308}
  {arXiv:hep-ph/9411308 [hep-ph]} \BibitemShut {NoStop}%
\bibitem [{\citenamefont {Li}\ \emph {et~al.}(2012)\citenamefont {Li},
  \citenamefont {Li}, \citenamefont {Lu},\ and\ \citenamefont
  {Su}}]{Li:2012xc}%
  \BibitemOpen
  \bibfield  {author} {\bibinfo {author} {\bibnamefont {Li}, \bibfnamefont
  {Xin-Qiang}}, \bibinfo {author} {\bibfnamefont {Yan-Min}\ \bibnamefont {Li}},
  \bibinfo {author} {\bibfnamefont {Gong-Ru}\ \bibnamefont {Lu}}, \ and\
  \bibinfo {author} {\bibfnamefont {Fang}\ \bibnamefont {Su}}} (\bibinfo {year}
  {2012}),\ \bibfield  {title} {\enquote {\bibinfo {title} {{$B_s^0-\bar B_s^0$
  mixing in a family non-universal $Z^{\prime}$ model revisited}},}\ }\href
  {\doibase 10.1007/JHEP05(2012)049} {\bibfield  {journal} {\bibinfo  {journal}
  {JHEP}\ }\textbf {\bibinfo {volume} {05}},\ \bibinfo {pages} {049}},\ \Eprint
  {http://arxiv.org/abs/1204.5250} {arXiv:1204.5250 [hep-ph]} \BibitemShut
  {NoStop}%
\bibitem [{\citenamefont {Ligeti}\ \emph {et~al.}(2010)\citenamefont {Ligeti},
  \citenamefont {Papucci}, \citenamefont {Perez},\ and\ \citenamefont
  {Zupan}}]{Ligeti:2010ia}%
  \BibitemOpen
  \bibfield  {author} {\bibinfo {author} {\bibnamefont {Ligeti}, \bibfnamefont
  {Zoltan}}, \bibinfo {author} {\bibfnamefont {Michele}\ \bibnamefont
  {Papucci}}, \bibinfo {author} {\bibfnamefont {Gilad}\ \bibnamefont {Perez}},
  \ and\ \bibinfo {author} {\bibfnamefont {Jure}\ \bibnamefont {Zupan}}}
  (\bibinfo {year} {2010}),\ \bibfield  {title} {\enquote {\bibinfo {title}
  {{Implication s of the dimuon CP asymmetry in $B_{d,s}$ decays}},}\ }\href
  {\doibase 10.1103/PhysRevLett.105.131601} {\bibfield  {journal} {\bibinfo
  {journal} {Phys. Rev. Lett.}\ }\textbf {\bibinfo {volume} {105}},\ \bibinfo
  {pages} {131601}},\ \Eprint {http://arxiv.org/abs/1006.0432} {arXiv:1006.0432
  [hep-ph]} \BibitemShut {NoStop}%
\bibitem [{\citenamefont {Ligeti}\ and\ \citenamefont
  {Robinson}(2015)}]{Ligeti:2015yma}%
  \BibitemOpen
  \bibfield  {author} {\bibinfo {author} {\bibnamefont {Ligeti}, \bibfnamefont
  {Zoltan}}, \ and\ \bibinfo {author} {\bibfnamefont {Dean~J.}\ \bibnamefont
  {Robinson}}} (\bibinfo {year} {2015}),\ \bibfield  {title} {\enquote
  {\bibinfo {title} {{Towards more precise determinations of the CKM phase
  $\beta$}},}\ }\href@noop {} {\ }\Eprint {http://arxiv.org/abs/1507.06671}
  {arXiv:1507.06671 [hep-ph]} \BibitemShut {NoStop}%
\bibitem [{\citenamefont {Liu}\ \emph {et~al.}(2014)\citenamefont {Liu},
  \citenamefont {Wang},\ and\ \citenamefont {Xie}}]{Liu:2013nea}%
  \BibitemOpen
  \bibfield  {author} {\bibinfo {author} {\bibnamefont {Liu}, \bibfnamefont
  {Xin}}, \bibinfo {author} {\bibfnamefont {Wei}\ \bibnamefont {Wang}}, \ and\
  \bibinfo {author} {\bibfnamefont {Yuehong}\ \bibnamefont {Xie}}} (\bibinfo
  {year} {2014}),\ \bibfield  {title} {\enquote {\bibinfo {title} {{Penguin
  pollution in $B\to J/\psi V$ decays and impact on the extraction of the
  $B_s-\bar B_s$ mixing phase}},}\ }\href {\doibase 10.1103/PhysRevD.89.094010}
  {\bibfield  {journal} {\bibinfo  {journal} {Phys. Rev.}\ }\textbf {\bibinfo
  {volume} {D89}}~(\bibinfo {number} {9}),\ \bibinfo {pages} {094010}},\
  \Eprint {http://arxiv.org/abs/1309.0313} {arXiv:1309.0313 [hep-ph]}
  \BibitemShut {NoStop}%
\bibitem [{\citenamefont {Liu}\ \emph {et~al.}(2008)\citenamefont {Liu} \emph
  {et~al.}}]{Liu:2008bta}%
  \BibitemOpen
  \bibfield  {author} {\bibinfo {author} {\bibnamefont {Liu}, \bibfnamefont
  {Y}},  \emph {et~al.} (\bibinfo {collaboration} {Belle})} (\bibinfo {year}
  {2008}),\ \bibfield  {title} {\enquote {\bibinfo {title} {{Search for $\Bd
  \to \jpsi \phi$ decays}},}\ }\href {\doibase 10.1103/PhysRevD.78.011106}
  {\bibfield  {journal} {\bibinfo  {journal} {Phys. Rev.}\ }\textbf {\bibinfo
  {volume} {D78}},\ \bibinfo {pages} {011106}},\ \Eprint
  {http://arxiv.org/abs/0805.3225} {arXiv:0805.3225 [hep-ex]} \BibitemShut
  {NoStop}%
\bibitem [{\citenamefont {Mannel}\ \emph {et~al.}(2007)\citenamefont {Mannel},
  \citenamefont {Pecjak},\ and\ \citenamefont {Pivovarov}}]{Mannel:2007am}%
  \BibitemOpen
  \bibfield  {author} {\bibinfo {author} {\bibnamefont {Mannel}, \bibfnamefont
  {T}}, \bibinfo {author} {\bibfnamefont {B.~D.}\ \bibnamefont {Pecjak}}, \
  and\ \bibinfo {author} {\bibfnamefont {A.~A.}\ \bibnamefont {Pivovarov}}}
  (\bibinfo {year} {2007}),\ \bibfield  {title} {\enquote {\bibinfo {title}
  {{Analyzing $\Bs - \barBs$ mixing: Non-perturbative contributions to bag
  parameters from sum rules}},}\ }\href@noop {} {\ }\Eprint
  {http://arxiv.org/abs/hep-ph/0703244} {arXiv:hep-ph/0703244 [HEP-PH]}
  \BibitemShut {NoStop}%
\bibitem [{\citenamefont {Mannel}\ \emph {et~al.}(2011)\citenamefont {Mannel},
  \citenamefont {Pecjak},\ and\ \citenamefont {Pivovarov}}]{Mannel:2011zza}%
  \BibitemOpen
  \bibfield  {author} {\bibinfo {author} {\bibnamefont {Mannel}, \bibfnamefont
  {T}}, \bibinfo {author} {\bibfnamefont {B.~D.}\ \bibnamefont {Pecjak}}, \
  and\ \bibinfo {author} {\bibfnamefont {A.~A.}\ \bibnamefont {Pivovarov}}}
  (\bibinfo {year} {2011}),\ \bibfield  {title} {\enquote {\bibinfo {title}
  {{Sum rule estimate of the subleading non-perturbative contributions to $\Bs
  - \barBs$- mixing}},}\ }\href {\doibase 10.1140/epjc/s10052-011-1607-4}
  {\bibfield  {journal} {\bibinfo  {journal} {Eur. Phys. J.}\ }\textbf
  {\bibinfo {volume} {C71}},\ \bibinfo {pages} {1607}}\BibitemShut {NoStop}%
\bibitem [{\citenamefont {Nierste}(2014)}]{Nierste:2014ckm}%
  \BibitemOpen
  \bibfield  {author} {\bibinfo {author} {\bibnamefont {Nierste}, \bibfnamefont
  {Ulrich}}} (\bibinfo {year} {2014}),\ \bibfield  {title} {\enquote {\bibinfo
  {title} {{Effects of Delta Gamma on the dimuon asymmetry in B decays}},}\
  }\href@noop {} {\bibinfo  {journal} {Talk at CKM 2014, Vienna}\ }\BibitemShut
  {NoStop}%
\bibitem [{\citenamefont {Norrbin}\ and\ \citenamefont
  {Vogt}(2000)}]{Norrbin:2000jy}%
  \BibitemOpen
\bibfield  {journal} {  }\bibfield  {author} {\bibinfo {author} {\bibnamefont
  {Norrbin}, \bibfnamefont {E}}, \ and\ \bibinfo {author} {\bibfnamefont
  {R.}~\bibnamefont {Vogt}}} (\bibinfo {year} {2000}),\ \bibfield  {title}
  {\enquote {\bibinfo {title} {{Bottom production asymmetries at the LHC}},}\
  }in\ \href
  {http://www.osti.gov/energycitations/product.biblio.jsp?osti$\_$id=822605}
  {\emph {\bibinfo {booktitle} {{Fifth Workshop on electronics for LHC
  experiments, Snowmass, CO, USA, 20-24 Sep 1999: Proceedings}}}},\ \Eprint
  {http://arxiv.org/abs/hep-ph/0003056} {arXiv:hep-ph/0003056 [hep-ph]}
  \BibitemShut {NoStop}%
\bibitem [{\citenamefont {Olive}\ \emph {et~al.}(2014)\citenamefont {Olive}
  \emph {et~al.}}]{Agashe:2014kda}%
  \BibitemOpen
  \bibfield  {author} {\bibinfo {author} {\bibnamefont {Olive}, \bibfnamefont
  {K~A}},  \emph {et~al.} (\bibinfo {collaboration} {Particle Data Group})}
  (\bibinfo {year} {2014}),\ \bibfield  {title} {\enquote {\bibinfo {title}
  {{Review of Particle Physics}},}\ }\href {\doibase
  10.1088/1674-1137/38/9/090001} {\bibfield  {journal} {\bibinfo  {journal}
  {Chin. Phys.}\ }\textbf {\bibinfo {volume} {C38}},\ \bibinfo {pages}
  {090001}}\BibitemShut {NoStop}%
\bibitem [{\citenamefont {Pivk}\ and\ \citenamefont
  {Le~Diberder}(2005)}]{Pivk:2004ty}%
  \BibitemOpen
  \bibfield  {author} {\bibinfo {author} {\bibnamefont {Pivk}, \bibfnamefont
  {Muriel}}, \ and\ \bibinfo {author} {\bibfnamefont {Francois~R.}\
  \bibnamefont {Le~Diberder}}} (\bibinfo {year} {2005}),\ \bibfield  {title}
  {\enquote {\bibinfo {title} {{SPlot: A Statistical tool to unfold data
  distributions}},}\ }\href {\doibase 10.1016/j.nima.2005.08.106} {\bibfield
  {journal} {\bibinfo  {journal} {Nucl. Instrum. Meth.}\ }\textbf {\bibinfo
  {volume} {A555}},\ \bibinfo {pages} {356--369}},\ \Eprint
  {http://arxiv.org/abs/physics/0402083} {arXiv:physics/0402083
  [physics.data-an]} \BibitemShut {NoStop}%
\bibitem [{\citenamefont {Sahoo}\ \emph {et~al.}(2011)\citenamefont {Sahoo},
  \citenamefont {Das},\ and\ \citenamefont {Maharana}}]{Sahoo:2011zz}%
  \BibitemOpen
  \bibfield  {author} {\bibinfo {author} {\bibnamefont {Sahoo}, \bibfnamefont
  {S}}, \bibinfo {author} {\bibfnamefont {C.~K.}\ \bibnamefont {Das}}, \ and\
  \bibinfo {author} {\bibfnamefont {L.}~\bibnamefont {Maharana}}} (\bibinfo
  {year} {2011}),\ \bibfield  {title} {\enquote {\bibinfo {title} {{The
  Prediction of Mass of $Z'$-Boson from $B_q^0-\bar{B}_q^0$ mixing}},}\ }\href
  {\doibase 10.1142/S0217751X11053936} {\bibfield  {journal} {\bibinfo
  {journal} {Int. J. Mod. Phys.}\ }\textbf {\bibinfo {volume} {A26}},\ \bibinfo
  {pages} {3347--3356}},\ \Eprint {http://arxiv.org/abs/1112.0460}
  {arXiv:1112.0460 [hep-ph]} \BibitemShut {NoStop}%
\bibitem [{\citenamefont {Sahoo}\ \emph {et~al.}(2013)\citenamefont {Sahoo},
  \citenamefont {Kumar},\ and\ \citenamefont {Banerjee}}]{Sahoo:2013vuf}%
  \BibitemOpen
  \bibfield  {author} {\bibinfo {author} {\bibnamefont {Sahoo}, \bibfnamefont
  {S}}, \bibinfo {author} {\bibfnamefont {M.}~\bibnamefont {Kumar}}, \ and\
  \bibinfo {author} {\bibfnamefont {D.}~\bibnamefont {Banerjee}}} (\bibinfo
  {year} {2013}),\ \bibfield  {title} {\enquote {\bibinfo {title} {{The effect
  of $Z'$ boson on same-sign dimuon charge asymmetry in $ B_q^0 - \bar{B}_q^0$
  system}},}\ }\href {\doibase 10.1142/S0217751X13500607} {\bibfield  {journal}
  {\bibinfo  {journal} {Int. J. Mod. Phys.}\ }\textbf {\bibinfo {volume}
  {A28}},\ \bibinfo {pages} {1350060}},\ \Eprint
  {http://arxiv.org/abs/1306.5087} {arXiv:1306.5087 [hep-ph]} \BibitemShut
  {NoStop}%
\bibitem [{\citenamefont {Sakharov}(1967)}]{Sakharov:1967dj}%
  \BibitemOpen
  \bibfield  {author} {\bibinfo {author} {\bibnamefont {Sakharov},
  \bibfnamefont {A~D}}} (\bibinfo {year} {1967}),\ \bibfield  {title} {\enquote
  {\bibinfo {title} {{Violation of CP Invariance, c Asymmetry, and Baryon
  Asymmetry of the Universe}},}\ }\href {\doibase
  10.1070/PU1991v034n05ABEH002497} {\bibfield  {journal} {\bibinfo  {journal}
  {Pisma Zh. Eksp. Teor. Fiz.}\ }\textbf {\bibinfo {volume} {5}},\ \bibinfo
  {pages} {32--35}},\ \bibinfo {note} {[Usp. Fiz.
  Nauk161,61(1991)]}\BibitemShut {NoStop}%
\bibitem [{\citenamefont {Salam}(1968)}]{Salam:1968rm}%
  \BibitemOpen
  \bibfield  {author} {\bibinfo {author} {\bibnamefont {Salam}, \bibfnamefont
  {Abdus}}} (\bibinfo {year} {1968}),\ \bibfield  {title} {\enquote {\bibinfo
  {title} {{Weak and Electromagnetic Interactions}},}\ }\bibfield  {booktitle}
  {\emph {\bibinfo {booktitle} {{8th Nobel Symposium Lerum, Sweden, May 19-25,
  1968}}},\ }\href@noop {} {\bibfield  {journal} {\bibinfo  {journal} {Conf.
  Proc.}\ }\textbf {\bibinfo {volume} {C680519}},\ \bibinfo {pages}
  {367--377}}\BibitemShut {NoStop}%
\bibitem [{\citenamefont {Shifman}\ and\ \citenamefont
  {Voloshin}(1985)}]{Shifman:1984wx}%
  \BibitemOpen
  \bibfield  {author} {\bibinfo {author} {\bibnamefont {Shifman}, \bibfnamefont
  {Mikhail~A}}, \ and\ \bibinfo {author} {\bibfnamefont {M.~B.}\ \bibnamefont
  {Voloshin}}} (\bibinfo {year} {1985}),\ \bibfield  {title} {\enquote
  {\bibinfo {title} {{Preasymptotic Effects in Inclusive Weak Decays of Charmed
  Particles}},}\ }\href@noop {} {\bibfield  {journal} {\bibinfo  {journal}
  {Sov. J. Nucl. Phys.}\ }\textbf {\bibinfo {volume} {41}},\ \bibinfo {pages}
  {120}},\ \bibinfo {note} {[Yad. Fiz.41,187(1985)]}\BibitemShut {NoStop}%
\bibitem [{\citenamefont {Stone}(2014)}]{Stone:2014pra}%
  \BibitemOpen
  \bibfield  {author} {\bibinfo {author} {\bibnamefont {Stone}, \bibfnamefont
  {Sheldon}}} (\bibinfo {year} {2014}),\ \bibfield  {title} {\enquote {\bibinfo
  {title} {{Lifetimes of some $b$-flavored hadrons}},}\ }in\ \href
  {http://inspirehep.net/record/1303109/files/arXiv:1406.6497.pdf} {\emph
  {\bibinfo {booktitle} {{Proceedings, 19th Real Time Conference (RT2014)}}}},\
  \Eprint {http://arxiv.org/abs/1406.6497} {arXiv:1406.6497 [hep-ex]}
  \BibitemShut {NoStop}%
\bibitem [{\citenamefont {Stone}\ and\ \citenamefont
  {Zhang}(2009)}]{Stone:2008ak}%
  \BibitemOpen
  \bibfield  {author} {\bibinfo {author} {\bibnamefont {Stone}, \bibfnamefont
  {Sheldon}}, \ and\ \bibinfo {author} {\bibfnamefont {Liming}\ \bibnamefont
  {Zhang}}} (\bibinfo {year} {2009}),\ \bibfield  {title} {\enquote {\bibinfo
  {title} {{S-waves and the Measurement of CP Violating Phases in $B_s$
  Decays}},}\ }\href {\doibase 10.1103/PhysRevD.79.074024} {\bibfield
  {journal} {\bibinfo  {journal} {Phys. Rev.}\ }\textbf {\bibinfo {volume}
  {D79}},\ \bibinfo {pages} {074024}},\ \Eprint
  {http://arxiv.org/abs/0812.2832} {arXiv:0812.2832 [hep-ph]} \BibitemShut
  {NoStop}%
\bibitem [{\citenamefont {Stone}\ and\ \citenamefont
  {Zhang}(2013)}]{Stone:2013eaa}%
  \BibitemOpen
  \bibfield  {author} {\bibinfo {author} {\bibnamefont {Stone}, \bibfnamefont
  {Sheldon}}, \ and\ \bibinfo {author} {\bibfnamefont {Liming}\ \bibnamefont
  {Zhang}}} (\bibinfo {year} {2013}),\ \bibfield  {title} {\enquote {\bibinfo
  {title} {{Use of $B\to J/\psi f_0$ decays to discern the $q \bar{q}$ or
  tetraquark nature of scalar mesons}},}\ }\href {\doibase
  10.1103/PhysRevLett.111.062001} {\bibfield  {journal} {\bibinfo  {journal}
  {Phys. Rev. Lett.}\ }\textbf {\bibinfo {volume} {111}}~(\bibinfo {number}
  {6}),\ \bibinfo {pages} {062001}},\ \Eprint {http://arxiv.org/abs/1305.6554}
  {arXiv:1305.6554 [hep-ex]} \BibitemShut {NoStop}%
\bibitem [{\citenamefont {Urban}\ \emph {et~al.}(1998)\citenamefont {Urban},
  \citenamefont {Krauss}, \citenamefont {Jentschura},\ and\ \citenamefont
  {Soff}}]{Urban:1997gw}%
  \BibitemOpen
  \bibfield  {author} {\bibinfo {author} {\bibnamefont {Urban}, \bibfnamefont
  {J}}, \bibinfo {author} {\bibfnamefont {F.}~\bibnamefont {Krauss}}, \bibinfo
  {author} {\bibfnamefont {U.}~\bibnamefont {Jentschura}}, \ and\ \bibinfo
  {author} {\bibfnamefont {G.}~\bibnamefont {Soff}}} (\bibinfo {year} {1998}),\
  \bibfield  {title} {\enquote {\bibinfo {title} {{Next-to-leading order QCD
  corrections for the $B_0 -\bar{B}_0$ mixing with an extended Higgs
  sector}},}\ }\href {\doibase 10.1016/S0550-3213(98)00043-1} {\bibfield
  {journal} {\bibinfo  {journal} {Nucl. Phys.}\ }\textbf {\bibinfo {volume}
  {B523}},\ \bibinfo {pages} {40--58}},\ \Eprint
  {http://arxiv.org/abs/hep-ph/9710245} {arXiv:hep-ph/9710245 [hep-ph]}
  \BibitemShut {NoStop}%
\bibitem [{\citenamefont {Voloshin}(2000)}]{Voloshin:2000zc}%
  \BibitemOpen
  \bibfield  {author} {\bibinfo {author} {\bibnamefont {Voloshin},
  \bibfnamefont {M~B}}} (\bibinfo {year} {2000}),\ \bibfield  {title} {\enquote
  {\bibinfo {title} {{Inclusive weak decay rates of heavy hadrons}},}\ }in\
  \href {http://alice.cern.ch/format/showfull?sysnb=2185755} {\emph {\bibinfo
  {booktitle} {{Workshop on New Strong Dynamics for Run II of the Fermilab
  Tevatron: Technicolor, Topcolor, New Dynamics at the TeV Scale Batavia,
  Illinois, October 30-31, 1998}}}},\ \Eprint
  {http://arxiv.org/abs/hep-ph/0004257} {arXiv:hep-ph/0004257 [hep-ph]}
  \BibitemShut {NoStop}%
\bibitem [{\citenamefont {Wang}\ \emph {et~al.}(2011)\citenamefont {Wang},
  \citenamefont {Xu}, \citenamefont {Chang},\ and\ \citenamefont
  {Yang}}]{Wang:2011ax}%
  \BibitemOpen
  \bibfield  {author} {\bibinfo {author} {\bibnamefont {Wang}, \bibfnamefont
  {Ru-Min}}, \bibinfo {author} {\bibfnamefont {Yuan-Guo}\ \bibnamefont {Xu}},
  \bibinfo {author} {\bibfnamefont {Qin}\ \bibnamefont {Chang}}, \ and\
  \bibinfo {author} {\bibfnamefont {Ya-Dong}\ \bibnamefont {Yang}}} (\bibinfo
  {year} {2011}),\ \bibfield  {title} {\enquote {\bibinfo {title} {{Studying of
  $B_s^0-\bar{B}_s^0$ mixing and $B_{s}\to K^{(*)-}K^{(*)+}$ decays within
  supersymmetry}},}\ }\href {\doibase 10.1103/PhysRevD.83.095010} {\bibfield
  {journal} {\bibinfo  {journal} {Phys. Rev.}\ }\textbf {\bibinfo {volume}
  {D83}},\ \bibinfo {pages} {095010}},\ \Eprint
  {http://arxiv.org/abs/1102.2031} {arXiv:1102.2031 [hep-ph]} \BibitemShut
  {NoStop}%
\bibitem [{\citenamefont {Wang}\ \emph {et~al.}(2010)\citenamefont {Wang},
  \citenamefont {Xu}, \citenamefont {Liu},\ and\ \citenamefont
  {Li}}]{Wang:2010vv}%
  \BibitemOpen
  \bibfield  {author} {\bibinfo {author} {\bibnamefont {Wang}, \bibfnamefont
  {Ru-Min}}, \bibinfo {author} {\bibfnamefont {Yuan-Guo}\ \bibnamefont {Xu}},
  \bibinfo {author} {\bibfnamefont {Mo-Lin}\ \bibnamefont {Liu}}, \ and\
  \bibinfo {author} {\bibfnamefont {Bing-Zhong}\ \bibnamefont {Li}}} (\bibinfo
  {year} {2010}),\ \bibfield  {title} {\enquote {\bibinfo {title}
  {{Reevaluating $R$-parity Violating Supersymmetry Effects in
  $B_s^0-\bar{B}_s^0$ Mixing}},}\ }\href {\doibase 10.1007/JHEP12(2010)034}
  {\bibfield  {journal} {\bibinfo  {journal} {JHEP}\ }\textbf {\bibinfo
  {volume} {12}},\ \bibinfo {pages} {034}},\ \Eprint
  {http://arxiv.org/abs/1007.2944} {arXiv:1007.2944 [hep-ph]} \BibitemShut
  {NoStop}%
\bibitem [{\citenamefont {Weinberg}(1967)}]{Weinberg:1967tq}%
  \BibitemOpen
  \bibfield  {author} {\bibinfo {author} {\bibnamefont {Weinberg},
  \bibfnamefont {Steven}}} (\bibinfo {year} {1967}),\ \bibfield  {title}
  {\enquote {\bibinfo {title} {{A Model of Leptons}},}\ }\href {\doibase
  10.1103/PhysRevLett.19.1264} {\bibfield  {journal} {\bibinfo  {journal}
  {Phys. Rev. Lett.}\ }\textbf {\bibinfo {volume} {19}},\ \bibinfo {pages}
  {1264--1266}}\BibitemShut {NoStop}%
\bibitem [{\citenamefont {Wolfenstein}(1983)}]{Wolfenstein:1983yz}%
  \BibitemOpen
  \bibfield  {author} {\bibinfo {author} {\bibnamefont {Wolfenstein},
  \bibfnamefont {Lincoln}}} (\bibinfo {year} {1983}),\ \bibfield  {title}
  {\enquote {\bibinfo {title} {{Parametrization of the Kobayashi-Maskawa
  Matrix}},}\ }\href {\doibase 10.1103/PhysRevLett.51.1945} {\bibfield
  {journal} {\bibinfo  {journal} {Phys. Rev. Lett.}\ }\textbf {\bibinfo
  {volume} {51}},\ \bibinfo {pages} {1945}}\BibitemShut {NoStop}%
\bibitem [{\citenamefont {Xie}(2009)}]{Xie:2009rka}%
  \BibitemOpen
  \bibfield  {author} {\bibinfo {author} {\bibnamefont {Xie}, \bibfnamefont
  {Yuehong}}} (\bibinfo {year} {2009}),\ \bibfield  {title} {\enquote {\bibinfo
  {title} {{sFit: a method for background subtraction in maximum likelihood
  fit}},}\ }\href@noop {} {\ }\Eprint {http://arxiv.org/abs/0905.0724}
  {arXiv:0905.0724 [physics.data-an]} \BibitemShut {NoStop}%
\bibitem [{\citenamefont {Xie}\ \emph {et~al.}(2009)\citenamefont {Xie},
  \citenamefont {Clarke}, \citenamefont {Cowan},\ and\ \citenamefont
  {Muheim}}]{Xie:2009fs}%
  \BibitemOpen
  \bibfield  {author} {\bibinfo {author} {\bibnamefont {Xie}, \bibfnamefont
  {Yuehong}}, \bibinfo {author} {\bibfnamefont {Peter}\ \bibnamefont {Clarke}},
  \bibinfo {author} {\bibfnamefont {Greig}\ \bibnamefont {Cowan}}, \ and\
  \bibinfo {author} {\bibfnamefont {Franz}\ \bibnamefont {Muheim}}} (\bibinfo
  {year} {2009}),\ \bibfield  {title} {\enquote {\bibinfo {title}
  {{Determination of $2\beta_s$ in $\Bs \to \jpsi K^+ K^-$ Decays in the
  Presence of a $K^+ K^-$ S-Wave Contribution}},}\ }\href {\doibase
  10.1088/1126-6708/2009/09/074} {\bibfield  {journal} {\bibinfo  {journal}
  {JHEP}\ }\textbf {\bibinfo {volume} {09}},\ \bibinfo {pages} {074}},\ \Eprint
  {http://arxiv.org/abs/0908.3627} {arXiv:0908.3627 [hep-ph]} \BibitemShut
  {NoStop}%
\bibitem [{\citenamefont {Yeh}\ and\ \citenamefont {Li}(1997)}]{Yeh:1997rq}%
  \BibitemOpen
  \bibfield  {author} {\bibinfo {author} {\bibnamefont {Yeh}, \bibfnamefont
  {Tsung-Wen}}, \ and\ \bibinfo {author} {\bibfnamefont {Hsiang-nan}\
  \bibnamefont {Li}}} (\bibinfo {year} {1997}),\ \bibfield  {title} {\enquote
  {\bibinfo {title} {{Factorization theorems, effective field theory, and
  nonleptonic heavy meson decays}},}\ }\href {\doibase
  10.1103/PhysRevD.56.1615} {\bibfield  {journal} {\bibinfo  {journal} {Phys.
  Rev.}\ }\textbf {\bibinfo {volume} {D56}},\ \bibinfo {pages} {1615--1631}},\
  \Eprint {http://arxiv.org/abs/hep-ph/9701233} {arXiv:hep-ph/9701233 [hep-ph]}
  \BibitemShut {NoStop}%
\bibitem [{\citenamefont {Zhang}\ and\ \citenamefont
  {Stone}(2013)}]{Zhang:2012zk}%
  \BibitemOpen
  \bibfield  {author} {\bibinfo {author} {\bibnamefont {Zhang}, \bibfnamefont
  {Liming}}, \ and\ \bibinfo {author} {\bibfnamefont {Sheldon}\ \bibnamefont
  {Stone}}} (\bibinfo {year} {2013}),\ \bibfield  {title} {\enquote {\bibinfo
  {title} {{Time-dependent Dalitz-plot formalism for $B_q \to \jpsi h^+
  h^-$}},}\ }\href {\doibase 10.1016/j.physletb.2013.01.035} {\bibfield
  {journal} {\bibinfo  {journal} {Phys. Lett.}\ }\textbf {\bibinfo {volume}
  {B719}},\ \bibinfo {pages} {383--387}},\ \Eprint
  {http://arxiv.org/abs/1212.6434} {arXiv:1212.6434} \BibitemShut {NoStop}%
\end{thebibliography}%

\end{document}